\documentclass[acmsmall,nonacm]{acmart}
\setcopyright{none}

\newif\ifACMConf
\ACMConftrue
\usepackage{env}
\usepackage{kvmacro}
\usepackage{kvtikz}
\newcommand{\RootPath}{.}

\newif\ifCommentEdits
\CommentEditsfalse
\usepackage{comment-box}
\ifCommentEdits
\fi

\newcommand{\spacingabovemypar}{0pt}
\newcommand{\scaleboxscaling}{.7}
\newcommand{\displaymathfont}{}
\begin{document}

\title{%
Data Consistency in Transactional Storage Systems: 
A Centralised Approach%
} 

\author{Shale Xiong}

\affiliation{
  \department{Department of Computing}              
  \institution{Imperial College London}            
  \country{UK}                    
}
\email{shale.xiong14@ic.ac.uk}          

\author{Andrea Cerone}

\affiliation{
  \institution{Football Radar}            
  \country{UK}                    
}
\email{andrea.cerone@footballradar.com}          

\author{Azalea Raad}

\affiliation{
  \institution{MPI-SWS}            
  \country{Germany}                    
}
\email{azalea@mpi-sws.org}          

\author{Philippa Gardner}

\affiliation{
  \department{Department of Computing}              
  \institution{Imperial College London}            
  \city{London}
  \postcode{SW7 2AZ}
  \country{UK}                    
}
\email{p.gardner@ic.ac.uk}          


\begin{abstract}

We introduce an interleaving operational semantics for describing the
client-observable behaviour of atomic transactions on distributed
key-value stores. Our semantics builds on abstract states comprising
centralised, global key-value stores and partial client views.  We provide
operational definitions of consistency models for our key-value stores which
are shown to be equivalent to the well-known declarative definitions
of consistency model for execution graphs. We explore  two
immediate applications of our semantics: specific protocols of 
geo-replicated databases (e.g. COPS) and partitioned databases
(e.g. Clock-SI) can be shown to be correct for a specific consistency
model by embedding them in our centralised semantics; 
programs can be directly shown to have invariant properties such as 
robustness results against a weak consistency model.

\end{abstract}

\maketitle

\section{Introduction}
\label{sec:intro}

Transactions are the \emph{de facto} synchronisation mechanism in
modern distributed databases. To achieve scalability and performance, distributed databases  
often use weak transactional consistency guarantees. 
These weak  guarantees
pose several challenges: the formalisation of  client-observable
behaviour; and the verification of database protocols and client applications.
Much work has been done to formalise the semantics of such consistency guarantees, both
declaratively and operationally.
On the declarative side, several {general} formalisms have been proposed, 
such as dependency graphs~\cite{adya} and abstract executions~\cite{ev_transactions}, to provide a unified
semantics for formulating different consistency models.  
On the operational side, the semantics of \emph{specific} consistency models have
been captured using reference implementations~\cite{si,PSI,PSI-RA}. 
However, unlike declarative approaches, there has been
little work on {general} operational semantics for describing a range
of consistency models.

We summarise briefly the existing work on general operational semantics;
further discussion can be found in \cref{sec:conclusions}.
\citet{alonetogether} propose an operational
semantics for SQL transaction programs under the consistency models
given by the standard ANSI/SQL isolation levels~\cite{si}. Their
operational semantics  
accesses a global, centralised store,  and is used to develop 
a program logic  and prototype tool for reasoning about client
programs. They capture consistency models such as snapshot
isolation (\SI)~\cite{si}, but not  weaker ones 
 such as parallel snapshot isolation (\PSI)~\cite{PSI} and causal consistency (\CC)~\cite{cops} 
 which are important 
for distributed databases. 
Following this work, \citet{sureshConcur} propose an operational semantics over abstract
execution graphs, rather than   a concrete centralised store, in order to
prove the robustness of applications against
a given consistency model. They are able to 
capture weaker consistency models
such as \(\PSI\) and \(\CC\) . However, although they focus on consistency models with snapshot 
property\footnote{\emph{Snapshot property}, also known as \emph{atomic visibility},
means that a transaction reads from an atomic snapshot of the database, and commits atomically.},
their semantics allows for the fine-grained interleaving of operations in different
transactions. We believe that this results  in an unnecessarily complicated semantics.
\citet{seebelieve} provide a trace semantics over a global
centralised store, where the behaviour of clients is formalised by the   
observations they make on the totally-ordered history of states. 
They 
use their semantics to demonstrate  the equivalence of several
implementation-specific definitions of
consistency model. 
However, the usefulness  of their approach for analysing client programs is not clear, 
since observations made by their clients  involve information that is not generally 
available to real-world clients,  such as the total order in which transactions commit.

We introduce an interleaving operational semantics for describing the
client-observable behaviour of atomic transactions on distributed
key-value stores (\cref{sec:overview}, \cref{sec:model}),
successfully abstracting from the internal details of protocols of
geo-replicated and partitioned databases.  In our semantics,
transactions execute atomically, preventing fine-grained interleaving
of the operations they perform.
Our semantics comprises a global, centralised key-value store (kv-store)
with {\em multi-versioning}, which records all the versions of a key,
and partial {\em client views}, which let clients see only a subset of
the versions. Our approach is partly  inspired by the views in the  C11 operational semantics
in~\cite{promises}. 
Our operational semantics is parametric in the notion of \emph{execution test},
which determines if a client with a given view is allowed to commit a transaction.
Just as with standard operational semantics, the  next transaction
step just depends on  the current kv-store and client view. 
Our execution tests resemble  an approach taken  in~\cite{seebelieve}, 
except that the next step requires an analysis of the whole trace. 
We  capture  most of the well-known consistency models in
a uniform way (\cref{sec:cm}): e.g,  \CC, \PSI, 
\SI\ and serialisability (\SER).
We also identify 
a new consistency model  that sits between \PSI\; and \SI\; and retains good properties of both.
We call this new consistency model \emph{weak snapshot isolation} (\WSI).
Since we focus on
snapshot property,  we are not able to capture popular 
  consistency models such as \emph{read committed}. 
Because our focus is on protocols and applications employed  by distributed databases, 
  most of which  guarantee snapshot property, we do not find this constraint to be a severe
  limitation.
%

We prove that our operational definitions of consistency models for
kv-stores are equivalent to the well-known declarative definitions
of consistency model for execution graphs.  Such results have not
been given in \cite{alonetogether}, but have been given in
\cite{seebelieve}.  We provide a general proof technique which
captures the correspondence between our execution tests and the
axiomatic specifications of consistency models for abstract
executions (\cref{sec:other_formalisms} and \cref{sec:et-sound-complete-constructor}).  Using this
technique, we
prove 
that our  definitions of consistency model for kv-stores are equivalent to the  declarative definitions of
consistency model  for
abstract executions~\cite{framework-concur}, and hence for  dependency
graphs~\cite{adya,laws}.

We explore two immediate applications of our operational semantics:
the establishment of invariant properties such as robustness for  simple client applications; and the correctness of
specific distributed protocols. By contrast, these tasks  tend to be carried out in
{different} declarative formalisms:  clients are analysed using dependency
graphs~\cite{fekete-tods,SIanalysis,giovanni_concur16,psi-chopping,sureshConcur};
protocols are verified using
abstract execution graphs~\cite{repldatatypes,framework-concur}; and equivalence results are used to move between
the two~\cite{laws}. 
We prove the robustness of a single
counter against \PSI\ and the robustness of multiple counters against
our new model \WSI\ (\cref{sec:applications}).
\citet{bank-example-wsi} present a banking example that is robust against \( \SI \).
We show that it is also robust against \( \WSI \) (\cref{sec:applications}).
%
To our knowledge, our robustness results are the first to take into
account client sessions. With sessions, we demonstrate that multiple
counters are not robust against \(\PSI\).  Interestingly, without
sessions, it can be shown that multiple counters are robust against
\(\PSI\) using static-analysis techniques~\cite{giovanni_concur16};
these techniques are known not to be applicable to sessions.  We also
establish the correctness of two database protocols against their
consistency models, demonstrating that they can be accurately
expressed in our centralised semantics: the COPS protocol for the
fully replicated kv-stores~\cite{cops} which satisfies \CC\ 
(\cref{sec:applications}); and the Clock-SI protocol for partitioned
kv-stores~\cite{clocksi} which satisfies $\SI$ (\cref{app:implementation-verification}). 

\sx{The reviewer from OOPSLA confused about client sessions here.
    Multi-counters are actually robust against \( \WSI \).
    It immediately gives some idea why \( \WSI \) is a interesting consistency model.
}

\section{Overview}
\label{sec:overview}

We motivate our key ideas, centralised kv-stores, partial client views and execution tests,
via an intuitive example.
We show that our interleaving semantics is an ideal middle point for proving invariant properties such as robustness, and verifying distributed protocols.

\mypar{Example} We use a simple transactional library, \(\mathsf{Counter}(\key)\), to
 introduce our operational semantics.  Clients of this counter library can manipulate the
value of key \(\key\) via two transactions:

\vspace{-5pt}
{%
\displaymathfont
\[%
\begin{array}{r @{\hspace{2pt}} l @{\hspace{18pt}} r @{\hspace{2pt}} l}
\ctrinc(\key) \defeq 
&
\begin{session}
\begin{transaction}
\plookup{\pv{x}}{\key}; \ 
\pmutate{\key}{\pv{x}+1}
\end{transaction}
\end{session}
&
\ctrread(\key) \defeq &
\begin{session}
\begin{transaction}
\plookup{\pv{x}}{\key}
\end{transaction}
\end{session}
\end{array}
\]%
}%
Command \( \plookup{\pv{x}}{\key} \) reads the value of key \( \key \) to
local variable \( \vx \); command \( \pmutate{\key}{\pv{x}+1} \)
writes the value of \( \pv{x}+1 \) to key \( \key \).  The code of each
operation is wrapped in square brackets, denoting that 
it must be executed \emph{atomically} as a transaction.  

Consider a replicated database where
a client only interacts with one replica.
For such a database, the 
correctness of atomic transactions is subtle, depending heavily on the
particular consistency model under consideration.  
Consider the client program
$\prog_{\mathsf{LU}} = \left(\cl_1 : \ctrinc(\key) \;|| \; \cl_2:
  \ctrinc(\key) \right)$, 
where we assume that the clients \( \cl_1 \) and \( \cl_2 \) work on different replicas and
the \(\key\) initially holds value \(0\) in all replicas.
Intuitively, since transactions are executed atomically, after both
calls to \(\ctrinc(\key)\) have terminated, the counter should hold 
the value \(2\).
Indeed, this is the only outcome allowed under 
\SER\, where transactions
appear to execute in a sequential (serial) order, one after another.
The implementation of  \SER\ in distributed kv-stores comes at a
significant performance cost. Therefore, implementers are content with
{weaker} consistency models~\cite{ramp,rola,cops,wren,redblue,PSI,NMSI,gdur,clocksi,distrsi}. 
For example, if the replicas provide no synchronisation mechanism for transactions,
then it is possible for both clients to read the same initial value \(0\) for \(\key\) at their
distinct replicas, update them to \(1\), and eventually propagate their updates to other replicas. 
Consequently, both
sites  are unchanged with value  \(1\) for \(\key\).
This weak behaviour is known as the \emph{lost update} anomaly, which
is  allowed under the consistency model called {\em causal consistency} \cite{cops,wren,redblue}.

\begin{figure*}[t]
\centering
\captionsetup[subfigure]{aboveskip=0pt, belowskip=5pt}
\begin{tabularx}{\textwidth}{@{} c @{} | @{} c @{} | @{} c @{} | c@{}}
\hline
\phantom{-}& \phantom{-}& \phantom{-}& \phantom{-}\\[-5pt]
\begin{subfigure}{0.18\textwidth}
\begin{centertikz}
\node(locx) {$\key \mapsto$};
\draw pic at ([xshift=\tikzkvspace]locx.east) {vlist={versionx}{%
    /$0$/$\txid_0$/$\emptyset$
}};

\end{centertikz}
\caption{Initial state}
\label{fig:counter_kv_initial}
\end{subfigure}
&
\begin{subfigure}{0.22\textwidth}
\begin{centertikz}

\node(locx) {$\key \mapsto$};
\draw pic at ([xshift=\tikzkvspace]locx.east) {vlist={versionx}{%
    /$0$/$\txid_0$/$\Set{\txid}$
    , /$1$/$\txid$/$\emptyset$
}};

\end{centertikz}
\caption{After \(\txid \)}
\label{fig:counter_kv_first_inc}
\end{subfigure}
&
\begin{subfigure}{0.28\textwidth}
\begin{centertikz}

\node(locx) {$\key \mapsto$};
\draw pic at ([xshift=\tikzkvspace]locx.east) {vlist={versionx}{%
    /$0$/$\txid_0$/$\Set{\txid}$
    , nonvisible/$1$/$\txid$/$\emptyset$
}};

\end{centertikz}
\caption{A possible view of \( \cl_2 \)}
\label{fig:counter_kv_view}
\end{subfigure} 
&
\begin{subfigure}{0.25\textwidth}
\begin{centertikz}
    
\node(locx) {$\key \mapsto$};
\draw pic at ([xshift=\tikzkvspace]locx.east) {vlist={versionx}{%
    /$0$/$\txid_0$/$\Set{\txid,\txid'}$
    , /$1$/$\txid$/$\emptyset$
    , /$1$/$\txid'$/$\emptyset$
}};

\end{centertikz}%
\caption{After \( \txid' \), lost update}
\label{fig:counter_kv_final}
\end{subfigure}\\
\hline
\end{tabularx}
\caption{Example key-value stores (\subref{fig:counter_kv_initial}, \subref{fig:counter_kv_first_inc}, \subref{fig:counter_kv_final}); a client view (\subref{fig:counter_kv_view})}
\end{figure*}

\mypar{Centralised Operational Semantics}

A well-known declarative approach for providing general reasoning
about clients of distributed kv-stores is to use  execution 
graphs~\cite{adya-icde,adya,framework-concur,ev_transactions},
where nodes are atomic transactions and edges describe the
known dependencies between transactions. The graphs capture the
behaviour of the whole program, with different consistency models
corresponding to different sets of axioms constraining the graphs. 
However, execution graphs provide little information about how the 
state of a kv-store evolves throughout the execution of a program.
By contrast, we provide an interleaving operational semantics based on an
abstract centralised state. The centralised state comprises a
centralised, multi-versioned kv-store, which is {\em global} in the
sense that it contains all the versions written by clients, and client views of the store,
which are {\em partial} in the sense that clients may see different 
subsets of the versions in the kv-store. Each update is given by either
a simple primitive command or an atomic transaction. The atomic
transaction steps are subject to an {\em execution test} which
analyses the state to determine whether the update is allowed by 
the associated  consistency model.

Let us introduce  our global kv-stores and partial client views by
showing that we can reproduce the lost update anomaly given by 
\(\prog_{\mathsf{LU}}\).
Our kv-stores are functions mapping keys to lists of versions, where
the versions  record all the values written to each key together with the
meta-data of the transactions that access it. 
In the \(\prog_{\mathsf{LU}}\) example, the initial kv-store comprises a single key \(\key\), with only one possible 
version \((0, \txid_{0}, \emptyset)\),  stating that \(\key\) holds value \(0\), 
that the version \emph{writer} is the initialising transaction
\(\txid_0\) (this version was written by \(\txid_0\)), 
and that the version \emph{reader set} is empty (no transaction has read this version as of yet). 
\Cref{fig:counter_kv_initial} depicts this initial kv-store, with the version
represented as a box sub-divided in three sections: the value \(0\);
the writer \(t_0\); and the reader set \(\emptyset\).


First, suppose that \(\cl_1\)  invokes \(\ctrinc\) on
\cref{fig:counter_kv_initial}. It does this by choosing a fresh
transaction identifier, \(\txid\), 
and then proceeds with \(\ctrinc(\key)\). It reads the initial version
of \(\key\) with value \(0\) 
and then writes a new value \(1\) for \(\key\). 
The resulting kv-store is depicted in \cref{fig:counter_kv_first_inc},
where  the initial version of \(\key\)  has been  updated to reflect that it
has been read by \(\txid\). 

Second, client \(\cl_2\) invokes \(\ctrinc\) on
\cref{fig:counter_kv_first_inc}.  As there are now two versions
available for \(\key\), we must determine the version from which
\(\cl_2\) fetches its value, before running \(\ctrinc(\key)\).  This is
where \emph{client views} come into play.  Intuitively, a view of
client \(\cl_2\) comprises those versions in the kv-store that are
\emph{visible} to \(\cl_2\), \ie those that can be read by
\(\cl_2\).  If more than one version is visible, then the newest
(right-most) version is selected, modelling the \emph{last-writer-wins}
resolution policy used by many distributed
kv-stores~\cite{vogels:2009:ec:1435417.1435432}.  In our example,
there are two view candidates for \(\cl_2\) when running
\(\ctrinc(\key)\) on \cref{fig:counter_kv_first_inc}: (1) one containing
only the initial version of \(\key\); (2) the other containing both
versions of \(\key\).%
\footnote{ As we explain in \cref{sec:mkvs-view}, we always require
  the view of a client to include the initial version of each key.}  
For (1), the view is depicted in
\cref{fig:counter_kv_view}.  Client \(\cl_2\) chooses a fresh
transaction identifier \(t'\), reads the initial value \(0\) and writes a
new version with value \(1\), as depicted in
\cref{fig:counter_kv_final}.  Such a kv-store does not contain a
version with value \(2\), despite two increments on \(\key\), producing
the lost update anomaly.  For (2), client \(cl_2\) reads the newest
value \(1\) and writes a new version with value \(2\).

To avoid undesirable behaviour, such as the lost update anomaly, we
use an {\em execution test} which restricts the possible update at the
point of the transaction commit.  One such test is to enforce a client
to commit a transaction writing to \(\key\) if and only if its view
contains all versions available in the global state for \(\key\).  This
prevents \(\cl_2\) from running \(\ctrinc(\key)\) on
\cref{fig:counter_kv_first_inc} if its view only contains the initial
version of \(\key\).  Instead, the \(\cl_2\) view must contain both
versions of \(\key\), thus enforcing \(\cl_2\) to write a version with
value \(2\) after running \(\ctrinc(\key)\). This particular test
corresponds to \emph{write-conflict-freedom} of distributed kv-stores:
at most one concurrent transaction can write to a key at any one time.
In \cref{sec:cm} we give many examples of execution tests and their
associated consistency models on kv-stores. In \cref{sec:other_formalisms}, we
develop a proof technique, which we use in \cref{app:et_sound_complete} 
to show the equivalence of our operational definitions of consistency models and the 
declarative ones based on  
execution graphs. 

\mypar{General Robustness Results} 
The first application of our operational semantics is to prove
general robustness results for clients with respect to specific
consistency models (\cref{sec:robustness}). 
Using our general operational semantics, we can prove invariant properties (\eg robustness)
of a program \(\prog\) under weak consistency models. 
That is, the invariant obtained by executing \(\prog\) under a weak consistency model can also be obtained under serialisability.
To demonstrate this, we prove the robustness of the single
counter library discussed above against \(\PSI\), and the robustness of a multi-counter library and the banking library of \citet{bank-example-wsi}
against our new proposed model \(\WSI\) and all stronger models such as \(\SI\).
The latter is done through general conditions on invariant which guarantees robustness against \( \WSI \).
Thanks to our operational semantics, our invariant-based approaches only need to work with single program steps rather than whole program traces.

\mypar{Verifying Implementation Protocols} 
The second application of our operational
semantics is for showing that implementations of distributed
kv-stores satisfy certain consistency models. 
Kv-stores and views provide a 
faithful abstraction of geo-replicated and partitioned
databases, and  execution tests provide a powerful abstraction of the synchronisation mechanisms 
enforced by these databases when committing a transaction. 
This then allows us to use our 
formalism to verify the correctness of distributed database protocols. 
To demonstrate this, we show that the
COPS protocol \citep{cops} for implementing a replicated database satisfies causal consistency  (\cref{sec:verify-impl}), 
and the Clock-SI protocol \citep{clocksi} for implementing a
partitioned database satisfies snapshot isolation (\cref{sec:clock-si}).

\sx{
\newpage 

BELOW GOES

However, the situation becomes more complicated when the kv-store contains multiple counters:  
since each client has its own view on the kv-store, and views of clients are independent from each other, it is possible for two 
clients to observe the increments on two distinct counters, \(\Counter(\key_1)\) and \(\Counter(\key_2)\), in different order. 
For instance, consider the following program:

\vspace{-5pt}
{%
\displaymathfont
\begin{align}
		\cl_0: 
		 \ctrinc(\key_1) ; \ctrinc(\key_2)
		 \;\; || \;\;  \cl_1: 
		 \ctrread(\key_1) ; \ctrread(\key_2)
		  \;\; || \;\;  \cl_2: 
		 \ctrread(\key_1); \ctrread(\key_2)
	\tag{\textsc{LF}}
	\label{prog:LF}
\end{align}	 
}%
Suppose that \(\cl_0\) executes first by incrementing \(\key_1\), \(\key_2\).
This results in \(\key_1\) and \(\key_2\) having two versions with values \(0\) and \(1\) each. 
Client \(\cl_1\) executes its transactions next, using a view that 
contains both versions of \(\key_1\), but only 
the initial version of \(\key_2\):  client \(\cl_1\) then reads \(1\) for \(\key_1\) and \(0\) for \(\key_2\); \ie \(\cl_1\) observes
the increment of \(\key_1\) 
happening before the increment of \(\key_2\). 
Finally, \(\cl_2\) executes its transactions using a view that contains both versions for \(\key_2\), but only 
the initial version of \(\key_1\): 
client \(\cl_2\) reads \(0\) for \(\key_1\) and \(1\) for \(\key_2\); 
\ie \(\cl_2\)
observes the increment of \(\key_2\) 
happening before the increment of \(\key_1\). 
This behaviour is known as the \emph{long fork} anomaly (\cref{fig:cp-disallowed}). 

The long fork anomaly is disallowed under strong models, \eg serialisability (\(\SER\)) and snapshot isolation (\(\SI\)), 
but is allowed under weaker models \eg parallel SI (\(\PSI\)) and causal consistency (\(\CC\)). 
To capture such consistency models and rule out the long fork anomaly as a possible result 
of \eqref{prog:LF}, we must strengthen the execution test associated with the kv-store. 
For \(\SER\), we strengthen the execution test by ensuring that a client can execute a transaction 
only if its view contains all versions available in the global state. 
For SI, the candidate execution test recovers the order in which 
updates of versions have been observed by different clients (\eg \(\cl_1\)), 
and allows a transaction to commit only if the observations made by the committing client (\eg \(\cl_2\)) are consistent with previous clients (\ie \(\cl_1\)): we give the formal definition of this execution test  in \cref{sec:cm}.
Under such strengthened execution tests, in the \eqref{prog:LF} example \(\cl_2\) cannot
observe \(1\) for \(\key_2\) after observing \(0\) for \(\key_1\); 
this is because \(\cl_1\) has already established that the increment on \(\key_2\) happens after 
the one of \(\key_1\). 


\newpage 
}

\section{Operational Model}
\label{sec:model}

We define an interleaving operational semantics for atomic transactions over
global, centralised kv-stores and partial client views. 

\subsection{Key-Value Stores and Client Views}
\label{subsec:kvstores}
\label{sec:mkvs-view}
Our global, centralised key-value stores (kv-store) and partial client views
provide the abstract machine states for our operational semantics. A
kv-store comprises key-indexed lists of versions which record
the history of the key with values and meta-data of the
transactions that accessed it: the writer and readers.

We assume a countably infinite set of \emph{client identifiers}\footnote{ We use the notation
 $\sort A \ni a$ to denote that elements of $\sort A$ are ranged over
  by $a$ and its variants such as $a', a_1, \cdots$.},
$\Clients \ni \cl$.
The set of \emph{transaction identifiers}, $\TxID \ni t$, 
 is defined by
$\TxID \defeq  \Set{\txid_{0}} \uplus \Set{ \txid_{\cl}^{n} \mid \cl
  \in \Clients \land n \geq 0 }$, 
where  $\txid_0$ denotes  the  \emph{initialisation transaction}
and $\txid_{\cl}^{n}$ identifies a transaction committed by client
$\cl$ with $n$  determining  the client session order: that is, $\SO \defeq \Set{ (\txid, \txid') \mid \exsts{ \cl, n,m } \txid =
\txid_{\cl}^{n} \land \txid' = \txid_{\cl}^{m} \land n < m}$.
Subsets of $\TxID$  are ranged over by $\txidset, \txidset', \cdots$. 
We let $\TxID_{0} \defeq \TxID \setminus \Set{\txid_0}$. 

\begin{definition}[Kv-stores]
\label{def:his_heap}
\label{def:mkvs}
Assume a countably infinite set of \emph{keys}, $\Keys \ni \key$, 
and a countably infinite set of  \emph{values}, $\Val \ni \val$, 
which includes the keys and an \emph{initialisation value} $\val_0$.
The set of \emph{versions}, $\Versions \ni \ver$, is defined by $\Versions \defeq \Val \times \TxID \times \pset{\TxID_{0}}$. 
A \emph{kv-store} 
is a function $\mkvs: \Keys \to \func{List}[\Versions]$, 
where $\func{List}[\Versions] \ni \vilist$ is the set of lists of versions. 
\end{definition}

Each version has the form 
$\ver {=} (\val, \txid, \txidset)$, where $\val$ is
a value, the \emph{writer} $\txid$ identifies the transaction that
wrote $\val$,  and the \emph{reader set} $\txidset$ identifies the
transactions that read $\val$. We use the notation 
$\valueOf(\ver)$,
$\wtOf(\ver)$ and $\rsOf(\ver)$ to project
the individual components of $\ver$.
Given a kv-store $\mkvs$ and a transaction $\txid$, we write 
$\txid \in \mkvs$ if $\txid$ 
is either the writer or 
one of the readers of a version included in $\mkvs$, 
$\lvert \mkvs(\key) \rvert$ for the length of the version
list $\mkvs(\key)$,
and write $\mkvs(\key, i)$ for the $i$\textsuperscript{th} version of $\key$, 
with $0 \leq i < \lvert \mkvs(\key) \rvert$.

We focus on kv-stores whose consistency model satisfies the
\emph{snapshot property}, ensuring that
a transaction reads and writes at most one version for each key.
This is a normal assumption for distributed databases, \eg in~\cite{ramp,rola,cops,wren,redblue,PSI,NMSI,gdur,clocksi,distrsi}.
We also assume that 
the version list for each key has an initial version 
carrying the initialisation value $\val_0$,  written by the 
initialisation transaction $\txid_0$ with an initial empty reader set.
Finally, we assume that the kv-store agrees with the session order of clients: 
a client cannot read a
version of a key that has been written by a future transaction within
the same session; and the order in which versions are written by a
client must agree with its session order.  A kv-store is
\emph{well-formed} if it satisfies these three assumptions, defined
formally in \cref{def:mkvs-appendix}.  Henceforth, we assume kv-stores
are well-formed, and write $\MKVSs$ to denote the set of well-formed
kv-stores.

A global kv-store provides an abstract centralised description of
updates associated with distributed kv-stores that is \emph{complete} in 
that no update has been lost in the description. By contrast, in
both replicated and partitioned distributed databases, a client may
have incomplete information about updates distributed between
machines.  We model this incomplete information by
defining a {\em view} of the kv-store which provides a {\em
  partial} record of the updates observed by a client. We require that a client view be {\em atomic} in that it can
see either all or none of the updates of a transaction.

\pg{This makes sense for replicated databases. It is not at all clear
  to me that this makes sense for partitioned databases.}

\begin{definition}[Views]
\label{def:view}
\label{def:cuts}
\label{def:views}
A \emph{view} of a kv-store $\mkvs \in \MKVSs$ is a function
$\vi \in \Views(\mkvs) \defeq \Keys \to\pset{\Nat}$ such that, for all $i, i', \key, \key'$:
%

\vspace{-5pt}
{%
\displaymathfont
\begin{align*}
    & 
    0 \in \vi(\key) 
    \land (i \in \vi(\key) \implies 0 \leq i < \abs{ \mkvs(\key) }) 
    \tag{well-formed}
    \label{eq:view.wf}\\
    & 
    \begin{array}{@{}l@{}}
	i \in \vi(\key)  
  	\land \wtOf(\mkvs(\key, i)) {=} \wtOf(\mkvs(\key', i'))  
  	\implies i' \in \vi(\key')
    \end{array}
	\tag{atomic}
	\label{eq:view.atomic}
\end{align*}
}%
Given two views $\vi, \vi' \in \Views(\mkvs)$, 
the order between them is defined by $\vi \viewleq \vi' \defiff \fora{\key \in \dom(\mkvs)} \vi(k) \subseteq \vi'(\key)$.
The set of views is $\Views \defeq \bigcup_{\mkvs \in \MKVSs} \Views(\mkvs)$.
\noindent The \emph{initial view}, $\vi_{0}$,  is defined by
$\vi_{0}(\key) = \Set{0}$ for every $\key \in \Keys$. 
\end{definition}

Our operational semantics updates {\em configurations},  which are pairs
comprising a kv-store and a function describing the
views of a finite set of clients. 

\begin{definition}[Configurations]
\label{def:configuration}
A \emph{configuration}, $\conf \in \Confs $,  is a pair $ (\mkvs, \vienv)$
with $\mkvs \in \MKVSs$ and
$\vienv : \Clients \parfinfun \Views(\mkvs)$. 
The set of \emph{initial configurations}, $\Confs_0 \subseteq \Confs$,
contains configurations of the form $ (\mkvs_0, \vienv_0)$, where
$\mkvs_0$ is the initial kv-store defined by
$\mkvs_{0}(\key)\defeq  (\val_0, \txid_0, \emptyset)$ for
all $\key \in \Keys$. 
\end{definition}

Given a configuration $(\mkvs, \vienv)$ and a client $\cl$, 
if $\vi = \vienv(\cl)$ is defined then, for each $k$,  the
configuration determines the sub-list of versions in $\mkvs$ that $\cl$ sees.
If $i,j \in \vi(\key)$ and $i < j$, then $\cl$ sees the values 
carried by versions $\mkvs(\key, i)$ and  $\mkvs(\key, j)$, 
and it also sees that the version 
$\mkvs(\key, j)$ is more 
up-to-date than $\mkvs(\key, i)$. 
It is therefore possible to associate a \emph{snapshot}
with the view $\vi$, 
which identifies, for each key $\key$, the last version included in the view. 
This definition assumes that the database satisfies the \emph{last-write-wins}
resolution policy, employed by many distributed kv-stores.
However, our formalism can be adapted straightforwardly  to capture other resolution policies. 

\begin{definition}[View Snapshots]
\label{def:snapshot}
Given $\mkvs \in \MKVSs$ and $\vi \in \Views(\mkvs)$, the \emph{snapshot} of $\vi$ in 
$\mkvs$ is a function, $\snapshot[\mkvs, \vi] : \Keys \to
\Val$,   defined by $\snapshot[\mkvs, \vi] \defeq \lambda \key \ldotp \valueOf(\mkvs(\key, \max_{<}(\vi(\key))))$, 
where $\max_{<}(\vi(\key))$ is the maximum element in $\vi(\key)$ with respect to the natural 
order $<$ over $\mathbb{N}$.
\end{definition}

\subsection{Operational Semantics}

\vspace{5pt}
\mypar{Programming Language}
A \emph{program} \( \prog \) comprises a finite number of clients,
where each client is associated with a unique identifier \( \cl \in \Clients \), 
and executes a sequential \emph{command} $\cmd$, given by the following grammar:

\vspace{-5pt}
{%
\[
\displaymathfont
\begin{aligned}
\cmd & ::=  
\pskip \!\mid\!
\cmdpri \!\mid\!  
\ptrans{\trans} \!\mid\! 
\cmd \pseq \cmd \!\mid\! 
\cmd \pchoice \cmd \!\mid\! 
\cmd \prepeat
&
 \cmdpri & ::=  
\passign{\var}{\expr} \!\mid\! 
\passume{\expr} 
\\
\trans & ::=
\pskip \!\mid\!
\transpri \!\mid\! 
\trans \pseq \trans \!\mid\!
\trans \pchoice \trans \!\mid\!
\trans\prepeat    
&
\transpri & ::= 
\cmdpri \!\mid\!
\plookup{\var}{\expr} \!\mid\!
\pmutate{\expr}{\expr} 
\end{aligned}%
\]
}%
Sequential commands $\cmd$ comprise $\pskip$, primitive commands
$\cmdpri$, atomic transactions $\ptrans{\trans}$, and standard
compound constructs: sequential composition (\( ; \)), non-deterministic
choice (\( + \)) and iteration (\( * \)). 
Primitive commands include variable assignment
$\passign{\var}{\expr}$ and assume statements $\passume{\expr}$
which can be used to encode conditionals. They are used for computations based on client-local variables and can hence be invoked
without restriction.  Transactional commands $\trans$ comprise
$\pskip$, primitive transactional commands $\transpri$, and the
standard compound constructs.  Primitive transactional commands comprise
primitive commands, lookup $\plookup{\var}{\expr}$ and mutation
$\pmutate{\expr}{\expr}$ used for reading and writing to kv-stores
respectively, which can only be invoked as part of an atomic
transaction.

A {\em program} is a finite partial function from client identifiers to sequential
commands.
For clarity, we often write \( \cmd_{1}\ppar \dots \ppar \cmd_{n}\) as syntactic sugar 
for a program \( \prog \) with $n$ clients associated with identifiers
$\cl_1 \dots \cl_n$, where each client $\cl_i$ executes
$\cmd_i$. 
Each client $\cl_i$ is associated with its own client-local  \emph{stack}, 
$\stk_i \in \Stacks \defeq \Vars \to \Val$,  mapping program variables
(ranged over by $\pvar{x}, \pvar{y}, \cdots$)
to values. 
We assume a language of expressions built from values \( \val \)
and program variables \( \var \):
$\expr ::= \val \mid \var \mid \expr + \expr \mid \cdots$.
The \emph{evaluation} $\evalE{\expr}$ of expression $\expr$ is parametric in
the client-local stack \( \stk \):%

\vspace{-5pt}
{%
\[
\displaymathfont
\begin{gathered}
\evalE{\val} \defeq
\val
\quad
\evalE{\var} \defeq
\stk(\var)
\quad
\evalE{\expr_{1} + \expr_{2}} \defeq
\evalE{\expr_{1}} + \evalE{\expr_{2}}
\quad
\dots
\end{gathered}%
\]
}%

\mypar{Transactional Semantics} 
In our operational semantics, transactions are executed
\emph{atomically}. It is still possible for an underlying
implementation, such as COPS, to update the distributed kv-store while
the transaction is in progress. It just means that, given the
abstractions captured by our global kv-stores and partial views, 
such an update is modelled as  an instantaneous  atomic
update.
Intuitively, given a configuration $\conf = (\mkvs, \vienv)$, 
when a client $\cl$ executes a transaction $\ptrans{\trans}$, 
it performs the following steps: 
\begin{enumerate*}
	\item it constructs an initial \emph{snapshot} $\sn$ of $\mkvs$ using its view $\vienv(\cl)$ as defined in \cref{def:snapshot};  
	\item it executes $\trans$ in isolation over $\sn$
        accumulating the effects (the reads and writes) of executing $\trans$; and
	\item it commits $\trans$ by incorporating these effects into $\mkvs$.
\end{enumerate*}

\begin{definition}[Transactional Snapshots]
\label{def:heaps}
A \emph{transactional snapshot}, \( \sn \in \Snapshots \defeq \Keys \to
\Val\),  is a function from keys to values. When the meaning is clear,
it is just called a {\em snapshot}. 
\end{definition}

The rules for transactional commands will be defined  using an arbitrary 
transactional snapshot. The rules for sequential
commands and programs will be defined  using a transactional
snapshot given by a view snapshot. 
To capture the effects of executing a transaction $\trans$ on a snapshot $\sn$ of kv-store $\mkvs$, 
we identify a \emph{fingerprint}  of $\trans$ on $\sn$ which captures
 the values $\trans$ reads from $\sn$, and
the values $\trans$ writes to $\sn$ and intends to commit to $\mkvs$. 
Execution of a transaction in a given configuration may result in more than one fingerprint due to non-determinism (non-deterministic choice). 

\begin{definition}[Fingerprints]
\label{beebop}
\label{def:fingerprint}
Let \( \Ops\) denote the set of read (\( \otR\)) and write (\(\otW\)) \emph{operations} defined by 
$\Ops \defeq \Set{(l, \key, \val) }[ l \in \Set{\otR, \otW} \land \key \in \Keys \land \val \in \Val ]$.
A \emph{fingerprint} $\fp$ is a set of operations, $\fp \subseteq \Ops$,
such that: 
$\fora{\key \in \Keys, l  \in \Set{\otR, \otW}}
	(l, \key, \val_1), (l, \key, \val_2) \in \fp \implies \val_1 = \val_2$.
\end{definition}
\noindent 
According to \cref{def:fingerprint}, a fingerprint contains at most one read operation and at most one write operation. 
This reflects our assumption regarding transactions that satisfy the snapshot property: reads are taken from a single snapshot of the kv-store;
and 
only the last write of a transaction to each key is committed to the kv-store.
\begin{figure*}[!t]
\begin{center}
\scalebox{.9}{%
\begin{tabular}{@{}c@{\hspace{20pt}}c@{}}
    $
    \inferrule[\rl{TPrimitive}]{%
        (\stk, \sn) \toLTS{\transpri} (\stk', \sn') 
       \\
        \op = \func{op}[\stk, \sn, \transpri]
    }{%
        (\stk, \sn, \fp) , \transpri  \toTRANS   (\stk', \sn', \fp \addO \op) , \pskip 
    }%
    $
    &
    $
 \begin{array}{ll}
    \fp \addO (\otR, \key, \val)  
    & \defeq \
    \begin{cases}
        \fp \cup \Set{(\otR, \key, \val)} & \text{if } \fora{l, v'} (l, \key, v') \! \notin \! \fp \\
        \fp &  \text{otherwise} \\
    \end{cases}  \\
    \fp \addO (\otW, \key, \val) 
    & \defeq \
    \left( \fp \! \setminus \! \Set{(\otW, \key, v')}[v' \in \Val] \right)  \!
    \cup \! \Set{(\otW, \key, \val)} \\
    \fp \addO \emptyop  & \defeq \ \fp 
\end{array}
$
    \\
    $
     \inferrule[\rl{TChoice}]{%
		i \in \Set{1,2}
    }{%
		(\stk, \sn, \fp) , \trans_{1} \pchoice \trans_{2}  \toTRANS  (\stk, \sn, \fp) , \trans_{i}
    }
    $
    &
    $
    \inferrule[\rl{TIter}]{ }{%
        (\stk, \sn, \fp),  \trans\prepeat \toTRANS  (\stk, \sn, \fp), \pskip \pchoice (\trans \pseq \trans\prepeat)
    }%
    $
    \\[20pt]
    $
    \inferrule[\rl{TSeqSkip}]{ }{%
        (\stk, \sn, \fp), \pskip \pseq \trans \toTRANS  (\stk, \sn, \fp), \trans
    }%
    $
    &
    $
    \inferrule[\rl{TSeq}]{%
		(\stk, \sn, \fp), \trans_{1} \toTRANS  (\stk', \sn', \fp'), \trans_{1}'
    }{%
		(\stk, \sn, \fp), \trans_{1} \pseq \trans_{2} \toTRANS  (\stk', \sn', \fp'), \trans_{1}' \pseq \trans_{2}
    }%
    $\\[10pt]
\end{tabular}
}

\hrulefill

\end{center}
\caption{Rules for transactional commands.}
\label{fig:semantics-trans}
\end{figure*}

\Cref{fig:semantics-trans} presents the rules for transactional commands. 
The only non-standard rule is \( \rl{TPrimitive} \), which updates 
the snapshot and the fingerprint of a transaction: the premise 
$(\stk, \sn) \toLTS{\transpri} (\stk', \sn')$ describes how executing
$\transpri$ affects the local state (the client stack and the snapshot)
of a transaction; and the premise $o =\func{op}(\stk, \sn, \transpri)$ identifies the operation on the 
kv-store associated with $\transpri$. 

\begin{definition}
\label{def:primitive_semantics}
The relation $\toLTS{\transpri}\; \subseteq (\Stacks \times \Snapshots) \times (\Stacks \times \Snapshots)$ 
is defined by\footnote{ 
For any function \( \mathsf f \), the new function \( \mathsf f
\rmto{\mathsf d}{\mathsf r}\) 
is defined by 
\( \mathsf f
\rmto{\mathsf d}{\mathsf r}
( {\mathsf d}) = {\mathsf r}$, and \( 
\mathsf f
\rmto{\mathsf d}{\mathsf r}
({\mathsf d}') = {\mathsf f}({\mathsf d}')$ if \( 
{\mathsf d}' \neq {\mathsf d}\). } :
{%
\[
\displaymathfont
    \begin{array}{@{} r @{\ } c @{\ } l @{\quad} r @{\ } c @{\ } l@{}}
(\stk, \sn)  & \toLTS{\passign{\var}{\expr}}
             & (\stk\rmto{\var}{\evalE{\expr}}, \sn) 
&
(\stk, \sn)  & \toLTS{\passume{\expr}}  
             & (\stk, \sn) \text{ where } \evalE{\expr} \neq 0
\\
(\stk, \sn)  & \toLTS{ \plookup{\var}{\expr} } 
             & (\stk\rmto{\var}{\sn(\evalE{\expr})}, \sn) 
&
(\stk, \sn) & \! \toLTS{\pmutate{\expr_{1}}{\expr_{2}}} \!
            & (\stk, \sn\rmto{\evalE{\expr_{1}}}{\evalE{\expr_{2}}}) \ 
\end{array}
\]%
}%
%
%

The function  $\mathsf{op}$, computing the fingerprint of primitive
transactional 
commands,  is defined by:
\[%
\displaymathfont
\begin{aligned}
    \func{op}[\stk, \sn, \passign{\var}{\expr}] & \defeq  \emptyop 
    & 
    \func{op}[\stk, \sn, \passume{\expr}] & \defeq \emptyop 
    \\
    \func{op}[\stk, \sn,  \plookup{\var}{\expr}] & \defeq (\otR, \evalE{\expr}, \sn(\evalE{\expr})) 
    &
    \func{op}[\stk,  \sn, \pmutate{\expr_{1}}{\expr_{2}}] & \defeq (\otW, \evalE{\expr_{1}}, \evalE{\expr_{2}})
\end{aligned}
\]%
The  empty operation $\emptyop$ is used for those primitive commands that do not
contribute to the fingerprint.
\end{definition}
The conclusion of the \( \rl{TPrimitive}\)  rule uses the \emph{combination operator} $\addO: 
\pset{\Ops} \times (\Ops \uplus \Set{\emptyop}) \to \pset{\Ops}$, defined 
in \cref{fig:semantics-trans}, to extend the fingerprint $\fp$ accumulated with
operation $o$ associated with $\transpri$, as
appropriate: it adds  a read from $\key$  if $\fp$ does not already
contain an entry for $\key$; and it always updates the  write for 
$\key$ to $\fp$, removing previous writes to $\key$.

\begin{definition}[Fingerprint Set]
Given a client stack $\stk$ and a snapshot $\sn$, the \emph{fingerprint set} of a transaction $\trans$ is:
\(
\Fingerprints(\trans) \defeq \Set{\fp \mid \exsts{\stk,\stk',\sn,\sn'} (\stk, \sn, \emptyset), \trans \toTRANS^* (\stk', \sn', \fp), \pskip }
\)
where $\toTRANS^*$ denotes the reflexive, transitive closure of $\toTRANS$ given in \cref{fig:semantics-trans}.  
A set $\fp \in \Fingerprints(\trans)$ is called a \emph{final fingerprint} of $\trans$. 
\end{definition}
\noindent 

\mypar{Operational Semantics}
We give the operational semantics of commands and programs in
\cref{fig:semantics-commands}.  
\begin{figure*}[!t]
\begin{center}
\scalebox{.9}{%
\begin{tabular}{@{}c@{\hspace{20pt}}c@{}}
    $
    \inferrule[\rl{CPrimitive}]{
		\stk \toLTS{\cmdpri} \stk'
    }{
        \cl \vdash 
        ( \mkvs, \vi, \stk ) , \cmdpri 
        \toCMD{(\cl,\iota)}_{\ET} 
        ( \mkvs, \vi, \stk' ) , \pskip
    }%
    $
    &
    $
 \begin{array}{r @{\ }c @{\ } l}
        \stk & \toLTS{\passign{\var}{\expr}} & \stk\rmto{\var}{\evalE{\expr}} \\
        \stk & \toLTS{\passume{\expr}} & \stk \ \text{where} \ \evalE{\expr} \neq 0
\end{array}
$
\\[30pt]
\multicolumn{2}{c}{
    $
     \inferrule[\rl{CAtomicTrans}]{%
        \vi \viewleq  \vi'' 
        \\
        \sn = \snapshot[\mkvs,\vi''] 
        \\
        (\stk, \sn, \emptyset), \trans \toTRANS^{*}   (\stk', \stub,
  \fp) , \pskip
  \\
	\cancommit{\mkvs}{\vi''}{\fp}
    \\\\
   \txid \in \nextTxid[\cl, \mkvs] 
\\
     \mkvs' = \updateKV[\mkvs, \vi'', \fp, \txid] 
\\
	\vshift{\mkvs}{\vi''}{\mkvs'}{\vi'}	
    }{%
        \cl \vdash 
        ( \mkvs, \vi, \stk ), \ptrans{\trans} 
        \toCMD{(\cl, \vi'', \fp)}_{\ET}
        (\mkvs',\vi', \stk' ) , \pskip
    }%
    $
    }

    \\[20pt]
    \multicolumn{2}{c}{
    $
    \inferrule[\rl{PProg}]{%
	    \vi = \vienv (\cl)
        \\
        \stk = \thdenv(\cl)
        \\
        \cmd = \prog(\cl)
        \\
		\cl \vdash 
		( \mkvs, \vi, \stk ) , \cmd
		\toCMD{\lambda}_{\ET} 
		( \mkvs', \vi', \stk' ) , \cmd'  
    }{%
		\vdash 
		(\mkvs, \vienv, \thdenv ), \prog
		\toCMD{\lambda}_{\ET} 
		( \mkvs', \vienv\rmto{\cl}{\vi'}, \thdenv\rmto{\cl}{\stk'} ) , \prog\rmto{\cl}{\cmd'} ) 
    }%
    $
    }\\[10pt]
\end{tabular}
}
\end{center}

\hrulefill

\caption{Rules for  sequential commands and programs.}
\label{fig:semantics-commands}
\end{figure*}
The command semantics describes transitions of the form
$\cl \vdash (\mkvs, \vi, \stk), \cmd \ \toCMD{\lambda}_{\ET} \ (\mkvs', \vi', \stk') ,
\cmd'$, stating that given the kv-store $\mkvs$, view $\vi$ and stack $\stk$, 
a client $\cl$ may execute command $\cmd$ for one step, updating 
the kv-store to $\mkvs'$, the stack to $\stk'$, and the command to its continuation $\cmd'$.
The label $\lambda$ is either of the form $(\cl, \iota)$ denoting that $\cl$ executed a primitive command
that required no access to $\mkvs$, 
or $(\cl, \vi'', \fp)$ denoting that $\cl$ committed an atomic transaction with final fingerprint $\fp$ under the view $\vi''$.
The semantics is parametric in the choice of \emph{execution test}
$\ET$ for kv-stores, which is used to generate 
the \emph{consistency model} on kv-stores
under which a 
transaction can execute.
In \cref{sec:cm}, we give many examples of execution tests for
well-known consistency models.
In \cref{sec:other_formalisms} and \cref{app:et_sound_complete}, we prove that our execution tests (using kv-stores) generate consistency models that are equivalent to existing definitions of
consistency models (using execution graphs). 

The rules for the compound constructs are straightforward and given in \cref{sec:full-semantics}.
The rule for primitive commands, $\rl{CPrimitive}$, 
depends on the 
transition system $\toLTS{\cmdpri} \subseteq \Stacks \times \Stacks$ 
which simply describes how the primitive command $\cmdpri$ affects the local state of a client.
The rule \rl{CAtomicTrans}  describes the execution of an atomic 
transaction under the execution test $\ET$.

We describe the \rl{CAtomicTrans} rule in detail. 
The first premise 
states that the current view $\vi$ of the executing command may be advanced to a newer  view $\vi''$ (see \cref{def:views}). 
Given the new view $\vi''$, the transaction obtains a snapshot $\sn$ of the kv-store $\mkvs$, 
and executes $\trans$ locally to completion ($\pskip$), updating the stack to $\stk'$, while accumulating the fingerprint $\fp$; 
this behaviour  is modelled in the second and third premises of \rl{CAtomicTrans}.
Note that the resulting snapshot is ignored 
as the effect of the transaction is recorded in the fingerprint $\fp$. 
The $\cancommit{\mkvs}{\vi''}{\fp}$ premise ensures that under the execution test $\ET$, 
the final fingerprint $\fp$ of the transaction is compatible with the (original) kv-store
$\mkvs$ and the client view $u''$, and thus the transaction \emph{can commit}. 
Note that the \cancommitname check is parametric in the execution test $\ET$.
This is because the conditions checked upon committing depend on the consistency model under which the transaction is to commit. 
In \cref{sec:cm}, we define \cancommitname for several execution tests associated with well-known consistency models.

Now we are ready for client $\cl$ to commit the transaction resulting 
in the kv-store $\mkvs'$ with the client view $\vi''$ \emph{shifting} to a new view $\vi'$: 
pick a fresh transaction identifier $\txid \in \nextTxid[\cl, \mkvs]$;
compute the new kv-store via $\mkvs' = \updateKV[\mkvs, \vi'', \fp, \txid]$; 
and 
check if the \emph{view shift} is permitted under execution test $\ET$ using $\vshift{\mkvs}{\vi''}{\mkvs'}{\vi'}$. 
Observe that as with \cancommitname, the \vshiftname check is parametric in the execution test $\ET$. 
Once again this is because the conditions checked for shifting the client view depend on the consistency model. 
In \cref{sec:cm} we define \vshiftname for several execution tests associated with well-known consistency models.
The set $\nextTxid[\cl, \mkvs]$ is defined by
\(
\nextTxid[\cl, \mkvs] \defeq 
\Set{\txid_{\cl}^{n}}[%
\fora{m} \txid_{\cl}^{m} \in \mkvs \implies m < n ]
\).
The function $\updateKV[\mkvs, \vi, \fp, \txid]$
describes how the fingerprint $\fp$ of transaction $\txid$ executed under view $\vi$ updates kv-store $\mkvs$:
for each read $(\otR, \key, \val) \in \fp$, it adds $\txid$ 
to the reader set of the last version of $\key$ in $\vi$; 
for each write $(\otW, \key, \val)$, it appends a new version $(\val, \txid, \emptyset)$ 
to $\mkvs(\key)$.

\begin{definition}[Transactional update]
\label{eq:updatekv}
\label{def:updatekv}
The function  $\updateKV[\mkvs, \vi, \fp, \txid]$,  is
defined by:

\vspace{-5pt}
{%
\displaymathfont
\[%
\begin{array}{lcl}
    \updateKV[\mkvs, \vi, \emptyset, \txid] & \defeq & \mkvs 
    \\
    \updateKV[\mkvs, \vi, \Set{(\otR, \key, \val)} \uplus \fp, \txid]
    & \defeq & \text{let } i = \max_{<}(\vi(\key)), (\val,\txid',\txidset) = \mkvs(\key,i) \text{ in } \\
    && \;\;\;\; \updateKV[\mkvs\rmto{\key}{\mkvs(\key)\rmto{i}{(\val, \txid', \txidset \uplus \Set{\txid})}},\vi, \fp, \txid] \\
	\updateKV[\mkvs, \vi, \Set{(\otW, \key, \val)} \uplus \fp, \txid]
    & \defeq & \text{let } \mkvs' = \mkvs\rmto{\key}{ \mkvs(\key) \lcat (\val, \txid, \emptyset) } \text{ in } \updateKV[\mkvs', \vi, \fp, \txid] 
\end{array}
\]%
}%
\noindent where, given a list of versions $\vilist = \ver_0 \lcat \cdots \lcat \ver_n$ 
and an index $i: 0 \leq i \leq n$, 
then $\vilist\rmto{i}{\ver} \defeq \ver_0 \lcat \cdots \lcat \ver_{i-1} \lcat \ver \lcat \ver_{i+1} \cdots \ver_{n}$,

\end{definition}

The last rule, \( \rl{PProg} \), captures the execution of a program step 
given a \emph{client environment} $\thdenv \in \ThdEnv$.
A client environment $\thdenv$ is a function from client identifiers to variable stacks, associating each client with its stack. 
We assume that the domain of client environments contains 
the domain of the program throughout the execution: 
$\dom(\prog) \subseteq \dom(\thdenv)$.
Program transitions are simply defined in terms of the transitions of
their constituent client commands. 
This yields an interleaving semantics for transactions of different clients:  
a client executes a transaction in an atomic step without
interference from the  other clients. 

\section{Consistency Models: Kv-stores}
\label{sec:cm}
We define what it means for a kv-store 
to be in a consistent state. Many different consistency models for
distributed databases have 
been proposed in the literature
\cite{principle-eventual-consistency,rola,redblue,PSI,si},
capturing different trade-offs 
between  performance and application
correctness: examples range from  \emph{serialisability}, a strong
consistency model which only allows kv-stores 
obtained  from a serial execution of transactions
with inevitable performance drawbacks, to  \emph{eventual consistency},  a weak consistency model
which imposes few conditions on the structure of a kv-store leading to
good performance but anomalous behaviour.
We define consistency models for our kv-stores,
by introducing the notion of 
\emph{execution test} which specifies  whether a client is allowed to commit a transaction in a given 
kv-store. Each execution test induces a consistency model as the set of kv-stores obtained 
by having clients non-deterministically commit transactions so long as  the constraints 
imposed by the execution test are satisfied.
We explore a range of execution tests  associated with well-known consistency models in the literature. 
In \cref{sec:other_formalisms},  we demonstrate that our operational
formulation of  consistency models over kv-stores using execution
tests are  equivalent to the established declarative definitions of
consistency models  over abstract executions \cite{ev_transactions,framework-concur} and dependency graphs \cite{adya}.

\begin{definition}[Execution tests]
\label{def:execution.test}
An \emph{execution test} is a set of tuples \(\ET \subseteq \MKVSs \times \Views \times \Fingerprints \times \MKVSs \times \Views\) 
such that, for all \((\mkvs, \vi, \fp, \mkvs', \vi') \in \ET\): 
\begin{enumerate*}
	\item \(\vi \in \Views(\mkvs)\) and \(\vi' \in \Views(\mkvs')\); 
	\item \(\cancommit \mkvs \vi \fp\); 
	\item \(\vshift \mkvs \vi {\mkvs'} {\vi'}\); and 
	\item for all \(\key \!\in\! \mkvs\) and \(\val \!\in\! \Val\), if \((\otR, \key, \val) \!\in\! \fp \) then \(	\mkvs(\key, \max{}_{<}(\vi(\key))) {=} \val   \).
\end{enumerate*}

\end{definition}
\noindent 
Intuitively, \((\mkvs, \vi, \fp, \mkvs', \vi') \in \ET\) means that, under the execution test \(\ET\),
a client with initial view \(\vi\) over a kv-store \(\mkvs\) can commit a transaction with 
fingerprint \(\fp\) to obtain the resulting kv-store \(\mkvs'\) (\cref{def:updatekv}) while shifting its view
to \(\vi'\). We adopt the 
notation \(\ET \vdash (\mkvs, \vi) \csat \fp: (\mkvs', \vi')\)  to
capture this intuition. 
Note that the last condition in \cref{def:execution.test} enforces the last-write-wins
policy~\cite{vogels:2009:ec:1435417.1435432}: 
a transaction always reads the most recent writes from the initial view \(\vi\).  

The largest execution test is denoted by \(\ET_{\top}\), where for all \(\mkvs, \mkvs', \vi, \vi, \fp\): 
\[
	\cancommit[\ET_{\top}] \mkvs \vi \fp \defiff \mathsf{true}
	\qquad  \text{and} \qquad 
	\vshift[\ET_{\top}] \mkvs \vi {\mkvs'} {\vi'} \defiff \mathsf{true}
\] 
In \cref{sec:other_formalisms}, we show that the consistency model induced by \(\ET_{\top}\) 
corresponds to the \emph{Read Atomic} model \cite{ramp}, a variant of \emph{Eventual 
Consistency} \cite{ev_transactions} for atomic transactions. 

In \cref{subsec:cm_examples}, we give many examples of execution tests. Here, we
explain how an execution test induces a consistency model. Given an execution test \(\ET\), 
we define a \(\ET\)-reduction as a labelled transition which captures how
client \(\cl\) updates a  configuration:  either  \(\cl\)
shifts its view to a more up-to-date one
with  the label \(\varepsilon\) denoting  that 
	there was no access to the kv-store; or  \(\cl\) 
	commits a transaction with the label \(\fp\)  denoting the
        fingerprint of the committed transaction.

\begin{definition}[ET-reduction]
An \emph{\(\ET\)-reduction}, \((\mkvs, \vienv) \toET{(\cl, \alpha)} (\mkvs', \vienv')\), is defined by:
\begin{enumerate}
    \item either \(\alpha = \varepsilon\), \(\mkvs' = \mkvs\) and $\vienv' =
      \vienv\rmto{\cl}{\vi}\) for some \(u\) such that \( \vienv(\cl) \sqsubseteq u\); or
\item \(\alpha = \fp\) for some \(\fp\), and \(\ET \vdash (\mkvs, \vi ) \csat \fp : (\mkvs', \vi')\), where \(\mkvs' = \updateKV[\mkvs, \vi, \fp, \txid]\) 
   for some \(\txid \in \nextTxid(\cl, \mkvs)\), \(\vienv' = \vienv\rmto{\cl}{\vi'}\).
\end{enumerate}
\ac{Not changing due to time constraints, but I believe the best thing to do is to change the type of execution tests to have a client identifier 
rather than \(\mkvs'\), since this is determined by \(\mkvs\) and \(\cl\). In fact, I believe that the best we can do is to change execution tests to 
have the form \(\ET \vdash (\mkvs, \vienv) {\csat}_{\cl} \fp : \vi'\); then one can lift this to a \(\ET\)-reduction in a style similar to monadic lifting.}
A finite sequence of \(\ET\)-reductions starting in an
initial configuration \(\conf_{0}\) is called  an \emph{\(\ET\)-trace}. 
\end{definition}
Each \(\ET\)-trace  starting with an initial configuration
(\cref{def:configuration}) terminates in a configuration \((\mkvs, \stub)\) where \(\mkvs\) is obtained as a result of several clients committing transactions under the 
execution test \(\ET\). The consistency model induced by \(\ET\), 
written \(\CMs(\ET)\), is the set of all such terminal kv-stores.
%
%

\begin{definition}[Consistency Model]
\label{def:cm}
The \emph{consistency model} induced by an execution test \(\ET\) is defined as 
\(
\CMs(\ET) \defeq 
\Set{\mkvs}[ 
\exsts{\conf_0 \in \Confs_0}
\conf_0 \toET{\stub}^{*} (\mkvs, \stub)
]
\).
\end{definition}
%

Note that in the definition of \(\ET\)-traces, the view-shifts and 
transaction commits are decoupled. This is in contrast to the
operational semantics (\cref{sec:model}, \cref{fig:semantics-commands}), 
where view-shifts and transaction commits are combined in a single transition of programs (\rl{CAtomicTrans}). 
The reason for this mismatch is best understood when looking at the
intended applications. 
ET-traces are useful for 
proving that a distributed transactional 
protocol implements a given consistency model: in this case, it is convenient to separate shifting a view from committing a transaction, 
as these two steps often take place separately in distributed
protocols. The operational semantics  is particularly useful for  reasoning about transactional 
programs: in this case, the treatment of the view-shifts and transaction commits as a single transition reduces the number of interleavings in programs.
The \(\ET\)-traces and operational semantics are equally expressive as
the following theorem states. 
\ac{One may wonder whether this difference in approach leads to a difference in expressive power between \(\ET\)-traces 
and the operational model: that is, if there exists a kv-store \(\mkvs \in \CMs(\ET)\) that can never be obtained as a 
result of executing an arbitrary program \(\prog\) under the execution test \(\ET\). The next result shows that 
this is not the case.}
%

\begin{theorem}
	\label{thm:ettraces2sem}
	Let \(\interpr{\prog}_{\ET}\) be the set of kv-stores reachable by executing \(\prog\) under the execution test \(\ET\). 
    Then for all \(\ET\), \(\CMs(\ET) = \bigcup_{\prog} \interpr{\prog}_{\ET}\).
\end{theorem}

\begin{figure}[t]
\small
\centering
\begin{tabular}{ @{} l @{\hspace{2pt}} || @{\hspace{2pt}} c | @{\hspace{2pt}} l @{\hspace{2pt}} | @{\hspace{2pt}}  c @{} }
\hline
	\ET 
	& $\cancommit \mkvs \vi \fp$
	& Closure Relation (where applicable)
    & $\vshift \mkvs \vi {\mkvs'} {\vi'}$ 
	\\
	\hline
	\MR 
	& \true 
	& 
	& $\vi \viewleq \vi'$
	\\ \hline  
	\RYW
	& \true
	& 
	& 
	\protect{$
	\begin{array}[t]{@{} l @{}}
		\fora{\txid \in \mkvs' \setminus \mkvs} \fora{\key, i} \\
		\;\;\wtOf(\mkvs'(\key, i) ) \toEDGE{\!\!\SO\rflx\!\!} \txid \implies i \!\in\! \vi'(\key) 
	\end{array}
	$}
	\\ \hline  
%
%
%
%
%
	\CC
	& $\closed(\mkvs, \vi, \rel_{\CC})$
	& $\rel_{\CC}   \defeq \SO \cup \WR_{\mkvs}$ 
	& $\vshift[\MR \cap \RYW] \mkvs \vi {\mkvs'} {\vi'}$
	\\ \hline  
	\UA 
	& $\closed(\mkvs, \vi, \rel_{\UA})$
	& $\rel_{\UA}  \defeq {\textstyle\bigcup_{(\otW, \key, \stub) \in \fp}} \WW^{-1}_{\mkvs}(\key) $ 
	& \true  
	\\ \hline  
	\PSI
	& $\closed(\mkvs, \vi, \rel_{\PSI})$
	& $\rel_{\PSI} \defeq \rel_{\UA} \cup \rel_{\CC} \cup \WW_\mkvs$ 
	& $\vshift[\MR \cap \RYW] \mkvs \vi {\mkvs'} {\vi'}$
	\\ \hline   
	\CP 
	& $\closed(\mkvs, \vi, \rel_{\CP})$
	& $\rel_{\CP} \defeq \SO;\RW\rflx_\mkvs \cup \WR_\mkvs;\RW\rflx_\mkvs  \cup \WW_\mkvs$ 	
	& $\vshift[\MR \cap \RYW] \mkvs \vi {\mkvs'} {\vi'}$
    \\ \hline 
	\SI
	& $\closed(\mkvs,\vi, \rel_{\SI})$
	& $  \rel_{\SI}  \defeq \rel_{\UA} \cup \rel_{\CP} \cup (\WW_\mkvs; \RW_\mkvs)$ 
	& $\vshift[\MR \cap \RYW] \mkvs \vi {\mkvs'} {\vi'}$
	\\ \hline  
%
	\SER
	& $\closed(\mkvs,\vi, \rel_{\SER})$
	&$\rel_{\SER} \defeq \WW^{-1}$
	& \true	
	\\ \hline
\end{tabular}
\vspace{0pt}
\caption{Execution tests of well-known consistency models, where \SER* denotes an alternative equivalent $\SER$ specification and $\SO$ is as given in \cref{subsec:kvstores}.
}
\label{fig:execution.tests}
\label{fig:execution_tests}
\label{fig:execution-tests}
\end{figure}

\subsection{Example Execution Tests}
\label{subsec:cm_examples}
We give several examples of execution tests which give rise to consistency
models on kv-stores.
Recall that the snapshot property and the last write wins are hard-wired into our model. 
This means that we can only define  consistency models that satisfy these two constraints. 
Although this forbids us to express interesting consistency models such as \emph{Read Committed}, we are able to express a large variety of consistency models employed by distributed kv-stores.
We proceed with a summary of our notational conventions.

\mypar{Notation}
Given relations \(\mathsf r, \mathsf r' \subseteq \sort A \times \sort A\),
we write:  \(\mathsf r\rflx\), \(\mathsf r^+\) and \(\mathsf r^*\) for its reflexive, transitive and reflexive-transitive closures of \(\mathsf r\), respectively;
\(\mathsf r^{-1}\) for its inverse;
\(a_1 \toEDGE{\mathsf r} a_2\) for \((a_1, a_2) \in \mathsf r\);
and \( \mathsf r; \mathsf r'\) for \( \Set{(a_1,a_2)}[\exsts{a} (a_1,a) \in \mathsf r\land (a,a_2) \in \mathsf r']\).

Recall from \cref{def:execution.test} that an execution test \(\ET\)
comprises  tuples of the form \((\mkvs,\vi, \fp, \mkvs', \vi')\) 
where  \(\cancommit \mkvs \vi \fp\) and \(\vshift \mkvs \vi {\mkvs'} {\vi'}\). 
We define \(\cancommitname\) and \(\vshiftname\) for several consistency
models, using a couple of auxiliary definitions.

\mypar{Prefix Closure}
Given a kv-store \(\mkvs\) and a view \(\vi\), the {\em set of visible
transactions} is: 
{%
\displaymathfont
\begin{align*}
\Tx[\mkvs, \vi] & \defeq
\Set{\wtOf[\mkvs(\key, i)] }[ i \in \vi(\key)] 
\end{align*}
}%
%
%
%
%
%
Given a binary relation on transactions, \(\rel \subseteq \TxID \times \TxID\), we say that a view \(\vi\) is closed with respect to a kv-store \(\mkvs\) and \(\rel\), written \(\closed[\mkvs,\vi,\rel]\), iff:  
{%
\displaymathfont
\begin{align*}
	\closed[\mkvs,\vi,\rel]
	\defiff
	\Tx(\mkvs, \vi) = 
	\left( (\rel^*)^{-1} \Tx(\mkvs, \vi) \right) \setminus \Set{\txid }[ \fora{\key,i} \txid \neq \wtOf[\mkvs(\key,i)] ]
\end{align*}
}%
That is, if transaction \(\txid\) is visible in \(\vi\), then all transactions that are \(\rel^*\)-before \(\txid\) are also visible in \(\vi\).

\mypar{Dependency Relations}
We define transaction dependency relations for kv-stores,  inspired by
analogous relations for  dependency graphs due to \citet{adya}.
Given a kv-store \(\mkvs\), a key \(\key\) and 
indexes \(i,j\) such that  \(0 \leq i < j < \abs{ \mkvs(\key) }\), 
if there exists \(\txid_i, \T_i, \txid\) such that 
\(\mkvs(\key, i)  {=} (\stub, \txid_{i}, \T_{i})\), \(\mkvs(\key,j) {=} (\stub, \txid_{j}, \stub)\)
and \(\txid \in \T_{i}\), 
then we say that there is:
\begin{enumerate} 
\item a \emph{Write-Read} dependency over 
\(\key\) from \(\txid_{i}\) to \(\txid\), written \((\txid_{i},\txid) \in \WR_{\mkvs}(\key)\);
\item a \emph{Write-Write} dependency over \(\key\) from \(\txid_{i}\) to \(\txid_{j}\), 
    written \((\txid_{i},\txid_{j}) \in \WW_{\mkvs}(\key) \); and 
\item a \emph{Read-Write} anti-dependency from \(\txid\) to \(\txid_{j}\), provided that 
\(\txid \neq \txid_{j}\), written \((\txid,\txid_j) \in \RW_{\mkvs}(\key)\).
\end{enumerate}
\noindent \cref{fig:dependencies} illustrates an example kv-store and
its transaction dependency relations.
%

We give several definitions of
execution tests using \vshiftname and \cancommitname in \cref{fig:execution_tests}. 
In \cref{sec:other_formalisms}, we demonstrate that the associated consistency
models on kv-stores correspond to well-known consistency models on 
execution graphs. We anticipate these results, by labelling the
execution test with their well-known consistency models.

\subsubsection{Monotonic Reads \((\MR)\)}
This consistency model states that when committing, a client
cannot lose information in that it can only see increasingly more up-to-date versions from a kv-store.
This prevents, for example, the kv-store of \cref{fig:mr-disallowed},
since client \(\cl\) first reads the latest version of \(\key\) in \(\txid_{\cl}^{1}\), 
and then reads the older, initial version of \(\key\) in \(\txid_{\cl}^{2}\).  
As such, the \(\vshiftname_{\MR}\) predicate in \cref{fig:execution_tests} ensures that clients  can only extend their views. 
When this is the case, clients can then \emph{always} commit their transactions, and thus \(\cancommitname_{\MR}\) is simply defined as \(\true\). 
%
%


\subsubsection{Read Your Writes \((\RYW)\)}
\begin{figure*}[t]
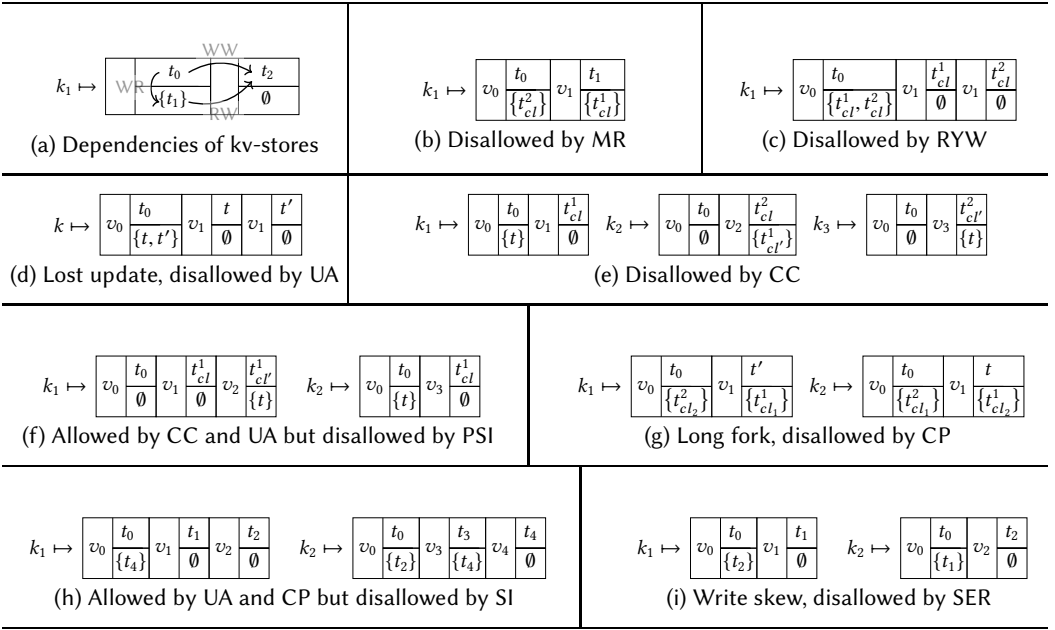

\newcommand{\SINGLEKV}{0.33\textwidth}
\newcommand{\RIGHTCOL}{0.61\textwidth}
\newcommand{\TWOKV}{0.49\textwidth}
\newcommand{\THREEKV}{0.61\textwidth}
\captionsetup[subfigure]{aboveskip=0pt, belowskip=5pt}

\begin{tabularx}{\textwidth}{@{} c @{}|@{} X @{}|@{} c @{} }
\hline
\phantom{-} & \phantom{-} & \phantom{-} 
\\[-5pt]
\begin{subfigure}{\SINGLEKV}
\vspace{-6pt}
\begin{centertikz}[.65]

\node(locx) {$\key_1 \mapsto$};
\draw pic at ([xshift=\tikzkvspace]locx.east) {vlist={versionx}{%
    /\;\;\;\;\;/\;\;\;\;\;\;$\txid_0$\;\;\;/\;\;\;\;$\Set{\txid_1}$\;\;\;\;
    , /\;/\;\;\;\;$\txid_2$\;\;\;\;/\;\;\;\;$\emptyset$\;\;\;\;\;
}};

\coordinate (A) at ([xshift=-25,yshift=3]versionx.center);
\coordinate (B) at ([xshift=0,yshift=-17]A.center);
\coordinate (C) at ([xshift=17,yshift=0]A.center);
\coordinate (D) at ([xshift=17,yshift=-17]A.center);
\coordinate (E) at ([xshift=37,yshift=0]C.center);
\coordinate (F) at ([xshift=0,yshift=-5]E.center);

\path[->, thick] ([yshift=5pt]A.center) edge[bend right=75] node[left,fill=white, opacity=.4, text opacity=1] {$\WR$} ([yshift=5pt]B.center)
([yshift=5pt]C.center) edge[bend left=30] node[above,fill=white, opacity=.4, text opacity=1] {$\WW$} ([yshift=5pt]E.center)
([yshift=5pt]D.center) edge[bend right=30] node[below, fill=white, opacity=.4, text opacity=1] {$\RW$} ([yshift=5pt]F.center);

\end{centertikz}%
\caption{Dependencies of kv-stores}
\label{fig:dependencies}
\end{subfigure}
&
\begin{subfigure}{\SINGLEKV}
\begin{centertikz}

\node(locx) {$\key_1 \mapsto$};
\draw pic at ([xshift=\tikzkvspace]locx.east) {vlist={versionx}{%
    /$\val_0$/$\txid_0$/$\Set{\txid_\cl^2}$
    , /$\val_1$/$\txid_1$/$\Set{\txid_\cl^1}$
}};

\end{centertikz}%
\caption{Disallowed by \(\MR\)}
\label{fig:mr-disallowed}
\end{subfigure}
&
\begin{subfigure}{\SINGLEKV}
\begin{centertikz}%

\node(locx) {$\key_1 \mapsto$};
\draw pic at ([xshift=\tikzkvspace]locx.east) {vlist={versionx}{%
    /$\val_{0}$/$\txid_0$/$\Set{\txid_\cl^1,\txid_\cl^2}$
    , /$\val_{1}$/$\txid_\cl^1$/$\emptyset$
    , /$\val_{1}$/$\txid_\cl^2$/$\emptyset$
}};
\end{centertikz}%
\caption{Disallowed by \(\RYW\)}
\label{fig:ryw-disallowed}
\end{subfigure}
\\
\hline
\\[-5pt]
\begin{subfigure}{\SINGLEKV}
\begin{centertikz}

\node(locx) {$\key \mapsto$};
\draw pic at ([xshift=\tikzkvspace]locx.east) {vlist={versionx}{%
    /$\val_{0}$/$\txid_0$/$\Set{\txid,\txid'}$
    , /$\val_{1}$/$\txid$/$\emptyset$
    , /$\val_{1}$/$\txid'$/$\emptyset$
}};

\end{centertikz}
\caption{Lost update, disallowed by \(\UA\)}
\label{fig:ua-disallowed}
\end{subfigure}

&

\multicolumn{2}{@{}c@{}}{%
\begin{subfigure}{\THREEKV}
\begin{centertikz}%

\node(locx) {$\key_1 \mapsto$};
\draw pic at ([xshift=\tikzkvspace]locx.east) {vlist={versionx}{%
        /$\val_0$/$\txid_0$/$\Set{\txid}$
    , /$\val_1$/$\txid^{1}_{\cl}$/$\emptyset$
}};

\path (versionx.east) + (0.75,0) node (locy) {$\key_2 \mapsto$};
\draw pic at ([xshift=\tikzkvspace]locy.east) {vlist={versiony}{%
    /$\val_0$/$\txid_0$/$\emptyset$
    , /$\val_2$/$\txid^{2}_{\cl}$/$\{\txid^{1}_{\cl'}\}$
}};

\path (versiony.east) + (0.75,0) node (locz) {$\key_3 \mapsto$};
\draw pic at ([xshift=\tikzkvspace]locz.east) {vlist={versionz}{%
    /$\val_0$/$\txid_0$/$\emptyset$
    , /$\val_3$/$\txid^{2}_{\cl'}$/$\Set{\txid}$
}};

\end{centertikz}%
\caption{Disallowed by \( \CC\)}
\label{fig:wr-wfr-allowed-but-cc}
\end{subfigure}%
}%
\end{tabularx}\\[-1pt]
\begin{tabularx}{\textwidth}{@{} c | X @{}}
\hline
\phantom{-}& \phantom{-} \\[-5pt]
\begin{subfigure}{\TWOKV}%
\begin{centertikz}%

\node(locx) {$\key_1 \mapsto$};
\draw pic at ([xshift=\tikzkvspace]locx.east) {vlist={versionx}{%
        /$\val_0$/$\txid_0$/$\emptyset$
        , /$\val_1$/$\txid^{1}_{\cl}$/$\emptyset$
        , /$\val_2$/$\txid^{1}_{\cl'}$/$\Set{\txid}$
}};

\path (versionx.east) + (1,0) node (locy) {$\key_2 \mapsto$};
\draw pic at ([xshift=\tikzkvspace]locy.east) {vlist={versiony}{%
    /$\val_0$/$\txid_0$/$\Set{\txid}$
    , /$\val_3$/$\txid^{1}_{\cl}$/$\emptyset$
}};

\end{centertikz}%
\caption{Allowed by \(\CC\) and \( \UA \) but disallowed by \( \PSI \)}
\label{fig:cc-ua-allowed-but-psi}
\end{subfigure}%
&
\begin{subfigure}{\TWOKV}%
\begin{centertikz}%
\node(locx) {$\key_1 \mapsto$};
\draw pic at ([xshift=\tikzkvspace]locx.east) {vlist={versionx}{%
    /$\val_0$/$\txid_0$/$\big\{\txid_{\cl_2}^2\big\}$
    , /$\val_1$/$\txid'$/$\big\{\txid_{\cl_1}^1\big\}$
}};

\path (versionx.east) + (0.75,0) node (locy) {$\key_2 \mapsto$};
\draw pic at ([xshift=\tikzkvspace]locy.east) {vlist={versiony}{%
    /$\val_0$/$\txid_0$/$\big\{\txid_{\cl_1}^2\big\}$
    , /$\val_1$/$\txid$/$\big\{\txid_{\cl_2}^1\big\}$
}};
\end{centertikz}%
\caption{Long fork, disallowed by \(\CP\)}
\label{fig:cp-disallowed-2}
\label{fig:cp-disallowed}
\end{subfigure}%
\end{tabularx}\\[-1pt]
\begin{tabularx}{\textwidth}{@{} c | X @{}}
\hline
\phantom{-}& \phantom{-} \\[-5pt]
\begin{subfigure}{0.54\textwidth}
\begin{centertikz}%
\node(locx) {$\key_1 \mapsto$};
\draw pic at ([xshift=\tikzkvspace]locx.east) {vlist={versionx}{%
    /$\val_0$/$\txid_0$/$\Set{\txid_4}$
    , /$\val_1$/$\txid_1$/$\emptyset$
    , /$\val_2$/$\txid_2$/$\emptyset$
}};

\path (versionx.east) + (1,0) node (locy) {$\key_2 \mapsto$};
\draw pic at ([xshift=\tikzkvspace]locy.east) {vlist={versiony}{%
    /$\val_0$/$\txid_0$/$\Set{\txid_2}$
    , /$\val_3$/$\txid_3$/$\Set{\txid_4}$
    , /$\val_4$/$\txid_4$/$\emptyset$
}};

\end{centertikz}
\caption{Allowed by \( \UA \) and \( \CP \) but disallowed by \(\SI\)}%
\label{fig:si-disallowed}%
\end{subfigure}%
&
\begin{subfigure}{0.45\textwidth}
\begin{centertikz}%

\node(locx) {$\key_1 \mapsto$};
\draw pic at ([xshift=\tikzkvspace]locx.east) {vlist={versionx}{%
    /$\val_0$/$\txid_0$/$\Set{\txid_2}$
    , /$\val_1$/$\txid_1$/$\emptyset$
}};

\path (versionx.east) + (1,0) node (locy) {$\key_2 \mapsto$};
\draw pic at ([xshift=\tikzkvspace]locy.east) {vlist={versiony}{%
    /$\val_0$/$\txid_0$/$\Set{\txid_1}$
    , /$\val_2$/$\txid_2$/$\emptyset$
}};

\end{centertikz}%
\caption{Write skew, disallowed by \(\SER\)}
\label{fig:ser-disallowed}
\end{subfigure}%
\\ \hline
\end{tabularx}
\caption{Behaviours disallowed under different consistency models. Sub-figure \ref{fig:dependencies} 
shows the dependencies of transactions in kv-stores, where values of versions have been 
removed for simplicity.}
\label{fig:anomalies}
\end{figure*}

This consistency model states that a client must always see all the versions written by the client itself. 
The \(\vshiftname_{\RYW}\) predicate thus states that after executing a transaction, a client 
contains all the versions it wrote in its view. This ensures that such versions will be included in the view of the client 
when committing future transactions.
Note that under \(\RYW\) the kv-store in \cref{fig:ryw-disallowed} is prohibited as
the initial version of \(\key\) holds value \(\val_0\) 
and client \(\cl\) tries to increment the value of \(\key\) twice.  
For its first transaction \( \txid_{\cl}^1\), it reads the initial value \(\val_0\) and then writes a new version with value \(\val_1\). 
For its second transaction \( \txid_{\cl}^2\), it reads the initial value \(\val_0\) again and write a new version with value \(\val_1\).
The \(\vshiftname_{\RYW}\) predicate rules out this example by requiring that
the client view, after it commits the transaction  \(\txid_{\cl}^{1}\), includes the version it wrote.  
When this is the case, clients can always commit their transactions, and thus \(\cancommitname_{\RYW}\) is simply \(\true\).

The \(\MR\) and \(\RYW\) models together with \emph{monotonic writes} (\MW) and \emph{write follows reads} (\WFR) models  are collectively known as \emph{session guarantees}. 
Due to space constraints, we omit the definitions associated with \(\MW\) and \(\WFR\), and refer the reader to \cref{sec:full-semantics}. 

\subsubsection{Causal Consistency \((\CC)\)}
Causal consistency subsumes the  four session guarantees discussed above. 
As such, the \(\vshiftname_\CC\) predicate is defined as the \emph{conjunction} of their associated \vshiftname predicates.
However, as shown in  \cref{fig:execution_tests}, it is sufficient to define \(\vshiftname_\CC\)
as the conjunction of the \(\MR\) and \(\RYW\) session guarantees alone, where for brevity we 
write \(\vshiftname_{\MR \cap \RYW}\) for  \(\vshiftname_{\MR} \land \vshiftname_{\RYW}\).
This is because 
as we demonstrate in \cref{sec:full-semantics},
the \(\vshiftname_{\MW}\) and \(\vshiftname_{\WFR}\) are defined simply as \( \true \), allowing us to remove them from \(\vshiftname_{\CC}\).

Additionally, \(\CC\) strengthens the session guarantees by requiring that if a client sees a version \(\ver\) prior to committing a transaction, then it must also see the versions 
on which \(\ver\) depends.
If \(\txid\) is the writer of \(\ver\), then 
\(\ver\) clearly depends on all versions that \(\txid\) reads. 
Moreover, if \(\ver\) is, or it depends on, a version \(\ver'\) accessed by 
a client \(\cl\), then it also depends on all versions that were previously 
read or written by \(\cl\). 
This is captured by the \(\cancommitname_{\CC}\) predicate in \cref{fig:execution_tests}, 
defined as \(\closed(\mkvs, u, \rel_{\CC})\) with \(\rel_\CC \defeq \SO \cup \WR_{\mkvs}\).
\azalea{without the definitions of \(\MW\) and \(\WFR\) we no longer can explain why this implies \(\MW\) and \(\WFR\).}
For example, the kv-store of 
\cref{fig:wr-wfr-allowed-but-cc} 
is disallowed by \(\CC\): the version of key \(\key_3\) carrying value \(\val_3\) depends on the version of key \(\key_1\) carrying value \(\val_1\). However, transaction \(\txid\) must have been committed by a client whose view included \(\val_3\), but not \(\val_1\).

\subsubsection{Update Atomic \((\UA)\)}
This consistency model has been proposed by \citet{framework-concur} 
and implemented by \citet{rola}.
\(\UA\) disallows concurrent transactions writing to the same key,
a property known as \emph{write-conflict freedom}:  
when two transactions write to the same key, one must see the version 
written by the other.
Write-conflict freedom is enforced by \(\cancommitname_{\UA}\) which allows a client to write to key \(\key\) only if its view includes all versions of \(\key\); 
\ie its view is closed with respect to the \(\WW^{-1}(\key)\) relation for all keys \(\key\) written in the fingerprint \(\fp\).
This prevents the kv-store of \cref{fig:ua-disallowed},
as \(\txid\) and \(\txid'\) concurrently increment the initial version of \(\key\) by \(1\).
As client views must include the initial versions, once \(\txid\) commits a new version \(\ver\) with value \(\val_1\) to \(\key\), then \(\txid'\) must include \(\ver\) in its view as there is a \(\WW\) edge from the initial version to \(\ver\). 
As such, when \(\txid'\) subsequently increments \(\key\), it must read from \(\ver\), and not the initial version as depicted in \cref{fig:ua-disallowed}.

\subsubsection{Parallel Snapshot Isolation \((\PSI)\)} 
This consistency model is defined as the conjunction of the guarantees provided by \(\CC\) and \(\UA\)~\cite{framework-concur}. 
As such, the \(\vshiftname_{\PSI}\) predicate is defined as the conjunction of the \(\vshiftname\) predicates for \(\CC\) and \(\UA\).
However, we cannot simply define \(\cancommitname_{\PSI}\) as the conjunction of the \(\cancommitname\) predicates for \(\CC\) and \(\UA\). 
This is for two reasons. 
First, their conjunction would only mandate that \(\vi\) be closed with respect to 
\(\rel_{\CC}\) and \(\rel_{\UA}\) \emph{individually}, but \emph{not} with respect to their \emph{union} (recall that closure is defined in terms of the transitive closure of a given relation and thus the closure of \(\rel_{\CC}\) and \(\rel_{\UA}\) is smaller than the closure of \(\rel_{\CC} \cup \rel_{\UA}\)).
As such, we define \(\cancommitname_{\PSI}\) as closure with respect to \(\rel_{\PSI}\) which must include \(\rel_{\CC} \cup \rel_{\UA}\).
Second, recall that \(\cancommitname_{\UA}\) requires that a transaction writing 
to a key \(\key\) must be able to see all previous versions of \(\key\), \ie all versions of \(\key\). 
That is, when write-conflict freedom is enforced, a version \(\ver\) of \(\key\) depends on all 
previous versions of \(\key\). 
This observation leads us to include write-write dependencies (\(\WW_{\mkvs}\)) in \(\rel_{\PSI}\). 
Observe that the kv-store in \cref{fig:cc-ua-allowed-but-psi} shows an example kv-store that satisfies \(\cancommitname_{\CC} \land \cancommitname_{\UA}\), 
but not \(\cancommitname_{\PSI}\).

\subsubsection{Consistent Prefix \((\CP)\)}
\label{para:cp}
If the total order in which transactions commit is known, \(\CP\)
can be described as a strengthening of \(\CC\): 
if a client sees the versions written by a transaction \(\txid\),
then it must also see all versions written by transactions that commit before \(\txid\). 
Although kv-stores only provide {\em partial} information about the transaction commit order via the dependency relations,
this is sufficient to formalise \emph{Consistent Prefix} \cite{laws}.

In practice, we approximate the order in which transactions 
commit in an \(\ET\)-trace that terminates in a configuration \((\mkvs, \_)\) via the \(\WR_{\mkvs}, \WW_{\mkvs}, \RW_{\mkvs}\) and \(\SO\)  relations. 
This approximation is best understood in terms of an idealised implementation of \(\CP\) on a centralised system,
where the snapshot of a transaction is determined at its start point and its effects are made visible to future transactions at its commit point.
With respect to this implementation, if \((\txid,\txid') \in \WR\), then 
\(\txid\) must commit before \(\txid'\) starts, and hence before \(\txid'\) commits.
Similarly, if \((\txid,\txid') \in \SO\), then \(\txid\) commits before \(\txid'\) starts, 
and thus before \(\txid'\) commits.
%
Recall that \((\txid'', \txid') \in \RW\)
denotes that \(\txid''\) reads a version that is later overwritten by \(\txid'\).
That is, \(\txid''\) cannot see the write of \(\txid'\), and thus \(\txid''\) must starts before 
\(\txid'\) commits. 
As such, if \(\txid\) commits before \(\txid''\) starts 
(\((\txid, \txid'') \in \WR\) or \((\txid,\txid'') \in \SO\)), 
and \((\txid'', \txid') \in \RW\), then \(\txid\) must commit before 
\(\txid'\) commits. 
In other words, if \((\txid,\txid') \in \WR;\RW\) or \((\txid,\txid') \in \SO;\RW\), then \(\txid\) commits before \(\txid'\).
Finally, if \((\txid,\txid') \in\WW\), then \(\txid\) must commit before \(\txid'\). 
We therefore define \(\rel_{\CP} \defeq (\WR_{\mkvs}; \RW_{\mkvs}\rflx \cup \SO;  \RW_{\mkvs}\rflx \cup \WW)\), approximating the order in which transactions commit. 
\citet{laws} show that the set \((\rel_{\CP}^{-1})^{+}(\txid)\) contains all transactions that must be observed by \(\txid\) under \(\CP\). 
We define \(\cancommitname_{\CP}\) by requiring that the client view be 
closed with respect to \(\rel_{\CP}\).

Consistent prefix disallows the \emph{long fork anomaly} shown in \cref{fig:cp-disallowed}, where clients \(\cl_1\) and \(\cl_2\) observe the updates to \(\key_1\) and \(\key_2\) 
in different orders. 
Assuming without loss of generality that \( \txid_{\cl_1}^{2} \) commits 
before \( \txid_{\cl_2}^{2} \), then prior to committing its transaction \(\cl_2\) sees 
the version of \(\key_1\) with value \(\val_0\). 
However, since \(\txid \xrightarrow{\WR_{\mkvs}} \txid_{\cl_{1}}^{1} 
\xrightarrow{\SO} \txid_{\cl_{1}}^{2} \xrightarrow{\RW} \txid' \xrightarrow{\WR} \txid_{\cl_{2}}^{1} \xrightarrow{\SO} 
\txid_{\cl_2}^{2}\), then \(\cl_2\) should also see the version of \(\key_1\) with 
value \(\val_2\), leading to a contradiction.

\subsubsection{Snapshot Isolation \((\SI)\)}
When the total order in which transactions commit is known,  
\(\SI\) can be defined compositionally from \(\CP\) and \(\UA\). 
As such, \(\vshiftname_{\SI}\) is defined as the conjunction of their associated \(\vshiftname\) predicates. 
However, as with \(\PSI\), we cannot define \(\cancommitname_{\SI}\) as the conjunction of their associated \(\cancommitname\) predicates. 
Rather, we define \(\cancommitname_{\SI}\) as closure with respect to \(\rel_{\SI}\), which includes \(\rel_\CP \cup \rel_{\UA}\).
Observe that the kv-store in \cref{fig:si-disallowed} shows an example kv-store that satisfies \(\cancommitname_{\UA} \land \cancommitname_{\CP}\), 
but not \(\cancommitname_{\SI}\).
Additionally, we include \(\WW;\RW\) in \(\rel_{\SI}\). 
This is because when the centralised \(\CP\) implementation (discussed above) is strengthened with write-conflict freedom, then a write-write dependency between two transactions \(\txid\) and \(\txid'\) 
does not only mandate that \(\txid\) commits before \(\txid'\) commits but also before \(\txid'\) starts. 
Consequently, if \((\txid, \txid') \in \WW ;\RW\), then \(\txid\) must commit 
before \(\txid'\) commit.

\subsubsection{(Strict) serialisability \((\SER)\)}
Serialisability is the strongest consistency model in the literature, requiring that transactions execute in a  total sequential order. 
The \(\cancommitname_{\SER}\) thus allows clients to commit transactions only when 
their view of the kv-store is complete, \ie the client view is closed with respect to \(\WW^{-1}\).
This requirement prevents the kv-store in  \cref{fig:ser-disallowed}: 
without loss of generality, suppose that \(\txid_1\) commits before \(\txid_2\). Then the client committing \(\txid_2\) must see the version of \(\key_1\) written by \(\txid_1\), 
and thus cannot read the outdated value \(\val_0\) for \(\key_1\). 
This example is allowed by all other execution tests in~\cref{fig:execution_tests}.

\subsubsection{Weak Snapshot Isolation \((\WSI)\): A New Consistency Model} 
\label{sec:new_cm}
Kv-stores and execution tests are useful for investigating new 
consistency models.  
One example is the consistency model induced by combining 
\(\CP\) and \(\UA\), which we refer to as \emph{Weak Snapshot Isolation} (\(\WSI\)). 
To justify this consistency model in full, it would be useful to explore its implementations. 
Here we focus on the benefits of implementing \(\WSI\).
Because \(\WSI\) is stronger than \(\CP\) and \(\UA\) by definition, 
it forbids all the  anomalies forbidden by these consistency models, \eg
the long fork (\cref{fig:cp-disallowed}) and the lost update (\cref{fig:ua-disallowed}). 
Moreover, \(\WSI\) is strictly weaker than \(\SI\). 
As such, \(\WSI\) allows all \(\SI\) anomalies, \eg the write skew (\cref{fig:ser-disallowed}), 
and allows behaviour not allowed under \(\SI\) such as that in \cref{fig:si-disallowed}.
We can construct a \((\ET_{\CP} \cap \ET_{\UA})\)-trace terminating in \((\mkvs, \_)\) by 
executing transactions \(\txid_{1}, \txid_{2}, \txid_{3}\) and \(\txid_{4}\) in this order. 
In particular, \(\txid_{4}\) is executed using \(\vi {=} [\key_{1} \mapsto \{0\}, \key_{2} \mapsto \{0,1\}]\). 
However, the same trace is not a valid \(\ET_{\SI}\)-trace. 
Under \(\SI\) transaction \(\txid_{4}\) cannot be executed using \(\vi\): 
\(\txid_{4}\) reads the version of \(\key_2\) written by \(\txid_3\), 
meaning that \(\vi\) must include the version written by 
\(\txid_{3}\). Since \((\txid_{2},\txid_{3}) \in \RW \)
and \((\txid_{1} ,\txid_2) \in \WW\), 
then \(\vi\) should contain the version of \(\key_{1}\) written by \(\txid_{1}\), 
contradicting the fact that \(\txid_{4}\) reads the initial version of \(\key_1\).

As \(\WSI\) is a weaker consistency model than \(\SI\), we believe that \(\WSI\) implementations would outperform known \(\SI\) implementations.
Nevertheless, the two consistency models are very similar in that 
many applications that 
are correct under \(\SI\) are also correct under \(\WSI\). We give an example of such an application in \cref{sec:program-analysis}.

\section{Consistency Models: Dependency Graphs and Abstract Executions}
\label{sec:other_formalisms}

We demonstrate that our consistency models for kv-stores
are equivalent to the declarative consistency models for 
dependency graphs \cite{adya} 
and abstract executions \cite{ev_transactions,framework-concur}. 
We outline our results here, and refer the reader
to \cref{sec:app-abstract-semantics-sound-complete,sec:et-sound-complete-constructor,app:et_sound_complete} for the full details.

\subsection{Relating KV-Stores and Dependency Graphs}
\label{sec:dep_graphs}
Dependency graphs \cite{adya-icde,adya} provide  perhaps the most
well-known 
formalism used for specifying transactional consistency models. 
A dependency graph \(\Gr\) is a directed, labelled graph whose 
nodes denote transactions, and whose edges denote \emph{dependencies} between transactions.  
More specifically, nodes are labelled with a transaction identifier
and the fingerprint associated with the  transaction. 
Edges are labelled with a dependency relation \(\SO, \WR, \WW, \RW\), in the 
same spirit of dependencies of transactions in kv-stores in \cref{sec:cm}.
An example of dependency graph is given in \cref{fig:dependency-graph}.
We give the formal definition of dependency graphs in \cref{app:depgraphs}.

We can always {extract} a dependency graph  from a kv-store, and vice-versa. 
For example, \cref{fig:dependency-graph} corresponds to the dependency graph extracted from the kv-store in \cref{fig:ser-disallowed}. 
\begin{theorem}
\label{thm:kv_graph_isomorph}
Dependency graphs are isomorphic to kv-stores.
\end{theorem}

\begin{figure*}[!t]
\captionsetup[subfigure]{aboveskip=0pt, belowskip=5pt}
\centering
\noindent
\begin{subfigure}{0.49\textwidth}
    \begin{centertikz}[.7]
\draw pic {transaction={t0}{%
        /$\otW$/$\key_1$/$\val_0$%
        , /$\otW$/$\key_2$/$\val'_0$%
}};
\path(t0.west) node[anchor=east] (t0lbl) {$\txid_0$};

\draw pic at ($(t0.north east) + (1.5,0.6)$) {transaction={t1}{%
        /$\otR$/$\key_2$/$\val'_0$%
        , /$\otW$/$\key_1$/$\val_1$%
}};
\path(t1.north) node[anchor=south] (t1lbl) {$\txid_1$};

\draw pic at ($(t0.south east) + (1.5,-0.3)$) {transaction={t2}{%
        /$\otR$/$\key_1$/$\val_0$%
        , /$\otW$/$\key_2$/$\val'_1$%
}};
\path(t2.south) node[anchor=north] (t2lbl) {$\txid_2$};

\path[->]
(t0.north) edge[bend left=30] node[above, yshift=3pt, xshift=-20pt, pos=0.3] {$\WR$} (t1.west)
(t0.south) edge[bend right=30] node[below, yshift=-3pt, xshift=-20pt, pos=0.3] {$\WR$} (t2.west)
([xshift=8pt]t1.south) edge[bend left=20] node[right] {$\RW$} ([xshift=8pt]t2.north)
([xshift=-16pt]t2.north) edge[bend left=20] node[right] {$\RW$} ([xshift=-16pt]t1.south);

\end{centertikz}
\caption{Dependency graph}
\label{fig:dependency-graph}
\end{subfigure}
\hfill
\begin{subfigure}{0.49\textwidth}
    \begin{centertikz}[.7]

\draw pic {transaction={t0}{%
        /$\otW$/$\key_1$/$\val_0$%
        , /$\otW$/$\key_2$/$\val'_0$%
}};
\path(t0.west) node[anchor=east] (t0lbl) {$\txid_0$};

\draw pic at ($(t0.north east) + (1.5,0.6)$) {transaction={t1}{%
        /$\otR$/$\key_2$/$\val'_0$%
        , /$\otW$/$\key_1$/$\val_1$%
}};
\path(t1.north) node[anchor=south] (t1lbl) {$\txid_1$};

\draw pic at ($(t0.south east) + (1.5,-0.3)$) {transaction={t2}{%
        /$\otR$/$\key_1$/$\val_0$%
        , /$\otW$/$\key_2$/$\val'_1$%
}};
\path(t2.south) node[anchor=north] (t2lbl) {$\txid_2$};

\path[->]
(t0.north) edge[bend left=30] node[above, yshift=3pt, xshift=-20pt, pos=0.3] {$\VIS, \AR$} (t1.west)
(t0.south) edge[bend right=30] node[below, yshift=-3pt, xshift=-20pt, pos=0.3] {$\VIS, \AR$} (t2.west)
([xshift=8pt]t1.south) edge[bend left=20] node[right] {$\AR$} ([xshift=8pt]t2.north);

\end{centertikz}
\caption{Abstract execution}
\label{fig:abstract_execution}
\end{subfigure}

\hrulefill

\caption{The dependency graph (\subref{fig:dependency-graph}) and abstract execution graph (\subref{fig:abstract_execution}) associated with the kv-store in \cref{fig:ser-disallowed}
}
\end{figure*}

Consistency models using dependency graphs can be specified by
constraining the shape of the graphs, typically by requiring the absence of certain cycles.  For example, strict serialisability is defined as
the set of dependency graphs with no cycles. 
We can immediately use such constraints to define execution tests on
kv-stores, and hence consistency models for kv-stores. However, to show 
that our consistency models over kv-stores given in~\cref{fig:execution-tests} are equivalent
to existing consistency model definitions using dependency graphs,
we first prove that our models are equivalent
to existing definitions using abstract executions, and then appeal 
to the results of \citet{laws} showing the equivalence between definitions using dependency graphs and those using abstract executions.

\subsection{Relating KV-Stores and Abstract Executions}
We compare our consistency model specifications using execution tests over kv-stores 
with an alternative, axiomatic specification style based on abstract 
executions \cite{framework-concur}, defined shortly. 
Our main contribution here is the development of a general proof technique for proving the equivalence of our execution-test-based specifications and abstract-execution-based specifications.
Our proof technique keeps the proof obligations (conditions that must be satisfied by its user) to a minimum. 
In particular, the user only needs to show that the constraints on client views in execution tests relate to analogous constraints on visibility edges in abstract executions.
We then provide a mapping between the \(\ET\)-traces to 
\(\mkvs\), to a set of abstract executions that satisfy the axiomatic specification corresponding to \(\ET\).
Here we use our proof technique to prove that the execution 
tests for serialisability (\(\SER\)) are equivalent to their 
axiomatic specifications. In \cref{sec:spec-proof} we apply our proof technique 
to show that all the execution tests from \cref{fig:execution.tests} are equivalent 
to their respective axiomatic specifications. 

\subsubsection{Abstract Executions}
Abstract executions are labelled graphs 
whose nodes comprise transaction identifiers and their fingerprints. 
These nodes may be connected either by a \emph{visibility edge}, \(\txid \xrightarrow{\VIS} 
\txid'\), when \(\txid'\) sees the updates of \(\txid\);  or an \emph{arbitration edge}, \(\txid \xrightarrow{\AR} 
\txid'\), when \(\txid'\) updates overwrite \(\txid\) updates.
\Cref{fig:abstract_execution} depicts an example abstract execution.
\begin{definition}[Abstract executions]
\label{def:main-body-absexec}
\label{def:main-body-aexec}
An {\em abstract execution} is a triple \(\aexec = (\TtoOp{T}, \VIS, \AR)\), where 
 \(\TtoOp{T}: \TxID \parfun \pset{\Ops}\) is a partial  
function mapping transaction identifiers to 
fingerprints, with \(\TtoOp{T}(\txid_{0}) = \Set{ (\otW, \key, \val_{0}) }[ \key \in \Keys]\), 
\(\VIS \subseteq \dom(\TtoOp{T}) \times \dom(\TtoOp{T})\) is an irreflexive relation 
such that, for any \(\txid \in \dom(\TtoOp{T})\), \(\txid_{0}
\xrightarrow{\VIS} \txid\) for the initial transaction \(\txid_0\), and 
\(\AR \subseteq \dom(\TtoOp{T}) \times \dom(\TtoOp{T})\) is a strict, total order 
such that \(\VIS \subseteq \AR\), \(\min_{\AR}(\dom(\TtoOp{T})) = \txid_{0}\)
and \(\txid_{\cl}^{n} \xrightarrow{\AR} \txid_{\cl}^{m}\) only if \(n < m\). 
\end{definition}
Given an abstract execution \(\aexec = (\TtoOp{T}, \VIS, \AR)\),  we let \(\TtoOp{T}_{\aexec} = \TtoOp{T}\), 
\(\VIS_{\aexec} = \VIS\) and \(\AR_{\aexec} = \AR\). 
We write \((l, \key, \val) \in_{\aexec} \txid\) as a shorthand for \((l, \key, \val) \in \TtoOp{T}_{\aexec}(\txid)\).
For \(\txid \in \T_{\aexec}\), 
we define its \emph{visible writes} in \(\aexec\) as 
$\visibleWrites_{\aexec}(\key, \txid) \defeq \VIS^{-1}_{\aexec}(\txid) \cap 
\{\txid' \mid (\otW, \key, \stub) \in_{\aexec} \txid'\}$. 
An abstract execution \(\aexec\) satisfies the \emph{last-writer-wins} policy: if
a transaction \( \txid \) reads key \( \key \), 
it must read from the latest transaction in the arbitration order that is visible to \(\txid\) and wrote to key \(\key\),
\ie $\forall \txid \in \T_{\aexec}.\; (\otR,\key,\val) \in_{\aexec} \txid 
\implies (\otW, \key, \val) \in \max_{\AR_{\aexec}}(\visibleWrites_{\aexec}(\key, \txid))$.
then \(\forall \key, \val.\; (\otR,\key,\val) \in_{\aexec} \txid \implies (\otW, \key, \val) \in \max_{\AR_{\aexec}}(\visibleWrites_{\aexec}(\key, \txid))\).
Henceforth we assume that abstract executions satisfy the last-writer-wins policy, 
and we let \(\Aexecs\) be the set of all such abstract executions.

%

Abstract-execution-based specifications of consistency models constrain the overall structure of abstract executions. 
For most consistency models \cite{laws,framework-concur,sureshConcur}, 
such constraints are over the set of transactions that  \textbf{must} be seen  by 
other transactions. For example, monotonic reads is specified by requiring 
that if a transaction \(\txid\) follows another transaction \(\txid'\) in the session order, 
then \(\txid\) must see all transactions that are seen by \(\txid'\).
Serialisability can be specified by requiring that 
a transaction \(\txid\) see all transactions preceding \(\txid\) in the arbitration order.

Formally, an axiomatic specification \(\Ax\) is a set of {\em axioms} \(\A : \Aexecs \to \pset{\TxID \times \TxID}\),
where  \(\forall \aexec.\;\A(\aexec) \subseteq \AR_{\aexec}\). 
We write \(\aexec \models \A\) when \(\A(\aexec) \subseteq \VIS_{\aexec}\).
We refer the reader to \cref{sec:abstract-execution} for details about abstract executions.

Returning to the monotonic reads (\MR) example, 
we define \(\Ax_{\MRd} \defeq \{\A_{\MRd}\}\), where \(\A_{\MRd}(\aexec) \defeq \VIS_{\aexec} ; \SO_{\VIS}\). 
By definition, for a given \(\aexec\), \(\aexec \models \A_{\MRd}\) if and only 
if \(\VIS_{\aexec} ; \SO_{\aexec} \subseteq \VIS_{\aexec}\). 
That is, whenever \(\txid'' \xrightarrow{\VIS_{\aexec}} \txid' \xrightarrow{\PO_{\aexec}} \txid \), 
then \(\txid'' \xrightarrow{\VIS_{\aexec}} \txid\).
Similarly, for serialisability (\SER) we define \(\Ax_{\SER} \defeq \{ \A_{\SER} \}\), where \(\A_{\SER}(\aexec) \defeq \AR_{\aexec}\), 
captures the constraint that a transaction \(\txid\) must see all transactions
preceding it in the arbitration order.

Any abstract executions \(\aexec\) can be mapped 
into an equivalent dependency graph \(\Gr_{\aexec}\) (\citet{laws}), and hence into a kv-store 
\(\hh_{\aexec}\) (\cref{thm:kv_graph_isomorph}). 
\sx{straightforward? Maybe better words like: 
we refer the readers to \cite{laws} for mapping an abstract execution \(\aexec\) into an
equivalent dependency graph \(\Gr_{\aexec}\). } 
We can then use this construction to define the consistency model induced by an abstract-execution-based specification 
\(\CMs(\Ax)\) by projecting abstract executions that satisfy the axioms in \(\Ax\) to kv-stores: 
\(\CMs(\Ax) \defeq \{ \hh_{\aexec} \mid \forall \A \in \Ax. \aexec \models \A \}\). 

In the remainder of this section we develop a proof techniques for showing  
that an execution test \(\ET\) and an axiomatic specification \(\Ax\) induce the same 
consistency model, \ie\(\CMs(\ET) {=} \CMs(\Ax)\). 
Due to space constraints, we focus only on \emph{soundness}, \ie on proving the left-to-right inclusion: \(\CMs(\ET) \subseteq \CMs(\Ax)\); 
we describe the other direction in full in \cref{sec:et-sound-complete-constructor}. 
The core of our proof technique lies in the soundness of the most permissive execution 
test, \(\CMs(\ET_{\top})\), with respect to the weakest axiomatic specification, given by 
the empty set of axioms, which we prove next.

\subsubsection{Equivalence of Read Atomic and \(\CMs(\ET_{\top})\)} 
\sx{high level comments: read atomic -> snapshot property? }
The weakest axiomatic specification, given by the empty set of 
axioms, corresponds to the \emph{Read Atomic} consistency model \cite{ramp}. 
To prove that the most permissive execution test \(\ET_{\top}\) is sound with 
respect to the weakest axiomatic specification, we map \(\ET_{\top}\)-traces 
terminating in a configuration of the form \((\hh, \stub)\), into the set of 
abstract executions whose underlying kv-store is \(\hh\). 

\begin{figure}[t]
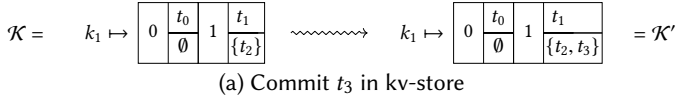
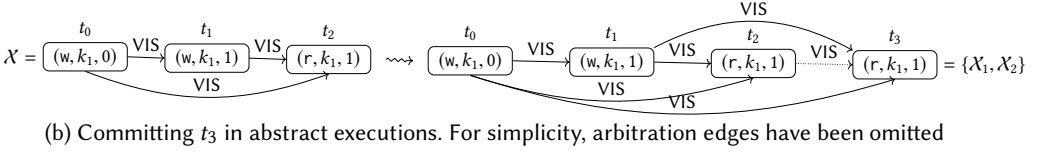

\captionsetup[subfigure]{aboveskip=0pt, belowskip=5pt}
\begin{subfigure}{0.95\textwidth}
\begin{centertikz}%
\node(locx) {$\key_1 \mapsto$};
\path (locx.west) + (-1,0) node (k) {$\hh = $};
\draw pic at ([xshift=\tikzkvspace]locx.east) {vlist={versionx}{%
    /$0$/$\txid_0$/$\emptyset$
    , /$1$/$\txid_1$/$\Set{\txid_2}$
}};

\path (versionx.east) + (3,0) node (locy) {$\key_1 \mapsto$};
\draw pic at ([xshift=\tikzkvspace]locy.east) {vlist={versiony}{%
    /$0$/$\txid_0$/$\emptyset$
    , /$1$/$\txid_1$/$\Set{\txid_2, \txid_3}$
}};

\draw[->,
line join=round,
decorate, decoration={
    zigzag,
    segment length=4,
    amplitude=.9,post=lineto,
    post length=2pt
}
] ($(versionx.east) + (0.5,0)$) -- ($(locy.west) + (-0.5,0)$);
\path (versiony.east) + (1,0) node (k1) {$= \hh'$};

\end{centertikz}%
\caption{Commit \( \txid_3 \) in kv-store}
\label{fig:et-sound-kv-store-update}
\end{subfigure}

\hrulefill

\begin{subfigure}{0.95\textwidth}
\begin{centertikz}[.66]%
\draw pic {transaction={t0}{%
        /$\otW$/$\key_1$/$0$%
}};
\path(t0.west) + (0,0) node[anchor=east] (a0) {$\aexec = $};
\path(t0.north) node[anchor=south] (t0lbl) {$\txid_0$};

\draw pic at ($(t0.east) + (1.6,0.15)$) {transaction={t1}{%
        /$\otW$/$\key_1$/$1$%
}};
\path(t1.north) node[anchor=south] (t1lbl) {$\txid_1$};

\draw pic at ($(t1.east) + (1.6,0.15)$) {transaction={t2}{%
        /$\otR$/$\key_1$/$1$%
}};
\path(t2.north) node[anchor=south] (t2lbl) {$\txid_2$};

\path[->]
(t0.east) edge node[above, yshift=0pt, xshift=0pt, pos=0.5] {$\VIS$} (t1.west)
(t0.south) edge[bend right=20] node[above, yshift=0pt, xshift=0pt, pos=0.5] {$\VIS$} (t2.south)
(t1.east) edge node[above, yshift=0pt, xshift=0pt, pos=0.5] {$\VIS$} (t2.west);

\draw pic at ($(t2.east) + (2,0.15)$) {transaction={t00}{%
        /$\otW$/$\key_1$/$0$%
}};
\path(t00.north) node[anchor=south] (t00lbl) {$\txid_0$};

\draw pic at ($(t00.east) + (2,0.15)$) {transaction={t11}{%
        /$\otW$/$\key_1$/$1$%
}};
\path(t11.north) node[anchor=south] (t11lbl) {$\txid_1$};

\draw pic at ($(t11.east) + (2,0.15)$) {transaction={t22}{%
        /$\otR$/$\key_1$/$1$%
}};
\path(t22.north) node[anchor=south] (t22lbl) {$\txid_2$};

\draw pic at ($(t22.east) + (2,0.15)$) {transaction={t3}{%
        /$\otR$/$\key_1$/$1$%
}};
\path(t3.east) + (0,0) node[anchor=west] (a1) {$= \{\aexec_1, \aexec_2\}$};
\path(t3.north) node[anchor=south] (t3lbl) {$\txid_3$};

\path[->]
(t00.east) edge node[above, yshift=0pt, xshift=0pt, pos=0.5] {$\VIS$} (t11.west)
(t00.south) edge[bend right=17] node[above, yshift=0pt, xshift=0pt, pos=0.5] {$\VIS$} (t22.south)
(t00.south) edge[bend right=17] node[above, yshift=0pt, xshift=0pt, pos=0.5] {$\VIS$} (t3.south)
(t11.north east) edge[bend left=28] node[above, yshift=0pt, xshift=0pt, pos=0.5] {$\VIS$} (t3.north west)
(t11.east) edge node[above, yshift=0pt, xshift=0pt, pos=0.5] {$\VIS$} (t22.west);

\path[densely dotted,->]
(t22.east) edge node[above, yshift=0pt, xshift=0pt, pos=0.5] {$\VIS$} (t3.west);

\draw[->,
line join=round,
decorate, decoration={
    zigzag,
    segment length=4,
    amplitude=.9,post=lineto,
    post length=2pt
}
] ($(t2.east) + (0.3,0)$) -- ($(t00.west) + (-0.3,0)$);

\end{centertikz}%
\vspace{-5pt}
\caption{Committing \( \txid_3 \) in abstract executions. For simplicity, arbitration edges have been omitted}
\label{fig:et-sound-aexec-update}
\end{subfigure}

\hrulefill

\caption{Correspondence between committing \( \txid_3 \) in kv-stores and abstract executions. 
The figure to the right denotes a set of abstract executions, which differ in the presence of the dashed visibility edge.}
\label{fig:et-sound-to-aexec}
\end{figure}

\begin{theorem}
\label{thm:kvtrace2aexec}
Given an \(\ET_{\top}\)-trace \(\tau\) terminating in \((\hh, \_)\), 
there exists a non-empty set of abstract executions \(\execs(\tau)\)
such that: \(\forall \aexec \in \execs(\tau).\ \hh_{\aexec} = \hh\), 
and the order in which transactions are executed in \(\tau\) is consistent with \(\AR_{\aexec}\). 
\end{theorem}
\noindent 
The proof of \cref{thm:kvtrace2aexec} is highly
non-trivial, and relies on the following intuition that drives the construction of the set \(\execs(\tau)\): 
whenever a client \(\cl\) in \(\tau\)  with view \(\vi\) commits a transaction \(\txid\), then in all 
abstract executions included in \(\execs(\tau)\), transaction \(\txid\) must see the writers  
of the versions included in \(\vi\), and it never sees the writers of versions not included in \(\vi\) (\cref{fig:et-sound-to-aexec}). 
These are defined by the set \(\Tx[\hh, \vi]\) (defined in \cref{subsec:cm_examples}). Furthermore, 
abstract executions in \(\execs(\tau)\) differ only in the set of read-only transactions (\ie those with no write operations) 
that transactions see. While mapping an \(\ET_{\top}\)-trace into multiple abstract executions is 
not strictly necessary for proving the soundness of \(\ET_{\top}\) with respect 
to the weakest axiomatic specification, it plays a significant role when proving the soundness 
of an arbitrary execution test \(\ET\) with respect to its counterpart in axiomatic specifications. 

The definition of \(\execs(\tau)\) is by induction on the length of the \(\ET_{\top}\)-trace
\(\tau\). For the base case with \(\tau_{0}\) consisting of a single configuration 
\((\hh_{0}, \stub)\), we define \(\execs(\tau_{0})\) to contain a single abstract execution with 
a single transaction \(\txid_{0}\) that initialises all the keys to the initial value \(\val_{0}\):
\(\execs(\tau_{0}) \defeq \{ ([\txid_{0} \mapsto \{ (\otW, \key, \val_{0} \mid \key \in \Keys)\}], \emptyset, \emptyset \}\). 
For the inductive case with \(\tau {=} \tau' \toET{(\cl, \alpha)} (\hh', \vienv')\), let \((\hh, \vienv)\) be the last 
configuration appearing in \(\tau'\). 
If \(\alpha {=} \varepsilon\), then \(\execs(\tau) \defeq \execs(\tau')\). 
If \(\alpha {=} \fp\) for some \(\fp\), we first determine the transaction identifier \(\txid'\) that was used to commit \(\fp\) in \(\hh'\), 
the view \(\vi' = \vienv'(\cl)\) of the client \(\cl\) when committing \(\txid'\), the 
set of transactions that \(\cl\) must see when committing \(\txid'\), given by 
\(\Tx[\hh', \vi']\), and the set of read-only transactions \(\txidset_{\rd}\) in \(\hh'\): 
the latter are those transactions that never appear as writers. 
Then, for all abstract execution \(\aexec' \in \execs(\tau')\), we define \(\extend[\aexec', \txid', \Tx[\mkvs',\vi'], \fp]\) as the largest set 
such that, whenever \(\aexec \in \extend[\aexec', \txid', \Tx[\mkvs',\vi'], \fp]\), then 
\begin{enumerate*}
\item \(\TtoOp{T}_{\aexec} = \TtoOp{T}_{\aexec'}\rmto{\txid'}{\fp}\);
\item  for all \(\txid' \in \T_{\aexec}\), 
\(\txid' \xrightarrow{\AR_{\aexec'}} \txid\); and 
\item if \(\txid' \xrightarrow{\VIS_{\aexec}} \txid\), 
then either \(\txid \in \Tx[\hh', \vi']\), or \(\txid \in \txidset_{\rd}\).  
\end{enumerate*}
Finally, we define \(\execs(\tau) \defeq \bigcup_{\aexec \in \execs(\tau')} \extend[\aexec, \txid, \Tx[\mkvs',\vi'], \txidset_{\rd}, \fp]\). 
In \cref{sec:kvtrace2aexec,sec:aexectrace2kv} we give the full details. 
To understand the construction outlined above, we 
illustrate one use of the function \(\extend\). The abstract 
execution \(\aexec\) to the left of \cref{fig:et-sound-aexec-update} has underlying kv-store \(\hh\), 
depicted to the left of \cref{fig:et-sound-kv-store-update}. If a client commits a transaction 
\(\txid_{3}\) that reads the last version of \(\key_1\), then the resulting kv-store \(\hh'\) would be the one 
to the right of \cref{fig:et-sound-kv-store-update}. This commit is simulated by the function 
\(\extend[\aexec', \txid_{3}, \Tx[\hh, \vi], \fp]\), where \(\vi, \fp\) are the view and fingerprint used to 
update \(\hh\) to \(\hh'\): the result of this function consists of two abstract executions \(\aexec_1, \aexec_2\), 
that only differ in read-only transactions (the right of \cref{fig:et-sound-kv-store-update}).
The visibility edges of \(\aexec_1\) are exactly the concrete edges in \cref{fig:et-sound-aexec-update}; however, 
\(\aexec_2\) has the extra dashed visibility edge \(\txid_{2} \xrightarrow{\VIS} \txid_{3}\). 
Note that \(\hh_{\aexec_2} = \hh_{\aexec_3} = \hh'\).

\begin{theorem}
\label{thm:aexec2kvtrace}
Given an abstract execution \(\aexec\), there exists a non-empty 
set of \(\ET_{\top}\)-traces \(\{\tau_{i}\}_{i \in I}\) such that, for each \(i \in I\), the last configuration of \(\tau_{i}\) is 
\((\hh_{\aexec}, \_)\), and \(\tau_{i}\) executes transactions in the order established by \(\AR_{\aexec}\). 
\end{theorem}
\noindent The proof of \cref{thm:aexec2kvtrace} is given 
in \cref{sec:kvtrace2aexec,sec:aexectrace2kv}. 
\Cref{thm:kvtrace2aexec,thm:aexec2kvtrace} together establish the
equivalence  of \(\ET_{\top}\) with the weakest axiomatic specification. 

\subsubsection{Equivalence of axiomatic specifications and execution tests}
We are now ready to present our proof technique for proving the soundness 
of an execution test \(\ET\) with respect to an axiomatic specification \(\Ax\).
It can be summarised as follows: 
the user considers an arbitrary 
\(\aexec: \aexec \models \Ax\) , and a tuple of the form 
\(\ET \vdash (\hh_{\aexec}, \vi) \csat \opset: (\hh', \vi')\). 
Then, it constructs a non-empty subset of \(\extend[\aexec, \txid, \Tx[\hh, \vi], \fp]\) 
whose elements satisfy the axioms \(\Ax\). Note that, because abstract executions in
\(\extend[\aexec, \txid, \Tx[\hh, \vi], \fp]\) differ only in visibility edges of the form \(\txid_{\mathsf{rd}}
\xrightarrow{\VIS} \txid\), where \(\txid_{\mathsf{rd}}\) is a read-only transaction, 
then constructing the set above amounts to identifying a suitable set of 
read-only transactions in \(\aexec\).

To see why our proof technique guarantees the soundness of \(\ET\) with respect 
to \(\Ax\) (\cref{thm:main-body-et_soundness}), we apply an inductive argument over the number of \(\ET\)-reductions \(n\) in a \(\ET\)-trace \(\tau\): 
first, if \(n = 0\) then \(\tau = (\mkvs_{0}, \vienv_{0})\), and \(\aexec_{0} \in \execs(\tau)\) trivially satisfies the axioms
\(\Ax\): \(\forall \A \in \Ax. \A(\aexec_{0}) \subseteq \AR_{\aexec_{0}} = \emptyset \subseteq \VIS_{\aexec_{0}}\). 
Otherwise, if \(\tau = \tau' \xrightarrow{(\cl, \fp)} (\mkvs, \vi)\), 
suppose that there exists an abstract execution \(\aexec' \in \execs(\tau')\) that satisfies 
the axioms \(\Ax\). If the proof obligations of our proof technique are satisfied, we can construct an abstract execution 
\(\aexec \in \extend[\aexec, \txid, \Tx[\hh, \vi], \fp] \subseteq \aexec(\tau)\) that satisfies \(\Ax\); furthermore \(\hh_{\aexec} = \mkvs\). 

In practice, our proof techniques allows defining a per-client invariant 
on the visibility relation of abstract executions, which must be proved to be preserved by \(\ET\)-reductions (\cref{def:main-body-et_sound}).
This invariant carries client-specific information that links to \( \vshiftname \) (defined in \cref{sec:cm}) in execution tests.
Defining the right invariant is crucial for proving the soundness of several execution-test-based specifications (see \cref{app:et_sound_complete}).

\begin{definition}
\label{def:main-body-et_sound}
An execution test \(\ET\) is sound with respect to an axiomatic 
specification \(\Ax\) if and only if
there exists an invariant condition \(I\) such that, 
for any \(\cl, \txid, \vi, \vi', \mkvs, \vi', \fp, \aexec\), if:
\begin{itemize}
    \item \(\ET \vdash (\mkvs, \vi) \csat \fp: (\mkvs',\vi')\), where \( \mkvs' = \updateKV[\mkvs, \vi ,\fp, \txid]\)
    \item  \(\mkvs_{\aexec} = \mkvs\) and \(I(\aexec, \cl) \subseteq \Tx[\mkvs, \vi]\),
\end{itemize}
then there exists a non-empty subset of \(\mathscr{X} \subseteq \extend[\aexec, \txid, \Tx[\hh, \vi], \fp]\) 
such that, for any \(\aexec' in \mathscr{X}\), \(\aexec \models \Ax\) and \(I(\aexec', \cl) \subseteq \Tx[\mkvs', \vi']\).
\end{definition}

\begin{theorem}
\label{thm:main-body-et_soundness}
If \(\ET\) is sound with respect to \((\Ax)\), then \(\CMs(\ET) \subseteq \CMs(\Ax)\).
\end{theorem}

We conclude this section by outlining how our proof technique can be applied to show that the 
execution test \(\ET_{\SER}\) defined for serialisability is sound with respect to the axiomatic 
specification \(\Ax_{\SER}\). Let \(\aexec\) be an abstract execution such that \(\aexec \models \Ax_{\SER}\), 
and suppose that the underlying 
kv-store \(\hh_{\aexec}\) is in \(\CMs(\ET_{\SER})\). An example is the abstract execution \(\aexec\) 
and kv-store \(\hh\) to the left of \cref{fig:et-sound-to-aexec}. 
We pick an invariant \( I_{\SER} \) that is always empty since \( \vshiftname_\SER \) is always true.
When a client \(\cl\) commits a transaction \(\txid\) 
with fingerprint \(\fp\) in \(\hh_{\aexec}\) under \(\ET_{\SER}\), then its view \(\vi\) must contain all the versions stored in \(\hh_{\aexec}\) (\( \cancommit[\SER]{\mkvs_\aexec}{\vi}{\fp} \)). 
This means that all the abstract executions \( \aexec'' \) in \(\extend[\aexec, \txid, \Tx[\hh_\aexec, \vi], \fp]\) are such that there is a visibility 
edge \(\txid' \xrightarrow{\VIS_{\aexec''}} \txid\) where \(\txid'\) is a writer transaction in \(\hh_{\aexec}\). 
It is easy to see there exists an singleton set \( \Set{\aexec'} \) 
that is a subset of \(\extend[\aexec, \txid, \Tx[\hh, \vi], \fp]\) and \( \aexec' \models \Ax_\SER \);
in particular, 
there is an edge \(\txid' \xrightarrow{\VIS_{\aexec'}} \txid\) for any transaction \(\txid' \in \T_{\aexec'}\).
For example, in \cref{fig:et-sound-to-aexec}, the possible abstract executions are \(\aexec_1, \aexec_2\).
We pick \(\Set{\aexec_2}\), because the new transaction \( \txid_3 \) in \( \aexec_2 \) sees all previous transactions including  
the visibility edge \(\txid_{2} \xrightarrow{\VIS} \txid_{3}\). 
Last, the invariant \( I_\SER(\aexec',\cl) \subseteq \Tx[\hh', \vi] \) given \( I_\SER(\aexec',\cl) = \emptyset \).

\section{Applications}
\label{sec:applications}
\label{sec:program-analysis}

To demonstrate the applications of our operational semantics, 
in \cref{sec:robustness} we use our formalism to prove the robustness of several transactional libraries.
In \cref{sec:verify-impl} we then use our formalism to verify several distributed protocols.

\subsection{Application: Robustness of Transactional Libraries}
\label{sec:robustness}
A transactional library, $L {=} \Set{\ptrans{\trans}_{i}}_{i \in I}$,
provides a set of transactional operations which can be used by its clients to access the underlying
kv-store\footnote{For simplicity, we model each operation as a single transaction; it is straightforward to extend this to multiple transactions.}. 
For instance, the set of operations of the counter library on key $\key$ in \cref{sec:overview} is: $\Counter(\key) \defeq \Set{\ctrinc(\key), \ctrread(\key)}$.
A program $\prog$ is a \emph{client program} of $L$ if the only transactional calls in $\prog$ are to operations of $L$.  
Let $\CMs(\ET, L)$ denote the set of kv-stores obtained by running $L$ clients under $\ET$.
A library $L$ is \emph{robust} against an execution test
$\ET$ if for all clients of $L$, the kv-stores obtained under $\ET$ can also be obtained under $\SER$; 
\ie $\CMs(\ET, L) \subseteq \CMs(\ET_{\SER})$.

\begin{theorem}
\label{thm:serialisable_nocycle}
For all kv-stores $\mkvs$, $\mkvs \in \CMs(\ET_{\SER})$ iff $(\SO \cup \WR_{\mkvs} 
\cup \WW_{\mkvs} \cup \RW_{\mkvs})^{+}$ is irreflexive.
\end{theorem}
This theorem is an adaptation of a well-known result on
dependency graphs~\cite{adya} stating that $\mkvs \in \CMs(\ET_{\SER})$ if and
only if its associated dependency graph is acyclic. 
Using this theorem, we prove the robustness of a single counter against $\PSI$.
As discussed in \cref{sec:overview}, the multi-counter library is not robust against $\PSI$. 
We thus prove the robustness of the multi-counter library and the banking library of \citet{bank-example-wsi} against $\WSI$ instead. 
While previous work on checking robustness \citep{giovanni_concur16,SIanalysis,laws,sureshConcur} uses
static-analysis techniques that cannot be extended to support client sessions, 
we give the first robustness proofs that take client sessions into account.

\subsubsection{Robustness of a Single Counter against $\PSI$}
In the single-counter library $\Counter(\key)$, 
a client reads from $\key$ by calling $\ctrread(\key)$, and writes to $\key$ by calling $\ctrinc(\key)$ which first reads the value of $k$ and subsequently increments it by one.
Pick an arbitrary key $\key$ and a kv-store $\mkvs \in \CMs(\ET_{\PSI}, \mathsf{Counter}(\key))$.
As $\PSI$ enforces write conflict freedom (\(\UA\)), we know that if a transaction $\txid$ updates $\key$ (by calling $\ctrinc(\key)$) and writes version $\ver$ to $\key$, then it must have read the version of $\key$ immediately preceding $\ver$:
$\fora{\txid, i > 0}\!\! \txid {=} \wtOf(\mkvs(\key, i)) \implies \txid \in \rsOf(\mkvs(\key, i{-}1))$. 
Moreover, as $\PSI$ enforces monotonic reads ($\MR$),
the order in which clients observe the versions of $\key$ (by calling $\ctrread(\key)$)
is consistent with the order of versions in $\mkvs(\key)$. 
As such, the kv-stores in $\CMs(\ET_{\PSI}, \mathsf{Counter}(\key))$ have the canonical structure depicted in  \cref{fig:prog_analysis} and defined below, where $\lcat$ denotes list concatenation, and 
$\{\txid^{\mathsf{}}_{i}\}_{i=1}^{n}$ and $\bigcup_{i=0}^{n} \txidset^{\mathsf{}}_{i}$ 
denote disjoint sets of transactions calling $\ctrinc(\key)$ and
$\ctrread(\key)$, respectively: 

\begin{figure}[!t]
\begin{centertikz}%
\node(locx) {$\key_1 \mapsto$};
\draw pic at ([xshift=\tikzkvspace]locx.east) {vlist={versionx}{%
    /$0$/$\txid_0$/$\Set{\txid_1} \cup \T_0$
    , /$1$/$\txid_1$/$\Set{\txid_2} \cup \T_1$
}};
\path(versionx.east)+(0.3,0) node (dots) {$\cdots$};
\draw pic at ([xshift=0.2]dots.east) {vlist= {versionxlast}{
	/$n-1$/$\txid_{n-1}$/$\Set{\txid_{n}} \cup \T_{n-1}$
	, /$n$/$\txid_{n}$/$\T_{n}$
}};
\coordinate  (A) at ([xshift=30,yshift=-14] versionx.north west);
\coordinate  (B) at ([xshift=17,yshift=-9] A.center);
\coordinate  (C) at ([xshift=24,yshift=10] B.center);
\coordinate  (D) at ([xshift=14,yshift=0] C.center);
\coordinate  (E) at ([xshift=19,yshift=-9] D.center);
\coordinate  (F) at ([xshift=73,yshift=9] E.center);
\coordinate  (G) at ([xshift=25,yshift=-9] F.center);
\coordinate  (H) at ([xshift=25,yshift=9] G.center);
\coordinate  (I) at ([xshift=14,yshift=0] H.center);
\coordinate  (J) at ([xshift=2,yshift=-15] I.center);

\path[->,dashed,thick] ([xshift=0,yshift=6]A) edge[bend left=30] ([xshift=-3,yshift=4]B)
([xshift=3,yshift=4]B) edge[bend left=30] ([xshift=2,yshift=6]C)
 ([xshift=0,yshift=6]D) edge[bend left=30] ([xshift=-3,yshift=4]E)
 ([xshift=0,yshift=6]F) edge[bend left=30] ([xshift=-9,yshift=3]G)
([xshift=3,yshift=4]G) edge[bend left=30] ([xshift=2,yshift=6]H)
([xshift=0,yshift=6]I) edge[bend left=60] ([xshift=-2,yshift=5]J);

%
%
\end{centertikz}

\hrulefill 

\caption{The canonical structure of a kv-store in $\CMs(\ET_{\PSI}, \mathsf{Counter})$, where each
$\txid_{i}$ is the result of a $\ctrinc(\key)$ operation, and each transaction in 
$\T_i$ is the result of a $\ctrread(\key)$ operation.} 
\label{fig:prog_analysis}
\end{figure}

\vspace{-7pt}
{%
\displaymathfont
\[%
\begin{aligned}
	\mkvs(\key) & = (0, \txid_{0}, \txidset^{\mathsf{}}_{0} \cup \Set{\txid^{\mathsf{}}_1}) 
	\lcat (1, \txid^{\mathsf{}}_{1}, \txidset^{\mathsf{}}_{1} \cup \Set{\txid^{\mathsf{}}_2}) 
	\lcat \cdots \lcat (n-1, \txid^{\mathsf{}}_{n-1}, \txidset^{\mathsf{}}_{n-1} \cup \Set{\txid^{\mathsf{}}_n})
	\lcat (n, \txid^{\mathsf{}}_n, \txidset^{\mathsf{}}_{n})
\end{aligned}%
\]
}%
%


We define the $\dashrightarrow$ relation depicted in \cref{fig:prog_analysis} by extending 
$\SO \cup 
\{
	(\txid_i, \txid_j) 
	\mid 
	\txid_j \in \T_i \lor 	
	(\txid_i \in \T_i \land j {=} i {+} 1)
\}$
to a strict total order (\ie is $\dashrightarrow$ irreflexive and transitive). 
Note that $\dashrightarrow$ contains $\SO \cup \WR_{\mkvs} \cup \WW_{\mkvs} \cup \RW_{\mkvs}$ and thus
$(\SO \cup \WR_{\mkvs} \cup \WW_{\mkvs} \cup \RW_{\mkvs})^+$ is irreflexive; 
\ie $\Counter(\key)$ is robust against $\PSI$.

Recall from \cref{sec:overview} that unlike in $\SER$ and $\SI$, under $\PSI$ clients can observe 
the increments on different keys in different orders (see \cref{fig:cp-disallowed}).
As such, multiple counters are not robust against $\PSI$. 

\subsubsection{Robustness Conditions against $\WSI$}
\sx{strictly no blind write, \SI also works for sure, etc.}
Several libraries in the literature that are robust against $\SI$ 
\citep{giovanni_concur16,bank-example-wsi} are also robust against $\WSI$.
The operations of these libraries all yield kv-stores that adhere to a particular pattern captured by the following definition.

\begin{definition}[\(\WSI\)-safe]
\label{def:main-body-wsi-safe}
    A kv-store \( \mkvs \) is \emph{\(\WSI\)-safe} if it is 
    reachable by executing a program \( \prog \) from an initial configuration \( \conf_0 \) under $\WSI$
   (\ie \( \conf_0, \prog \toPROG{}_{\ET_\WSI} (\mkvs, \stub), \stub\)), and for all $\txid, \key, i$:
    \begin{align}
         & \txid \in \rsOf[\mkvs(\key,i)] \land \txid \neq \wtOf[\mkvs(\key,i)]  \implies \fora{\key',j} \txid \neq \wtOf[\mkvs(\key',j)] \label{equ:main-wsi-safe-read-only} \\
         & \txid \neq \txid_0 \land \txid = \wtOf[\mkvs(\key,i)] \implies \exsts{j} \txid \in \rsOf[\mkvs(\key,j)] \label{equ:main-wsi-safe-write-must-read} \\
         & \txid \neq \txid_0 \land \txid = \wtOf[\mkvs(\key,i)] \land \exsts{k', j} \txid \in \rsOf[\mkvs(\key',j)] \implies \txid = \wtOf[\mkvs(\key',j)] \label{equ:main-wsi-safe-all-write}
    \end{align}
\end{definition}

This definition states that a kv-store $\mkvs$ is \WSI-safe if for each transaction $\txid$: 
\begin{enumerate*} 
    \item if $\txid$ reads from $\key$ without writing to it then $\txid$ must be a read-only transaction \eqref{equ:main-wsi-safe-read-only}; 
    \item if \( \txid \) writes to $\key$, then it must also read from it \eqref{equ:main-wsi-safe-write-must-read}, a property known as \emph{strictly-no-blind writes}; and
	\item  if \( \txid \) writes to $\key$, then it must also write to all keys it reads \eqref{equ:main-wsi-safe-all-write}.
\end{enumerate*}
It is straightforward to see that the version $j$ read by \( \txid \) in \eqref{equ:main-wsi-safe-write-must-read} must be written immediately before the version $i$ written by \( \txid \), \ie \( i {=} j + 1 \).

\begin{theorem}[\( \WSI \) robustness]
 \label{thm:main-wsi-robust}
    If a kv-store \( \mkvs \) is \(\WSI\)-safe, then it is robust against \(\WSI\).   
\end{theorem}

\noindent From \cref{thm:serialisable_nocycle} it suffices to prove that $(\SO \cup \WR_{\mkvs} \cup \WW_{\mkvs} \cup \RW_{\mkvs})^{+}$ is irreflexive.
We proceed by contradiction and assume that there exists $\txid_1$ such that $\txid_1 \toEDGE{(\SO \cup \WR_{\mkvs} \cup \WW_{\mkvs} \cup \RW_{\mkvs})^{+}} \txid_1$. 
Since \( \mkvs \) is reachable under \( \WSI \) and thus also reachable under \( \CC \),
this cycle is of the form:
\[
    \txid_1 \toEDGE{\rel^*} \txid_2 \toEDGE{\RW} \txid_3 \toEDGE{\rel^*} \cdots \toEDGE{\rel^*} \txid_{n-2} \toEDGE{\RW} \txid_{n-1} \toEDGE{\rel^*} \txid_n = \txid_1
\]
\noindent where \( \rel \defeq \WR \cup \SO \cup \WW \).
From \cref{equ:main-wsi-safe-write-must-read,equ:main-wsi-safe-all-write} we know that 
an \( \RW \) edge from a writing transaction can be replaced by a \( \WW \) edge.
Moreover, \( \WW \) edges can be replaced by \( \WR^* \) edges since \( \mkvs \) is reachable under \( \UA \).
We thus have:
\[
    \txid_1 \toEDGE{{\rel'}^*} \txid'_2 \toEDGE{\RW} \txid'_3 \toEDGE{{\rel'}^+} \cdots \toEDGE{{\rel'}^+} \txid'_{m-2} \toEDGE{\RW} \txid'_{m-1} \toEDGE{{\rel'}^*} \txid'_m = \txid_n = \txid_1
\]
\noindent where \( \rel' \defeq \WR \cup \SO \).
That is, \( \txid_1 \toEDGE{( (\WR \cup \SO); \RW^? )^* } \txid_n \).
This however leads to a contradiction as \( (\WR \cup \SO); \RW^? \subseteq \rel_\CP \) and 
\( \WSI \) requires views to be closed under \( \rel_\CP \) (see \cref{fig:execution_tests}). 

\subsubsection{Robustness of Multiple Counters against $\WSI$} 
\label{sec:multi-counter-robust}
A multi-counter library on a set of keys \( \codeFont{\keyset} \) is: 
\( \mathsf{Counters(\keyset)} \defeq \bigcup_{\codeFont{\key} \in \codeFont{\keyset} } \Counter(\codeFont{\key}) \).
We next show that a multi-counter library is \( \WSI \)-safe, and is therefore robust against \( \WSI \) and all stronger models such as \( \SI \).

\begin{theorem}
  A multi-counter library \( \mathsf{Counters(\keyset)}  \) is robust against \( \WSI \).
\end{theorem}

It is sufficient to show that a kv-store obtained by executing arbitrary transactional calls from the library are \( \WSI \)-safe.
We proceed by induction on the length of traces.
Let \( \conf_0 = (\mkvs_0, \vienv_0) \) be an initial configuration and \( \prog_0 \) be a program such that \( \dom(\prog_0) \subseteq \dom(\vienv_0) \).
The initial kv-store trivially satisfies \eqref{equ:main-wsi-safe-read-only}, \eqref{equ:main-wsi-safe-write-must-read} and \eqref{equ:main-wsi-safe-all-write} above. 
Let \( \mkvs_i \) be the resulting kv-store after \( i \) steps of execution under \( \WSI \).
The next transaction \( \txid_{i+1} \) may then be a call to either \( \ctrinc(\pv{\key}) \) or \( \ctrread(\pv{\key}) \).
If \( \txid_{i+1} \) is a \( \ctrread(\pv{\key}) \) call,
then the resulting kv-store is: 
\[ 
	\mkvs_{i+1} = \mkvs_{i}\rmto{\pv{\key}}{\mkvs_i(\pv{\key})\rmto{j}{(\val, \txid, \txidset \uplus \Set{\txid_{i+1}})}} 
\]
for some \( j \) and \( \mkvs_i(\pv{\key},j) {=} (\val,\txid,\txidset)\).
Since \( \txid_{i+1} \) is a read-only transaction,
then  \eqref{equ:main-wsi-safe-read-only}, \eqref{equ:main-wsi-safe-write-must-read} and \eqref{equ:main-wsi-safe-all-write} immediately hold.
On the other hand, if \( \txid_{i+1} \) is an \( \ctrinc(\codeFont{\key}) \) call, 
then the resulting kv-store is: 
\[
    \mkvs_{i+1} = \mkvs_{i}\rmto{\pv{\key}}{( \mkvs_i(\pv{\key})\rmto{j}{(\val, \txid, \txidset \uplus \Set{\txid_{i+1}})} ) \lcat (\val+1, \txid_{i+1}, \emptyset) } 
\]
where \( j  {=} \abs{\mkvs_i(\pv{\key})} \) and \( \mkvs_i(\pv{\key},j) {=} (\val,\txid,\txidset)\).
As \( \txid_{i+1} \) reads the latest version of \( \pv{\key} \) and writes a new version to \( \pv{\key} \),
the new kv-store \( \mkvs_{i+1} \) satisfies \eqref{equ:main-wsi-safe-read-only}, \eqref{equ:main-wsi-safe-write-must-read} and \eqref{equ:main-wsi-safe-all-write}.

\subsubsection{Robustness of a Banking Library against $\WSI$}

\citet{bank-example-wsi} present a banking library
which is robust against  \( \SI \).
Here, we show that this library is also robust against \( \WSI \).
The banking example is based on relational databases and has three tables: account, saving and checking.
The account table maps customer names to customer IDs (\( \codeFont{Account(\underline{Name}, CID )} \)); 
the saving table maps customer IDs to their saving balances (\( \codeFont{Saving(\underline{CID}, Balance )} \)); and
the checking table maps customer IDs to their checking balances (\( \codeFont{Checking(\underline{CID}, Balance )} \)).

For simplicity, we encode the saving and checking tables as a kv-store,
and forgo the account table as it is an immutable lookup table.
We model a customer ID as an integer \( n \in \mathbb{N}\), and assume that balances are integer values. 
We then define the key associated with customer $n$ in the checking table as 
$n_c \defeq 2 n$; 
and define the key associated with $n$ in the saving table as 
$n_s \defeq 2n {+} 1$. 
That is, \( \Keys \defeq \bigcup_{n \in \Nat} \Set{n_c, n_s} \).
Moreover, if \( n \) identifies a customer, \ie $(\stub, n) \in \codeFont{Account(\underline{Name}, CID )}$,
then:
\small
\[ 
	(n, \valueOf[\mkvs(n_s, \abs{\mkvs(n_s)})]) \in \codeFont{Saving(\underline{CID}, Balance )} 
	\qquad 
	(n, \valueOf[\mkvs(n_s, \abs{\mkvs(n_c)})]) \in \codeFont{Checking(\underline{CID}, Balance )}
\]
\normalsize
The banking library provides five transactional operations for accessing the database:
\small
\begin{align*}
    \codeFont{balance(n)} & \defeq
    \begin{transaction}
    \plookup{\pv{x}}{\pv{n}_c}; \ 
    \plookup{\pv{y}}{\pv{n}_s}; \ 
    \passign{\ret}{\pv{x}+\pv{y}}
    \end{transaction} \\
    \codeFont{depositChecking(n,v)} & \defeq
    \begin{transaction}
    \pifs{\pv{v} \geq 0} \ 
    \plookup{\pv{x}}{\pv{n}_c}; \ 
    \pmutate{\pv{n}_c}{\pv{x} + \pv{v}}; \ 
    \pife
    \end{transaction}  \\
    \codeFont{transactSaving(n,v)} & \defeq
    \begin{transaction}
    \plookup{\pv{x}}{\pv{n}_s}; \ 
    \pifs{\pv{v} + \pv{x} \geq 0} \ 
    \pmutate{\pv{n}_s}{\pv{x} + \pv{v}}; \ 
    \pife
    \end{transaction} \\
	 \codeFont{amalgamate(n,n')} & \defeq
    \begin{transaction}
    \plookup{\pv{x}}{\pv{n}_s}; \ 
    \plookup{\pv{y}}{\pv{n}_c}; \ 
    \plookup{\pv{z}}{\pv{n'}_c}; \ 
    \pmutate{\pv{n}_s}{0}; \ 
    \pmutate{\pv{n}_c}{0}; \ 
    \pmutate{\pv{n'}_c}{\pv{x} + \pv{y} + \pv{z}}; 
    \end{transaction} \\
    \codeFont{writeCheck(n,v)} & \defeq
    \begin{transaction}
    \plookup{\pv{x}}{\pv{n}_s}; \ 
    \plookup{\pv{y}}{\pv{n}_c}; \\
    \pifs{\pv{x} + \pv{y} < \pv{v} } \
        \pmutate{\pv{n}_c}{\pv{y} - \pv{v} - 1 }; \
    \pifm \
        \pmutate{\pv{n}_c}{\pv{y} - \pv{v} }; \ 
    \pife \ 
    \pmutate{\pv{n}_s}{\pv{x}}; 
    \end{transaction}     
\end{align*}
\normalsize
The \( \codeFont{balance(n)} \) operation returns the total balance of customer \codeFont{n} in  \( \ret \).
The \( \codeFont{depositChecking(n,v)} \) operation deposits \codeFont{v} to the checking account of customer \codeFont{n} when \codeFont{v} is non-negative; otherwise the checking account remains unchanged.
When $\codeFont{v} \geq 0$,  \( \codeFont{transactSaving(n,v)} \) deposits \codeFont{v} to the saving account of \codeFont{n}.
When $\codeFont{v} < 0$, \( \codeFont{transactSaving(n,v)} \) withdraws \codeFont{v} from the saving account of \codeFont{n} only if the resulting balance is non-negative; otherwise the saving account remains unchanged.
The \( \codeFont{amalgamate(n,n')} \) operation moves the combined checking and saving funds of consumer \codeFont{n} to the checking account of customer $\codeFont{n}'$.
Lastly, \( \codeFont{writeCheck(n,v)} \) cashes a cheque of customer \codeFont{n} in the amount  \codeFont{v} by deducting \codeFont{v} from its checking account.
If \codeFont{n} does not hold sufficient funds (\ie the combined checking and saving balance is less than \codeFont{v}), customer \codeFont{n} is penalised by deducting one additional pound. 
\citet{bank-example-wsi} argue that to make the banking library robust against \( \SI \),
the \( \codeFont{writeCheck(n,v)} \) operation must be strengthened by writing back the balance to the saving account 
(via \(\pmutate{\pv{n}_s}{\pv{x}} \)),
even though the saving balance is unchanged.


The banking library is more complex than the multi-counter library discussed in \cref{sec:multi-counter-robust}.
Nevertheless, all banking transactions are either read-only or
satisfy the strictly-no-blind-writes property; \ie the banking library is \WSI-safe.
As such, we can prove its robustness against $\WSI$ in a similar fashion to that of the multi-counter library.
More concretely, given a \( \WSI \)-safe kv-store \( \mkvs\),
we show that the kv-store resulting from executing a banking operation on \( \mkvs \) remains \WSI-safe. 

As \(\codeFont{balance(n)} \) is read-only, it immediately satisfies \eqref{equ:main-wsi-safe-read-only}, \eqref{equ:main-wsi-safe-write-must-read} and \eqref{equ:main-wsi-safe-all-write}.
When $\codeFont{v} \geq 0$, then \(\codeFont{depositChecking(n,v)} \) both reads and writes \( n_c \), and thus preserves  
\eqref{equ:main-wsi-safe-read-only}, \eqref{equ:main-wsi-safe-write-must-read} and \eqref{equ:main-wsi-safe-all-write}.
When $\codeFont{v} < 0$, then  \(\codeFont{depositChecking(n,v)} \) leaves the kv-store unchanged and thus \eqref{equ:main-wsi-safe-read-only}, \eqref{equ:main-wsi-safe-write-must-read} and \eqref{equ:main-wsi-safe-all-write} are trivially preserved.
Lastly, the
\( \codeFont{transactSaving(n,v)}, \codeFont{amalgamate(n,n')} \) and \( \codeFont{writeCheck(n,v)}\) operations
always read and write the keys they access, thus satisfying \eqref{equ:main-wsi-safe-read-only}, \eqref{equ:main-wsi-safe-write-must-read} and \eqref{equ:main-wsi-safe-all-write}.

\subsection{Verifying Database Protocols}
\label{sec:verify-impl}
Kv-stores and views faithfully abstract the state of geo-replicated and partitioned
databases, and  execution tests provide a powerful abstraction of the synchronisation mechanisms  enforced by these databases when committing a transaction. This makes it
possible to use our 
semantics to verify the correctness of distributed database protocols. 
We  demonstrate this by showing that  the replicated database 
COPS~\cite{cops} satisfies causal consistency  (this section and \cref{sec:cops}) and
the partitioned database Clock-SI~\cite{clocksi} satisfies snapshot isolation (\cref{sec:clock-si});

\begin{figure*}[!t]
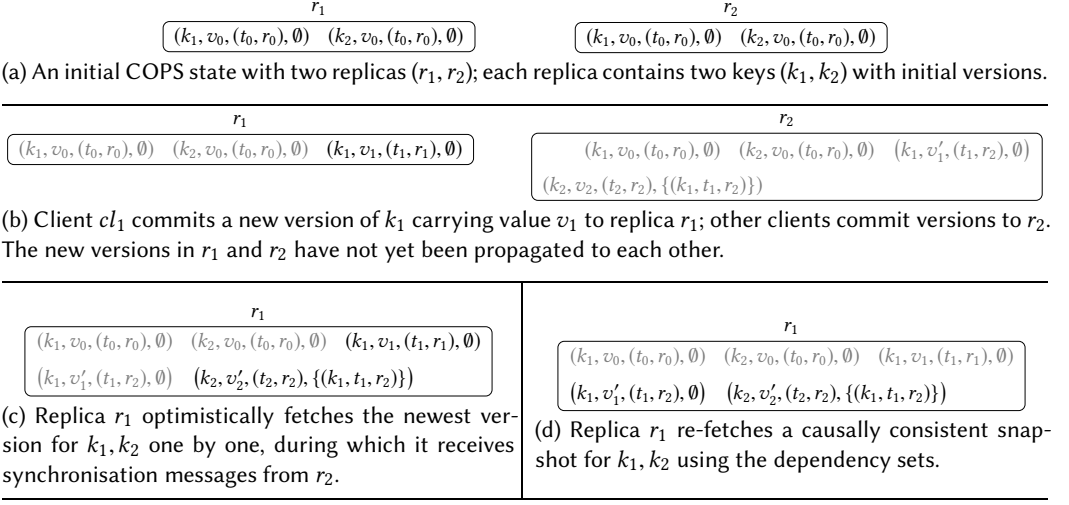

\captionsetup[subfigure]{aboveskip=0pt, belowskip=5pt}


\begin{subfigure}{\textwidth}
\begin{centertikz}
\draw pic {transaction={r1}{%
        /$\key_1$/$\val_0$/${(t_0,\repl_0),\emptyset}$%
        , /$\key_2$/$\val_0$/${(t_0,\repl_0),\emptyset}$%
}};
\path(r1.north) node[anchor=south] (r1lb) {$\repl_1$};

\draw pic at ($(r1.south east) + (3.4,0.45)$) {transaction={r2}{%
        /$\key_1$/$\val_0$/${(t_0,\repl_0),\emptyset}$%
        , /$\key_2$/$\val_0$/${(t_0,\repl_0),\emptyset}$%
}};
\path(r2.north) node[anchor=south] (r2lb) {$\repl_2$};

\end{centertikz}
\caption{An initial COPS state with two replicas ($\repl_1, \repl_2$); 
each replica contains two keys ($\key_1, \key_2$) with initial versions.}
\label{fig:initial-cops}
\end{subfigure}

\hrulefill 

\begin{subfigure}{\textwidth}
\begin{centertikz}
\draw pic {transaction={r1}{%
        nonvisible/$\key_1$/$\val_0$/${(t_0,\repl_0),\emptyset}$%
        , nonvisible/$\key_2$/$\val_0$/${(t_0,\repl_0),\emptyset}$%
        , /$\key_1$/$\val_1$/${(t_1,\repl_1),\emptyset}$%
}};
\path(r1.north) node[anchor=south] (r1lb) {$\repl_1$};

\draw pic at ($(r1.south east) + (3.4,0.45)$) {transaction={r2}{%
        nonvisible/$\key_1$/$\val_0$/${(t_0,\repl_0),\emptyset}$%
        , nonvisible/$\key_2$/$\val_0$/${(t_0,\repl_0),\emptyset}$%
        , nonvisible/$\key_1$/$\val'_1$/${(t_1,\repl_2),\emptyset}$%
        , nonvisible/$\key_2$/$\val_2$/${(t_2,\repl_2),\Set{(\key_1,t_1,\repl_2)}}$%
}};
\path(r2.north) node[anchor=south] (r2lb) {$\repl_2$};

\end{centertikz}
\caption{Client \( \cl_1 \) commits a new version of \( \key_1 \) carrying value \( \val_1 \) to replica \( \repl_1 \);
other clients commit versions to $\repl_2$. The new versions in $\repl_1$ and $\repl_2$ have not yet been propagated to each other.}
\label{fig:cops-after-write-transaction}
\end{subfigure}




\begin{tabularx}{\textwidth}{@{} X | c @{}}
\hline\\[-5pt]
\begin{subfigure}{0.49\textwidth}
\begin{centertikz}
\draw pic {transaction={r1}{%
        nonvisible/$\key_1$/$\val_0$/${(t_0,\repl_0),\emptyset}$%
        , nonvisible/$\key_2$/$\val_0$/${(t_0,\repl_0),\emptyset}$%
        , /$\key_1$/$\val_1$/${(t_1,\repl_1),\emptyset}$%
        , nonvisible/$\key_1$/$\val'_1$/${(t_1,\repl_2),\emptyset}$%
        , /$\key_2$/$\val'_2$/${(t_2,\repl_2),\Set{(\key_1,t_1,\repl_2)}}$%
}};
\path(r1.north) node[anchor=south] (r1lb) {$\repl_1$};

\end{centertikz}
\caption{Replica $\repl_1$ optimistically fetches the newest version for \( \key_1,\key_2 \) one by one, during which it receives synchronisation messages from \( \repl_2 \).}
\vspace{-15pt}%
\label{fig:cops-request-values}
\end{subfigure}

& 

\begin{subfigure}{0.49\textwidth}
\begin{centertikz}
\draw pic {transaction={r1}{%
        nonvisible/$\key_1$/$\val_0$/${(t_0,\repl_0),\emptyset}$%
        , nonvisible/$\key_2$/$\val_0$/${(t_0,\repl_0),\emptyset}$%
        , nonvisible/$\key_1$/$\val_1$/${(t_1,\repl_1),\emptyset}$%
        , /$\key_1$/$\val'_1$/${(t_1,\repl_2),\emptyset}$%
        , /$\key_2$/$\val'_2$/${(t_2,\repl_2),\Set{(\key_1,t_1,\repl_2)}}$%
}};
\path(r1.north) node[anchor=south] (r1lb) {$\repl_1$};

\end{centertikz}%
\caption{Replica \( \repl_1 \) re-fetches a causally consistent snapshot for \( \key_1,\key_2 \) using the dependency sets.}
\vspace{-15pt}%
\label{fig:cops-re-read-values}
\end{subfigure}%
\\ \hline
\end{tabularx}

\caption{COPS protocol}
\label{fig:cops-digraph}
\end{figure*}

COPS is a fully replicated database, with each replica storing multiple versions of each key as illustrated in \cref{fig:initial-cops,fig:cops-after-write-transaction}. 
Each COPS version \( \ver \),
\eg \( (\key_1,\val_1,(\txid_1,\repl_1), \emptyset) \) in \cref{fig:cops-after-write-transaction},
contains a key (\eg $\key_1$), a value (\eg $\val_1$), a time-stamp denoting when a client first wrote the version to the replica (\eg $(\txid_1, \repl_1)$), and a set of dependencies, written $\depOf[\ver]$ (\eg $\emptyset$). 
A time-stamp associated with a version $\ver$ is of the form $(\txid, \repl)$, where $\repl$ identifies the replica that committed $\ver$, and $\txid$ denotes the local, real time when $\repl$ committed $\ver$. 
Each dependency in $\depOf[\ver]$ comprises a key and the time-stamp of the version of that key on which $\ver$ directly depends.  
COPS assumes a total order among replica identifiers. Thus,  time-stamps can be
totally ordered lexicographically over time-stamps. 

The COPS API provides two operations: one for writing to a single
key; and another for atomically reading from a set of keys. 
Each call to a COPS operation is processed by a single replica. 
Each client maintains a \emph{context}, which is a set of dependencies
tracking the versions the client observes.  

We demonstrate how a client \( \cl \) interacts with a replica through the following example:
\begin{align}
    \label{equ:client-repl-1}
    \cl: \begin{transaction} \pmutate{\key_1}{\val_1}; \end{transaction} ; \ 
    \begin{transaction} \pderef{\vx}{\key_1}; \ \pderef{\vy}{\key_2}; \end{transaction}
    \tag{\textsc{cops-cl}}
\end{align}
For brevity, we assume that there are two keys ($\key_1$ and $\key_2$) and two replicas ($\repl_1$ and $\repl_2$) 
as shown in \cref{fig:initial-cops}, where $\repl_1 < \repl_2$.
Initially, client \( \cl \) connects to replica \( \repl_1 \) and initialises its local context as $ctx {=} \emptyset$.
To execute its first single-write transaction, client $\cl$ requests to write $\val_1$ to $\key_1$ by sending the message $(\key_1, \val_1, ctx)$ to its associated replica $\repl_1$. 
It then awaits a reply from $\repl_1$.
Upon receiving the message, $\repl_1$ produces a monotonically increasing local time $\txid_1$, and uses it to install  a new version $\ver {=} (\key_1,\val_1, (\txid_1,\repl_1), ctx)$, as shown in \cref{fig:cops-after-write-transaction}.
Note that the dependency set of $\ver$ is the $\cl$ context ($ctx {=} \emptyset$).
Replica $\repl_1$ then sends the time-stamp $(\txid_1,\repl_1)$ back to client $\cl_1$, and $\cl_1$ in turn  incorporates $(\key_1, \txid_1,\repl_1)$ in its local context; \ie 
$\cl$ observes its own write. 
Finally, replica $\repl_1$ propagates the written version to other replicas asynchronously by sending a \emph{synchronisation message} 
using \emph{causal delivery} as follows. 
In the general case, when a replica $\repl'$ receives a version $\ver'$ from another replica $\repl$, it 
first waits for all dependencies of $\ver$ to be committed to $\repl'$, and then commits $\ver$.
As such, the set of versions contained in each replica is closed with respect to causal dependencies.
In the example above, when other replicas receive version $\ver$ from $\repl_1$, they can immediately commit $\ver$ as its dependency set is empty. 
Note that replicas may accept new versions from different clients in parallel (see \cref{fig:cops-after-write-transaction}).

To execute its second multi-read transaction,
client  \( cl \) requests to read from the $\key_1, \key_2$ keys by sending the message 
\( \Set{\key_1, \key_2} \) to replica $\repl_1$ and awaits a reply.
Upon receiving this message, replica $\repl_1$ builds a causally consistent snapshot (a mapping from $\Set{\key_1, \key_2}$ to values) in two rounds as follows. 
First, $\repl_1$ optimistically fetches the most recent versions for $\key_1$ and $\key_2$,
one at a time. 
This process may be interleaved with other writes and synchronisation messages (propagated from other replicas). 
For instance, \cref{fig:cops-request-values} depicts a scenario where \( \repl_1 \)
\begin{enumerate*}
	\item first fetches \( (\key_1,\val_1,(\txid_1,\repl_1), \emptyset) \) for $\key_1$ (highlighted); 
	\item then receives two synchronisation messages from \( \repl_2 \), 
containing versions \( ( \key_1, \val'_1, (\txid_1,\repl_2),\emptyset ) \) and \( ( \key_2, \val'_2, (t_2,\repl_2),\Set{(\key_1,\txid_1,\repl_2)} ) \); and
	\item finally fetches \( ( \key_2, \val'_2, (t_2,\repl_2),\Set{(\key_1,\txid_1,\repl_2)} ) \) for $\key_2$ (highlighted).
\end{enumerate*}
As such, the versions fetched for \( \Set{\key_1,\key_2}\) are not causally consistent: 
\( ( \key_2, \val'_2, (t_2,\repl_2),\Set{(\key_1,\txid_1,\repl_2)} ) \) depends on a $\key_1$ version with time-stamp $(\txid_1, \repl_2)$ which is bigger than that fetched for $\key_1$, namely $(\txid_1, \repl_1)$.
To remedy this, after the first round of optimistic reads,
$\repl_1$ collects all dependency sets and uses them as an lower-bound in the second round
to re-fetch the most recent version (with the biggest time-stamp) of each key from the dependency sets.
For instance, in \cref{fig:cops-request-values} replica $\repl_1$ re-fetches the newer version \( ( \key_1, \val'_1, (\txid_1,\repl_2),\emptyset ) \) for \( \key_1 \), as depicted in \cref{fig:cops-re-read-values}.
The snapshot obtained after the second round is thus causally consistent. 
Finally, the snapshot and the dependencies of each version read are sent to the client, and subsequently added to the client context.
\ac{I will leave this to myself for future reference: why the hell does COPS do this in two rounds? It looks like you can do 
this in a single round, by simply taking the time-stamp of the first version you read as an upper bound for future reads.
Update: Went to check directly on the COPS protocol, and indeed COPS really does something more - but not much more - complicated. 
The main aspect from which COPS differs from Shale's idealisation of it is that (1) reads happen concurrently, and the 
one with highest time-stamp is used to fix the bound on versions in the first round, and (2) in the second round not all keys 
are re-fetched, but only those that are really needed to recover a causally consistent snapshot.}

\begin{figure*}[!t]
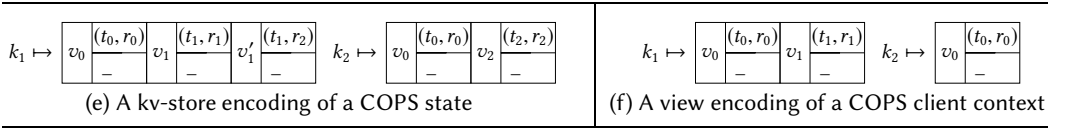

\captionsetup[subfigure]{aboveskip=0pt, belowskip=5pt}

\begin{tabularx}{\textwidth}{@{} X | c @{}}
\hline\\[-5pt]
\begin{subfigure}{0.53\textwidth}
\begin{centertikz}
\node(locx) {$\key_1 \mapsto$};
\draw pic at ([xshift=\tikzkvspace]locx.east) {vlist={versionx}{%
    /$\val_0$/${(\txid_0,\repl_0)}$/$\stub$
    , /$\val_1$/${(\txid_1,\repl_1)}$/$\stub$
    , /$\val'_1$/${(\txid_1,\repl_2)}$/$\stub$
}};

\path (versionx.east) + (0.75,0) node (locy) {$\key_2 \mapsto$};
\draw pic at ([xshift=\tikzkvspace]locy.east) {vlist={versiony}{%
    /$\val_0$/${(\txid_0,\repl_0)}$/$\stub$
    , /$\val_2$/${(\txid_2,\repl_2)}$/$\stub$
}};

\end{centertikz}
\caption{A kv-store encoding of a COPS state}
\vspace{-15pt}%
\label{fig:encode-mkvs}
\end{subfigure}

& 

\begin{subfigure}{0.42\textwidth}
\begin{centertikz}
\node(locx) {$\key_1 \mapsto$};
\draw pic at ([xshift=\tikzkvspace]locx.east) {vlist={versionx}{%
    /$\val_0$/${(\txid_0,\repl_0)}$/$\stub$
    , /$\val_1$/${(\txid_1,\repl_1)}$/$\stub$
}};

\path (versionx.east) + (0.75,0) node (locy) {$\key_2 \mapsto$};
\draw pic at ([xshift=\tikzkvspace]locy.east) {vlist={versiony}{%
    /$\val_0$/${(\txid_0,\repl_0)}$/$\stub$
}};

\end{centertikz}
\caption{A view encoding of a COPS client context}
\vspace{-15pt}%
\label{fig:encode-view}
\end{subfigure}%
\\ \hline
\end{tabularx}

\caption{COPS encoding}
\label{fig:cops-encode}
\end{figure*}

To prove that COPS satisfies causal consistency,
we encode the state of the system 
(comprising the state of all replicas and clients) as a
configuration in  our operational semantics. 
As each replica stores a set of versions (for each key in COPS) and their dependencies, 
we can project the state of  COPS replicas into a kv-store 
by mapping  COPS versions into our kv-store versions.
The writer of a mapped version is uniquely
determined by the time-stamp of the corresponding COPS version.
The reader set of the mapped version 
can be recovered by annotating read-only transactions.
For example, the COPS state in \cref{fig:cops-after-write-transaction} can be encoded as the kv-store depicted in \cref{fig:encode-mkvs}.
Similarly, as the context of a client $\cl$ identifies the set of COPS versions that $\cl$ sees, 
we can project COPS client contexts to our client views over kv-stores. 
For example, the context of \( \cl \) after committing its first single-write transaction in \eqref{equ:client-repl-1} is encoded as the client view depicted in \cref{fig:encode-view}.

We next map the execution of a COPS transaction into an $\ET_{\CC}$ reduction between the configurations 
obtained from encoding the COPS states before and after executing the transaction.
Note that existing verification techniques \cite{framework-concur,seebelieve} require examining the \emph{entire} sequence of operations of a protocol to show that it implements a consistency model.
By contrast, we only need to look at how the system evolves after a \emph{single} transaction is executed.
In particular, we check the client views (obtained from COPS client contexts) over kv-stores after executing a single transaction as follows.
Intuitively, we observe that when a COPS client $\cl$ executes a transaction then:
\begin{enumerate*} 
\item the $\cl$ context grows, and thus we obtain a more up-to-date view of the associated kv-store; \ie $\vshiftname_{\MR}$ holds;
\item the $\cl$ context always includes the time-stamp of the versions written by itself, and thus the 
corresponding client view always includes the versions $\cl$ has written; \ie $\vshiftname_{\RYW}$ holds; and
\item the $\cl$ context always contains the dependencies corresponding to versions it has 
either written or read from other transactions, and thus the corresponding client view is always closed-down 
with respect to the relation $\SO \cup \WR_{\mkvs}$; \ie $\closed[\mkvs, \vi, R_{\CC}])$ holds.
\end{enumerate*}
As such, from the definition of $\CC$ in \cref{fig:execution_tests} we know that COPS satisfies causal consistency ($\CC$).
We refer the reader to  \cref{sec:cops} for further details and the full soundness proof.


\section{Conclusions and Related Work}
\label{sec:conclusions}
\sx{A cite here \cite{surech-session-guarantee} who mentions
  composition of 4 session guarantees.
  \cite{principle-eventual-consistency} mentions causal consistency is
  combination of session guarantees, yet they only define 3 types of
  session guarantees.  }

We have introduced a simple interleaving semantics for atomic
transactions, based on a global, centralised kv-stores and partial
client views. It is expressive enough to capture the anomalous
behaviour of many weak consistency models.  We have demonstrated that
our semantics can be used to both verify protocols of distributed databases
and analyse client programs.

We have defined a large variety of consistency models for kv-stores
based on execution tests, and have shown these models  to be equivalent to
well-known declarative consistency models for dependency graphs and
abstract executions. We do not know of an appropriate consistency
model that we cannot express with our semantics, bearing in mind the
constraints that our transactions satisfy snapshot property and the last-write-wins policy. We have
identified a new consistency model, called weak snapshot
isolation, which lies between $\PSI$ and $\SI$ and inherits many of
the good properties of $\SI$. 
We have shown that examples are robust against \( \WSI \).
We would need to provide an implementation of 
this model to justify it in full in the future. 
We have proved the correctness of two real-world protocols employed by distributed 
databases: COPS~\cite{cops}, a 
protocol for replicated databases that satisfies causal consistency;
and Clock-SI~\cite{clocksi}, a protocol for partitioned databases that satisfies 
snapshot isolation. We have also demonstrated the usefulness of our framework
for proving invariant properties: the robustness of simple transactional 
libraries against different consistency models. 

In future, we aim to extend our framework to handle other 
weak consistency models. For example, we believe that, by introducing promises 
in the style of \cite{promises}, we can capture  consistency models such 
as \emph{Read Committed}. We also plan to validate further the usefulness of our framework
by: verifying other well-known protocols of 
distributed databases, e.g. Eiger~\cite{eiger}, Wren~\cite{wren} and
Red-Blue~\cite{redblue};  exploring robustness results for OLTP
workloads  such as TPC-C~\cite{tpcc} and RUBiS~\cite{rubis}; 
and exploring other program analysis techniques such as 
transaction chopping \cite{chopping,psi-chopping}, invariant checking 
\cite{cise,repliss} and program logics \cite{alonetogether}. 

\mypar{Related Work} 
In the introduction (\cref{sec:intro}), we highlight three general operational semantics for 
distributed transactional databases. We discuss these semantics in more detail here,
and also give some additional related work on program analysis. 

\citet{alonetogether} propose an operational semantics of SQL transactional programs 
under the consistency models given by the standard ANSI/SQL isolation levels \cite{si}.
In their  framework, transactions work on a local copy of the global state 
of the system, and the local effects of a transaction are committed to the  
system state when it terminates. Because state changes 
are made immediately available to all clients of a system, this model 
is not suitable to capture weak consistency models such as \(\PSI\) or \(\CC\). 
They introduce a program logic and prototype verification tool for reasoning 
about client programs. However, their definitions of consistency models 
are not validated against previously known 
formal definitions.

\citet{sureshConcur} propose an operational semantics for weak consistency 
based on abstract executions. Their semantics 
is parametric in the declarative definition of a consistency model. 
They introduce a tool for checking the robustness of transactional  
libraries.
They focus on consistency models with snapshot property, but confusingly allow 
the interleaving of fine-grained operations between transactions. 
This results in an unnecessary explosion of the space of traces generated by 
the program. In our semantics, the interleaving is between transactions.

\citet{seebelieve} propose a state-based formal framework for weak
consistency models 
that employs concepts  similar to execution tests and views, called commit tests and read states 
respectively.
They prove that consistency models previously thought to be different 
are in fact equivalent in their semantics.
They capture 
a wide range of consistency models including read committed which we cannot do. 
In their semantics, one-step trace reduction is determined by the whole previous history of the trace. 
In contrast, our reduction step only depends on the current configuration (kv-store and view).
They do not consider program analysis. Their notion of commit tests and read states requires 
the knowledge of information that is not known to clients of the system, i.e. the total order of system changes that happened in the database 
prior to committing a transaction. For this reason, we believe that
their framework is not suitable for the development of techniques for
analysing client programs. 

\citet{op-semantics-c11-rar} develop an operational semantics for release-acquire fragment of
C11 memory model, an variant of causal consistency.
Their semantics is based on an variant of dependency graph where nodes and edges 
are tailored for C11 operations.
They introduce per-thread observations, and they must be compatible for executing next operations;
this is similar to our views and execution tests.
We believe we can also model release-acquire fragment of C11.

Several other works have focused on program analysis for transactional systems. 
\citet{dias-tm} developed a separation logic for
the robustness of applications against \(\SI\). \citet{fekete-tods} derived 
a static analysis check for \(\SI\) based on dependency graph. 
\citet{giovanni_concur16}
developed a static analysis check for several consistency models with snapshot property. 
\citet{snapshot-isolation-robust-tool} developed a tool based on Lipton's reduction \cite{Lipton-reduction} 
for checking robustness against \( \SI \).
\citet{laws} investigated the relationship between abstract 
executions and dependency graphs from an algebraic perspective, and applied it to infer 
robustness checks for several consistency models.


\bibliography{main}

\newpage
\onecolumn
\begin{appendices}
\section{Operational Semantics on KV-Stores}
\label{sec:full-semantics}

\begin{definition}[Multi-version Key-value Stores]
\label{def:mkvs-appendix}
Assume a countably infinite set of \emph{keys} $\Keys \ni \key$, 
and a countably infinite set of  \emph{values} $\Val \ni \val$, 
including an \emph{initialisation value} $\val_0 $.
The set of \emph{versions}, $\Versions \ni \ver$, is: $\Versions \defeq \Val \times \TxID \times \pset{\TxID_{0}}$. 
A \emph{kv-store} is a function $\mkvs: \Keys \to \func{List}[\Versions]$, 
where $\func{List}[\Versions] \ni \vilist$ is the set of lists of versions $\Versions$. 
Well-formed key-values store satisfy:
\begin{align}
& \fora{\key, i, j} 
\rsOf(\mkvs(\key, i)) \cap \rsOf(\mkvs(\key, j)) \neq \emptyset \lor
\wtOf(\mkvs(\key, i)) = \wtOf(\mkvs(\key, j))
\implies i = j  \\
& \fora{\key} \mkvs(\key, 0) = (\val_0, \txid_0, \stub) \\
& \fora{ \key, \cl, i,j, n, m} 
\txid_{\cl}^{n} = \wtOf(\mkvs(\key,i)) \land \txid_{\cl}^{m} \in
\Set{\wtOf(\mkvs(\key,j))} \cup \rsOf(\mkvs(\key, i)) \implies n < m
\end{align}
\end{definition}

The full semantics is in \cref{def:program_semantics} and the full definition of consistency models is in  \cref{fig:app-execution-tests}.

\begin{figure*}[!t]
\begin{align*}
\toTRANS & : ((\Stacks \times \Snapshots \times \Fingerprints) \times \Transactions) \times ((\Stacks \times \Snapshots \times \Fingerprints) \times \Transactions)
\end{align*}
\begin{mathpar}
    \inferrule[\rl{TPrimitive}]{%
        (\stk, \sn) \toLTS{\transpri} (\stk', \sn') 
        \\
        \op = \func{op}[\stk, \sn, \transpri]
    }{%
        (\stk, \sn, \fp) , \transpri  \toTRANS   (\stk', \sn', \fp \addO \op) , \pskip 
    }%
    \\
     \inferrule[\rl{TChoice}]{%
		i \in \Set{1,2}
    }{%
		(\stk, \sn, \fp) , \trans_{1} \pchoice \trans_{2}  \toTRANS  (\stk, \sn, \fp) , \trans_{i}
    }
    \and
    \inferrule[\rl{TIter}]{ }{%
        (\stk, \sn, \fp),  \trans\prepeat \toTRANS  (\stk, \sn, \fp), \pskip \pchoice (\trans \pseq \trans\prepeat)
    }%
    \and
    \inferrule[\rl{TSeqSkip}]{ }{%
        (\stk, \sn, \fp), \pskip \pseq \trans \toTRANS  (\stk, \sn, \fp), \trans
    }%
    \and
    \inferrule[\rl{TSeq}]{%
		(\stk, \sn, \fp), \trans_{1} \toTRANS  (\stk', \sn', \fp'), \trans_{1}'
    }{%
		(\stk, \sn, \fp), \trans_{1} \pseq \trans_{2} \toTRANS  (\stk', \sn', \fp'), \trans_{1}' \pseq \trans_{2}
    }%
\end{mathpar}
\hrulefill
\begin{align*}
	\toCMD{}  & : 
    \begin{multlined}[t]
    \Clients \; \times \;
	\left( ( \MKVSs \times \Views \times \Stacks ) \times \Commands \right)  
    \; \times\; \ETs \;\times \sort{Labels} \times \;
	\left( ( \MKVSs \times \Views \times \Stacks ) \times \Commands \right) 
    \end{multlined}
\end{align*}
\begin{mathpar}
     \inferrule[\rl{CAtomicTrans}]{%
        \vi \viewleq  \vi'' 
        \\
        \sn = \snapshot[\mkvs,\vi''] 
        \\
        (\stk, \sn, \emptyset), \trans \toTRANS^{*}   (\stk', \stub,
  \fp) , \pskip
  \\
   \txid \in \nextTxid[\cl, \mkvs] 
    \\\\
     \mkvs' = \updateKV[\mkvs, \vi'', \fp, \txid] 
\\
	\cancommit{\mkvs}{\vi''}{\fp}
\\
	\vshift{\mkvs}{\vi''}{\mkvs'}{\vi'}	
    }{%
        \cl \vdash 
        ( \mkvs, \vi, \stk ), \ptrans{\trans} 
        \toCMD{(\cl, \vi'', \fp)}_{\ET}
        (\mkvs',\vi', \stk' ) , \pskip
    }%
    \and
    \inferrule[\rl{CPrimitive}]{
		\stk \toLTS{\cmdpri} \stk'
    }{
        \cl \vdash 
        ( \mkvs, \vi, \stk ) , \cmdpri 
        \toCMD{(\cl,\iota)}_{\ET} 
        ( \mkvs, \vi, \stk' ) , \pskip
    }%
    \and
    \inferrule[\rl{CChoice}]{%
        i \in \Set{1,2}
    }{%
        \cl \vdash ( \mkvs, \vi, \stk ) , \cmd_{1} \pchoice \cmd_{2} \ \toCMD{(\cl,\iota)}_{\ET} \  ( \mkvs, \vi, \stk ) , \cmd_{i}
    }
    \and
    \inferrule[\rl{CIter}]{ }{%
        \cl \vdash ( \mkvs, \vi, \stk ) , \cmd\prepeat \ \toCMD{(\cl,\iota)}_{\ET} \  ( \mkvs, \vi, \stk ) , \pskip \pchoice (\cmd \pseq \cmd\prepeat)
    }
    \and
    \inferrule[\rl{CSeqSkip}]{ }{%
        \cl \vdash ( \mkvs, \vi, \stk ) , \pskip \pseq \cmd \ \toCMD{(\cl,\iota)}_{\ET} \  ( \mkvs, \vi, \stk ) , \cmd
    }
    \and
    \inferrule[\rl{CSeq}]{%
        \cl \vdash ( \mkvs, \vi, \stk ) , \cmd_{1} \ \toCMD{(\cl,\iota)}_{\ET} \  ( \mkvs, \vi', \stk' ) , {\cmd_{1}}' 
    }{%
        \cl \vdash ( \mkvs, \vi, \stk ) , \cmd_{1} \pseq \cmd_{2} \ \toCMD{(\cl,\iota)}_{\ET} \ ( \mkvs, \vi', \stk' ) , {\cmd_{1}}' \pseq \cmd_{2}
    }
\end{mathpar}

\hrulefill

\[
	\toPROG{} : 
    ( \Confs \times \ThdEnv \times \Programs) 
    \;\times\; \ETs \;\times \sort{Label} \times \;
    ( \Confs \times \ThdEnv \times \Programs) 
\]
\begin{mathpar}
    \inferrule[\rl{PProg}]{%
        \cl \vdash ( \mkvs, \vienv(\cl), \thdenv(\cl) ), \prog(\cl), \ \toCMD{\lambda}_{\ET} \  ( \mkvs', \vi', \stk' ) , \cmd'  
    }{%
    ( \mkvs, \vienv, \thdenv ) , \prog \ \toPROG{\lambda}_{\ET} \  ( \mkvs', \vienv\rmto{\cl}{\vi'}, \thdenv\rmto{\cl}{\stk'}) , \prog\rmto{\cl}{\cmd'} 
    }
\end{mathpar}
\hrulefill
\caption{Operational Semantics on Key-value Store}
\label{fig:transaction_semantics}
\label{def:thread_semantics}
\label{fig:thread_semantics}
\label{def:thread_pool_semantics}
\label{fig:thread_pool_semantics}
\label{def:program_semantics}
\label{fig:program_semantics}
\label{fig:full-semantics}
\end{figure*}

\begin{figure*}[t]
\small
\centering
\scalebox{.8}{%
\begin{tabular}{ @{} l @{\hspace{2pt}} || @{\hspace{2pt}} c | @{\hspace{2pt}} l @{\hspace{2pt}} | @{\hspace{2pt}}  c @{} }
\hline
	\ET 
	& $\cancommit \mkvs \vi \fp$
	& Closure Relation (where applicable)
    & $\vshift \mkvs \vi {\mkvs'} {\vi'}$ 
	\\
	\hline
	\MR 
	& \true 
	& 
	& $\vi \viewleq \vi'$
	\\ \hline  
	\RYW
	& \true
	& 
	& 
	\protect{$
	\begin{array}[t]{@{} l @{}}
		\fora{\txid \in \mkvs' \setminus \mkvs} \fora{\key, i} \\
		\;\;\wtOf(\mkvs'(\key, i) ) \toEDGE{\!\!\SO\rflx\!\!} \txid \implies i \!\in\! \vi'(\key) 
	\end{array}
	$}
	\\ \hline  
    \MW 
    & \( \closed(\mkvs, \vi, \rel_{\MW} ) \)
    & \(\rel_{\MW} \defeq \SO \cap \WW_\mkvs\)
    & \true  
    \\ \hline
    \WFR
    & $\closed(\mkvs, \vi, \rel_{\WFR})$
    & $\rel_{\WFR}  \defeq \WR_{\mkvs} ; (\SO \cup \RW_\mkvs)\rflx $ 
    & \true \\ \hline
	\CC
	& $\closed(\mkvs, \vi, \rel_{\CC})$
	& $\rel_{\CC}   \defeq \SO \cup \WR_{\mkvs}$ 
	& $\vshift[\MR \cap \RYW] \mkvs \vi {\mkvs'} {\vi'}$
	\\ \hline  
	\UA 
	& $\closed(\mkvs, \vi, \rel_{\UA})$
	& $\rel_{\UA}  \defeq {\textstyle\bigcup_{(\otW, \key, \stub) \in \fp}} \WW^{-1}_{\mkvs}(\key) $ 
	& \true  
	\\ \hline  
	\PSI
	& $\closed(\mkvs, \vi, \rel_{\PSI})$
	& $\rel_{\PSI} \defeq \rel_{\UA} \cup \rel_{\CC} \cup \WW_\mkvs$ 
	& $\vshift[\MR \cap \RYW] \mkvs \vi {\mkvs'} {\vi'}$
	\\ \hline   
	\CP 
	& $\closed(\mkvs, \vi, \rel_{\CP})$
	& $\rel_{\CP} \defeq \SO;\RW\rflx_\mkvs \cup \WR_\mkvs;\RW\rflx_\mkvs  \cup \WW_\mkvs$ 	
	& $\vshift[\MR \cap \RYW] \mkvs \vi {\mkvs'} {\vi'}$
    \\ \hline 
	\WSI
	& $\closed(\mkvs,\vi, \rel_{\WSI})$
	& $  \rel_{\SI}  \defeq \rel_{\UA} \cup \rel_{\CP}$ 
	& $\vshift[\MR \cap \RYW] \mkvs \vi {\mkvs'} {\vi'}$
	\\ \hline  
	\SI
	& $\closed(\mkvs,\vi, \rel_{\SI})$
	& $  \rel_{\SI}  \defeq \rel_{\UA} \cup \rel_{\CP} \cup (\WW_\mkvs; \RW_\mkvs)$ 
	& $\vshift[\MR \cap \RYW] \mkvs \vi {\mkvs'} {\vi'}$
	\\ \hline  
	\SER
	& $\closed(\mkvs,\vi, \rel_{\SER})$
	&$\rel_{\SER} \defeq \WW^{-1}$
	& \true	
	\\ \hline
    \SER*
    & $ \closed(\mkvs,\vi, \rel_{\SER^*})$
    & $\rel_{\SER^*} \defeq \rel_{\UA} \cup \SO \cup \WW_\mkvs \cup \WR_{\mkvs} \cup \RW_\mkvs$ 
    & $\vshift[\MR \cap \RYW] \mkvs \vi {\mkvs'} {\vi'}$
    \\ \hline  
\end{tabular}%
}
\vspace{0pt}
\caption{Execution tests of well-known consistency models, where \SER* denotes an alternative equivalent $\SER$ specification and $\SO$ is as given in \cref{subsec:kvstores}.
}
\label{fig:app-execution-tests}
\end{figure*}


\section{Relations to Dependency Graphs}
\label{app:depgraphs}
\label{sec:dependent-graph}
\emph{Dependency graphs} were introduced by Adya to define consistency models of transactional databases \cite{adya}. 
They are directed graphs consisting of transactions as nodes, 
each of which is labelled with transaction identifier and a set of read and write operations,
and labelled edges between transactions for describing how information flows between nodes. 
Specifically, a transaction $\txid$ reads a version for a key $\key$ that has been written by another transaction $\txid'$ 
(\emph{write-read dependency} \( \WR\)), overwrites a version of $\key$ written by $\txid'$ (\emph{write-write dependency} \( \WW \)),
or reads a version of $\key$ that is later overwritten by $\txid'$ (\emph{read-write anti-dependency} \( \RW \)). 
Note that we have named dependencies in kv-stores after the labelled edges of dependency graph. 
The main result of this Section shows that kv-stores are in fact isomorphic to dependency graphs, 
and dependencies in a kv-store naturally translates into a labelled edge in the associated dependency graph.

\begin{definition}
\label{def:dgraph}
A \emph{dependency graph} is a quadruple $\Gr = (\TtoOp{T}, \WR, \WW, \RW)$, where
\begin{itemize}
\item 
    $\TtoOp{T}: \TxID \parfun \pset{\Ops}$ is a partial mapping from transaction identifiers 
    to the set of operations, where there are at most one read operation and one write operation per key, 
    and such that $\TtoOp{T}(\txid_{0}) = \{(\otW, \key, \val_{0} \mid \key \in \Keys \}$; furthermore, 
    $\txid_{0} \in \dom(\TtoOp{T}$, and $\TtoOp{T}(\txid_{0}) = \{(\otW, \key, \val_{0}) \mid \key \in \Keys\}$, 
\item
    $\WR : \Keys \to \pset{\dom(\TtoOp{T}) \times \dom(\TtoOp{T})}$ is a function that 
maps each key $\key$ into a relation between transactions, such that for any $\txid, \txid_1, \txid_2, 
\key, \cl, m, n$: 
\begin{itemize}
\item if $(\otR, \key, \val) \in \TtoOp{T}(\txid)$, 
there exists $\txid' \neq \txid$ such that $(\otW, \key, \val) \in \TtoOp{T}(\txid')$, and $\txid' \toEDGE{\WR(\key)} \txid$, 
\item if $\txid_1 \toEDGE{\WR(\key)} \txid$ and $\txid_2 \toEDGE{\WR(\key)} \txid$, then 
$\txid_1 = \txid_2$.
\item if $\txid_{\cl}^{m} \toEDGE{\WR(\key)} \txid_{\cl}^{n}$, then $m < n$.
\end{itemize}
\item $\VO: \Keys \to \pset{\dom(\TtoOp{T}) \times \dom(\TtoOp{T})}$ is a function 
that maps each key into an irreflexive relation between transactions, such that for any $\txid, \txid', \key, \cl, m, n$, 
\begin{itemize}
\item if $\txid \toEDGE{\WW(\key)} \txid'$, then $(\otW, \key, \_) \in \TtoOp{T}(\txid), (\otW, \key, \_) \in \TtoOp{T}(\txid')$, 
\item if $(\otW, \key, \_) \in \TtoOp{T}(\txid), (\otW, \key, \_) \in \TtoOp{T}(\txid')$, then either $\txid = \txid'$, 
$\txid \toEDGE{\WW(\key)} \txid'$, or $\txid' \toEDGE{\WW(\key)} \txid$; furthermore, if $\txid = \txid_{0}$, 
then it must be the case that $\txid \toEDGE{(\WW(\key))} \txid'$,
\item if $\txid_{\cl}^{m} \toEDGE{\WW(\key)} \txid_{\cl}^{n}$, then $m < n$, 
\end{itemize}
\item $\AD: \Keys \to \pset{\dom(\TtoOp{T}) \times \dom(\TtoOp{T})}$ is defined 
by letting $\txid \toEDGE{\RW(\key)} \txid'$ if and only if $\txid'' \toEDGE{\WR(\key)} \txid$, 
$\txid'' \toEDGE{\WW(\key)} \txid'$ for some $\txid''$.
\end{itemize}
Let $\Dgraphs$ be the set of all dependency graphs.
\end{definition}

Given a dependency graph $\Gr = (\TtoOp{T}, \WR, \WW, \RW)$, we 
let $\WR_{\Gr} = \WR$, and similarly for $\WW$ and $\RW$. We also let $
\T_{\Gr} = \dom(\TtoOp{T})$, and write $(l, \key, \val) \in_{\Gr} \txid$ if 
$(;, \key, \val) \in \TtoOp{\Gr}(\txid)$. We
often 
commit an abuse of notation and use $\WR$ to denote the relation 
$\bigcup_{\key \in \Keys} \WR(\key)$; a similar notation is adopted for $\WW, \RW$. 
It will always be clear from the context whether the symbol $\WR$ refers to a function 
from keys to relations, or to a relation between transactions. 

As stated above, kv-stores are isomorphic to dependency graphs. The proof 
of this result is the topic of this Section. 

\begin{theorem}
\label{thm:kv2graph}
There is a one-to-one map between kv-stores and dependency graphs.
\end{theorem}
The proof structure of \cref{thm:kv2graph} is standard in its nature. 
We first how to encode a kv-store into a dependency graph. Then we 
show how to encode a dependency graph into a kv-store. Finally, 
we prove that the two constructions are one the inverse of the other: 
if we convert a kv-store $\mkvs$ into a dependency graph $\Gr_{\mkvs}$, 
then back to a kv-store $\mkvs_{\Gr_{\mkvs}}$, we obtain the initial kv-store.

To convert a kv-store $\mkvs$ into a dependency graph, we first define how 
to extract a fingerprint of a transaction identifier $\txid$ appearing in $\mkvs$:
\begin{definition}
\label{def:mkvs_fingerprint}
Let $\mkvs$ be a kv-store. For any transaction identifier $\txid$, we define 
$\fp_{\mkvs}(\txid)$ to be the smallest set such that whenever 
$\mkvs(\key, \_) = (\val, \txid, \_)$ then $(\otW, \key, \val) \in \txid$, and 
whenever $\mkvs(\key, \_) = (\val, \_, \{\txid\} \cup \_)$, then $(\otR, \key, \val) \in \txid$. 
\end{definition}
\begin{proposition}
\label{prop:mkvs_fp_welldefined}
For any $\mkvs, \txid$, the fingerprint $\fp_{\mkvs}(\txid)$ is well defined. 
That is, whenever $(\otW,\key,\val_1), (\otW,\key,\val_2) \in \fp_{\mkvs}(\txid)$, 
then $\val_1 = \val_2$, and whenever $(\otR, \key, \val_1), (\otR,\key, \val_2) \in \fp_{\mkvs}(\txid)$, 
then $\val_1 = \val_2$.
\end{proposition}

\begin{proof}
Suppose that $(\otW, \key, \val_1), (\otW,\key,\val_2) \in \fp_{\mkvs}(\txid)$ for some $key, 
\val_1, \val_2$. That is, there exist two indexes $i_1, i_2$ such that 
$\mkvs(\key, i_1) = (\val_1, \txid, \_)$, and $\mkvs(\key, i_2) = (\val_2, \txid, \_)$. 
That is, $\wtOf(\mkvs(\key, i_1)) = \wtOf(\mkvs(\key, i_2))$, and it follows 
from \cref{def:mkvs-appendix} that $i_1 = i_2$. In particular, this implies that $\val_1 = \val_2$. 

A similar argument can be used to prove that if $(\otR, \key, \val_1), (\otR,\key, \val_2) \in \fp_{\mkvs}(\txid)$, 
then $\val_1 = \val_2$. In this case, in fact, we have that there exist two indexes $i_1, i_2$ such that 
$\mkvs(\key, i_1) = (\val_1, \_,\{\txid\} \cup \_)$, and $\mkvs(\key, i_2) = (\val_2, \_, \{\txid\} \cup \_)$. 
Equivalently, $\txid \in \rsOf(\mkvs(\txid, i_1)) \cap \rsOf(\mkvs(\txid, i_2))$, and from 
\cref{def:mkvs-appendix} it must be the case that $i_1 = i_2$, hence $\val_1 = \val_2$.
\end{proof}

Using \cref{def:mkvs_fingerprint}, conerting a kv-store $\mkvs$  into a dependency graph is immediate, as the following 
definition shows: 

\begin{definition}
\label{def:kv2graph}
Given a kv-store $\mkvs$, the \emph{dependency graph} $\Gr_{\mkvs} = (\TtoOp{T}_{\mkvs}, \WR_{\mkvs}, 
\WW_{\mkvs}, \RW_{\mkvs})$ is defined by letting  $\TtoOp{T}_{\mkvs}(\txid)$ be defined if and only if
$\fp_{\mkvs}(\txid) \neq \emptyset$, in which case we let $\TtoOp{T}_{\mkvs}(\txid) = \fp_{\mkvs}(\txid)$. 
The relations $\WR_{\mkvs}, \WW_{\mkvs}, \RW_{\mkvs}$ are inherited directly from the transactional 
dependencies defined for $\mkvs$.
\end{definition}

\begin{definition}
\label{def:dependency-to-kv-store}
Given a dependency graph $\Gr = (\TtoOp{T}, \WR, \WW, \RW)$, we define the kv-store $\mkvs_{\Gr}$ as follows: 
\begin{enumerate}
\item for any transaction $\txid \in \dom(\TtoOp{T})$ such that $(\otW, \key, \val) \in \TtoOp{T}(\txid)$, 
    let $\txidset = \Set{ \txid' }[ \txid \toEDGE{\WR(\key)} \txid']$, and let $\ver(\txid, \key) = (\val, \txid, \txidset)$, 
\item For each key $\key$, let $\ver_{\key}^{0} = (\val_0, \txid_0, \txidset_k^{0})$, where $\txidset_{k}^{0} = \Set{ \txid }[ (\otR, \key, \stub) \in 
\TtoOp{T}(\txid) \land \fora{ \txid' } \neg( \txid' \toEDGE{\WR(\key)} \txid ]$. 
Let also $\Set{ \ver_{\key}^{i} }_{i = 1}^{n}$ be the ordered set of versions such that, for any 
$i=1,\cdots,n$, $\ver_{\key}^{i} = \ver(\txid, \key)$ for some $\txid$ such that $(\otW, \key, \_) \in \TtoOp{T}(\txid)$, 
and such that for any $i, j: 1 \leq i < j \leq n$, $\wtOf(\ver_{\key}^{i}) \toEDGE{\WW(\key)} \wtOf(\ver_{\key}^{j})$. 
Then we let $\mkvs_{\Gr}= \lambda \key. \prod_{i=0}^{n} \ver_{\key}^{i}$.
\end{enumerate}
\end{definition}

\begin{proposition}
\label{prop:well-formed-kv-store-to-dependency}
Let $\mkvs$ be a well-formed kv-store. Then $\Gr_{\mkvs}$ is a well-formed dependency graph.
\end{proposition}

\begin{proof}
Let $\mkvs$ be a (well-formed) kv-store. We need to show that 
$\Gr_{\mkvs} = (\TtoOp{T}_{\mkvs}, \WR_{\mkvs}, \WW_{\mkvs}, \RW_{\mkvs})$ is a dependency graph. 
As a first step, we show that $\Gr_{\mkvs}$ is a dependency graph, 
i.e. it satisfies all the constraints placed by \cref{def:dgraph}.

\begin{itemize}
\item Let $\txid \in \dom(\TtoOp{T}_{\mkvs})$, and suppose that $(\otR, \key, \val) \in \TtoOp{T}_{\mkvs}(\txid)$. 
We need to prove that there exists a transaction $\txid' \in \dom(\TtoOp{T}_{\mkvs})$ such
Because $(\otR, \key, \val) \in \TtoOp{T}_{\mkvs}(\txid)$, there 
must exist an index $i: 0 \leq i < \lvert \mkvs(\key) \rvert$ such that $\mkvs(\key, i) = (\val, \txid', \Set{\txid } \cup \_ )$ 
for some $\txid' \in \TxID$.  In this case we have that $\txid' \toEDGE{\WR_{\mkvs}(\key)} \txid$, 
and by \cref{def:mkvs_fingerprint} we have that $(\otW, \key, \val) \in_{\Gr_{\mkvs}} \txid'$.
\item Let $\txid \in \dom(\TtoOp{T}_{\mkvs})$, and suppose that there exist $\txid_1, \txid_2$ such that 
$\txid_{1} \toEDGE{\WR_{\key}(\mkvs)} \txid$, $\txid_{2} \toEDGE{\WR_{\key}(\mkvs)} \txid$. 
By \cref{def:kv2graph}, there exist two indexes $i, j: 0 \leq i, j < \lvert \mkvs(\key) \rvert$, such that 
$\mkvs(\key, i) = (\_, \txid_1, \Set{\txid} \cup \_)$, $\mkvs(\key, j) = (\_, \txid_2, \Set{\txid} \cup \_)$. 
We have that $\txid \in \rsOf(\mkvs(\key, i)) \cap \rsOf(\mkvs(\key, j))$, i.e. 
$\rsOf(\mkvs(\key,i)) \cap \rsOf(\mkvs(\key, j)) \neq \emptyset$. Because we are assuming 
that $\mkvs$ is well-formed, then it must be the case that $i = j$. This implies that $\txid_1 = \txid_2$.
\item Let $\cl \in \Clients$, $m, n \in \Nat$ and $\key \in \Keys$ be such that 
$\txid_{\cl}^{n} \toEDGE{\WR_{\mkvs}(\key)} \txid_{\cl}^{m}$.  We prove that 
$n < m$. By \cref{def:kv2graph}, it must be the case that 
there exists an index $i : 0 \leq i < \lvert \mkvs(\key) \rvert$ such that $\mkvs(\key, i) = 
(\_, \txid_{\cl}^{n}, \Set{\txid_{\cl}^{m}} \cup \_)$. Because $\mkvs$ is well-formed, 
it must be the case that $n < m$.
\item Let $\txid \in \dom(\TtoOp{T}_{\mkvs})$. We show that $\neg (\txid \toEDGE{\WW_{\mkvs}} \txid)$. 
We prove this fact by contradiction: suppose that $\txid \toEDGE{\WW_{\mkvs}(\key)} \txid$ for some key $\key$. By \cref{def:kv2graph}, 
there must exist two indexes $i,j: 0 \leq i < j < \lvert \mkvs(\key) \rvert$ such that $\txid = \wtOf(\mkvs(\key,i))$ and 
$\txid = \wtOf(\mkvs(\key, j))$. Because we are assuming that $\mkvs$ is well-formed, then it must be the 
case that $i = j$, contradicting the statement that $i < j$. 
\item Let $\txid, \txid'$ be such that $\txid' \toEDGE{\WW_{\key}(\mkvs)} \txid$. 
We must show that  $(\otW, \key, \_) \in \TtoOp{T}_{\mkvs}(\txid')$, and $(\otW, \key, \_) \in \TtoOp{T}_{\mkvs}(\txid)$.
By \cref{def:kv2graph}, there exist $i, j: 0 \leq i,j < \lvert \mkvs(\key) \rvert$ such that 
$\mkvs(\key, i) = (\val', \txid', \_)$ and $\mkvs(\key, j) = (\val, \txid, \_)$, for some 
$\val, \val' \in \Val$. \cref{def:kv2graph} also ensures that $(\otW, \key, \val') \in 
\TtoOp{T}_{\mkvs}(\txid')$, and $(\otW, \key, \val) \in \TtoOp{T}_{\mkvs}(\txid)$.
\item Let $\txid, \txid'$ be such that $(\otW, \key, \_) \in \TtoOp{T}_{\mkvs}(\txid)$ 
and $(\otW, \key, \_) \in \TtoOp{T}_{\mkvs}(\txid')$. We need to prove that 
either $\txid = \txid', \txid \toEDGE{\WW_{\mkvs}(\key)} \txid'$, or $\txid' \toEDGE{\WW_{\mkvs}(\key)} \txid$. 
By \cref{def:kv2graph} there exist two indexes $i, j: 0 < i,j< \lvert \mkvs(\key) \rvert$ such that 
$\mkvs(\key, i) = (\_, \txid, \_)$ and $\mkvs(\key, j) = (\_, \txid', \_)$. If $i = j$, then $\txid = \txid'$ 
and there is nothing left to prove. Otherwise, suppose without loss of generality that 
$i < j$. Then \cref{def:kv2graph} ensures that $\txid \toEDGE{\WW_{\mkvs}(\key)} \txid'$. 
\item Suppose that $\txid_{\cl}^{m} \toEDGE{\WW_{\mkvs}(\key)} \txid_{\cl}^{n}$ for 
some $\cl \in \Clients$ and $m, n \in \Nat$. We need to show that $m < n$. 
By \cref{def:kv2graph}, because  $\txid_{\cl}^{m} \toEDGE{\WW_{\mkvs}(\key)} \txid_{\cl}^{n}$ 
there exist two indexes $i,j: 0 < i,j < \lvert \mkvs(\key) \rvert$ such that $\wtOf(\mkvs(\key,i)) = \txid_{\cl}^{m}$ 
and $\wtOf(\mkvs(\key, j)) = \txid_{\cl}^{n}$. From the assumption that $\mkvs$ is well-formed, it 
follows that $n < m$.
\end{itemize}
\end{proof}

Next, we show how to convert a dependency graph $\Gr$ into a kv-store $\mkvs$. 
The main idea is that any transaction $\txid \in \T_{\Gr}$ induces a set of versions, and 
for each key $\key$, the write-write-dependency order $\WW_{\Gr}(\key)$ determines 
the order of these versions in $\mkvs_{\Gr}$. 

\begin{definition}
\label{def:kv-store-to-dependency-graph}
Let $\Gr$ be a dependency graph. Given a key $\key$, let $n_{\key}$, 
$\{\val_{i}^{\key}\}_{i=0}^{n_{\key}}$
$\{\txid^{\key}_{i}\}_{i=0}^{n_{\key}}$ be such that 
$\{\txid^{\key}_{i}\}_{i=0}^{n_{\key}}= \{ \txid \mid (\otW, \key, \val_{i}^{\key}) \in_{\Gr} \txid \}$, 
where the index set $\{1,\cdots,n_{\key}\}$  is chosen to be consistent
with $\WW_{\Gr}(\key)$: that is, $\txid_{i} \xrightarrow{\WW(\key)} \txid_{j}$ if 
and only if $i < j$. Given a key $\key$ and an index $i=1,\cdots, n_{\key}$, we also 
let $\T_{i}^{\key} = \{ \txid \mid \txid_{i}^{\key} \xrightarrow{\WR(\key)}\} \txid$. Note that 
this set is possibly empty. Finally, we let $\mkvs_{\Gr}$ be such that, for any $\key \in \Keys$, 
$\lvert \mkvs_{\Gr}(\key) \rvert = n_{\key}$, and for any $i=0,\cdots,n$, $\mkvs_{\Gr}(\key,i) = 
(\val_{i}^{\key}, \txid_{i}^{\key}, \T_{i}^{\key})$.
\end{definition}

\begin{proposition}
\label{prop:dependency-to-kv-store}
For any dependency graph $\Gr$, $\mkvs_{\Gr}$ is a (well-formed) kv-store.
\end{proposition}

\begin{proof}
We show that $\mkvs_{\Gr}$ satisfies all the constraints fromf \cref{def:mkvs-appendix}. Throughout 
the proof, we adopt the same notation of \cref{def:kv-store-to-dependency-graph}.

Let $\key \in \Keys$, and let $i,j$ be such that $\rsOf(\mkvs_{\Gr}(\key,i)) \cap \rsOf(\mkvs_{\Gr}(\key,j)) \neq \emptyset$, 
that is there exists a transaction $\txid \in \rsOf(\mkvs_{\Gr}(\key, i)) \cap \rsOf(\mkvs_{\Gr}(\key,j))$. We show that $i = j$. 
By definition, $\rsOf(\mkvs_{\Gr}(\key, i)) = \T_{\key}^{i}$, and $\rsOf(\mkvs_{\Gr}(\key, j)) = \T_{\key}^{j}$. 
\cref{def:kv-store-to-dependency-graph} ensures that $\txid^{\key}_{i} \xrightarrow{\WR_{\Gr}(\key)} \txid$, 
and $\txid^{\key}_{j} \xrightarrow{\WR_{\Gr}(\key)} \txid$. By definition of dependency graph, it must be the 
case that $\txid^{\key}_{i} = \txid^{\key}_{j}$, and because the order of writers 
transactions in versions in $\mkvs_{\Gr}(\key)$ 
is defined to be consistent with $\WW_{\Gr}(\key)$, then it must also be the case that $i = j$. 

Suppose no that $\key, i, j$ are such that $\wtOf(\mkvs_{\Gr}(\key, i)) = \wtOf(\mkvs_{\Gr}(\key, j))$. By 
definition $\wtOf(\mkvs_{\Gr}(\key, i)) = \txid_{i}^{\key}$, and $\wtOf(\mkvs_{\Gr}(\key, j) = \txid^{\key}_{j}$. 
That is, $\txid_{i}^{\key} = \txid_{j}^{\key}$. Because the order of writer transactions in $\mkvs_{\Gr}(\key)$ 
is consistent with $\WW_{\Gr}(\key)$, we also have that $i = j$.

Next, note that for any key $\key$, $\txid^{\key}_{0} = \txid_{0}$. In fact, because $\txid_{0} \in_{\Gr} 
(\otW, \key, \val_{0})$, we have that $\txid_{0} = \txid^{\key}_{i}$ for some $i=0,\cdots, n_{\key}$.
Also, because whenever $\txid$ is such that $(\otW, \key, \_) \in_{\Gr} \txid$, then 
it must be the case that $\txid_{0} \xrightarrow{\WW(\key)} \txid$, then it must be 
the case that $i = 0$. It follows that, for any $\key \in \Keys$, $\mkvs_{\Gr}(\key, 0) = (\val_{0}, \txid_{0}, \_)$.

Finally, suppose that $\txid_{\cl}^{n} = \wtOf(\mkvs_{\Gr}(\key, i))$, $\txid_{\cl}^{m}  = \wtOf(\mkvs_{\Gr}(\key, j))$ 
for some $i, j$ such that $i < j$. In this case we have that $\txid_{\cl}^{n} = \txid_{i}^{\key}$, $\txid_{\cl}^{m} = 
\txid_{j}^{\key}$, and because $i < j$ it must be the case that $\txid_{\cl}^{n} \xrightarrow{\WW_{\Gr}(\key)} 
\txid_{\cl}^{m}$. The definition of dependency graph ensures then that it must $n < m$. A similar 
argument shows that, if $\txid_{\cl}^{n} \in \wtOf(\mkvs_{\Gr}(\key, i)), \txid_{cl}^{m} \in \rsOf(\mkvs_{\Gr}(\key, i))$, 
then it must be the case that $\txid_{\cl}^{n} \xrightarrow{\WR_{\Gr}(\key)} \txid_{\cl}^{m}$, and therefore 
$n < m$.
\end{proof}

Finally, we need to show that the two constructions outlined in \cref{def:kv-store-to-dependency-graph} and 
\cref{def:dependency-to-kv-store} are one the inverse of the other.

\begin{proposition}
\label{prop:kv-store-back-and-forth}
For any kv-store $\mkvs$, $\mkvs_{\Gr_{\mkvs}} = \mkvs$.
\end{proposition}

\begin{proof}
We prove that for any $\key \in \Keys$, 
$\mkvs(\key) = \mkvs_{\Gr_{\mkvs}}(\key)$.

Let then $\key \in \Keys$, and suppose that $\mkvs(\key) = (\val_{0}, \txid_{0}, \T_{0}) \cdots (\val_{n}, \txid_{n}, \T_{n})$. 
By construction, in $(\otW, \key, \val_{i}) \in_{\Gr_{\mkvs}} \txid_{i}$, and whenever there is a transaction 
$\txid$ such that $(\otW, \key, \txid) \in_{\Gr_{\mkvs}} \txid$, then $\txid = \txid_{i}$ for some $i=0,\cdots, n$. 
In particular, we have that $\txid_{0} \xrightarrow{\WW_{\mkvs}(\key)} \cdots \xrightarrow{\WW_{\mkvs}(\key)} \txid_{n}$ 
completely characterises the write-write-dependency relation $\WW_{\mkvs}(\key)$ over $\mkvs_{\Gr}$ 
 (recall that, by \cref{def:kv-store-to-dependency-graph}, $\WR_{\Gr_{\mkvs}} = \WR_{\mkvs}$).
By definition, we have that $\mkvs_{\Gr_{\mkvs}} = (\val_{0}, \txid_{0}, \T'_{0}) \cdots (\val_{n}, \txid_{n}, \T'_{n})$. 

It remains to prove that, for any $i=0,\cdots, n$, $\T'_{i} = \T_{i}$.
For any $i=0,\cdots, n$, and transaction $\txid \in \T_{i}$, \cref{def:kv-store-to-dependency-graph} ensures that 
$\txid_{i} \xrightarrow{\WR_{\mkvs}} \txid$, 
and by \cref{def:dependency-to-kv-store} it must be the case that $\txid \in \T'_{i}$.
Furthermore, if $\txid' \in \T'_{i}$, then from \cref{def:dependency-to-kv-store} it must be 
the case that $\txid_{i} \xrightarrow{\WR_{\Gr_{mkvs}}(\key)} \txid'$, 
or equivalently $\txid_{i} \xrightarrow{\WR_{\mkvs}(\key)} \txid'_{i}$.
(\cref{def:kv-store-to-dependency-graph}). Then it must be the case that $\txid' \in \T_{i}$. 
\end{proof}

\begin{proposition}
\label{prop:dependency-back-and-forth}
For any dependency graph $\Gr$, $\Gr_{\mkvs_{\Gr}} = \Gr$. 
\end{proposition}

\begin{proof}
The proof of this claim is similar to \cref{prop:kv-store-back-and-forth}, and therefore omitted.
\end{proof}

\section{Operational Semantics of Abstract Executions}
\label{sec:abstract-execution}

Abstract executions are a framework originally introduced in \cite{ev_transactions} 
to capture the run-time behaviour of clients interacting with a database. 
In abstract execution, two relations between transactions are introduced: 
the \emph{visibility} relation establishes when a transaction observes the effects of another transaction; 
and the \emph{arbitration} relation helps to determine the value of a key $\key$ read by a transaction, 
in the case that the transaction observes multiple updates to $\key$ performed by different transactions. 

\begin{definition}
\label{def:absexec}
\label{def:aexec}
An abstract execution is a triple $\aexec = (\TtoOp{T}, \VIS, \AR)$, where 
\begin{itemize}
    \item $\TtoOp{T}: \TxID \parfun \pset{\Ops}$ is a partial, 
finite function mapping transaction identifiers to the set of operations that they perform, 
with $\TtoOp{T}(\txid_{0}) = \{ (\otW, \key, \val_{0} \mid \key \in \Keys\}$,
\item $\VIS \subseteq \dom(\TtoOp{T}) \times \dom(\TtoOp{T})$ is an irreflexive relation, 
called \emph{visibility}, 
\item $\AR \subseteq \dom(\TtoOp{T}) \times \dom(\TtoOp{T})$ is a strict, total order 
such that $\VIS \subseteq \AR$, and whenever $\txid_{\cl}^{n} \toEDGE{\AR} 
\txid_{\cl}^{m}$, then $n < m$.
\end{itemize} 
The set of abstract executions is denoted by $\aeset$.
\end{definition}

Given an abstract execution $\aexec = (\TtoOp{T}, \VIS, \AR)$, 
the notation $\TtoOp{T}_{\aexec} = \TtoOp{T}$,
$\txidset_{\aexec} = \dom(\TtoOp{T})$, $\VIS_{\aexec} = \VIS$ 
and $\AR_{\aexec} = \AR$. 
The session order for a client \( \SO_{\aexec}(\cl) \)  and 
then the overall session order \( \SO_\aexec\) are defined as the following:
\[
    \SO_{\aexec}(\cl) = \Set{(\txid_{\cl}^{n}, \txid_{\cl}^{m})}[ \cl \in \Clients 
    \land \txid_{\cl}^{n} \in \txidset_{\aexec} \land \txid_{\cl}^{m} \in \txidset_{\aexec} \land n < m]
\]
and 
\[
    \SO_{\aexec} = \bigcup_{\cl \in \Clients} \SO_{\aexec}(\cl)
\]
The notation $(\otR, \key, \val) \in_{\aexec} \txid$ denotes $(\otR, \key, \val) \in \TtoOp{T}_{\aexec}(\txid)$, 
and similarly for write operations \( (\otW, \key, \val) \in_{\aexec} \txid \).
Given an abstract execution $\aexec$, a transaction $\txid \in \txidset_{\aexec}$, and a key $\key$, 
the visible writers set 
$\visibleWrites_{\aexec}(\key, \txid) \defeq \Set{ \txid' }[ \txid' \toEDGE{\VIS_{\aexec}} \txid \land (\otW, \key, \stub) \in_{\aexec} \txid']$.

The operational semantics on abstract executions (\cref{fig:aexec.semantics}) is parametrised in the axiomatic definition $(\RP, \Ax)$ of a consistency model:
transitions take the form $(\aexec, \ThdEnv, \prog) \toAEXEC{\_}_{(\RP, \Ax)} (\aexec', \ThdEnv', \prog')$. 
An axiomatic definition of a consistency model is given by a pair $(\RP, \Ax)$, 
where $\RP$ is a resolution policy (\cref{def:rp}) 
and $\Ax$ is a set of axioms for visibility relation (\cref{def:aexec-axioms}).
An abstract execution $\aexec$ satisfies the consistency model, 
written $\aexec \models (\RP, \Ax)$ if it satisfies its individual components. 
The set of abstract executions induced by an axiomatic definition is given 
by $\CMa(\RP, \Ax) = \Set{ \aexec }[ \aexec \models (\RP, \Ax)]$.

We first introduce a notation of two abstract executions \emph{agree}.
Given two abstract executions $\aexec_1, \aexec_2 \in \aeset$ and set of transactions $\txidset \subseteq \txidset_{\aexec_1} \cap \txidset_{\aexec_2}$,
 $\aexec_1$ and $\aexec_2$ \emph{agree} on $\txidset$ if and only if for any transactions \( \txid \) \( \txid' \) in \( \txidset \):
\[
\begin{array}{l}
    \TtoOp{T}_{\aexec_1}(\txid) = \TtoOp{T}_{\aexec_2}(\txid) \land 
((\txid \toEDGE{\VIS_{\aexec_1}} \txid') \iff (\txid \toEDGE{\VIS_{\aexec_2}} \txid'))
\land ((\txid \toEDGE{\AR_{\aexec_1}} \txid') \iff (\txid \toEDGE{\AR_{\aexec_2}} \txid'))
\end{array}
\]
\begin{definition}
\label{def:rp}
A resolution policy $\RP$ is a function $\RP: \aeset \times \pset{\TxID} \rightarrow \pset{\Snapshots}$ 
such that, for any $\aexec_1, \aexec_2$ that agree on a subset of transactions $\txidset$, then 
$\RP(\aexec_1, \txidset) = \RP(\aexec_2, \txidset)$.
An abstract execution $\aexec$ satisfies the execution policy $\RP$ if, 
\[
    \fora{\txid \in \txidset_{\aexec} } 
    \exsts{ \sn \in \RP(\aexec, \VIS_{\aexec}^{-1}(\txid)) }
    \fora{ \key,\val } (\otR, \key, \val) \in_{\aexec} \txid 
    \implies \sn(\key) = \val
\]
\end{definition}

\begin{definition}
\label{def:aexec-axioms}
An axiom $\A$ is a function from abstract executions to relations between 
transactions, $\A: \aeset \rightarrow \pset{\TxID \times \TxID}$, 
such that whenever $\aexec_1, \aexec_2$ agree on a subset of 
transactions $\txidset$, then $\A(\aexec_1) \cap (\txidset \times \txidset) \subseteq \A(\aexec_2)$.
\end{definition}

Axioms of a consistency model are constraints of the form $\A(\aexec) \subseteq \VIS_{\aexec}$. 
For example, if we require $A(\aexec) = \AR_{\aexec}$, 
then the corresponding axiom is given by $\AR_{\aexec} \subseteq \VIS_{\aexec}$,
thus capturing the serialisability of transactions,
\ie this axiom is equivalent to require that $\VIS_{\aexec}$ is a total order.
The requirement on subsets of transactions on which 
abstract executions agree will be needed later, 
when we define an operational semantics of transactions where 
clients can append a new transaction $\txid$ at the tail of an abstract execution $\aexec$,
which satisfies an axiom $\A$. This requirement ensures that 
we only need to check that the axiom is $\A$ is satisfied by the pre-visibility 
and pre-arbitration relation of the transaction $\txid$ in $\aexec'$. 
In fact, the resulting abstract execution $\aexec'$ agrees with $\aexec$ on the set $\txidset_{\aexec}$: 
in this case we'll note that we can rewrite 
$\A(\aexec') = \A(\aexec') \cap ((\txidset_{\aexec} \times \txidset_{\aexec}) ) \cup (\txidset_{\aexec} \times \Set{\txid}))$.
Then
$\A(\aexec') \cap ((\txidset_{\aexec} \times \txidset_{\aexec})) \subseteq \A(\aexec) \cap (\txidset_{\aexec} \times \txidset_{\aexec}) \subseteq \VIS_{\aexec} \cap (\txidset_{\aexec} \times \txidset_{\aexec}) \subseteq \VIS_{\aexec'}$, 
hence we only need to check that $\A(\aexec') \cap (\txidset_{\aexec} \times \Set{\txid}) \subseteq \VIS_{\aexec'}$.

We say that an abstract execution $\aexec$ satisfies an axiom $\A$, 
if $\A(\aexec) \subseteq \VIS_{\aexec}$. 
An abstract execution $\aexec$ satisfies $(\RP,\Ax)$, written $\aexec \models (\RP,\Ax)$,  
if the abstract execution \( \aexec \) satisfies \( \RP \) and \( \Ax \).

\begin{definition}[Abstract executions induced by axiomatic definition]
\label{def:axiom-to-aexec}
The set of all abstract executions induced by an axiomatic definition, \( \CMa(\RP,\Ax)\) is defined as \( \CMa(\RP,\Ax) \defeq \Set{\aexec}[\aexec \models (\RP,\Ax)]\).
\end{definition}

\begin{figure}[t]

\hrulefill

\begin{align*}
	\toAEXEC{}  & : 
    \begin{multlined}[t]
    \Clients \; \times \; 
	\left( ( \Aexecs \times \Stacks ) \times \Commands \right) 
    \; \times\; \ETs \; \times \; \sort{Label} \;\times  
	\left( ( \Aexecs \times \Stacks ) \times \Commands \right) 
    \end{multlined}
\end{align*}
\begin{mathpar}
    \inferrule[\rl{AAtomicTrans}]{
        \txidset \subseteq \txidset_{\aexec} \qquad \sn \in \RP(\aexec, \txidset) \qquad
		(\stk, \sn, \emptyset), \trans \ \toTRANS^{*} \  (\stk', \stub,  \fp) , \pskip \\\\
		\txid \in \nextTxid(\txidset_{\aexec}, \cl) \qquad \aexec' = \extend(\aexec, \txid, \txidset, \fp) \qquad 
    \fora{ A \in \Ax } \Set{\txid' }[ (\txid', \txid) \in \A(\aexec') ] \subseteq \txidset
    }{
    \cl \vdash ( \aexec, \stk ), \ptrans{\trans} \ \toAEXEC{(\cl, \txidset,\fp)}_{(\RP, \Ax)} \ ( \aexec', \stk' ) , \pskip
    }
    \and
    \inferrule[\rl{APrimitive}]{
        \stk \toLTS{\cmdpri} \stk'
    }{%
    \cl \vdash ( \aexec, \stk ) , \cmdpri \ \toAEXEC{(\cl,\iota)}_{\ET} \  ( \aexec, \stk' ) , \pskip
    }
    \and
    \inferrule[\rl{AChoice}]{
        i \in \Set{1,2}
    }{%
        \cl \vdash ( \aexec, \stk ) , \cmd_{1} \pchoice \cmd_{2} \ \toAEXEC{(\cl,\iota)}_{\ET} \  ( \aexec, \stk ) , \cmd_{i}
    }
    \quad
    \inferrule[\rl{AIter}]{ }{%
        \cl \vdash ( \aexec, \stk ) , \cmd\prepeat \ \toAEXEC{(\cl,\iota)}_{\ET} \  ( \aexec, \stk ) , \pskip \pchoice (\cmd \pseq \cmd\prepeat)
    }
    \and
    \inferrule[\rl{ASeqSkip}]{ }{%
        \cl \vdash ( \aexec, \stk ) , \pskip \pseq \cmd \ \toAEXEC{(\cl,\iota)}_{\ET} \  ( \aexec, \stk ) , \cmd
    }
    \quad
    \inferrule[\rl{ASeq}]{%
        \cl \vdash ( \aexec, \stk ) , \cmd_{1} \ \toAEXEC{(\cl,\iota)}_{\ET} \  ( \aexec, \stk' ) , {\cmd_{1}}' 
    }{%
        \cl \vdash ( \aexec,\stk ) , \cmd_{1} \pseq \cmd_{2} \ \toAEXEC{(\cl,\iota)}_{\ET} \ ( \aexec, \stk' ) , {\cmd_{1}}' \pseq \cmd_{2}
    }
\end{mathpar}

\hrulefill

\[
	\toAEXEC{} : 
    ( \Aexecs \times \ThdEnv \times \Programs) 
    \;\times\; \ETs \; \times \sort{Label} \times \;
    ( \Aexecs \times \ThdEnv \times \Programs) 
\]
\[
    \inferrule[\rl{AProg}]{%
         \cl \vdash ( \aexec, \thdenv(\cl) ) , \prog(\cl), \ \toAEXEC{\lambda}_{(\RP, \Ax)} \  ( \aexec', \stk' ) , \cmd'  
    }{%
         (\aexec, \thdenv ), \prog  \ \toAEXEC{\lambda}_{(\RP, \Ax)} \  ( \aexec', \thdenv\rmto{\cl}{\stk'} ) , \prog\rmto{\cl}{\cmd'} ) 
    }
\]
\hrulefill
\caption{Operational Semantics on Abstract Executions}
\label{fig:aexec.semantics}
\end{figure}

The \cref{fig:aexec.semantics} presents all rules of the operational semantics of programs based on abstract executions. 
The\rl{ACommit} rule is the abstract execution counterpart of rule \rl{PCommit} for kv-stores.
The\rl{ACommit} models how an abstract execution $\aexec$ evolves 
when a client wants to execute a transaction whose code is $\ptrans{\trans}$. 
In the rule, $\txidset$ is the set of transactions of $\aexec$ that are visible to the client $\cl$ that wishes to execute $\ptrans{\trans}$.
Such a set of transactions is used to determine a snapshot $\sn \in \RP(\aexec, \txidset)$ that 
the client $\cl$ uses to execute the code $\ptrans{\trans}$, and obtain a fingerprint $\fp$. 
This fingerprint is then used to extend abstract execution $\aexec$ 
with a transaction from the set $\nextTxid(\txidset_{\aexec}, \cl)$.
Similar \rl{PProg} rule, the \rl{AProg} rule in \cref{fig:aexec.semantics}
models multi-clients concurrency in an interleaving fashion. 
All the rest rules of the abstract operational semantics in \cref{fig:aexec.semantics}
have a similar counterpart in the kv-store semantics.

Note that \rl{AAtomicTrans} is more general than Rule \rl{PAtomicTrans} in the kv-store semantics.
In the latter, the snapshot of a transaction is uniquely determined from a view of the client,
in a way that roughly corresponds to the last write wins policy in the abstract execution framework. 
In contrast, the snapshot of a transaction used in \rl{AAtomicTrans}
is chosen non-deterministically from those made available to the client by 
the resolution policy $\RP$, which may not necessarily be last-write-win. 

Throughout this report we will work mainly with the \emph{Last Write Wins} resolution policy (\cref{def:lww}).
When discussing the operational semantics of transactional programs, 
we will also introduce the \emph{Anarchic} resolution policy.

\begin{definition}
\label{def:lww}
The Last Write Wins resolution policy $\RP_{\LWW}$ is defined as 
$\RP_{\LWW}(\aexec, \txidset) \defeq \Set{\sn}$ where
\[
\sn = \lambda \key. \text{let} \ \txidset_{\key} = ( \txidset \cap \Set{\txid }[ (\otW,\key, \stub) \in_{\aexec} \txid])  \text{ in }
\begin{cases}
    \val_{0} & \text{ if } \txidset_{\key} =  \emptyset\\
\val & \text{ if } (\otW, \key, \val) \in_{\aexec} \max_{\AR_{\aexec}}(\txidset_{\key})
\end{cases}
\]
\end{definition}

\section{Relationship between kv-stores and abstract execution}
\label{sec:app-abstract-semantics-sound-complete}
\subsection{KV-Store to Abstract Executions}
\label{app:aexec2kv}
\label{sec:thm:aexec2kv-compatible-proof}

We introduce the definition of the dependency graph induced an abstract execution:

\begin{definition}
\label{def:aexec2graph}
Given an abstract execution \(\aexec\) that satisfies the last write wins policy,
the dependency graph $\graphOf[\aexec] \defeq (\TtoOp{T}_{\aexec}, \WR_{\aexec}, 
\WW_{\aexec}, \RW_{\aexec})$ is defined by letting
\begin{itemize}
\item \(\txid \toEDGE{\WR_{\aexec}(\key)} \txid'\) if and only if 
\(\txid = \max_{\AR_{\aexec}}(\visibleWrites_{\aexec}(\key, \txid'))\), 
\item \(\txid \toEDGE{\WW_{\aexec}(\key)} \txid'\) if and only 
\(\txid, \txid' \in_{\aexec} (\otW, \;\key, \stub)\) 
and \(\txid \toEDGE{\AR_{\aexec}} \txid'\),
\item \(\txid \toEDGE{\RW_{\aexec}(\key)} \txid'\) if and only if either 
\((\otR, \key, \stub) \in_{\aexec} \txid, (\otW, \key, \stub) \in_{\aexec} \txid'\) and 
whenever \(\txid'' \toEDGE{\WR_{\aexec}(\key)} \txid\), 
then \(\txid'' \toEDGE{\WW_{\aexec}(\key)} \txid'\).
\end{itemize}
\end{definition}

Note that each abstract execution \(\aexec\) determines a kv-store \(\mkvs_{\aexec}\),
as a result of \cref{def:aexec2graph} and \cref{thm:kv2graph}. 
Let \(\mkvs\) be the unique kv-store such that \(\Gr_{\mkvs} = \graphOf[\aexec]\), then \(\mkvs_{\aexec} = \mkvs\). 
As we will discuss later in this Section,
this mapping \(\mkvs_{(\stub)}\) is NOT a bijection, 
in that several abstract executions may be encoded in the same kv-store.
Because kv-stores abstract away the total arbitration order of transactions.

Upon the relation \( \mkvs_{\aexec} = \mkvs \),
there is a deeper link between kv-store plus views and abstract exertions.
This notion, named \emph{compatibility}, bases on the intuition that 
clients can make observations over kv-stores and abstract executions, in terms of snapshots.

In kv-stores, observations are snapshots induced by views. 
While in abstract executions, observations correspond to the snapshots induced by the visible transactions.
Note that it is under the condition that the abstract execution adopts \(\RP_{\LWW}\) resolution policy.
This approach is analogous to the one used by operation contexts in \cite{repldatatypes}.
Thus, a kv-store \(\mkvs\) is \emph{compatible} with an abstract execution \(\aexec\), written \( \mkvs \compatible \aexec \)
if any observation made on \(\mkvs\) can be replicated by an observation made on \(\aexec\), and vice-versa. 

\begin{definition}
\label{def:compatible}
Given a kv-store \(\mkvs\),
an abstract execution \(\aexec\) is compatible with \(\mkvs\), written 
\(\aexec \compatible \mkvs\), if and only if there exists a  mapping 
\(f: \pset{\txidset_{\aexec}} \rightarrow \Views(\mkvs)\)
such that  
\begin{itemize}
\item for any subset \(\txidset \subseteq \txidset_{\aexec}\), then \(\RP_{\LWW}(\aexec, \txidset) = \Set{\snapshot[\mkvs, f(\txidset)]}\); 
\item for any view \(\vi \in \Views(\mkvs)\), there exists a subset \(\txidset \subseteq \txidset_{\aexec}\) 
such that \(f(\txidset) = \vi\), and \(\RP_{\LWW}(\aexec, \txidset) = \Set{\snapshot[\mkvs_{\aexec}, \vi]}\).
\end{itemize}
\end{definition}

The function \(\getView[\aexec, \txidset]\) defines the view on \( \mkvs_\aexec \) that corresponds to \( \txidset \) as the following:
\[
    \getView[\aexec, \txidset] \defeq \lambda \key. \Set{0} \cup \Set{i}[\wtOf(\mkvs_{\aexec}(\key, i)) \in \txidset]
\]
Inversely, the function \( \Tx[\mkvs, \vi] \) converts a view to a set of observable transactions:
\[
    \Tx[\mkvs, \vi] \defeq \Set{\wtOf(\mkvs(\key, i))}[\key \in \Keys \land i \in \vi(\key)]
\]
Given \( \getView \), \( \Tx \), \cref{def:compatible}, 
it follows \( \aexec \compatible \mkvs_{\aexec} \) shown in \cref{thm:aexec2kv.compatible}.

\begin{theorem}
\label{thm:aexec2kv.compatible}
For any abstract execution \(\aexec\) that satisfies the last write wins policy, \(\aexec \compatible \mkvs_{\aexec}\).
\end{theorem}
\begin{proof}
Given the function \(\getView[\aexec, \cdot]\) from \(\pset{\txidset_{\aexec}}\) to \(\Views(\mkvs_{\aexec})\),
we prove it satisfies the constraint of \cref{def:compatible}.
Fix a set of transitions \( \txidset \).
By the \cref{prop:getview.valid}, the view \(\getView[\aexec, \txidset]\)  on \( \mkvs_\aexec \) is a valid view,
that is, \( \getView[\aexec, \txidset] \in \Views(\mkvs_\aexec) \).
Given that it is a valid view, the \cref{prop:compatible.aexec2kv} proves:
\begin{equation}
    \label{equ:visible-trans-to-view}
    \RP_{\LWW}(\aexec, \txidset) = \Set{\snapshot[\mkvs_{\aexec}, \getView[\aexec, \txidset]]} 
\end{equation}

The another way round is more subtle,
because \( \txidset \) contains any read only transaction.
By \cref{prop:getview.tx}, it is safe to erase read only transactions from \( \txidset \),
when calculating the view \( \getView[\aexec, \txidset] \).
Last, by \cref{prop:compatible.kv2aexec}, we prove the following:
\begin{equation}
    \label{equ:view-to-visible-trans}
    \RP_{\LWW}(\aexec, \txidset) = \snapshot[\mkvs_{\aexec}, \vi]
\end{equation}
By \cref{equ:visible-trans-to-view} and \cref{equ:view-to-visible-trans},
it follows \( \aexec \compatible \mkvs_{\aexec} \).
\end{proof}

\begin{proposition}[Valid views]
\label{prop:getview.valid}
For any abstract execution \(\aexec\), and \(\txidset \subseteq \txidset_{\aexec}\), 
\(\getView[\aexec, \txidset] \in \Views(\mkvs_{\aexec})\).
\end{proposition}
\begin{proof}
Assume an abstract execution \(\aexec\), a set of transactions \(\txidset \subseteq \txidset_{\aexec}\), and a key \( \key \).
By the definition of \(\getView[\aexec, \txidset]\), 
then \(0 \in \getView[\aexec, \txidset](\key)\), and 
\(0 \leq i < \abs{ \mkvs_{\aexec}(\key) }\) for any index \( i \) such that \(i \in \getView[\aexec, \txidset](\key)\).
Therefore we only need to prove that \(\getView[\aexec, \txidset]\) satisfies \eqref{eq:view.atomic}.
Let \(j \in \getView[\aexec, \txidset](\key)\) for some key \(\key\), and let \(\txid = 
\wtOf(\mkvs_{\aexec}(\key, j))\). Let also \(\key', i\) be such that 
\(\wtOf(\mkvs_{\aexec}(\key', i)) = \txid\). We need to show that 
\(i \in \getView[\aexec, \txidset](\key')\). Note that it \(\txid = \txid_{0}\) 
then \(\wtOf(\mkvs_{\aexec}(\key', i)) = \txid\) only if \(i = 0\), and 
\(0 \in \getView[\aexec, \txidset](\key')\) by definition. 
Let then \(\txid \neq \txid_{0}\). Because \(\wtOf(\mkvs_{\aexec}(\key, j)) = \txid\) 
and \(j \in \getView[\aexec, \txidset]\), then it must be the case that \(\txid \in \txidset\). 
Also, because \(\wtOf(\mkvs_{\aexec}(\key', i)) = \txid\), then \((\otW, \key, \stub) \in 
\TtoOp{T}_{\aexec}(\txid)\). It follows that there exists an index \(i' \in \getView[\aexec, \txid](\key')\) 
such that \(\wtOf(\mkvs_{\aexec}(\key', i')) = \txid\). By definition of 
\(\mkvs_{\aexec}\), if \(\wtOf(\mkvs_{\aexec}(\key', i')) = \txid\), then it must 
be \(i' = i\), and therefore \(i \in \getView[\aexec, \txid](\key')\).
\end{proof}

\begin{proposition}[Visible transactions to views]
\label{prop:compatible.aexec2kv}
For any subset \(\txidset \subseteq \txidset_{\aexec}\), \(\RP_{\LWW}(\aexec, \txidset) = \Set{\snapshot[\mkvs_{\aexec}, \getView[\aexec, \txidset]]}\).
\end{proposition}

\begin{proof}
Fix \(\txidset \subseteq \aexec\), and let \(\Set{\mkvs} = \RP_{\LWW}(\aexec, \txidset)\). We prove that, for any \(\key \in \Keys\), 
\(\mkvs(\key) = \snapshot[\getView[\aexec, \txidset]](\key)\). There are two different cases: 
\begin{enumerate}
    \item \(\txidset \cap \Set{ \txid }[ (\otW, \key, \stub) \in_{\aexec} \txid ] = \emptyset\). 
In this case \(\mkvs(\key) = \val_0\). 
We know that \(\graphOf[\aexec]\) satisfies all the constraints required by the definition of dependency graph 
(\cite{laws}). Together with \cref{thm:kv2graph} it follows that \(\mkvs_{\aexec}(\key, 0) = (\val_0, \txid_0, \stub)\).
We prove that \(\getView[\aexec, \txidset](\key) = \Set{0}\), 
hence 
\[ 
\snapshot[\mkvs_{\aexec}, \getView[\aexec, \txidset]](\key) = \valueOf(\mkvs_{\aexec}(\key, 0)) = \val_{0}
\]
Note that whenever \((\otW, \key, \stub) \in_{\aexec} \txid\) for some \(\txid\), then 
\(\txid \notin \txidset\). Therefore, whenever \((\val, \txid, \stub) = \mkvs_{\aexec}(\key, i)\) for some \(i \geq 0\), then 
\(\txid \notin \txidset\).
\[
\getView[\aexec, \txidset](\key) = \Set{0} \cup \Set{i }[ \wtOf(\mkvs_{\aexec}(\key, i)) \in \txidset)] = \Set{0} \cup \emptyset = \Set{0}
\]
\item Suppose now that \(\txidset \cap \Set{ \txid }[ (\otW, \key, \stub) \in_{\aexec} \txid ] \neq \emptyset\). 
Let then \(\txid = \max_{\AR_{\aexec}}(\txidset \cap \Set{\txid }[ (\otW, \key, \stub) \in_{\aexec} \txid])\). 
Then \((\otW, \key, \val) \in_{\aexec} \txid\) for some \(\val \in \Val\). Furthermore, \(\RP_{\LWW}(\aexec, \txidset)(\key) = \val\).
By definition, \(\txid' \in \txidset \cap \Set{ \txid }[ (\otW, \key, \stub) \in_{\aexec} \txid]\), 
then either \(\txid' = \txid\) or \(\txid' \toEDGE{\AR_{\aexec}} \txid\). The definition of 
\(\graphOf[\aexec]\) gives that \(\txid' \toEDGE{\WW_{\aexec}(\key)} \txid\). 
Because \((\otW, \key, \val) \in_{\aexec} \txid\), then there exists an index 
\(i \geq 0\) such that \(\mkvs_{\aexec}(\key, i) = (\val, \txid, \stub)\). Furthermore, 
whenever \(\wtOf(\key, j) = \txid'\) for some \(\txid'\) and \(j > i\), then it must 
be the case that \(\txid \toEDGE{\WW_{\aexec}(\key)} \txid'\), and because 
\(\WW_{\aexec}(\key)\) is transitive and irreflexive, it must be that  
\(\neg( \txid' \toEDGE{\WW_{\aexec}(\key)} \txid)\) and \(\txid \neq \txid'\): this implies that 
\(\txid' \notin \txidset\). It follows that \(\max(\getView[\aexec, \txidset](\key)) = i\), hence 
\(\snapshot[\mkvs_{\aexec}, \getView[\aexec, \txidset]] = \valueOf(\mkvs_{\aexec}(\key, i)) = \val\).
\end{enumerate}
\end{proof}

\begin{proposition}[Read-only transactions erasing]
\label{prop:getview.tx}
Let \(\vi \in \Views(\mkvs_{\aexec})\), and let \(\txidset \subseteq \txidset_{\aexec}\) be a 
set of read-only transactions in \(\aexec\). Then 
\(\getView[\aexec, \txidset \cup \Tx[\mkvs_{\aexec}, \vi]] = \vi\). 
\end{proposition}

\begin{proof}
Fix a key \(\key\). Suppose that \(i \in \getView[\aexec, \txidset \cup \Tx[\mkvs_{\aexec}, \vi]](\key)\). 
By definition, \(\mkvs_{\aexec}(\key, j) = (\stub, \txid, \stub)\) for some \(\txid \in \txidset \cup \Tx[\mkvs_{\aexec}, \vi]\). 
Because \(\txidset\) only contains read-only transactions, by definition of \(\mkvs_{\aexec}\) there exists 
no index \(j\) such that \(\mkvs_{\aexec}(\key, j) = (\stub, \txid', \stub)\) for some \(\txid' \in \txidset\), 
hence it must be the case that \(\txid \in \Tx[\mkvs_{\aexec}, \vi]\). By definition of \(\Tx\), 
this is possible only if there exist a key \(\key'\) and an index \(j\) such that \(\mkvs_{\aexec}(\key', \vi) = (\stub, \txid, \stub)\). 
Because \(\vi\) is atomic by definition, and because \(\mkvs_{\aexec}(\key, i) = (\stub, \txid, \stub)\), then we have that \(i \in \vi(\key)\). 

Now suppose that \(i \in \vi(\key)\), and let \(\mkvs_{\aexec}(\key, i) = (\stub, \txid, \stub)\) for some \(\txid\). 
This implies that \((\otW, \key, \stub) \in_{\aexec} \txid\).
By definition \(\txid \in \Tx[\mkvs_{\aexec}, \vi]\), hence \(\txid \in \txidset \cup \Tx[\mkvs_{\aexec}, \vi)]\). 
Because \(\txid \in \txidset \cup \Tx[\mkvs_{\aexec}, \vi]\), then for any key \(\key'\) such that 
\((\otW, \key', \stub) \in_{\aexec} \txid\), there exists an index \(j \in \getView[\aexec, \txidset \cup \Tx[\mkvs_{\aexec}, \vi]]\) 
\(\mkvs(\key', j) = (\stub, \txid, \stub)\); because kv-stores only allow a transaction to write at most one version 
per key, then the index \(j\) is uniquely determined. In particular, we know that \((\otW, \key, \stub) \in_{\aexec} \txid\), 
and \(\mkvs_{\aexec}(\key, i) = (\stub, \txid, \stub)\), from which it follows that \(i \in \getView[\aexec, \txidset \cup \Tx[\mkvs_{\aexec}, \vi]](\key)\).
\end{proof}

\begin{proposition}[Views to visible transactions]
\label{prop:compatible.kv2aexec}
Given a view \(\vi \in \Views(\mkvs_{\aexec})\), there exists \(\txidset \subseteq \txidset_{\aexec}\) 
such that \(\getView[\aexec, \txidset] = \vi\), and \(\RP_{\LWW}(\aexec, \txidset) = \snapshot[\mkvs_{\aexec}, \vi]\).
\end{proposition}

\begin{proof}
We only need to prove that, for any \(\vi \in \Views(\mkvs_{\aexec})\), there exists \(\txidset \subseteq \txidset_{\aexec}\) such 
that \(\getView[\aexec, \txidset] = \vi\). Then it follows from \cref{prop:compatible.aexec2kv} that 
\(\RP_{\LWW}(\aexec, \txidset) = \snapshot[\mkvs_{\aexec}, \vi]\). 
It suffices to choose \(\txidset = \bigcup_{\key \in \Keys}(\Set{\wtOf(\mkvs_{\aexec}(\key, i))}[ i > 0 \land i \in \vi(\key)])\).
Fix a key \(\key\), and let \(i \in \vi(\key)\). We prove that \(i \in \getView[\aexec, \txidset]\). 
If \(i = 0\), then \(i \in \getView[\aexec, \txidset]\) by definition. 
Therefore, assume that \(i > 0\). Let \(\txid = \wtOf(\mkvs_{\aexec}(\key, i))\).
It must be the case that \(\txid \in \txidset\) and \(i \in \getView[\aexec, \txidset](\key)\).

Next, suppose that \(i \in \getView[\aexec, \txidset](\key)\). We prove that \(i \in \vi(\key)\).
Note that if \(i = 0\), then \(i \in \vi(\key)\) because of the 
definition of views. Let then \(i > 0\). Because \(i \in \getView[\aexec, \txidset](\key)\), we have that 
\(\wtOf(\mkvs_{\aexec}(\key, i)) \in \txidset\).  Let \(\txid = \wtOf(\mkvs_{\aexec}(\key, i))\). Because \(i > 0\), 
it must be the case that \(\txid \neq \txid_0\).
By definition, \(\txid \in \txidset\) only if there 
exists an index \(j\) and key \(\key'\), possibly different from \(\key\), such that \(\wtOf(\mkvs_{\aexec}(\key', j)) = \txid\) and \(j \in \vi(\key')\). 
Because \(\txid \neq \txid_0\) we have that \(j > 0\). Finally, because \(\vi\) is atomic by definition, \(j \in \vi(\key')\)
\(\wtOf(\mkvs_{\aexec}(\key', j)) = \txid = \wtOf(\mkvs_{\aexec}(\key, i))\), then it must be the case 
that \(i \in \vi(\key)\), which concludes the proof.
\end{proof}

\subsection{KV-Store Traces to Abstract Execution Traces}
\label{sec:kvtrace2aexec}

To prove our definitions using execution test on kv-stores 
is sound and complete with respect with the axiomatic definitions on abstract executions (\cref{sec:kv-sound-complete-proof}),
we need to prove trace equivalent between these two models.

In this section, we only consider the trace that does not involve \( \prog \) but only committing fingerprint and view shift.
In \cref{sec:et-sound-complete-constructor}, we will go further and discuss the trace installed with \( \prog \).

Let $\ET_\top$ be the most permissive execution test.
That is $\ET_\top \vdash (\mkvs, \vi) \csat \fp: (\mkvs',\vi')$ 
such that whenever $\vi(\key) \neq \vi'(\key)$ then either $(\otW, \key, \stub) \in \fp$ or $(\otR, \key, \stub) \in \fp$.
We will relate $\ET_{\top}$-traces to abstract executions that satisfy the last write wins resolution policy, \ie \( (\RP_{\LWW}, \emptyset) \).

To bridge $\ET_{\top}$-traces to abstract executions, 
The \aeset(\tr) function converse the trace of \( \ET_\top \) to set of possible abstract executions (\cref{def:kvtrace2aexec}).
In fact, for any trace \( \tr \) and abstract execution $\aexec \in \aeset(\tr)$, 
the last configuration of $\tr$ is $(\mkvs_{\aexec}, \stub)$ (\cref{prop:kvtrace2aexec}).
We often use \( \aexec_\tr \) for \( \aexec \in \aeset(\tr) \).

\begin{definition}
\label{def:kvtrace2aexec}
Given a kv-store $\mkvs$, a view $\vi$, 
an initial abstract execution $\aexec_0 = ( [ ], \emptyset, \emptyset)$, 
an abstract execution $\aexec$, a set of transactions  
$\txidset \subseteq \txidset_{\aexec}$, a transaction identifier $\txid$ and a set of operations $\fp$,
the \( \extend \)  function defined as the follows:
\begin{align*}
\extend[\aexec, \txid, \txidset, \fp] & \defeq 
\begin{cases}
\text{undefined} & \text{if} \ \txid = \txid_{0}\\
\left(\TtoOp{T}_{\aexec} \uplus \Set{\txid \mapsto \fp}, \VIS', \AR' \right) & \text{if} \ \dagger \\
\end{cases} \\
\dagger & \equiv 
\begin{multlined}[t]
\txid = \txid_{\cl}^{n}
\land \VIS' = \VIS_{\aexec} \uplus \Set{(\txid', \txid)}[\txid \in \txidset]  \\
{} \land \AR' = \AR_{\aexec} \uplus \Set{(\txid', \txid)}[\txid' \in \txidset_{\aexec}]
\end{multlined}
\end{align*}
Given a $\ET_{\top}$ trace $\tr$, let $\lastConf(\tr)$ be the last configuration appearing in $\tr$.
The set of abstract executions $\aeset(\tr)$ is defined as the smallest set such that:
\begin{itemize}
\item $\aexec_{0} \in \aeset((\mkvs_{0}, \vienv_{0}))$, 
\item if $\aexec \in \aeset(\tr)$, then $\aexec \in \aeset\left(\tr \toET{(\cl, \varepsilon)}[\ET_{\top}] (\mkvs, \vienv) \right)$, 
\item if $\aexec \in \aeset(\tr)$, then $\aexec \in \aeset\left(\tr \toET{(\cl, \emptyset)}[\ET_{\top}] (\mkvs, \vienv) \right)$, 
\item 
    let $(\mkvs', \vienv') = \lastConf(\tr)$; 
    if $\aexec \in \aeset(\tr)$, $\fp \neq \emptyset$,
    and $\txidset = \Tx[\mkvs, \vienv'(\cl)] \cup \txidset_\rd$ where \( \txidset_\rd \) is a set of \emph{read-only transactions}
    such that $(\otW, \key, \val) \notin_{\aexec} \txid'$ for all keys \( \key \) and values \( \val \) and transactions \( \txid' \in \txidset_\rd\),
    and if the transaction $\txid$ is the transaction appearing in $\lastConf(\tr)$ but not in $\mkvs$, 
    then $\extend[\aexec, \txid, \txidset, \fp] \in \aeset\left(\tr \toET{(\cl, \fp)}[\ET_{\top}] (\mkvs, \vienv) \right)$.
\end{itemize}
\end{definition}

\begin{proposition}[Trace of \( \ET \) to abstract executions]
\label{prop:kvtrace2aexec}
For any $\ET_{\top}$-trace $\tr$, 
the abstract execution $\aexec \in \aeset(\tr)$ satisfies the last write wins policy,
and $(\mkvs_{\aexec}, \stub) = \lastConf(\tr)$.
\end{proposition}
\begin{proof}
Fix a $\ET_{\top}$-trace $\tau$. 
We prove by induction on the number of transitions $n$ in $\tau$. 
\begin{itemize}
\item \caseB{$n = 0$}
It means $\tr = (\mkvs_{0}, \stub)$.
It follows from \cref{def:kvtrace2aexec} that $\aexec_{\tau} = ([], \emptyset, \emptyset)$. 
This triple satisfies the constraints of \cref{def:aexec}, as well as the resolution policy $\RP_{\LWW}$. 
It is also immediate to see that $\graphOf[\aexec] = ([], \emptyset, \emptyset, \emptyset)$.
In particular, $\txidset_{\graphOf[\aexec]} = \emptyset$, 
and the only kv-store $\mkvs$ such that $\txidset_{\Gr_{\mkvs}} = \emptyset$ 
is given by $\mkvs = \mkvs_{0}$. 
By definition, $\mkvs_{\aexec_{\tr}} = \mkvs_{0}$, as we wanted to prove.

\item \caseI{$n > 0$} In this case, we have that $\tr = \tr' \toET{(\cl, \mu)} (\mkvs, \vienv)$ 
for some $\cl, \mu, \mkvs, \vienv$. The $\ET_{\top}$-trace $\tau'$ contains exactly $n-1$ transitions, 
so that by induction we can assume that $\aexec_{\tau'}$ is a valid abstract execution that satisfies 
$\RP_{\LWW}$. and $\lastConf(\tau') = (\mkvs_{\aexec_{\tau'}}, \vienv')$ for some $\vienv'$. 

We perform a case analysis on $\mu$. 
If $\mu = \varepsilon$, then it follows that $\mkvs = \mkvs_{\aexec_{\tau'}}$, 
and $\aexec_{\tau} = \aexec_{\tau'}$ by \cref{def:kvtrace2aexec}. 
Then by the inductive hypothesis $\aexec_{\tau}$ is an abstract execution that satisfies $\RP_{\LWW}$,
$\lastConf(\tau) = (\mkvs, \stub)$, and $\mkvs_{\aexec_{\tau}} = \mkvs_{\aexec_{\tau'}} = \mkvs$, 
and there is nothing left to prove. 

Suppose now that $\mu = \fp$, for some $\fp$. In this case we have that  
$\mkvs \in \updateKV[\mkvs_{\aexec_{\tau'}}, \vienv'(\cl), \fp, \cl]$. Note that if 
$\fp = \emptyset$, then $\mkvs = \mkvs_{\aexec_{\tau'}}$ and $\aexec_{\tau} = \aexec_{\tau'}$. 
By the inductive hypothesis, $\aexec_{\tau}$ is an abstract execution that satisfies 
$\RP_{\LWW}$, and $\mkvs = \mkvs_{\aexec_{\tau'}} = \mkvs_{\aexec_{\tau}}$. 
Assume then that $\fp \neq \emptyset$. 
By definition, $\mkvs = \updateKV[\mkvs_{\aexec_{\tau'}}, \vienv'(\cl), \fp, \txid]$ 
for some $\txid \in \nextTxid(\cl, \mkvs_{\aexec_{\tau}})$. It follows that $\txid$ 
is the unique transaction such that $\txid \notin \mkvs_{\aexec_{\tau'}}$, and $\txid \in \mkvs$ 
(the fact that $\txid \in \mkvs$ follows from the assumption that $\fp \neq \emptyset$). Let 
$\txidset = \Tx[\mkvs_{\aexec_{\tau'}}, \vienv'(\cl)]$; then $\aexec_{\tau} = \extend[\mkvs_{\aexec_{\tau'}}, \txid, \txidset, \fp]$. 
Note that $\aexec_{\tau}$ satisfies the constraints of abstract execution required by \cref{def:aexec}:
\begin{itemize}
\item  Because $\txid \in \nextTxid(\cl, \mkvs_{\aexec_{\tau}})$, it must be the case that $\txid = \txid_{\cl}^{m}$ for some 
$m \geq 1$; we have that $\TtoOp{T}_{\aexec_{\tau}} = \TtoOp{T}_{\aexec_{\tau'}}\rmto{\txid_{\cl}^{m}}{\fp}$, 
from which it follows that 
\[
\txidset_{\aexec_{\tau}} = \dom(\TtoOp{T}_{\aexec_{\tau}}) = \dom(\TtoOp{T}_{\aexec_{\tau'}}) \cup 
\Set{\txid_{\cl}^{m} } = \txidset_{\aexec_{\tau'}} \cup \Set{\txid_{\cl}^{m} }
\]
By inductive hypothesis, $\txid_0 \notin \txidset_{\aexec_{\tau'}}$, and therefore $\txid_{0} \notin 
\txidset_{\aexec_{\tau'}} \cup \Set{\txid_{\cl}^{m} } = \txidset_{\aexec}$.

\item \( \VIS_{\aexec_{\tau}} \subseteq \AR_{\aexec_{\tau}} \).
    Let $(\txid' ,\txid'') \in \VIS_{\aexec_{\tau}}$. Then either $\txid'' = \txid_{\cl}^{m}$ and $\txid' \in \txidset$, or $(\txid', \txid'') \in 
\VIS_{\aexec_{\tau'}}$. In the former case, we have that $(\txid', \txid_{\cl}^{m}) \in \AR_{\aexec_{\tau}}$ by definition; 
in the latter case, we have that $(\txid', \txid'') \in \AR_{\aexec_{\tau'}}$ because $\aexec_{\tau'}$ is a valid 
abstract execution by inductive hypothesis, and therefore $(\txid', \txid'') \in \AR_{\aexec_{\tau}}$ by definition. 
This concludes the proof that $\VIS_{\aexec_{\tau}} \subseteq \AR_{\aexec_{\tau}}$. 
\item \( \VIS_{\aexec_\tr} \) is irreflexive.
Assume $(\txid', \txid'') \in \VIS_{\aexec_{\tau}}$, then either 
$(\txid' \txid'') \in \VIS_{\aexec_{\tau'}}$, and because $\VIS_{\aexec_{\tau'}}$ is irreflexive by the inductive hypothesis, 
then $\txid' \neq \txid''$; 
or $\txid'' = \txid_{\cl}^{m}$, $\txid' \in \txidset \subseteq \txidset_{\aexec_{\tau'}}$, 
and because $\txid_{\cl}^{m} \notin \mkvs_{\aexec_{\tau'}}$, then $\txid' \neq \txid_{\cl}^{m}$. 

\item $\AR_{\aexec_{\tau}}$ is total. Let $(\txid', \txid'') \in \txidset_{\aexec_{\tau}}$. 
Suppose that $\txid' \neq \txid''$.
\begin{enumerate}
\item If $\txid' \neq 
\txid_{\cl}^{m}$, $\txid'' \neq \txid_{\cl}^{m}$, then it must be the case that $\txid', \txid'' \in \txidset_{\aexec_{\tau'}}$; 
this is because we have already argued that $\txidset_{\aexec_{\tau}} = \txidset_{\aexec_{\tau'}} \cup \Set{\txid_{\cl}^{m}}$. 
By the inductive hypothesis, we have that either $(\txid', \txid'') \in \AR_{\aexec_{\tau'}}$, or 
$(\txid'', \txid') \in \AR_{\aexec_{\tau'}}$. Because $\AR_{\aexec_{\tau'}} \subseteq \AR_{\aexec_{\tau}}$, 
then either $(\txid', \txid'') \in \AR_{\aexec_{\tau'}}$ or $(\txid'', \txid') \in \AR_{\aexec_{\tau}}$. 
\item if $\txid'' = \txid_{\cl}^{m}$, then it must be $\txid' \in \txidset_{\aexec_{\tau'}}$. By definition, 
$(\txid', \txid_{\cl}^{m}) \in \AR_{\aexec_{\tau}}$. Similarly, if $\txid' = \txid_{\cl}^{m}$, we 
can prove that $(\txid'', \txid_{\cl}^{m}) \in \AR_{\aexec_{\tau}}$.
\end{enumerate}
\item  $\AR_{\aexec_{\tau}}$ is irreflexive. It follows is the same as the one of $\VIS_{\aexec_{\tau}}$.
\item \( \AR_{\aexec_{\tau}} \) is transitive.
Assume $(\txid', \txid'') \in \AR_{\aexec_{\tau}}$ and $(\txid'', \txid''') \in \AR_{\aexec_{\tau}}$. 
Note that it must be the case that $\txid', \txid'' \in \txidset_{\aexec_{\tau'}}$ by the definition of 
$\AR_{\aexec}$, and in particular $(\txid', \txid'') \in \AR_{\aexec_{\tau'}}$. 
For $\txid'''$, we have two possible cases. 
\begin{enumerate}
\item Either $\txid''' \in \txidset_{\aexec_{\tau}}$, from 
which it follows that $(\txid'', \txid''') \in \AR_{\aexec_{\tau'}}$; because
of $\AR_{\aexec_{\tau'}}$ is transitive by the inductive hypothesis, then 
$(\txid', \txid''') \in \AR_{\aexec_{\tau'}}$, and therefore $(\txid' ,\txid''') \in 
\AR_{\aexec_{\tau}}$.
\item Or $\txid''' = \txid_{\cl}^{m}$, and because $\txid' \in \txidset_{\aexec_{\tau'}}$, then 
$(\txid', \txid_{\cl}^{m}) \in \AR_{\aexec_{\tau}}$ by definition. 
\end{enumerate}
\item \( \SO_{\aexec_{\tau}} \subseteq \AR_{\aexec_{\tau}} \).
Let $\cl'$ be a client such that $(\txid_{\cl'}^{i}, \txid_{\cl'}^{j}) \in \AR_{\aexec_{\tau}}$. 
If $\cl' \neq \cl$, then it must be the case that $\txid_{\cl'}^{i}, \txid_{\cl'}^{j} \in \txidset_{\aexec_{\tau'}}$, 
and therefore $(\txid_{\cl'}^{i}, \txid_{\cl'}^{j}) \in \AR_{\aexec_{\tau'}}$. By the inductive hypothesis, 
it follows that $i < j$. If $\cl' = \cl$, then by definition of $\AR_{\aexec_{\tau}}$ it must be  $i \neq m$. 
If $j \neq m$ we can proceed as in the previous case to prove that $i < j$. If $j = m$, then 
note that $\txid_{\cl}^{i} \in \txidset_{\aexec_{\tau}}$ only if $\txid_{\cl}^{i} \in \mkvs_{\aexec_{\tau'}}$. 
Because $\txid_{\cl}^{m} \in \nextTxid(\mkvs_{\aexec_{\tau'}}, \cl)$, then we have that $i < m$, 
as we wanted to prove.
\end{itemize}

Next, we prove that $\aexec_{\tau}$ satisfies the last write wins policy. 
Let $\txid' \in \txidset_{\aexec_{\tau}}$, and suppose that $(\otR, \key, \val) \in_{\aexec_{\tau}} \txid'$. 
\begin{itemize} 
\item If $\txid' \neq \txid$, then we have that $\txid \in \txidset_{\aexec_{\tau'}}$. We also have that 
$\VIS^{-1}_{\aexec_{\tau}}(\txid') = \VIS^{-1}_{\aexec_{\tau'}}(\txid')$, $\AR^{-1}_{\aexec_{\tau}}(\txid') 
= \AR^{-1}_{\aexec_{\tau'}}(\txid')$; finally, for any $\txid'' \in \txidset_{\aexec_{\tau'}}$, 
$(\otW, \key, \val') \in_{\aexec_{\tau}} \txid''$ if and only if $(\otW, \key, \val') \in_{\aexec_{\tau'}} 
\txid''$. Therefore, let $\txid_{r} = \max_{\AR_{\aexec_{\tau}}}(\VIS^{-1}_{\aexec_{\tau}}(\txid') \cap 
\Set{\txid'' }[ (\otW, \key, \stub) \in_{\aexec_{\tau}} \txid''])$. We have that 
\[
    \txid_{r} = \max_{\AR_{\aexec_{\tau'}}}(\VIS^{-1}_{\aexec_{\tau'}}(\txid) 
    \cap \Set{ \txid'' }[ (\otW, \key, \stub ) \in_{\aexec_{\tau'}} \txid''])
\]
and because $\aexec_{\tau'}$ satisfies the last write 
wins resolution policy, then $(\otW, \key, \val) \in_{\aexec_{\tau'}} \txid_{r}$. This also implies that 
$(\otW, \key, \val) \in_{\aexec_{\tau}} \txid_{r}$. 

\item Now, suppose that $\txid' = \txid$. Suppose that $(\otR, \key, \val) \in_{\aexec_{\tau}} \txid'$. 
By definition, we have that $(\otR, \key, \val) \in \fp$. Recall that $\tau = \tau' \toET{(\cl, \fp)}[\ET_{\top}] (\mkvs, \vienv)$, 
and $\lastConf(\tau') = (\mkvs_{\aexec_{\tau'}}, \vienv')$ for some $\vienv'$. 
That is, 
\[
    (\mkvs_{\aexec_{\tau'}}, \vienv') \toET{(\cl, \fp)}[\ET_{\top}] (\mkvs, \vienv)
\]
which in turn implies that $\ET_{\top} \vdash (\mkvs_{\aexec_{\tau'}}, \vienv'(\cl)) \csat \fp : (\mkvs,\vienv(\cl) )$. 
Let then $r = \max\Set{i}[ i \in \vienv'(\cl)(\key)]$. 
By definition of execution test, and because $(\otR, \key, \val) \in \fp$, then it must be the case that 
$\mkvs_{\aexec_{\tau'}}(\key, r) = (\val, \txid'', \stub)$ for some $\txid''$. 

We now prove that 
$\txid'' = \max_{\AR_{\aexec_{\tau}}}(\VIS^{-1}_{\aexec_{\tau}}(\txid) \cap \Set{ \txid'' }[ (\otW, \key, \stub) \in_{\aexec_{\tau}} \txid''])$. 
First we have
\[ 
\begin{array}{l}
\VIS^{-1}_{\aexec_{\tau}}(\txid) = 
\Tx[\mkvs_{\aexec_{\tau'}}, \vienv'(\cl)] = 
\Set{\wtOf(\mkvs_{\aexec_{\tau'}}(\key',  i)) }[ \key' \in \Keys \land  i \in \vienv'(\cl)(\key')]
\end{array}
\]
Note that $r \in \vienv'(\cl)(\key)$, and $\txid'' = \wtOf(\mkvs_{\aexec_{\tau'}}(\key, r))$. 
Therefore, $\txid'' \in \VIS^{-1}_{\aexec_{\tau}}(\txid)$. 
Because $\mkvs = \updateKV[\mkvs_{\aexec_{\tau'}}, \vienv'(\cl), \fp, \txid]$, it 
must be the case that $\wtOf(\mkvs(\key, r)) = \txid''$. Also, because $\wtOf(\mkvs_{\aexec_{\tau'}}(\key, r)) = \txid''$, 
then $(\otW, \key, \stub) \in_{\aexec_{\tau''}} \txid''$, or equivalently $(\otW, \key, \stub) \in \TtoOp{T}_{\aexec_{\tau'}}(\txid'')$. 
We have already proved that $\VIS_{\aexec_{\tau}}$ is irreflexive, hence it must be the case that $\txid'' \neq \txid$. 
In particular, because $\aexec_{\tau} = \extend[\aexec_{\tau'}, \txid, \stub, \stub]$, then we have that 
$\TtoOp{T}_{\aexec_{\tau}}(\txid'') = \TtoOp{T}_{\aexec_{\tau'}}\rmto{\txid}{\fp}(\txid'') = 
\TtoOp{T}_{\aexec_{\tau'}}(\txid'')$, hence $(\otW, \key, \stub) \in \TtoOp{T}_{\aexec_{\tau}}(\txid'')$. Equivalently, 
$(\otW, \key, \stub) \in_{\aexec_{\tau}} \txid''$. We have proved that $\txid'' \in \VIS^{-1}_{\aexec_{\tau}}(\txid)$, 
and $(\otW, \key, \stub) \in_{\aexec_{\tau}} \txid''$. 

Now let $\txid'''$ be such that $\txid''' \in \VIS^{-1}_{\aexec_{\tau}}(\txid)$, and $(\otW, \key \stub) \in_{\aexec_{\tau}} \txid'''$. 
Note that $\txid''' \neq \txid$ because $\VIS_{\aexec_{\tau}}$ is irreflexive.
We show that either $\txid''' = \txid''$, or $\txid''' \toEDGE{\AR_{\aexec_{\tau}}} \txid''$. 
Because $\txid''' \in \VIS^{-1}_{\aexec_{\tau}}(\txid)$, then there exists a key $\key'$ and an index $i \in \vienv'(\cl)$ 
such that $\wtOf(\mkvs_{\aexec_{\tau'}}(\key', i)) = \txid'''$. Because $(\otW, \key, \stub) \in_{\aexec_{\tau}} \txid'''$, 
and because $\txid''' \neq \txid$, then $(\otW, \key, \stub) \in_{\aexec_{\tau'}} \txid'''$, and therefore there exists 
an index $j$ such that $\wtOf(\mkvs_{\aexec_{\tau'}}(\key, j)) = \txid'''$. We have that $\wtOf(\mkvs_{\aexec_{\tau'}}(\key, j) = 
\wtOf(\mkvs_{\aexec_{\tau'}}(\key', i))$, and $i \in \vienv'(\cl)$. By \cref{eq:view.atomic}, it must be $j \in \vienv'(\cl)$. 
Note that $r = \max\Set{i}[ i \in \vienv'(\cl)]$, hence we have that $j \leq r$. If $j = r$, then $\txid''' = \txid''$ and 
there is nothing left to prove. If $j < r$, then we have that $(\txid''', \txid'') \in \AR_{\aexec_{\tau'}}$, and 
therefore $(\txid''', \txid'') \in \AR_{\aexec_{\tau}}$.
\end{itemize}
Finally, we need to prove that $\mkvs = \mkvs_{\aexec_{\tau}}$.
Recall $\mkvs = \updateKV[\mkvs_{\aexec_{\tau'}}, \vienv'(\cl), \fp, \txid]$, 
and $\aexec_{\tau} = \extend[\aexec_{\tau'}, \Tx[\mkvs_{\aexec_{\tau'}}, \vienv'(\cl)], \txid, \fp]$. 
The result follows then from \cref{prop:extend.update.sameop}. 
\end{itemize}
\end{proof}

\begin{proposition}[\( \extend \) matching \( \updateKV\)]
\label{prop:extend.update.sameop}
Given an abstract execution $\aexec$, a set of transactions $\txidset \subseteq \txidset_{\aexec}$,
a transaction $\txid \notin \txidset_{\aexec}$, and a fingerprint $\fp \subseteq \pset{\Ops}$,
if the new abstract execution $\aexec' = \extend[\aexec, \txidset, \txid, \fp]$,
and the view $\vi = \getView[\mkvs_{\aexec}, \txidset]$,
then $\updateKV[\mkvs_{\aexec}, \vi, \fp, \txid] = \mkvs_{\aexec'}$.
\end{proposition}

\begin{proof}
Let $\Gr = \Gr_{\updateKV[\mkvs_{\aexec}, \vi, \fp, \txid]}$, $\Gr' = \graphOf[\aexec']$. 
Note that $\mkvs_{\aexec'}$ is the unique kv-store such that $\Gr_{\mkvs_{\aexec'}} = \graphOf[\aexec'] = \Gr'$. 
It suffices to prove that $\Gr = \Gr'$. Because the function $\Gr_{\cdot}$ is injective, it follows that 
$\updateKV[\mkvs_{\aexec}, \vi, \fp, \txid] = \mkvs_{\aexec'}$, as we wanted to prove.  

The proof is a consequence of \cref{lem:graph.extend} and \cref{lem:graph.update}. 
Consider the dependency graph $\Gr_{\mkvs_{\aexec}}$.
Recall that $\mkvs_{\aexec}$ is the unique kv-store such that $\Gr_{\mkvs_{\aexec}} = \graphOf[\aexec]$. 
We prove that $\TtoOp{T}_{\Gr} = \TtoOp{T}_{\Gr'}$, $\WR_{\Gr} = \WR_{\Gr'}$ and 
$\WW_{\Gr} = \WW_{\Gr'}$ (from the last two it follows that $\RW_{\Gr} = \RW_{\Gr'}$). 
\begin{itemize}
\item It is easy to see $\TtoOp{T}_{\Gr} = \TtoOp{T}_{\Gr'}$.

\item $\WR_{\Gr} = \WR_{\Gr'}$.
Let \( \mkvs  = \mkvs_\aexec \).
Suppose that $\txid' \toEDGE{\WR_{\Gr}(\key)} \txid''$ for some $\txid', \txid''$. 
By \cref{lem:graph.update} we have that either $\txid' \toEDGE{\WR_{\Gr_\mkvs}(\key)} \txid''$, 
or $\txid'' = \txid$, $(\otR, \key, \stub) \in \fp$, $\txid' = \max_{\WW_{\Gr_\mkvs}(\key)}\Set{\wtOf(\key, i) }[i \in \vi(\key)]$. 

\begin{itemize}
\item If $\txid' \toEDGE{\WR_{\Gr_\mkvs}(\key)} \txid''$, then because 
$\Gr_\mkvs = \graphOf[\aexec]$, we have that $\txid' \toEDGE{\WR_{\graphOf[\aexec]}(\key)} \txid''$. 
Recall that $\Gr' = \graphOf[\extend[\aexec, \txidset, \txid, \fp]]$, hence by \cref{lem:graph.extend} 
we obtain that $\txid' \toEDGE{\WR_{\Gr'}(\key)} \txid''$. 

\item If $\txid'' = \txid$, $(\otR, \key, \stub) \in \fp$, and $\txid' = \max_{\WW_{\Gr_\mkvs}(\key)} \Set{\wtOf(\mkvs_{\aexec}(\key, i))}[i \in \vi(\key)]$, 
    then we also have that $\txid' = \max_{\WW_{\graphOf[\aexec]}(\key)} (\txidset \cap \Set{\txid'''}[(\otW, \key, \stub) \in_{\aexec} \txid''']) $. 
This is because of the assumption that 
\begin{align*}
    \Set{\wtOf(\mkvs_{\aexec}(\key, i))}[i \in \vi(\key)]
    & = \Set{\wtOf(\mkvs_{\aexec}(\key', i))}[\key' \in \Keys \land i \in \vi(\key')] \cap \Set{\wtOf(\mkvs_{\aexec}(\key, \stub)} \\
    & = \Tx[\mkvs_{\aexec}, \vi] \cap \Set{\wtOf(\mkvs_{\aexec}(\key, \stub)}  \\
    & = \txidset \cap \Set{\txid'''}[(\otW, \key, \stub) \in_{\aexec} \txid''']
\end{align*}
Again, it follows from \cref{lem:graph.extend} that $\txid' \toEDGE{\WR_{\Gr'}(\key)} \txid''$. 
\end{itemize}
\item \( \WW_{\Gr} = \WW_{\Gr'}\). The \( \WW_{\Gr} = \WW_{\Gr'} \) follows the similar reasons as $\WR_{\Gr} = \WR_{\Gr'}$.
\end{itemize}
\end{proof}

\begin{lemma}[Graph to abstract execution extension]
\label{lem:graph.extend}
Let $\aexec$ be an abstract execution, 
$\txid \notin \txidset_{\aexec} \cup \Set{\txid_0}$ be a transaction identifier $\txidset_{\aexec}$, and $\fp \in \txidset_{\aexec}$. 
Let $\txidset \subseteq \txidset_{\aexec}$ be a set of transaction identifiers.
Let $\Gr = \graphOf[\aexec], \Gr' = \graphOf[\extend[\aexec, \txid, \txidset, \fp]]$. 
We have the following: 
\begin{enumerate}
\item for any $\txid' \in \txidset_{\Gr'}$, either $\txid' \in \txidset_{\Gr}$ and $\TtoOp{T}_{\Gr}(\txid') = \TtoOp{T}_{\Gr'}(\txid')$, 
or $\txid' = \txid$ and $\TtoOp{T}_{\Gr'}(\txid) = \fp$.
\item $\txid' \toEDGE{\WR_{\Gr'}(\key)} \txid''$ if and only if either 
$\txid' \toEDGE{\WR_{\Gr}(\key)_{\Gr}} \txid''$, or $(\otR, \key, \stub) \in \fp$, $\txid'' = \txid$ and 
$\txid' = \max_{\WW_{\Gr}(\key)}(\txidset)$, 
\item $\txid' \toEDGE{\WW_{\Gr'}(\key)} \txid''$ if and only if 
either $\txid' \toEDGE{\WW_{\Gr}(\key)} \txid''$, or $(\otW, \key, \stub) \in \fp$, $\txid'' = \txid$, 
and $(\otW, \key, \stub) \in_{\Gr} \txid'$.
\end{enumerate}
\end{lemma}

\begin{proof}
Fix a key $\key$. Let $\aexec' = \extend[\aexec, \txid, \txidset, \fp]$. Recall that $\Gr' = \graphOf[\aexec']$.

\begin{enumerate}
\item By definition of $\extend$, and 
because $\txid \notin \txidset_{\aexec}$, we have that 
$\txidset_{\aexec'} = \txidset_{\aexec} \uplus \Set{\txid}$. Furthermore, $\TtoOp{T}_{\aexec'}(\txid) = \fp$, 
from which it follows that $\TtoOp{T}_{\Gr'}(\txid) = \fp$.
For all $\txid' \in \txidset_{\aexec}$, we have that $\TtoOp{T}_{\aexec'}(\txid') = 
\TtoOp{T}_{\aexec}(\txid') = \TtoOp{T}_{\Gr}(\txid')$.
\item
There are two cases that either the \( \txid'' \) already exists in the dependency graph before,
or it is the newly committed transaction.
\begin{itemize}
\item Suppose that $\txid' \toEDGE{\WR(\key)_{\Gr}} \txid''$ for some $\txid', \txid'' \in \txidset_{\Gr}$. 
By definition, $(\otR, \key, \stub) \in_{\aexec} \txid''$,  
and $\txid' = \max_{\AR_{\aexec}}(\VIS_{\aexec}^{-1}(\txid'') \cap \Set{\txid'''}[(\otW, \key, \stub) \in_{\aexec} \txid'''])$. 
Because $\txid'' \in \txidset_{\Gr} = \txidset_{\aexec}$, it follows that $\txid'' \neq \txid$. By definition, 
$\VIS^{-1}_{\aexec'}(\txid'') = \VIS^{-1}_{\aexec}(\txid)$: also, whenever 
$\txid_{a}, \txid_{b} \in \VIS^{-1}_{\aexec'}(\txid)$ we have that $\txid_{a}, \txid_{b} \in \txidset_{\aexec}$, 
and therefore if $\txid_{a} \toEDGE{\AR_{\aexec'}} \txid_{b}$, then it must be the case 
that $\txid_{a} \toEDGE{\AR_{\aexec}} \txid_b$; also, $\TtoOp{T}_{\aexec}(\txid_{a}) = \TtoOp{T}_{\aexec'}(\txid_{a})$. 
As a consequence, we have that 
\[
    \begin{array}{l}
        \max{}_{\AR_{\aexec'}}(\VIS^{-1}_{\aexec'}(\txid) \cap \Set{ \txid'''}[(\otW, \key, \stub) \in_{\aexec'} \txid''']) =
        \max{}_{\AR_{\aexec}}(\VIS^{-1}_{\aexec}(\txid) \cap \Set{\txid'''}[(\otW, \key, \stub) \in_{\aexec} \txid''']) = \txid'
    \end{array}
\] 
and therefore $\txid' \toEDGE{\WR_{\Gr'}} \txid$. 

\item Suppose now that $(\otR,\key, \stub) \in \fp$, and $\txid' = \max_{\WW(\key)_{\Gr}}(\txidset)$. 
    By Definition, $\txid' = \max_{\AR_{\aexec}}(\txidset) \cap \Set{\txid'''}[(\otW, \key, \stub) \in_{\aexec} \txid''']$, 
and, $\txidset = \VIS^{-1}_{\aexec'}(\txid)$.
Because $\txidset \subseteq \txidset_{\aexec}$, we have 
that for any $\txid_{a}, \txid_{b}$, if $\txid_{a} \toEDGE{\AR_{\aexec}} \txid_{b}$, 
then $\txid_{a} \toEDGE{\AR_{\aexec'}} \txid_{b}$; and $\TtoOp{T}_{\aexec'}(\txid_{a}) = 
\TtoOp{T}_{\aexec}(\txid_a)$. Therefore, 
\[
    \txid' = \max{}_{\AR_{\aexec'}}(\VIS^{-1}_{\aexec'}(\txid) \cap \Set{\txid'''}[(\otW, \key, \stub) \in_{\aexec'} \txid'''], 
\] 
from which it follows that $\txid' \toEDGE{\WR_{\Gr'}(\key)}\txid$.

Now, suppose that $\txid' \toEDGE{\WR_{\Gr'}(\key)} \txid''$ for some $\txid', \txid'' \in \txidset_{\Gr'} = 
\txidset_{\aexec'}$. We have that $ (\otR, \key, \stub) \in_{\aexec'} \txid''$, 
$(\otW, \key, \stub) \in_{\aexec'} \txid'$, and $\txid'' = \max_{\AR_{\aexec'}}(\VIS_{\aexec'}^{-1}(\txid'') 
\cap \Set{\txid'''}[(\otW, \key, \stub) \in_{\aexec'} \txid''']$. 
We also have that $\txidset_{\aexec'} = \txidset_{\aexec} \uplus \Set{\txid}$. We perform a case 
analysis on $\txid''$. 

\begin{itemize}
\item If $\txid'' = \txid$, then by definition of $\extend$ we have that 
$\VIS^{-1}_{\aexec'}(\txid) = \txidset$. Note that $\txidset \subseteq \txidset_{\aexec}$, so that 
for any $\txid_{a}, \txid_{b} \in \txidset_{\aexec}$, we have that $\txid_{a} \toEDGE{\AR_{\aexec'}} \txid_{b}$ 
if and only if $\txid_{a} \toEDGE{\AR_{\aexec}} \txid_{b}$, 
and $(\otW, \key, \val) \in_{\aexec'} \txid_{a}$ if and only if $(\otW, \key, \val) \in_{\aexec} \txid_{a}$. 
Thus, $\txid' = \max_{\AR_{\aexec}}(\txidset 
\cap \Set{\txid'''}[(\otW, \key, \stub) \in_{\aexec} \txid''']) = \max_{\WW_{\Gr}(\key)}(\txidset)$. 

\item If $\txid'' \in \txidset_{\aexec}$, then it is the case that 
    $\txid' = \max_{\AR_{\aexec'}}(\VIS^{-1}_{\aexec'}(\txid'') \cap \Set{ \txid'''}[(\otW, \key, \stub) \in_{\aexec'} \txid''']$. 
Similarly to the case above, we can prove that $\VIS^{-1}_{\aexec'}(\txid'') = \VIS^{-1}_{\aexec}(\txid)$, 
for any $\txid_{a}, \txid_{b} \in \VIS^{-1}_{\aexec}(\txid)$, $(\otW, \key, \val) \in_{\aexec'} \txid_{a}$ 
implies $(\otW, \key, \val) \in_{\aexec} \txid_{a}$, and $\txid_{a} \toEDGE{\AR_{\aexec'}} \txid_{b}$ 
implies $\txid_{a} \toEDGE{\AR_{\aexec}} \txid_{b}$, from which it follows that 
$\txid' = \max_{\AR_{\aexec}}(\VIS^{-1}_{\aexec}(\txid'') \cap \Set{\txid'''}[(\otW, \key \stub) \in_{\aexec} \txid'''])$, 
and therefore $\txid' \toEDGE{\WR_{\Gr}(\key)} \txid''$.
\end{itemize}
\end{itemize}

\item 
Similar to \( \WR(\key)_{\Gr} \), there are two cases that either the \( \txid'' \) already exists in the dependency graph before,
or it is the newly committed transaction.
\begin{itemize}
\item Suppose that $\txid' \toEDGE{\WW_{\Gr}(\key)} \txid''$ for some $\txid', \txid'' \in \txidset_{\aexec}$. 
Then $(\otW,\key,\stub) \in_{\aexec} \txid', (\otW, \key, \stub) \in_{\aexec} \txid''$, and $\txid' \toEDGE{\AR_{\aexec}} \txid''$. 
By definition of $\extend$, it follows that $\txid' \toEDGE{\AR_{\aexec'}} \txid''$, and because 
$\txid', \txid'' \in \txidset_{\aexec}$, hence $\txid', \txid'' \neq \txid$, then 
$(\otW,\key, \stub) \in_{\aexec'} \txid'$, $(\otW, \key, \stub) \in_{\aexec'} \txid''$. By definition, 
we have that $\txid' \toEDGE{\WW_{\aexec'}(\key)} \txid''$.

\item Suppose that $(\otW, \key, \stub) \in_{\aexec} \txid'$, $(\otW, \key, \stub) \in \fp$. Because $\txid' \in \txidset_{\aexec}$, 
we have that $\txid' \neq \txid$, hence $(\otW, \key, \stub) \in_{\aexec' }\txid'$. By definition, 
$\TtoOp{T}_{\aexec'}(\txid) = \fp$, hence $(\otW, \key, \stub) \in_{\aexec'} \txid$. Finally, 
the definition of $\extend$ ensures that $\txid' \toEDGE{\AR_{\aexec'}} \txid$. Combining 
these three facts together, we obtain that  
$\txid' \toEDGE{\WW_{\Gr'}(\key)} \txid$. 

Now, suppose that $\txid' \toEDGE{\WW_{\Gr'}(\key)} \txid''$ for some $\txid', \txid'' \in \txidset_{\aexec}$. 
Then $\txid' \toEDGE{\AR_{\aexec'}} \txid''$, $(\otW, \key, \stub) \in_{\aexec'} \txid'$, $(\otW, \key, \stub) 
\in_{\aexec'} \txid''$. 
Recall that $\txidset_{\Gr'} = \txidset_{\aexec'} = \txidset_{\aexec} \uplus \Set{ \txid }$. We perform a case analysis on $\txid''$. 

\begin{itemize}

\item If $\txid'' = \txid$, then the definition of $\extend$ ensures that $\txid' \toEDGE{\AR_{\aexec'}} \txid$ 
implies that $\txid \in \txidset_{\aexec}$, hence $\txid' \neq \txid$. 
Together with $(\otW, \key, \stub) \in_{\aexec'} 
\txid'$, this leads to $(\otW, \key, \stub) \in_{\aexec} \txid'$. 

\item If $\txid'' \in \txidset_{\aexec}$, then $\txid'' \neq \txid$. The definition of $\extend$ ensures that $\txid' \toEDGE{\AR_{\aexec}} \txid''$. 
This implies that $\txid', \txid'' \in \txidset_{\aexec}$, hence $\txid', \txid'' \neq \txid$, and $\TtoOp{T}_{\aexec'}(\txid') = \TtoOp{T}_{\aexec}(\txid')$, 
$\TtoOp{T}_{\aexec'}(\txid'') = \TtoOp{T}_{\aexec}(\txid'')$. It follows that $(\otW, \key, \stub) \in_{\aexec} \txid'$, 
$(\otW, \key, \stub) \in_{\aexec} \txid''$, and therefore $\txid' \toEDGE{\WW_{\Gr}(\key)} \txid''$.

\end{itemize}
\end{itemize}
\end{enumerate}
\end{proof}

\begin{lemma}[Graph to kv-store update]
\label{lem:graph.update}
Let $\mkvs$ be a kv-store, and $\vi \in \Views(\mkvs)$. Let $\txid \notin \mkvs$, and 
$\fp \subseteq \pset{\Ops}$, and let $\mkvs' = \updateKV[\mkvs, \vi, \fp, \txid]$. 
Let $\Gr = \Gr_{\mkvs}$, $\Gr' = \Gr_{\mkvs'}$; then for all $\txid', \txid'' \in \txidset_{\Gr'}$ and keys $\key$, 
\begin{itemize}
\item $\TtoOp{T}_{\Gr'} = \TtoOp{T}_{\Gr}\rmto{\txid}{\fp}$, 
\item $\txid' \toEDGE{\WR_{\Gr'}(\key)} \txid''$ if and only if either 
$\txid' \toEDGE{\WR_{\Gr}(\key)} \txid''$, or $(\otR, \key, \stub) \in \fp$ and 
\[\txid' = \max_{\WW_{\Gr}(\key)}(\Set{\wtOf(\mkvs(\key, i)) }{ i \in \vi(\key)})\]
\item $\txid' \toEDGE{\WW_{\Gr'}(\key)} \txid''$ if and only if either 
$\txid' \toEDGE{\WW_{\Gr}(\key)} \txid''$, or $(\otW, \key, \stub) \in \fp$ 
and $\txid' = \wtOf(\mkvs(\key, \stub))$. 
\end{itemize}
\end{lemma}

\begin{proof}
Fix $\key \in \Keys$. Because $\txid \notin \mkvs$, then $\txid \notin \txidset_{\Gr}$, 
and by definition of $update$ we obtain that $\Set{\txid'}[\txid' \in \mkvs'] = 
\Set{\txid'}[\txid' \in \mkvs] \cup \Set{\txid}$. It follows that $\txidset_{\Gr'} = \txidset_{\Gr} \uplus \Set{\txid }$.

\begin{enumerate}
\item Suppose that $(\otR, \key, \val) \in_{\Gr} \txid'$. By definition, 
there exists an index $i$ such 
that $\mkvs(\key, i) = (\val, \stub, \Set{\txid'} \cup \stub)$. Because $\mkvs' = \updateKV[\mkvs, \vi, \fp, \txid]$, 
it is immediate to observe that $\mkvs'(\key, i) = (\val, \stub, \Set{\txid'} \cup \stub)$, and therefore 
$(\otR,\key, \val) \in_{\Gr'} \txid'$. Conversely, note that if $(\otR, \key, \val) \in_{\Gr'} \txid$, 
then there exists an index $i = 0,\cdots, \lvert \mkvs'(\key) \rvert - 1$ such that 
$\mkvs'(\key, i) = (\val, \stub, \Set{\txid'} \cup \stub)$. 
it follows that it must be the case that $i \leq \lvert \mkvs(\key) \rvert - 1$, and because 
$\txid' \neq \txid$, we have that $\mkvs(\key, i) = (\val, \stub, \Set{\txid'} \cup \stub)$. Therefore 
$(\otR, \key, \val) \in_{\Gr} \txid'$. 

Similarly, if $(\otW, \key, \val) \in_{\Gr} \txid'$, 
then there exists an index $i=0,\cdots, \lvert \mkvs(\key) \rvert - 1$ such that 
$\mkvs(\key, i) = (\val, \txid', \val)$. It follows that $\mkvs'(\key, i) = (\val, \txid', \stub)$, hence 
$(\otW, \key, \val) \in_{\Gr'} \txid'$. If $(\otW, \key, \val) \in \fp$, then we 
have 
that $\mkvs'(\key, \lvert \mkvs'(\key) \rvert - 1) = (\val, \txid', \stub)$, 
hence $(\otW, \key, \val) \in_{\Gr'} \txid'$. 
Conversely, if $(\otW, \key, \val) \in_{\Gr'} \txid'$, then there exists an index 
$i = 0, \cdots, \lvert \mkvs'(\key) \rvert - 1$ such that $\mkvs(\key, i) = (\val, \txid', \stub)$. 
We have two possible cases: either $i < \lvert \mkvs'(\key, i) \rvert - 1$, leading to  
$\txid' \neq \txid$ and $\mkvs(\key, i) = (\val, \txid', \stub)$, or equivalently 
$(\otR,\key, \val) \in_{\Gr} \txid'$; or $i = \lvert \mkvs'(\key, i) \rvert - 1$, 
leading to $\txid' = \txid$, and $\mkvs(\key, i) = (\val, \txid, \emptyset)$ 
for some $\val$ such that $(\otW, \key, \val) \in \fp$. 

Putting together the facts above, we obtain that $\TtoOp{T}_{\Gr'} = 
\TtoOp{T}_{\Gr}\rmto{\txid}{\fp}$, as we wanted to prove.

\item There are two cases that either the \( \txid'' \) already exists in the dependency graph before,
or it is the newly committed transaction.
\begin{itemize}
\item Suppose that $\txid' \toEDGE{\WR_{\Gr}(\key)} \txid''$. 
By definition, there exists an index $i = 0,\cdots, \lvert \mkvs(\key) \rvert - 1$ 
such that $\mkvs(\key, i) = (\stub, \txid', \Set{\txid''} \cup \stub)$. It is immediate 
to observer, from the definition of $\updateKV$, that $\mkvs'(\key, i) = (\stub, \txid', \Set{\txid''} \cup \stub)$, 
and therefore $\txid' \toEDGE{\WR_{\Gr'}(\key)} \txid''$. 

\item Next, suppose that $(\otR, \key, \stub) \in \fp$, and $\txid' = \max_{\WW_{\Gr}(\key)}(\Set{\wtOf(\mkvs(\key, i))}[i \in \vi(\key)]$. 
By Definition, $\mkvs(\key, i) = (\stub, \txid', \stub)$, where $i = \max(\vi(\key))$. This is because 
$\txid' \rightarrow{\WW_{\Gr}(\key)} \txid''$ if and only if $\txid' = \wtOf(\mkvs(\key, j_1)), \txid'' = 
\wtOf(\mkvs(\key, j_2))$ for some $j_1, j_2$ such that $j_1 < j_2$. 
The definition of $\updateKV$ now ensures that $\mkvs'(\key, i) = (\stub, \txid', \Set{\txid } \cup \stub)$, 
from which it follows that $\txid' \toEDGE{\WR_{\Gr'}(\key)} \txid$.

Conversely, suppose that $\txid' \toEDGE{\WR_{\Gr'}(\key)} \txid''$. 
Recall that $\txidset_{\Gr'} = \txidset_{\Gr} \cup \Set{ \txid }$, hence either 
$\txid'' \in \txidset_{\Gr}$ or $\txid'' = \txid$. 

\begin{itemize}
\item If $\txid'' = \txid$, then it must be the case that there exists an index $i = 0,\cdots, \lvert \mkvs'(\key) \rvert - 1$ 
such that $\mkvs'(\key, i) = (\stub, \txid', \Set{\txid } \cup \stub)$. Note that if $\mkvs'(\key, \lvert \mkvs'(\key) \rvert -1)$ is 
defined, then it must be the case that $\mkvs'(\key, \lvert \mkvs'(\key) \rvert -1) = (\stub, \txid, \emptyset)$, 
hence it must be the case that $i < \lvert \mkvs'(\key) \rvert - 1$. Because $\txid \notin \mkvs$, 
then by the definition of $\updateKV$ it must be the case that $(\otR, \key, \stub) \in \fp$, 
$\mkvs(\key, i) = (\stub, \txid', \stub)$ and $i = \max(\vi(\key))$; this also implies that $\txid' = 
\max_{\WW(\key)}\Set{\wtOf(\mkvs(\key, i))}[i \in \vi(\key)]$. 

\item If $\txid'' \in \txidset_{\Gr}$, then  it must be the case that $\txid'' \neq \txid$. 
In this case, it also must exist an index $i = 0,\cdots, \lvert \mkvs'(\key) \rvert - 1$ 
such that $\mkvs'(\key, i) = (\stub, \txid', \Set{\txid''} \cup \stub)$. As in the previous 
case, we note that $i < \lvert \mkvs'(\key) \rvert - 1$, which together 
with the fact that $\txid'' \neq \txid$ leads to $\mkvs(\key, i) = (\stub, \txid', \Set{\txid''} \cup \stub)$. 
It follows that $\txid' \toEDGE{\WR_{\Gr}(\key)} \txid''$.
\end{itemize}
\end{itemize}

\item 
Similar to \( \WR(\key)_{\Gr} \), there are two cases that either the \( \txid'' \) already exists in the dependency graph before,
or it is the newly committed transaction.
\begin{itemize}
\item Suppose that $\txid' \toEDGE{\WW_{\Gr}(\key)} \txid''$. 
By definition, there exist two indexes $i, j$ such that 
$\mkvs(\key, i) = (\stub, \txid', \stub)$, $\mkvs(\key, j) = (\stub, \txid'', \stub)$ 
and $i < j$. The definition of $\updateKV$ ensures that 
$\mkvs'(\key, i) = (\stub, \txid', \stub)$, $\mkvs'(\key, j) = (\stub, \txid'', \stub)$, 
and because $i < j$ we obtain that $\txid' \toEDGE{\WW_{\Gr'}(\key)} \txid''$. 

\item Suppose that $(\otW, \key, \stub) \in \fp$. Then $\mkvs'(\key, \lvert \mkvs(\key) \rvert) = (\stub, \txid, \stub)$.
Let $\txid' \in \txidset_{\Gr}$; by definition there exists an index $i = 0,\cdots, \lvert \mkvs(\key) \rvert$ 
such that $\mkvs(\key, i) = (\stub, \txid', \stub)$. It follows that $\mkvs'(\key, i) = (\stub, \txid', \stub)$, and 
because $i < \lvert \mkvs(\key) \rvert$, then we have that $\txid' \toEDGE{\WW_{\Gr'}(\key)} \txid$. 

Conversely, suppose that $\txid' \toEDGE{\WW_{\Gr'}(\key)} \txid''$. Because 
$\txidset_{\Gr'} = \txidset_{\Gr} \cup \Set{ \txid }$, we have two possibilities. Either $\txid'' = \txid$, 
or $\txid'' \in \txidset_{\Gr}$. 

\begin{itemize}
\item If $\txid'' = \txid$, then it must be the case that $(\otW, \key, \stub) \in_{\Gr'} \txid$, 
or equivalently there exists an index $i=0,\cdots, \lvert \mkvs'(\key) \rvert -1 $ such that 
$\mkvs'(\key, i) = (\stub, \txid, \stub)$. Because $\txid \notin \mkvs$, and because for any 
$i = 0, \cdots, \lvert \mkvs(\key) \rvert - 1$, $\mkvs'(\key, i) = (\stub, \txid, \stub) \implies 
\mkvs(\key, i) = (\stub, \txid, \stub)$, then it necessarily has to be $i = \mkvs'(\key) \rvert - 1$. 
According to the definition of $\updateKV$, this is possible only if $(\otW,\key, \stub) \in \fp$. 
Finally, note that because $\txid' \toEDGE{\WW_{\Gr'}(\key)} \txid$, then 
there exists an index $j < \lvert \mkvs'(\key, i) \rvert - 1$ such that 
$\mkvs'(\key, j) = (\stub, \txid' ,\stub)$. The fact that $j < \lvert \mkvs'(\key, i) \rvert - 1$ 
we obtain that $\mkvs(\key, j) = (\stub, \txid', \stub)$, 
or equivalently $\txid' = \wtOf(\mkvs(\key, \stub))$. 

\item If $\txid'' \in \txidset_{\Gr}$, then there exist two indexes $i,j$ such that 
$j < \lvert \mkvs'(\key, j) \rvert - 1$, $\mkvs'(\key, j) = (\stub, \txid'', \stub)$, 
$i < j$, and $\mkvs'(\key, i) = (\stub, \txid', \stub)$. It is immediate to observe 
that $\mkvs(\key, i) = (\stub, \txid', \stub)$, $\mkvs(\key, j) = (\stub, \txid'', \stub)$, 
from which $\txid' \toEDGE{\WW_{\Gr}(\key)} \txid''$ follows. 
\end{itemize}
\end{itemize}

\end{enumerate}
\end{proof}

\subsection{Abstract Execution Traces to KV-Store Traces}
\label{sec:aexectrace2kv}

We show to construct, given an abstract execution $\aexec$, 
a set of $\ET_{\top}$-traces $\KVtrace(\ET_{\top}, \aexec)$ in normal form such that for any 
$\tr \in \KVtrace(\ET_{\top}, \aexec)$, the trace \( \tr \) satisfies $\lastConf(\tr) = (\mkvs_{\aexec}, \stub)$. 
We first define the \( \cut[\aexec,n] \) function in \cref{def:aexec.inductive} 
which gives the prefix of the first \( n \) transactions of the abstract execution \( \aexec \).
The  \( \cut[\aexec,n] \) function is very useful for later discussion.

\begin{definition}
\label{def:aexec.inductive}
Let $\aexec$ be an abstract execution, let $n = \lvert \txidset_{\aexec} \rvert$, and let 
$\Set{\txid_{i}}_{i=1}^{n} \subseteq \txidset_{\aexec}$ be such that $\txid_{i} \toEDGE{\AR_{\aexec}} \txid_{i+1}$. 
The \emph{cut} of the first \( n \) transactions from an abstract execution \( \aexec \) is defined as the follows:
\begin{align*}
\cut[\aexec, 0] & \defeq ([], \emptyset, \emptyset)\\
\cut[\aexec , i+1] & \defeq \extend[\cut[\aexec, i], \txid_{i+1}, \VIS^{-1}_{\aexec}(\txid_{i+1}), \TtoOp{T}_{\aexec}(\txid_{i+1})]
\end{align*}
\end{definition}

\begin{proposition}[Well-defined \( \cut \)]
\label{prop:aexec.inductive}
For any abstract execution $\aexec$, $\aexec = \cut[\aexec, \abs{ \txidset_{\aexec} }]$.
\end{proposition}
\begin{proof}
    This is an instantiation of \cref{lem:cut.explicit} by choosing $i = \lvert \txidset_{\aexec} \rvert$. 
\end{proof}

\begin{lemma}[Prefix]
\label{lem:cut.explicit}
For any abstract execution $\aexec$, and index $i: i \leq j \leq \lvert \txidset_{\aexec} \rvert$, 
if $\txidset_{\aexec} = \Set{\txid_{i}}_{i=1}^{n}$ be such that 
$\txid_{i} \toEDGE{\AR_{\aexec}} \txid_{i+1}$, 
then $\cut[\aexec, i] = \aexec_{i}$ where 
\begin{align*}
\TtoOp{T}_{\aexec_{i}}(\txid) &=
\begin{cases}
\TtoOp{T}_{\aexec}(\txid) & \text{if } \exists j \leq i.\; \txid = \txid_{j}\\
\text{undefined} & \text{otherwise}\\
\end{cases} \\
\VIS_{\aexec_{i}} &= \Set{ (\txid, \txid') \in \txidset_{\aexec_{i}} }[ \txid \toEDGE{\VIS_{\aexec}} \txid'] \\
\AR_{\aexec_{i}} &= \Set{ (\txid, \txid') \in \txidset_{\aexec_{i}} }[ \txid \toEDGE{\AR_{\aexec}} \txid']
\end{align*}
\end{lemma}

\begin{proof}
Fix an abstract execution $\aexec$. We prove by induction on $i = \lvert \txidset_{\aexec} \rvert$.
\begin{itemize}
\item \caseB{$i = 0$} Then $\TtoOp{T}_{\aexec'} = [], \VIS_{\aexec'} = \emptyset$, 
$\AR_{\aexec'} = \emptyset$, which leads to $\aexec' = \cut[\aexec, 0]$. 
\item \caseI{$i = i' + 1$} 
Assume that $\cut[\aexec, i'] = \aexec_{i'}$. 
We prove the following: 
\begin{itemize}
\item $\TtoOp{T}_{\cut[\aexec, i]} = \TtoOp{T}_{\aexec_i}$. 
By definition, 
\[
    \begin{array}{l}
\TtoOp{T}_{\cut[\aexec,i]} = \TtoOp{T}_{\cut[\aexec, i']}\rmto{\txid_{i}}{\TtoOp{T}_{\aexec}(\txid_{i})} 
\TtoOp{T}_{\aexec_{i'}}\rmto{\txid_{i}}{\TtoOp{T}_{\aexec}}(\txid_{i}) = \TtoOp{T}_{\aexec_{i}}
\end{array}
\]
\item $\VIS_{\cut[\aexec, i]} = \VIS_{\aexec_{i}}$. 
Note that, by inductive hypothesis, $\txidset_{\cut[\aexec, i']} = \txidset_{\aexec_{i'}} = \Set{\txid_{j}}_{j=1}^{i'}$. 
We have that  
\[
\begin{array}{l}
    \VIS_{\cut[\aexec, i]}
    \begin{array}[t]{l}
    = \VIS_{\cut[\aexec, i']} \cup \Set{(\txid_j, \txid_{i}) \in \VIS_{\aexec}}[1 \leq j \leq i'] \\ 
    = \VIS_{\aexec_{i'}} \cup \Set{(\txid_{j}, \txid_{i}) \in \VIS_{\aexec} }[ 1 \leq j \leq i'] \\ 
    = \Set{(\txid_{j'}, \txid_{j}) \in \VIS_{\aexec}}[1 \leq j \leq i'] \cup \Set{(\txid_{j}, \txid_{i}) \in \VIS_{\aexec}}[1 \leq j \leq i'] \\
    = \Set{(\txid_{j'}, \txid_{j}) \in \VIS_{\aexec}}[1 \leq j \leq i'] \\
    = \VIS_{\aexec_{i}}
    \end{array}
\end{array}
\]
\item $\AR_{\cut[\aexec, i]} = \AR_{\aexec_{i}}$. It follows the same way 
as the above. 
\end{itemize}
\end{itemize}
\end{proof}

Let $\Clients(\aexec) \defeq \Set{\cl}[\exsts{ n } \txid_{\cl}^{n} \in \txidset_{\aexec}]$.
Given an abstract execution $\aexec$, a client $\cl$ and an index $i : 0 \leq i < \abs{\txidset_\aexec}$,
the function $\nextTxid[\aexec, \cl, i] \defeq \min_{\AR_{\aexec}} \Set{\txid_{\cl}^{j} }[ \txid_{\cl}^{n} \notin \txidset_{\cut[\aexec, i]}]$. 
Note that $\nextTxid[\aexec, \cl, i]$ could be undefined. 
The conversion from abstract execution tests to \( \ET \) traces is in \cref{def:aexec2kvtrace}.

\begin{definition}
\label{def:aexec2kvtrace}
Given an abstract execution $\aexec$ and an index $i : 0 \leq i < \abs{\txidset_\aexec}$, 
the function $\KVtrace(\ET_{\top}, \aexec, i)$ is defined as the smallest set such that:
\begin{itemize}
\item 
$(\mkvs_{0}, \lambda \cl \in \Clients(\aexec). \lambda \key.\Set{0}) \in \KVtrace(\ET_{\top}, \aexec, 0)$, 
\item suppose that $\tr \in \KVtrace(\ET_{\top}, \aexec, i)$ for some $i$.  
Let
\begin{itemize} 
\item $\txid = \min_{\AR_{\aexec}}(\txidset_{\aexec} \setminus T_{\cut[\aexec, i]})$, 
\item  $\cl, n$ be such that $\txid = \txid_{\cl}^{n}$, 
\item  $\vi = \getView[\aexec, \VIS^{-1}_{\aexec}(\txid_{\cl}^{n})]$, 
\item $\vi' = \getView[\aexec, \txidset]$, where $\txidset$ is an arbitrary subset of $\txidset_{\aexec}$ if 
$\nextTxid[\aexec, \cl, i+1]$ is undefined, or is such that 
$\txidset \subseteq (\AR_{\aexec}^{-1})\rflx(\txid) \cap \VIS^{-1}_{\aexec}(\nextTxid[\cl, i+1])$, 
\item $\fp = \TtoOp{T}_{\aexec}(\txid)$, 
\item $(\mkvs_{\tr}, \vienv_{\tr}) = \lastConf(\tr)$, and
\item $\mkvs = \updateKV[\mkvs_{\tr}, \vi, \fp, \txid]$.
\end{itemize}
Then
\[
\left( 
\begin{array}{l}
\tr \toET{(\cl, \varepsilon)}[\ET_{\top}] (\mkvs_{\tr}, \vienv_{\tr}\rmto{\cl}{\vi}) 
\toET{(\cl, \fp)}[\ET_{\top}] (\mkvs, \vienv_{\tr}\rmto{\cl}{\vi'}) 
\end{array}
\right) \in \KVtrace(\ET_{\top}, \aexec, i+1)
\]
\end{itemize}
Last, the function $\KVtrace(\ET_{\top}, \aexec) \defeq \KVtrace(\ET_{\top}, \aexec, \lvert \txidset_\aexec \rvert)$.
\end{definition}

\begin{proposition}[Abstract executions to trace \( \ET_\top \)]
\label{prop:aexec2kvtrace}
Given an abstract execution $\aexec$ satisfying $\RP_{\LWW}$, 
and a trace $\tr \in \KVtrace(\ET_{\top}, \aexec)$,
then $\lastConf(\tr) = (\mkvs_{\aexec}, \stub)$ and $\mkvs_{\aexec} \in \CMs(\ET_{\top})$. 
\end{proposition}
\begin{proof}
Let $\aexec$ be an abstract execution that satisfies the last write wins policy. 
Let $n = \lvert \txidset_{\aexec} \rvert$. Fix $i =0,\cdots, n$, 
and let $\tr \in \KVtrace(\ET_{\top}, \aexec, i)$. We prove, by 
induction on $i$, that $\tr \in \CMs(\ET_{\top})$, and 
$\lastConf(\tr) = (\mkvs_{(\cut[\aexec, i]}, \stub)$. 
Then the result follows from  \cref{prop:aexec.inductive}.

\begin{itemize}
\item \caseB{$i = 0$} By definition, $\tr = (\mkvs_{0}, \vienv_{0})$, 
where $\vienv_{0} = \lambda \cl \in \Clients(\aexec). \lambda \key.\Set{0}$. 
Clearly, we have that $\tr \in \CMs(\ET_{\top})$. 
\item \caseI{$i = i'+1$} Let $\txid_{i} = \min_{\AR_{\aexec}}(\txidset_{\aexec} \setminus \txidset_{\cut[\aexec, i']})$, 
and suppose that $\txid_{i} = \txid_{\cl}^{m}$ for some client $\cl$ and index $m$. 
Fix $\vi = \getView[\aexec, \VIS_{\aexec}^{-1}(\txid_{i})]$, and  $\fp = \TtoOp{T}_{\aexec}(\txid_{i})$.
We prove that there exists a trace $\tr' \in \KVtrace(\ET_{\top}, \aexec, i')$ and a set 
$\txidset$ such that: 
\begin{enumerate}
\item if $\nextTxid[\cl, \aexec, i]$ is undefined then $\txidset \subseteq \txidset_{\aexec}$, otherwise 
\[
    \txidset \subseteq \VIS^{-1}_{\aexec}(\nextTxid[\cl, \aexec, i]) \cap (\AR_{\aexec}^{-1})\rflx(\txid_{i})
\]
\item the new trace \( \tr \) such that
\[
    \tr = \tr' \toET{(\cl, \varepsilon)}[] (\mkvs_{\tr'}, \vienv_{\tr'}\rmto{\cl}{\vi}) \toET{(\cl, \fp)}[] 
(\mkvs,  \vienv_{\tr'}\rmto{\cl}{\vi'})
\]
where $(\mkvs_{\tr'}, \vienv_{\tr'}) = \lastConf(\tr')$, and $\mkvs = \updateKV[\mkvs_{\tr'}, \vi, \fp, \txid_{i}]$, 
and $\vi' = \getView[\aexec, \txidset]$.
\end{enumerate}
By inductive hypothesis, we may assume that $\tr' \in \CMs(\ET_{\top})$, and $\mkvs_{\tr'} = \mkvs_{\cut[\aexec, i']}$. 
We prove the following facts: 
\begin{enumerate}
\item $\mkvs = \mkvs_{\cut[\aexec, i]}$. 
Because of \cref{prop:extend.update.sameop} and \cref{prop:aexec.inductive},
we obtain 
\[
\begin{array}{l}
\mkvs = \updateKV[\mkvs_{\tr'}, \vi, \fp, \txid_{i}] \\
\quad = \updateKV[\mkvs_{\cut[\aexec, i']}, \getView[\aexec, \VIS^{-1}_{\aexec}(\txid_{i})], \TtoOp{T}_{\aexec}(\txid_{i}), \txid_{i}] \\
\quad = \mkvs_{\extend[\cut[\aexec, i'], \VIS^{-1}_{\aexec}(\txid_{i}), \txid_{i}, \TtoOp{T}_{\aexec}(\txid_{i})] } \\
\quad = \mkvs_{\extend[\cut[\aexec, i]]}
\end{array}
\]

\item $(\mkvs_{\tr'}, \vienv_{\tr'}) \toET{(\cl, \varepsilon)}[] (\mkvs_{\tr'}, \vienv_{\tr'}\rmto{\cl}{\vi})$. 
It suffices to prove that $\vienv_{\tr'}(\cl) \viewleq \vi$ for any key $\key$.
By \cref{lem:cut.explicit} we have that $\txidset_{\cut[\aexec, i']} = \Set{\txid_{j}}_{j=1}^{i'}$, for 
some $\txid_{1},\cdots, \txid_{i'}$ such that whenever $1 \leq j < j' \leq i'$, then 
$\txid_{j} \toEDGE{\AR_{\aexec}} \txid_{j'}$. We consider two possible cases: 

\begin{itemize}
\item For all $j : 1 \leq j \leq i'$, and $h \in \Nat$, then $\txid_{j} \neq \txid_{\cl}^{h}$.
In this case we have that no transition contained in $\tr'$ has the form 
$(\stub, \stub) \toET{(\cl, \stub)}[.] (\stub, \stub)$, from which it is possible to infer 
that  $\vienv_{\tr'}(\cl) = \lambda \key. \Set{0}$. Because $\vi = \getView[\aexec, \VIS^{-1}_{\aexec}(\txid_{i})]$, 
then by definition we have that $0 \in \vi(\key)$ for all keys $\key \in \Keys$. It follows that 
$\vienv_{\tr'}(\cl) \viewleq \vi$. 

\item There exists an index $j : 1 \leq j \leq i'$ and an integer $h \in \Nat$ such that $\txid_{j} = \txid_{\cl}^{h}$. 
Without loss of generality, let $j$ be the largest such index. 
It follows that the last transition in $\tr'$ of the form $(\stub, \stub) \toEDGE{(\cl, \fp_{j})} (\stub, \vienv_{\mathsf{pre}})$ 
is such that $\vienv_{\mathsf{pre}}(\cl) = \getView[\aexec, \txidset_{\mathsf{pre}}]$, 
for some $\txidset_{\mathsf{pre}} \subseteq \VIS^{-1}_{\aexec}(\txid_{i}) \cap (\AR^{-1}_{\aexec})\rflx(\txid_{j})$.
This is because $\nextTxid[\cl, \aexec, j]$  is defined and equal to $\txid_{i}$. 
Furthermore, because the trace $\tr'$ is in normal form by construction, 
in $\tr'$ a transition of the form $(\stub, \stub) \toET{(\cl, \varepsilon)}[\ET_{\top}] (\stub, \stub)$ 
is always followed by a transition of the form $(\stub, \stub) \toET{(\cl, \fp')}[\ET_{\top}] (\stub, \stub)$. 
Because we assume that the last transition where client $\cl$ executes a transaction in $\tr'$ 
has the form $(\stub, \stub) \toET{(\cl, \fp_{j})}[\ET_{\top}] (\stub, \vienv_{\mathsf{pre}})$, 
then the latter is also the last transition for client $\cl$ in $\tr'$ 
(i.e. including both execution of transactions and view updates). 
It follows that $\vienv_{\tr'}(\cl) = \vienv_{\mathsf{pre}}(\cl)$, and in particular 
$\vienv_{\tr'}(\cl) = \getView[\aexec, \txidset_{\mathsf{pre}}]$. By definition, 
$\txidset_{\mathsf{pre}} \subseteq  \VIS^{-1}_{\aexec}(\txid_{i}) \cap (\AR^{-1}_{\aexec})\rflx(\txid_{j}) 
\subseteq \VIS^{-1}_{\aexec}(\txid_{i})$. By  \cref{lem:getView.monotone}, 
we have that $\vienv_{\tr'}(\cl) = \getView[\aexec, \txidset_{\mathsf{pre}}] \viewleq 
\getView[\aexec, \VIS^{-1}_{\aexec}(\txid_{i})] = \vi$, as we wanted to prove.
\ac{Note: this is more a sketch, rather than a real proof. A Proposition giving an explicit form to the 
structure of any $\tr \in \KVtrace(\ET_{\top}, \aexec)$ would be helpful for a more rigorous proof here.}
\end{itemize}

\item $(\mkvs_{\tr'}, \vienv_{\tr'}\rmto{\cl}{\vi}) \toET{(\cl, \fp)}[\ET_{\top}] (\mkvs,  \vienv_{\tr'}\rmto{\cl}{\vi'})$. 
    It suffices to show that $\ET_{\top} \vdash (\mkvs_{\tr'}, \vi) \csat \fp: (\mkvs,\vi')$. 
That is, it suffices to show that $\vi \in \Views(\mkvs_{\tr'})$, $\vi' \in \Views(\mkvs)$, 
and whenever $(\otR, \key, \val) \in \fp$, then $\max_{<}(\vi(\key)) = (\val, \stub, \stub)$. 
The first two facts are a consequence of \cref{lem:cut.views}, $\mkvs_{\tr'} = \mkvs_{\cut[\aexec, i']}$, and  $\mkvs_{\cut[\aexec, i]}$. 
The last one that if $(\otR, \key, \val) \in \fp$ then $\max_{<}(\vi(\key)) = (\val, \stub, \stub)$ follows the fact that 
$\aexec$ satisfies the last write wins policy and the fact that $\vi = \getView[\VIS^{-1}_{\aexec}(\txid_{i})]$.
\ac{Again, the proof is really loose here, mostly because I got bored.}
\end{enumerate} 

\end{itemize}
\end{proof}

\begin{lemma}[Monotonic \( \getView \)]
\label{lem:getView.monotone}
Let $\aexec$ be an abstract execution, and let $\txidset_1 \subseteq \txidset_2 \subseteq \txidset_{\aexec}$. 
Then $\getView[\aexec, \txidset_1] \viewleq \getView[\aexec, \txidset_2]$.
\end{lemma}
\begin{proof}
Fix $\key \in \Keys$. By definition  
\[
\begin{array}{l}
    \getView[\aexec, \txidset_1](\key) = \Set{0} \cup \Set{i}[\wtOf(\mkvs_{\aexec}(\key, i)) \in \txidset_1] \\
    \quad \subseteq \Set{0} \cup \Set{i}[\wtOf(\mkvs_{\aexec}(\key, i)) \in \txidset_2] \\
\quad = \getView[\aexec, \txidset_2](\key)
\end{array}
\]
then it follows that  $\getView[\aexec, \txidset_1] \viewleq \getView[\aexec, \txidset_2]$.
\end{proof}

\begin{lemma}[Valid view on cut of abstract execution]
\label{lem:cut.views}
Let $\aexec$ be an abstract execution, with $\txidset_{\aexec} = \Set{\txid_{i}}_{i = 1}^{n}$ for 
$n = \lvert \txidset_{\aexec} \rvert$, and \( i : 0 \leq i < n\) such that $\txid_{i} \toEDGE{\AR_{\aexec}} \txid_{i+1}$.
Assuming $\txidset_{0} = \emptyset$, and $\txidset_{i} \subseteq (\AR^{-1})\rflx(\txid_{i})$ for $i : 0 \leq i \leq n$,
then $\getView[\aexec, \txidset_{i}] \in \Views(\mkvs_{\cut[\aexec, i]})$.
\end{lemma}

\begin{proof}
We prove by induction on the index $i$. 
\begin{itemize}
\item \caseB{$i = 0$} It follows $\txidset_{0} = \emptyset$, and $\getView[\aexec, \txidset_{0}] = \lambda \key. \Set{0}$. 
We also have that $\mkvs_{\cut[\aexec, 0]} = \lambda \key \ldotp (\val_0, \txid_{0}, \emptyset)$, hence 
it is immediate to see that $\getView[\aexec, \txidset_{0}] \in \Views(\mkvs_{\cut[\aexec, 0]})$.

\item \caseI{$i = i'+1$}
Suppose that for any $\txidset \subseteq (\AR_{\aexec}^{-1})\rflx(\txid_{i'})$, 
then $\getView[\aexec, \txidset] \in \Views(\mkvs_{\cut[\aexec, i]})$. 
Let consider the set $\txidset_{i}$.
Note that, because of \cref{prop:extend.update.sameop}, we have that
\[
\begin{array}{l}
\mkvs_{\cut[\aexec, i]} =
\mkvs_{\extend[\cut[\aexec, i'], \txid_{i}, \VIS^{-1}_{\aexec}(\txid_{i}), \TtoOp{T}_{\aexec}(\txid_{i})]} 
= \updateKV[\mkvs_{\cut[\aexec, i']}, \getView[\VIS^{-1}_{\aexec}(\txid_{i})], \TtoOp{T}_{\aexec}(\txid_{i}), \txid_{i}]
\end{array}
\]
There are two possibilities:
\begin{itemize}
\item $\txid_{i} \notin \txidset_{i}$, where case $\txidset_{i} \subseteq (\AR_{\aexec}^{-1})\rflx(\txid_{i'})$.
From the inductive hypothesis we get $\getView[\aexec, \txidset_{i}] \in \Views(\mkvs_{\cut[\aexec, i']})$. 
Note that $\mkvs_{\cut[\aexec, i']}$ only contains the transactions identifiers from $\txid_{1}$ to $\txid_{i'}$;
in particular, it does not contain $\txid_{i}$. 
Because $\mkvs_{\cut[\aexec, i]} = \updateKV[\mkvs_{\cut[\aexec, i']}, \stub, \stub, \txid_{i}]$, 
then by \cref{lem:updatekv.preserveviews} we have that $\getView[\aexec, \txid_{i}] \in \Views(\mkvs_{\cut[\aexec, i]})$.

\item $\txid \in \txidset_{i}$. Note that for any key $\key$ such that 
$(\otW, \key, \stub) \notin \TtoOp{T}_{\aexec}(\txid_{i})$, then 
$\getView[\aexec, \txidset_{i}](\key) = \getView[\aexec, \txidset_{i} \setminus \Set{\txid_{i}}](\key)$; 
and for any key $\key$ such that $(\otW, \key, \stub) \in \TtoOp{T}_{\aexec}(\txid_{i})$, 
then $\getView[\aexec, \txidset_{i}](\key) = \getView[\aexec, \txidset_{i} \setminus \Set{\txid_{i}}](\key) 
\cup \Set{j}[\wtOf(\mkvs_{\aexec}(\key, i)) = \txid_{j}]$. 
In the last case, the index $j$ must be such that $j < \lvert \mkvs_{\cut[\aexec, i]} \rvert - 1$, 
because we know that $\txid_{i} \in \mkvs_{\cut[\aexec, i]}$. 
It follows from this fact and the inductive hypothesis, 
that $\getView[\aexec, \txidset_{i}] \in \Views(\mkvs_{\cut[\aexec, i]})$.
\ac{This is a loose proof sketch.} 
\end{itemize}
\end{itemize}
\end{proof}

\begin{lemma}[\(\updateKV \) preserving views]
\label{lem:updatekv.preserveviews}
Given a kv-store $\mkvs$, a transactions $\txid \notin \mkvs$, views $\vi, \vi' \in \Views(\mkvs)$, 
and set of operations $\fp$, then $\vi \in \updateKV[\mkvs, \vi', \fp, \txid]$.
\end{lemma}

\begin{proof}
Immediate from the definition of $\updateKV$. Note that $\txid \notin \mkvs$ ensures that 
$\vi$ still satisfies \eqref{eq:view.atomic} with respect to the new kv-store $\updateKV[\mkvs, \vi', \fp, \txid]$.
\end{proof}

\section{The Sound and Complete Constructors of the KV-Store Semantics with Respect to Abstract Executions}
\label{sec:et-sound-complete-constructor}
In this Section we first define the set of $\ET$-traces generated by a program $\prog$. 
Then we prove correctness our semantics on kv-stores,
meaning that if a program $\prog$ executing under the execution 
test $\ET$ terminates in a state $(\mkvs, \_)$, then $\mkvs \in \CMs(\ET)$. 
\subsection{Traces of Programs under KV-Stores}
\label{sec:kv-sound-complete-theorem}

The \( \Ptraces(\ET, \prog) \) is the set of all possible traces generated by the program \( \prog \)
starting from the initial configuration \( ( \mkvs_0, \vienv_0 ) \).

\begin{definition}
Given an execution test $\ET$ a program $\prog$ and a state 
$(\mkvs, \vienv, \thdenv)$, the  $\Ptraces(\ET, \prog, \mkvs, \vienv, \thdenv)$ function
is defined as the smallest set such that:
\begin{itemize}
\item $(\mkvs, \vienv) \in \Ptraces(\ET, \prog, \mkvs, \vienv, \thdenv)$
\item if $\tr \in \Ptraces(\ET, \prog', \mkvs', \vienv',\thdenv')$
and $((\mkvs, \vienv, \thdenv) , \prog) \toCMD{(\cl, \iota)}_{\ET} (\mkvs', \vienv', \thdenv')$, 
then 
\[\tr \in \Ptraces(\ET, \prog, \mkvs, \vienv, \thdenv')\]
\item if $\tr \in \Ptraces(\ET, \prog', \mkvs', \vienv', \thdenv')$ and 
$(\mkvs, \vienv, \thdenv), \prog) \toCMD{(\cl, \vi, \fp)} ((\mkvs', \vienv', \thdenv'), \prog')$,  
then 
\[
\left( 
\begin{array}{l}
(\mkvs, \vienv) \toET{(\cl, \varepsilon)}
(\mkvs, \vienv\rmto{\cl}{\vi}) \toET{(\cl, \fp)} \tr 
\end{array}
\right) \in \Ptraces(\ET, \prog, \mkvs, \vienv, \thdenv)
\]
\end{itemize}
The set of traces generated by a program $\prog$ under the execution test $\ET$ is 
then defined as $\Ptraces(\ET, \prog) \defeq \Ptraces(\ET, \prog, \mkvs_{0}, \vienv_{0}, \thdenv_{0})$, 
where $\vienv_{0} = \lambda \cl \in \dom(\prog).\lambda \key.\Set{0}$, and 
$\thdenv_{0} = \lambda \cl \in \dom(\prog).\lambda a.0$.
\end{definition}

\begin{proposition}
\label{prop:program-trace-in-et-trace}
For any program $\prog$ and execution test $\ET$, 
$\Ptraces(\ET, \prog) \subseteq \confOf[\ET]$ and $\tr \in \Ptraces(\ET, \prog)$ is in normal form. 
\end{proposition}
\begin{proof}
    First, by the definition of \( \Ptraces \), 
    it only constructs trace in normal form.
    It is easy to prove that for any trace \( \tau \) in \( \Ptraces(\ET, \prog) \), by induction on the trace length,
    the trace is also in \( \confOf[\ET] \).
\end{proof}

\begin{corollary}
If a trace in the following form
\[
    (\mkvs_{0}, \vienv_{0}, \thdenv_{0}), \prog) \toTRANS_{\ET} \cdots \toTRANS_{\ET} 
    (\mkvs, \vienv, \thdenv, \lambda \cl \in \dom(\prog). \pskip)
\]
then $\mkvs \in \CMs(\ET)$.
\end{corollary}
\begin{proof}
    By the definition of \( \Ptraces \), 
    there exists a corresponding trace \( \tau \in \Ptraces(\ET, \prog) \).
    By \cref{prop:program-trace-in-et-trace}, such trace \( \tau \in \confOf[\ET] \),
    therefore \( \mkvs \in \CMs(\ET)\) by definition of \( \CMs(\ET) \).
\end{proof}

Similar to \( \interpr{\prog}_{(\RP,\Ax)} \) 
, the function \( \interpr{\prog}_{\ET} \) is defined as the following:
\[
    \interpr{\prog}_{\ET} = \Set{ \mkvs }[ (\mkvs_{0}, \vienv_0, \thdenv_{0} ), \prog \toPROG{\stub}_{\ET}^{\ast} (\mkvs, \stub, \stub ), \prog_{f}) ]
\]
where $\thdenv_{0} = \lambda \cl \in \dom(\prog).\lambda \vx.0$ and $\prog_{f} = \lambda \cl \in \dom(\prog).\pskip$.

\begin{proposition}
    \label{thm:consistency-intersect-permissive}
    For any program $\prog$ and execution test $\ET$:
    \( \interpr{\prog}_{\ET} = \interpr{\prog}_{\ET_\top}  \cap \CMs(\ET) \).
\end{proposition}
\begin{proof}
    We prove a stronger result that for any program $\prog$ and execution test $\ET$, $\Ptraces(\ET, \prog) = \Ptraces(\ET_{\top}, \prog) \cap \confOf[\ET]$.
    It is easy to see \(\Ptraces(\ET, \prog) \subseteq \Ptraces(\ET_\top, \prog) \).
    By \cref{prop:program-trace-in-et-trace}, we know \( \Ptraces(\ET, \prog) \subseteq \confOf[\ET]\).
    Therefore \(  \Ptraces(\ET, \prog) \subseteq \Ptraces(\ET_\top, \prog) \cap \confOf[\ET] \).

    Let consider a trace \( \tau \) in \( \Ptraces(\ET_\top, \prog) \cap \confOf[\ET] \).
    By inductions on the length of trace, 
    every step that commits a new transaction  must satisfy \( \ET \) as \( \tau \in \confOf[\ET] \).
    It also reduce the program \( \prog \) since \( \tau \in \Ptraces(\ET_\top, \prog) \).
    By the definition \( \Ptraces(\ET, \prog) \), we can construct the same trace \( \tau \) so that \( \tau \in \Ptraces(\ET, \prog) \).
\end{proof}

\subsection{Adequate of KV-Store Semantic}
\label{sec:adequate}

Our main aim is to prove that for any program $\prog$, 
the set of  kv-stores generated by $\prog$ under $\ET$ 
corresponds to all the possible abstract executions that 
can be obtained by running $\prog$ on a database that satisfies the axiomatic definition $\Ax$. 
In this sense, we aim to establish that our operational semantics is \emph{adequate}.

More precisely, suppose that a given execution test $\ET$ captures precisely 
a consistency model defined in the axiomatic style, using a set of 
axioms $\Ax$ and a resolution policy $\RP$ over abstract executions.
That is, for any abstract execution $\aexec$ that satisfies 
the axioms $\Ax$ and the resolution policy $\RP$, then $\KVtrace(\ET_{\top}, \aexec) \cap \CMs(\ET) \neq \emptyset$; 
and for any $\tr \in \CMs(\ET)$, there exists an abstract execution 
$\aexec \in \aeset(\tr)$ that satisfies the axioms $\Ax$ and the resolution policy $\RP$. 
%

We now consider the program \( \prog \).
The \cref{prop:kv2aexec_transition} and \cref{prop:aexec2kv_transition} show 
the connection between reduction steps between 
the last write win resolution policy \( (\RP_{\LWW}, \emptyset) \) 
and the most permissive execution test \( \ET_\top \).

\begin{proposition}[Permissive execution test to last write win]
\label{prop:kv2aexec_transition}
Suppose that 
\[(\mkvs, \vienv, \thdenv), \prog \toCMD{(\cl, \vi, \fp)}_{\ET_{\top}} (\mkvs', \vienv', \thdenv'), \prog'\]
Assuming an abstract execution $\aexec$ 
such that $\mkvs_{\aexec} = \mkvs$, and a set of read-only transactions $\txidset \subseteq \txidset_{\aexec}$,
then there exists an abstract execution $\aexec'$ such that $\mkvs_{\aexec'} = \mkvs'$, and 
\[
(\aexec, \thdenv), \prog \toAEXEC{(\cl, \txidset \cup \Tx[\mkvs, \vi], \fp)}_{(\RP_{\LWW}, \emptyset)}
(\aexec', \thdenv'), \prog'
\]
\end{proposition}
\begin{proof}
Suppose that $(\mkvs, \vienv, \thdenv), \prog \toCMD{(\cl, \vi, \fp)}_{\ET_{\top}} (\mkvs', \vienv', \thdenv), \prog'$. 
This transition can only be inferred by applying Rule \rl{PSingleThread}, meaning that 
\begin{itemize}
\item $\prog(\cl) \mapsto \cmd$ for some command $\cmd$, 
\item $\cl \vdash (\mkvs, \vienv(\cl), \thdenv(\cl)), \cmd \toCMD{(\cl, \vi, \fp)}_{\ET_{\top}} (\mkvs', \vi', \stk'), \cmd'$ 
for some $\vi', \stk'$, and 
\item $\vienv' = \vienv\rmto{\cl}{\vi'}$, $\thdenv' = \thdenv\rmto{\cl}{\stk'}$ and $\prog' = \prog\rmto{\cl}{\cmd'}$. 
\end{itemize}
Let $\aexec$ be such that $\mkvs_{\aexec} = \mkvs$, and let $\txidset \subseteq \txidset_{\aexec}$ be a set of read-only transactions in $\aexec$. 
It suffices to show that there exists an abstract execution $\aexec'$ such that 
$\mkvs_{\aexec'} = \mkvs'$, and 
\[
    \cl \vdash (\aexec, \thdenv(\cl)), \cmd \toAEXEC{(\cl, \txidset \cup \Tx[\mkvs, \vi], \fp)}_{(\RP_{\LWW}, \emptyset)} (\aexec', \stk'), \cmd'.
\]
By the \rl{ASingleThread} rule, we obtain 
\[ 
    (\aexec, \thdenv) \prog, \toAEXEC{(\cl, \txidset \cup \Tx[\mkvs,\vi], \fp)}_{(\RP_{\LWW}, \emptyset)} (\aexec' ,\thdenv'), \prog'
\]
Now we perform a rule induction on the derivation of the transition 
\[
    \cl \vdash (\mkvs, \vienv(\cl), \thdenv(\cl)), \cmd \toCMD{(\cl, \vi, \fp)}_{\ET_{\top}} (\mkvs', \vi', \sn'), \cmd'
\]
\caseB{\rl{PCommit}}
This implies that 
\begin{itemize}
\item $\cmd = \ptrans{\trans}$ for some $\trans$, and $\cmd' = \pskip$,
\item $\vienv(\cl) \viewleq \vi$, 
\item let $\sn = \snapshot[\mkvs, \vi]$; then $(\thdenv(\cl), \sn, \emptyset) \toTRANS^{\ast} (\stk', \_, \fp)$, 
\item $\mkvs' = \updateKV[\mkvs, \vi, \fp, \txid]$ for some $\txid \in \nextTxid(\mkvs, \cl)$, and
\item $\ET_{\top} \vdash (\mkvs, \vi) \csat \fp: (\mkvs',\vi')$.
\end{itemize}
Choose an arbitrary set of of read-only transactions $\txidset \subseteq \txidset_{\aexec}$.
We observe that $\getView[\aexec, \txidset \cup \Tx[\mkvs, \vi]] = \vi$ since $\mkvs_{\aexec} = \mkvs$ and \cref{prop:getview.tx}.
We can now apply \cref{prop:compatible.aexec2kv} and ensure that $\RP_{\LWW}(\aexec, \txidset \cup \Tx[\mkvs, \vi]) = \Set{\sn}$.
Let $\aexec' = \extend[\aexec, \txid, \txidset \cup \Tx[\mkvs, \vi]]$. 
Because $\getView[\aexec, \txidset \cup \Tx[\mkvs, \vi]] = \vi$, $\mkvs_{\aexec} = \mkvs$,
then by \cref{prop:extend.update.sameop} we have that $\mkvs_{\aexec'} = \updateKV[\mkvs, \vi, \txid, \fp] = \mkvs'$. 
To summarise, we have that $\txidset \cup \Tx[\mkvs, \vi] \subseteq \txidset_{\aexec}$, $\sn \in \RP_{\LWW}(\aexec, \txidset \cup \Tx[\mkvs, \vi])$,
$(\thdenv(\cl), \sn, \emptyset) \toTRANS^{\ast} (\stk', \_, \fp)$ and $\txid \in \nextTxid(\txidset_{\aexec}, \cl)$. 
Now we can apply \rl{ACommit} and infer
\[
\cl \vdash (\aexec, \thdenv(\cl)), \ptrans{\trans} \toAEXEC{\cl, \txidset \cup \Tx[\mkvs_{\aexec}, \vi]}_{(\RP_{\LWW}, \emptyset)} 
(\aexec', \stk'), \pskip
\]
which is exactly what we wanted to prove. 

\caseB{\rl{PPrimitive},\rl{PChoice},\rl{PIter},\rl{PSeqSkip}}
These cases are trivial since they do not alter the state of \( \mkvs \).
\caseI{\rl{PSeq}}
It is derived by the \ih
\end{proof}

\begin{proposition}[Last write win to permissive execution test]
\label{prop:aexec2kv_transition}
Suppose that 
\[(\aexec, \thdenv), \prog \toAEXEC{(\cl, \txidset, \fp)}_{(\RP_{\LWW}, \emptyset)} (\aexec', \thdenv'), \prog'\]
Then for any $\vienv$ and $\vi \in \Views(\mkvs_{\aexec})$ such that $\vi \viewleq \getView[\aexec, \txidset]$, 
the following holds:
\[
    (\mkvs_{\aexec}, \vienv\rmto{\cl}{\vi}, \thdenv), \prog 
    \toAEXEC{(\cl, \getView[\aexec, \txidset], \fp)}_{\ET_{\top}} (\mkvs_{\aexec'}, \vienv, \thdenv'), \prog'
\]
\end{proposition}
\begin{proof}
Suppose that \[(\aexec, \thdenv), \prog \toAEXEC{(\cl, \txidset, \fp)}_{(\RP_{\LWW}, \emptyset)} (\aexec', \thdenv'), \prog'\]
Fix a function $\vienv$ from clients in $\dom(\prog)$ to views in $\Views(\mkvs)$, and a view $\vi \viewleq \getView[\aexec, \txidset]$.
We show that 
$(\mkvs_{\aexec}, \vienv\rmto{\cl}{\vi}, \thdenv) \toCMD{(\cl, \getView[\aexec, \txidset], \fp)}_{\ET_{\top}} (\mkvs_{\aexec'}, 
\vienv, \thdenv'), \prog'$. 

Note that the transition 
$\aexec, \thdenv, \prog \toAEXEC{(\cl, \txidset, \fp)}_{(\RP_{\LWW}, \emptyset)} (\aexec', \thdenv'), \prog'$ 
can only be inferred using \rl{ASingleThread} rule, 
from which it follows that 
\[
    \cl \vdash (\aexec, \thdenv(\cl)), \prog(\cl) 
    \toAEXEC{(\cl, \txidset, \fp)}_{(\RP_{\LWW}, \emptyset)} (\aexec' ,\stk'), \cmd'
\]
for some $\stk'$ such that $\thdenv' = \thdenv\rmto{\cl}{\stk'}$ 
and $\cmd'$ such that $\prog' = \prog\rmto{\cl}{\cmd'}$.
It suffices to show that 
\[
    \cl \vdash (\mkvs_\aexec, \vi, \thdenv(\cl)), \prog(\cl) 
   \toCMD{(\cl, \getView[\mkvs_{\aexec}, \txidset], \fp)}_{\ET_{\top}} 
    (\mkvs_{\aexec'}, \vienv(\cl), \stk'), \cmd'
\]
Then by applying \rl{PSingleThread} we obtain 
\[
    (\mkvs_{\aexec}, \vienv\rmto{\cl}{\vi}, \thdenv), \prog 
    \toCMD{(\cl, \getView[\mkvs_{\aexec}, \txidset], \fp)}_{\ET_{\top}} 
    (\mkvs_{\aexec'}, \vienv, \thdenv'), \prog'
\]
The rest of the proof is performed by a rule induction on the derivation to inter 
\[ 
    \cl \vdash (\aexec, \thdenv(\cl)), \prog(\cl) 
    \toAEXEC{(\cl, \txidset, \fp)}_{(\RP_{\LWW}, \emptyset)} (\aexec', \stk'), \cmd'
\]
\caseB{\rl{ACommit}}
In this case we have that 
\begin{itemize}
    \item $\prog = \ptrans{\trans}$, 
    \item $\prog' = \pskip$, 
    \item $(\thdenv(\cl), \sn, \emptyset), \trans \toTRANS^{\ast} (\stk', \_, \fp), \pskip$ for an index $\sn \in \RP_{\LWW}(\aexec, \txidset)$, and 
    \item $\aexec' = \extend[\aexec, \txid, \txidset, \fp]$ for some $\txid \in \nextTxid(\aexec, \cl)$. 
\end{itemize}
Furthermore, it is easy to see by induction on the length of the derivation 
$(\thdenv(\cl), \sn, \emptyset), \trans \toTRANS^{\ast} (\stk', \_, \fp), \pskip$, 
that whenever $(\otR, \key, \val) \in \fp$ then $\sn(\key) = \val$.
Note that $\snapshot[\mkvs_{\aexec}, \getView[\aexec, \txidset]] = \sn$ by \cref{prop:compatible.aexec2kv}.
Also, if $(\otR,\key, \val) \in \fp$ then $\sn(\key) = \val$, which is possible only if  
$\mkvs_{\aexec}(\key, \max_{<}(\getView[\aexec, \txidset](\key))) = (\val, \_, \_)$.
This ensures that $\ET_{\top} \vdash (\mkvs_{\aexec}, \getView[\aexec, \txidset]) \csat \fp: \vienv(\cl)$. 
\ac{There should be a condition here that $\vienv(\cl)(\key)$ is the same as $\getView[\aexec, \txidset](\key)$ 
for any $\key$ that is neither read nor written by $\fp$.} 
We can now combine all the facts above to apply rule \rl{PCommit}
\[
    \cl \vdash (\mkvs_{\aexec}, \vi, \thdenv(\cl)), \ptrans{\trans}
    \toCMD{(\cl, \getView[\mkvs_{\aexec}, \txidset], \fp)}_{\ET_{\top}} 
    (\mkvs', \vienv(\cl), \stk'), \pskip, 
\] 
where $\mkvs' = \updateKV[\mkvs_{\aexec}, \txid, \getView[\aexec, \txidset], \fp]$. 
Recall that $\aexec' = \extend[\aexec, \txidset, \txid, \fp]$. 
Therefore by \cref{prop:extend.update.sameop} we have that $\mkvs' = \mkvs_{\aexec'}$, 
which concludes the proof of this case.

\caseB{\rl{APrimitive},\rl{AChoice},\rl{AIter},\rl{ASeqSkip}}
These cases are trivial since they do not alter the state of \( \aexec \).
\caseI{\rl{ASeq}}
It is derived by the \ih
\end{proof}

\begin{corollary}
For any program $\prog$, 
\[
\interpr{\prog}_{\ET_{\top}} = \Set{\mkvs_{\aexec} }[ \aexec \in \interpr{\prog}_{(\RP_{\LWW}, \emptyset)}]
\]
\end{corollary}
\begin{proof}
    It can be derived by \cref{prop:aexec2kv_transition} and \cref{prop:kv2aexec_transition}.
\end{proof}

\subsection{Soundness and Completeness Constructor}
\label{sec:kv2aexec-sound-complete}

We now show how all the results illustrated so far 
can be put together to show that the kv-store operational semantics 
is sound and complete with respect to abstract execution operational semantics.

\subsubsection{Soundness}
Recall that in the abstract execution operational semantics,
a client \( \cl \) loses information of the visible transactions immediately after it commits a transaction.
Yet such information is indirectly presented when the next transaction from the same client is committed.
To define the soundness judgement (\cref{def:et_sound}), we introduce a notation of \emph{invariant} (\cref{def:invariant-for-clients})
to encore constraints on the visible transactions after each commit.

\ac{The idea behind client-based invariant being that \(I(\aexec, \cl)\) represents 
the minimal set of transactions that \(\cl\) must see in \(\aexec\), before 
updating the view and performing a transaction. Such a set of transaction 
roughly correspond to the view of the client before performing a 
sequence of \emph{update view+execute transaction} operations, 
or equivalently from the view obtained after the execution of the 
last transaction from that client.}

\begin{definition}[Invariant for clients]
\label{def:invariant-for-clients}
A \emph{client-based invariant condition}, or simply \emph{invariant}, is a 
function \(I : \Aexecs \times \Clients \to \pset{\TxID}\) 
such that for any \(\cl\) we have that \(I(\aexec, \cl) \subseteq \txidset_{\aexec}\), and 
for any  \(\cl'\) such that \(\cl' \neq \cl\) we have that 
\(I(\extend[\aexec, \txid_{\cl'}^{\cdot}, \stub, \stub], \cl) = I(\aexec, \cl)\).
\end{definition}

\begin{definition}[Soundness judgement]
\label{def:et_sound}
An execution test \(\ET\) is sound with respect to an axiomatic 
definition \((\RP_{\LWW}, \Ax)\) if and only if
there exists an invariant condition \(I\) such that 
if assuming that
\begin{itemize}
    \item a client \( \cl \) having an initial view \( \vi \), 
        commits a transaction \( \txid \) with a fingerprint \( \fp \) and updates the view to \( \vi' \), 
        which is allowed by \( \ET \) \ie \(\ET \vdash (\mkvs, \vi) \csat \fp: (\mkvs',\vi')\) where \( \mkvs' = \updateKV[\mkvs, \vi ,\fp, \txid]\),
    \item a \(\aexec\) such that \(\mkvs_{\aexec} = \mkvs\) and \(I(\aexec, \cl) \subseteq \Tx[\mkvs, \vi]\),
\end{itemize}
then there exist a set of read-only transactions \(\txidset_{\rd}\) such that 
\begin{itemize}
    \item the new abstract execution \( \aexec'  = \extend[\aexec,\txid,\Tx[\mkvs,\vi] \cup \txidset_\rd, \fp]\),
    \item the view \( \vi \) satisfies \( \Ax \), \ie \(\fora{ \A \in \Ax } \Set{\txid' }[ (\txid', \txid) \in \A(\aexec')] \subseteq \Tx[\mkvs, \vi] \cup \txidset_{\rd}\), 
    \item the invariant is preserved, \ie \(I(\aexec', \cl) \subseteq \Tx[\mkvs', \vi']\).
\end{itemize}
\end{definition}

\begin{theorem}[Soundness]
\label{thm:et_soundness}
If \(\ET\) is sound with respect to \((\RP_{\LWW}, \Ax)\), then 
\[
    \CMs(\ET) \subseteq \Set{ \mkvs }[ \exsts{ \aexec \in \CMa(\RP_{\LWW}, \Ax)) }\mkvs_{\aexec} = \mkvs]
\]
\end{theorem}
\begin{proof}
Let \(\ET\) be an execution test that is sound with respect to an 
axiomatic definition \((\RP_{\LWW}, \Ax)\). Let \(I\) be 
the invariant that satisfies \cref{def:et_sound}. 
Let consider an \(\ET\)-trace \(\tr\).
We can assume that \(\tr\) is in normal form, 
a trace that every view shift of a client \( \cl \) is followed by a transaction from \( \cl \),
and any transaction from \( \cl \) must be after a view shift of \( \cl \).
Without lose generality, we can also assume that the trace does not have transitions labelled as \((\stub, \emptyset)\).
Thus we have that the following trace \( \tr \):
\begin{align*}
\tr & =  (\mkvs_{0}, \vienv_{0}) \toET{(\cl_{0}, \varepsilon)} (\mkvs_{0}, \vienv_{0}') 
\toET{(\cl_{0}, \fp_{0})} 
(\mkvs_1, \vienv_{1}) \toET{(\cl_1, \varepsilon)}  \cdots
\toET{(\cl_{n-1}, \fp_{n-1})} (\mkvs_{n}, \vienv_{n})
\end{align*}
For any \(i : 0 \leq i \leq n\), let \(\tr_{i}\) be the prefix of \(\tr\) that 
contains only the first \(2i\) transitions. 
Clearly \(\tr_{i}\) is a valid \(\ET\)-trace, and it is also a \(\ET_{\top}\)-trace. 
By \cref{prop:kvtrace2aexec}, 
any abstract execution \(\aexec_{i} \in \aeset(\tr_{i})\) satisfies the last write wins policy. 
We show by induction on \(i\) that we can always find 
an abstract execution \(\aexec_{i} \in \aeset(\tr_{i})\) such that \(\aexec_i \models \Ax\) and \(I(\aexec_{i}, \cl) \subseteq \txidset^{i}_{\cl}\)
for any client \(\cl\) and set of transactions 
\(\txidset^{i}_{\cl} = \Tx[\aexec_{i}, \vienv_{i}(\cl)] \cup \txidset^{i}_\rd\), 
and read-only transactions \(\txidset_\rd^{i}\) in \(\aexec_{i}\).
If so, because \(\aexec_{i}\) satisfies the last write wins policy,
then it must be the case that \(\aexec_{i} \models (\RP_{\LWW}, \Ax)\). 
Then by choosing \(i = n\), we will obtain that \(\aexec_{n} \models (\RP_{\LWW}, \Ax)\). 
Last, by \cref{prop:kvtrace2aexec}, \(\mkvs_{\aexec_{n}} = \mkvs_{n}\), and there is nothing left to prove.
Now let prove such \(\aexec_{i} \in \aeset(\tr_{i})\) always exists.

\caseB{\(i = 0\)} 
Let \(\aexec_{0}\) be the only abstract execution included in \(\aeset(\tr_{0})\), 
that is \(\aexec_{0} = ([], \emptyset, \emptyset)\). 
For any \(\A \in \Ax\), it must be the case that 
\(\A(\aexec_{0}) \subseteq \txidset_{\aexec_{0}} = \emptyset\), 
hence the inequation \(\A(\aexec_{0}) \subseteq \VIS_{\aexec_{0}}\) is trivially satisfies.
Furthermore, for the client invariant \(I\) we also require that \(I(\aexec_{0}, \stub) \subseteq \txidset_{\aexec_{0}} = \emptyset\); 
for any client \(\cl\) we can choose \(\txidset_{\cl}^{0} = \Tx[\mkvs_{\aexec_{0}},\vienv_{0}(\cl)] \cup \emptyset = \emptyset\). 
Therefore \(I(\aexec_{0}, \cl) = \emptyset \subseteq \emptyset = \txidset_{\cl}^{0}\).

\caseI{\(i' = i + 1\) where \(i < n\)}
By the inductive hypothesis, there exists an abstract execution \(\aexec_i\) such that  
\begin{itemize}
\item \(\aexec_{i} \models \A\) for all \(\A \in \Ax\), and 
\item \(I(\aexec, \cl) \subseteq \txidset_{\cl}^{i}\) for any client \(\cl\) and set of transactions \(\txidset_{\cl}^{i} = \Tx[\mkvs_{i}, \vienv_{i}(\cl)]\).
\end{itemize}

We have two transitions to check, the view shift and committing a transaction.
\begin{itemize}
\item the view shift transition \((\mkvs_{i}, \vienv_{i}) \toET{(\cl_{i}, \varepsilon)} (\mkvs_{i}, \vienv'_{i})\). 
By definition, it must be the case that \(\vienv'_{i} = \vienv_{i}\rmto{\cl}{\vi'_{i}}\) 
for some \(\vi'_{i}\) such that \(\vienv_{i}(\cl) \viewleq \vi'_{i}\).
Let \((\txidset_{\cl}^{i})' = \Tx[\mkvs_{i}, \vi'_{i}]\); then we have 
\(
\txidset_{\cl}^{i} = \Tx[\mkvs_{i}, \vienv_{i}(\cl)] \subseteq \Tx[\mkvs_{i}, \vi'_{i}] = (\txidset_{\cl}^{i})' \)
As a consequence, \(I(\aexec, \cl) \subseteq \txidset_{\cl}^{i} \subseteq (\txidset_{\cl}^{i})'\).

\item the commit transaction transition $(\mkvs_{i}, \vienv_{i}') \toET{(\cl_{i}, \fp_{i})}
(\mkvs_{i+1}, \vienv_{i+1})$.
A necessary condition for this transition 
to appear in \(\tr\) is that \(\ET \vdash (\mkvs_{i}, \vienv(\cl)) \csat \fp_{i}: (\mkvs_{i+1},\vienv_{i+1}(\cl))\). 
Because \(I\) is the invariant to derive that \(\ET\) is sound with respect to \(\Ax\), 
and because \(I(\aexec_{i}, \cl_{i}) \subseteq (\txidset^{i}_{\cl})'\), 
then by \cref{def:et_sound} we have the following:
\begin{itemize}
\item there exists a set of read-only transactions \(\txidset_\rd\) 
such that 
\[
    \Set{\txid' }[ (\txid', \txid_{(\cl, i)}) \in \A(\aexec_{i+1})] \subseteq {\txidset^{i}_{\cl}}' \cup \txidset_\rd
\]
where 
\(\txid_{(\cl, i)} \in \nextTxid[\mkvs_{i}, \cl]\)
and \(\aexec_{i+1} = \extend[\aexec_{i}, \txid_{(\cl, i)}, (\txidset^{i}_{\cl})' \cup \txidset_\rd, \fp_{i}]\),
\item  \(I(\aexec_{i+1}, \cl) \subseteq \Tx[\mkvs_{i+1}, \vienv_{i+1}(\cl)]\).
\end{itemize} 
Because \(\aexec_{i} \in \aeset(\tr_{i})\), by definition of \(\aeset(\stub)\) we have that 
\(\aexec_{i+1} \in \aeset(\tr)\) (under the assumption that \(\fp_{i} \neq \emptyset\)), 
and because \(\lastConf(\tr_{i+1}) = (\mkvs_{i+1}, \stub)\), then \(\mkvs_{\aexec_{i+1}} = \mkvs_{i+1}\). 

Now we need to check if \( \aexec_{i+1} \) satisfies \( \Ax\) and the invariant \( I \) is preserved.
\begin{itemize}
\item \(\A(\aexec_{i+1}) \subseteq \VIS_{\aexec}^{i+1}\) for any \(\A \in \Ax\).
Fix \(\A \in \Ax\) and \((\txid', \txid) \in \A(\aexec_{i+1})\). 
Because \(\aexec_{i+1} = \extend[\aexec_{i}, \txid_{(\cl, i)}, (\txidset_{\cl}^{i})' \cup \txidset_\rd, \fp_{i}]\), 
we distinguish between two cases.
\begin{itemize}
\item If \(\txid = \txid_{(\cl, i)}\), then it must be the case that \(\txid' \in (\txidset^{i}_{\cl})' \cup \txidset_\rd\), 
and by definition of \(\extend\) we have that \((\txid' ,\txid_{(\cl, i)}) \in \VIS_{\aexec_{i+1}}\). 
\item If \(\txid \neq \txid_{(\cl, i)}\), then we have that \(\txid, \txid' \in \txidset_{\aexec_{i}}\). 
Because \(\aexec_{i}\) and \(\aexec_{i+1}\) agree on \(\txidset_{\aexec_{i}}\), then \((\txid', \txid) \in \A(\aexec_{i})\).
Because \(\aexec_{i} \models \A\), then \((\txid', \txid) \in \VIS_{\aexec_{i}}\). 
By definition of \(\extend\), it follows that \((\txid', \txid) \in \VIS_{\aexec_{i+1}}\).
\end{itemize}

\item Finally, we show the invariant is preserved.
Fix a client \(\cl'\). 
\begin{itemize}
\item If \(\cl' = \cl\), then we have already proved that 
\(I(\aexec_{i+1}, \cl) \subseteq \txidset_{\cl}^{i+1}\). 
\item if \(\cl' \neq \cl\), then note that \(\vienv_{i}(\cl') = \vienv'_{i}(\cl') = \vienv_{i+1}(\cl')\), 
and in particular \((\txidset^{\cl'}_{i})' = \Tx[\aexec_{i}, \vienv'_{i}(\cl')] = \Tx[\aexec_{i+1}, \vienv_{i+1}(\cl')] =  \txidset_{\cl'}^{i+1}\).
By the inductive hypothesis we know that \(I(\aexec_{i}, \cl) \subseteq \txidset_{\cl'}^{i}\), 
and by the definition of invariant, we have \(I(\aexec_{i+1}, \cl) \subseteq \txidset_{\cl'}^{i} = \txidset_{\cl'}^{i+1}\). 
\end{itemize}
\end{itemize}
\end{itemize}
\end{proof}

\begin{corollary}
\label{cor:et-soundness}
If \(\ET\) is sound with respect to \((\RP_{\LWW}, \Ax)\), then 
for any program \(\prog\), \(\interpr{\prog}_{\ET} \subseteq \Set{ \mkvs_{\aexec} }[ \aexec \in \interpr{\prog}_{(\RP_{\LWW}, \Ax)} ]\).
\end{corollary}
\begin{proof}
\begin{align*}
\interpr{\prog}_{\ET} 
& \stackrel{
}{=} 
\interpr{\prog}_{\ET_\top} \cap \CMs(\ET) \\
& \stackrel{
}{=} 
\Set{\mkvs_{\aexec} }[ \aexec \text{ satisfies } \RP_{\LWW}] \cap \CMs(\ET) \\
& \stackrel{\cref{thm:et_soundness}}{\subseteq} 
\Set{\mkvs_{\aexec} }[ \aexec \text{ satisfies } \RP_{\LWW} \land \aexec \in \CMa(\RP_{\LWW}, \Ax) ] \\
& \stackrel{
}{=}
\Set{ \mkvs_{\aexec} }[ \aexec \in \interpr{\prog}_{(\RP_{\LWW}, \Ax)} ]
\end{align*}
\end{proof}

\subsubsection{Completeness}
The Completeness judgement is in \cref{def:et_complete}.
Given a transaction \( \txid_i \) from client \( \cl \), it converts the visible transactions \( \VIS_{\aexec}^{-1}(\txid_{i}) \) into view  and such view should satisfy the \( \ET \).
Note that \( \aexec \) does not contain precise information about final view after update,
yet the visible transactions of the immediate next transaction from the same client \( \cl \) include those information.

\begin{definition}
\label{def:et_complete}
An execution test \(\ET\) is \emph{complete} with respect 
to an axiomatic definition \((\RP_{\LWW}, \Ax)\) if, for any abstract execution \(\aexec \in \CMa(\RP_{\LWW}, \Ax)\) 
and index \( i : 1 \leq i < \abs{\txidset_{\aexec}}\) such that \( \txid_{i} \toEDGE{\AR_{\aexec}} \txid_{i+1} \), there exist an initial view \(\vi_{i}\) and a final view \(\vi_{i}'\) where 
\begin{itemize}
\item \(\vi_{i} = \getView[\aexec, \VIS_{\aexec}^{-1}(\txid_{i})]\), 
\item let \(\txid_{i} = \txid_{\cl}^{n}\) for some \(\cl, n\); 
    \begin{itemize}
        \item if the transaction \(\txid_{i}' = \min_{\SO_{\aexec}}\Set{\txid' }[ \txid_i \toEDGE{\SO_{\aexec}} \txid']\) is defined, then \(\vi' = \getView[\aexec, \txidset_{i}]\) where \(\txidset_{i} \subseteq (\AR_{\aexec}^{-1})\rflx(\txid_{i}) \cap \VIS_{\aexec}^{-1}(\txid_{i}'))\); 
        \item otherwise \(\vi' = \getView[\aexec, \txidset_{i}]\) where \(\txidset_{i} \subseteq (\AR_{\aexec}^{-1})\rflx(\txid_{i})\), 
    \end{itemize}
\item \(\ET \vdash (\mkvs_{\cut[\aexec, i-1]}, \vi_{i}) \csat \TtoOp{T}_{\aexec}(\txid_{i}) : (\mkvs_{\cut[\aexec, i]},\vi_{i}')\).
\end{itemize}
\end{definition}

\begin{theorem}
\label{thm:et_complete}
Let \(\ET\) be an execution test that is complete with respect to an axiomatic definition \((\RP_{\LWW}, \Ax)\). 
Then \(\CMa(\RP_{\LWW}, \Ax) \subseteq \CMs(\ET)\).
\end{theorem}
\begin{proof}
Fix an abstract execution \(\aexec \in \CMa(\RP_{\LWW}, \Ax)\). 
For any \(i : 1 \leq i < \abs{\txidset_\aexec} \), suppose that \( \txid_i \) that is the i-\emph{th} transaction follows the arbitrary order, \ie \(\txid_{i} \toEDGE{\AR_{\aexec}} \txid_{i+1}\) 
and let \(\cl_{i}\) be the client of the i-\emph{th} step, \ie \(\txid_{i} = \txid_{\cl_{i}}^{\stub}\).
Because \(\ET\) is complete with respect to \((\RP_{\LWW}, \Ax)\), 
for any step \(i\) we can find an initial views \(\vi_i\),and a final view \(\vi'_{i}\) such that 
\begin{itemize}
\item \(\vi_i = \getView[\aexec, \VIS^{-1}_{\aexec}(\txid_{i})]\), 
\item there exists a set of transactions \(\txidset_{i}\) such that \(\getView[\aexec, \txidset_{i}] = \vi'_{i}\), and 
either \(\min_{\SO_{\aexec}}\Set{\txid' }[ \txid_{i} \toEDGE{\SO_{\aexec}} \txid']\) is 
is defined and \(\txidset_{i} \subseteq (\AR_{\aexec}^{-1})\rflx(\txid_{i}) \cap \VIS^{-1}_{\aexec}(\txid')\), 
or \(\txidset_{i} \subseteq (\AR_{\aexec}^{-1})\rflx(\txid_{i})\), 
\item \(\ET \vdash (\mkvs_{\cut[\aexec, i-1]}, \vi_i) \csat \TtoOp{T}_{\aexec}(\txid_{i}): (\mkvs_{\cut[\aexec, i]}, \vi'_{i})\).
\end{itemize}
Given above, let \(\mkvs_{i} = \cut[\aexec, i]\) and \(\fp_{i} = \TtoOp{T}_{\aexec}(\txid_{i})\). Define the views for clients as 
\[
\vienv_{0} = \lambda \cl \in \Set{\cl' }[ \exsts{ \txid \in \txidset_{\aexec} } \txid = \txid_{\cl'}] \ldotp \lambda \key \ldotp \Set{0}
\quad \vienv'_{i-1} = \vienv_{i}\rmto{\cl_{i}}{ \vi_i}
\quad \vienv_{i} = \vienv'_{i-1}\rmto{\cl_{i} }{\vi'_{i}}
\]
and the ke-stores as
\[
\mkvs_{0} = \lambda \key.(\val_{0}, \txid_{0}, \emptyset)
\quad \mkvs_{i} = \updateKV[\mkvs_{i-1}, \vi_i, \fp_{i}, \txid_{i}]
\]
Now by \cref{prop:aexec2kvtrace} we have that the following sequence of \(\ET_{\top}\)-reductions 
\[
\begin{array}{l}
(\mkvs_{0}, \vienv_{0}) \toET{(\cl_{1}, \varepsilon)}[\ET_{\top}] (\mkvs_{0}, \vienv'_{0}) 
\toET{(\cl_{1}, \fp_{1})}[\ET_{\top}] (\mkvs_{1}, \vienv_{1}) 
\toET{(\cl_{2}, \varepsilon)}[\ET_{\top}]
\cdots \toET{(\cl_{n}, \fp_{n})}[\ET_{\top}] (\mkvs_{n}, \vienv_{n})
\end{array}
\]
Note that \(\mkvs_{i} = \mkvs_{\cut[\aexec, i]}\). 
Because \(\ET \vdash ( \mkvs_{\cut[\aexec,i-1]}, \vi_i ) \csat \fp_{i} : (\mkvs_{i}, \vi'_{i})\), 
or equivalently \(\ET \vdash ( \mkvs_{\cut[\aexec, i-1]}, \vienv'_{i-1}(\cl_{i}) ) \csat \fp_{i} : ( \mkvs_{\cut[\aexec, i-1]}, \vienv_{i}(\cl_{i}) )\), therefore 
\[
\begin{array}{l}
(\mkvs_{0}, \vienv_{0}) \toET{(\cl_{1}, \varepsilon)} (\mkvs_{0}, \vienv'_{0}) 
\toET{(\cl_{1}, \fp_{1})} (\mkvs_{1}, \vienv_{1})
\toET{(\cl_{2}, \varepsilon)}
\cdots \toET{(\cl_{n}, \fp_{n})} (\mkvs_{n}, \vienv_{n})
\end{array}
\]
It follows that \(\mkvs_{n} \in \CMs(\ET)\) then \(\mkvs_{n} = \mkvs_{\cut[\aexec, n]} = \mkvs_{\aexec}\), and there is nothing left to prove.
\end{proof}

\begin{corollary}
\label{cor:et-completeness}
If \(\ET\) is complete with respect to \((\RP_{\LWW}, \Ax)\), then 
for any program \(\prog\), 
\[\Set{ \mkvs_{\aexec} }[ \aexec \in \interpr{\prog}_{(\RP_{\LWW}, \Ax)} ] \subseteq \interpr{\prog}_{\ET}\]
\end{corollary}
\begin{proof}
\begin{align*}
\Set{ \mkvs_{\aexec} }[ \aexec \in \interpr{\prog}_{(\RP_{\LWW}, \Ax)} ]
& \stackrel{
}{=} 
\Set{\mkvs_{\aexec} }[ \aexec \text{ satisfies } \RP_{\LWW} \land \aexec \in \CMa(\RP_{\LWW}, \Ax) ] \\
& \stackrel{\cref{thm:et_complete}}{\subseteq} 
\Set{\mkvs_{\aexec} }[ \aexec \text{ satisfies } \RP_{\LWW}] \cap \CMs(\ET) \\
& \stackrel{
}{=} 
\interpr{\prog}_{\ET_\top} \cap \CMs(\ET) \\
& \stackrel{
}{=} 
\interpr{\prog}_{\ET} 
\end{align*}
\end{proof}

\section{The Soundness and Completeness of Execution Tests}
\label{app:et_sound_complete}
\label{app:et-sound-complete}
\label{sec:kv-sound-complete-proof}
We use \cref{def:et_sound,def:et_complete} to prove the soundness and completeness of execution tests with respect to axiomatic definitions.
It is sufficient to match these two definition, 
then by \cref{cor:et-soundness,cor:et-completeness} we have \( \CMs(\ET) = \Set{\mkvs_\aexec}[\aexec \in \CMa(\RP_{\LWW},\Ax)] \).
\label{sec:spec-proof}

We first prove the \cref{thm:view-vis-relation}, which states that the least fix point of view matches 
 the constraint on the visibility relation on abstract execution.

\begin{theorem}[View closure to visibility closure]
    \label{thm:view-vis-relation}
    Assume \( \mkvs \) and \( \aexec \) such that \( \mkvs = \mkvs_\aexec \), 
    and \( \rel_\mkvs \) and \( \rel_\aexec \) such that \( \rel_\mkvs = \rel_\aexec \).
    For any \(\txid, \fp \),
    if there is a view \( \vi = \getView[\mkvs,\left(\rel_\mkvs^{-1}\right)^{*}(\Tx[\mkvs,\vi])] \),
    then the new abstract execution \( \aexec' = \extend[\aexec, \txid, \fp, \left(\rel_\mkvs^{-1}\right)^{*}(\Tx[\mkvs,\vi])] \)
    satisfies \( \rel_\aexec^{-1}(\VIS_{\aexec'}^{-1}(\txid)) \subseteq \VIS_{\aexec'}^{-1}(\txid) \).
    Conversely,
    If there a new abstract execution \( \aexec' = \extend[\aexec, \txid, \fp, \txidset] \) for some \( \txidset \)
    that satisfies \( \rel_\aexec^{-1}(\txidset) \subseteq \txidset \),
    and if a view \( \vi = \getView[\mkvs,\txidset] \),
    then the view \( \vi = \getView[\mkvs,\left(\rel_\mkvs^{-1}\right)^{*}(\Tx[\mkvs,\vi])] \).
\end{theorem}
\begin{proof}
    Assume \( \mkvs \) and \( \aexec \) such that \( \mkvs = \mkvs_\aexec \), 
    and \( \rel_\mkvs \) and \( \rel_\aexec \) such that \( \rel_\mkvs = \rel_\aexec \),
    Assume \(\txid, \fp \).
    Let \( \txidset  =  \left(\rel_\mkvs^{-1}\right)^{*}(\Tx[\mkvs,\vi]) \).
    Assume that a view satisfies \( \vi = \getView[\mkvs,\txidset] \).
    By the definition of \( \extend \),  the visible transactions 
    \( \VIS_{\aexec'}^{-1}(\txid) = \txidset \).
    Let consider transactions \( \txid', \txid'' \) such that \( \txid' \toEDGE{\rel_{\aexec}} \txid'' \toEDGE{\VIS_{\aexec'}} \txid \).
    This means there exists a natural number \( n \) such that 
    \( \txid'' \in \left(\rel_\mkvs^{-1}\right)^{n}(\Tx[\mkvs,\vi])\).
    Given that \( \rel_\mkvs = \rel_\aexec \), it follows \( \txid' \in \left(\rel_\mkvs^{-1}\right)^{n + 1}(\Tx[\mkvs,\vi])\), 
    then \( \txid' \in \txidset \) and so \( \txid' \toEDGE{\VIS_{\aexec'}} \txid \).

    Assume there a new abstract execution \( \aexec' = \extend[\aexec, \txid, \fp, \txidset] \),
    that satisfies \( \rel_\aexec^{-1}(\txidset) \subseteq \txidset \).
    Assume \( \vi = \getView[\mkvs,\txidset] \).
    Note that \( \rel_\aexec = \rel_\mkvs \).
    It suffices to prove
    \( \Set{\txid' \in \txidset }[\txid' \ \text{has writes}] = \Set{ \txid' \in \left(\rel_\mkvs^{-1}\right)^{*}(\Tx[\mkvs,\vi]) }[\txid' \ \text{has writes}]\).

    \begin{itemize}
    \item Assume a transaction \( \txid' \in \txidset \) that has writes.
    It is easy to see there are \( \key,i \) such that \( i \in \vi(\key) \) and
    \( \wtOf[\mkvs(\key,i)] \in \txidset \).
    This means \( \txid' \in \Tx[\mkvs,\vi] \).
    \item Assume a transaction \( \txid' \in \Tx[\mkvs,\vi] \),
    we now prove \( \left(\rel_\mkvs^{-1}\right)^{n}(\txid') \subseteq \txidset \) for all \( n \).
    \begin{itemize}
        \item \caseB{n = 0} It trivially holds that  \( \txid' \in \Tx[\mkvs,\vi]  \subseteq \txidset \).
        \item \caseI{n + 1} 
            Assume a transaction \( \txid''' \in \left(\rel_\mkvs^{-1}\right)^{n + 1}(\txid') \).
            It means there is a \( \txid'' \in \left(\rel_\mkvs^{-1}\right)^{n}(\txid') \) such that \( \txid''' \toEDGE{\rel_\mkvs} \txid'' \).
            By \ih, \( \txid'' \in \txidset \).
            Given \( \rel_\mkvs = \rel_\aexec \) and \( \rel_\aexec^{-1}(\txidset) \subseteq \txidset\),
            it is known that \( \txid''' \in \txidset \).
    \end{itemize}
    \end{itemize}
\end{proof}

\subsection{Monotonic Read \( \MR \)}
\label{sec:sound-complete-mr}

The execution test $\ET_\MR$ is sound with respect to the axiomatic definition \cite{surech-session-guarantee}
\[(\RP_{\LWW}, \Set{\lambda \aexec. \VIS_{\aexec} ; \SO_{\aexec} })\] 
We choose an invariant as the following,  
\[
    I(\aexec, \cl) = \left( \bigcup_{\Set{\txid_{\cl}^{n} \in \txidset_{\aexec} }[ n \in \Nat]} \VIS_{\aexec}^{-1}(\txid^n_\cl) \right) \setminus \txidset_\rd
\]
where \( \txidset_\rd \) is all the read-only transactions in 
\( \bigcup_{\Set{\txid_{\cl}^{n} \in \txidset_{\aexec} }[ n \in \Nat]} \VIS_{\aexec}^{-1}(\txid^n_\cl) \).
Assume a kv-store $\mkvs$, an initial and a final view $\vi, \vi'$  a fingerprint $\fp$ 
such that $\ET_{\MR} \vdash (\mkvs, \vi) \csat \fp: (\mkvs',\vi')$. 
Also choose an arbitrary $\cl$, a transaction identifier $\txid \in \nextTxid(\mkvs, \cl)$, 
and an abstract execution $\aexec$ such that $\mkvs_{\aexec} = \mkvs$ and 
\begin{equation}
I(\aexec, \cl) \subseteq \Tx[\mkvs, \vi]
\label{eq:mr_invariant}
\end{equation}
Let \( \aexec' = \extend[\aexec, \txid, \fp, \Tx[\mkvs, \vi] \cup \txidset_\rd] \).
We now check if \( \aexec' \) satisfies the axiomatic definition and the invariant is preserved:
\begin{itemize}
    \item $\Set{\txid' }[ (\txid', \txid) \in \VIS_{\aexec'} ; \SO_{\aexec'} ] \subseteq \Tx[\mkvs, \vi] \cup \txidset_\rd$. 
Suppose that $\txid' \toEDGE{\VIS_{\aexec'}} \txid'' \toEDGE{\SO_{\aexec'}} \txid$ 
for some $\txid', \txid''$. We show that $\txid' \in I(\aexec, \cl)$, and then \cref{eq:mr_invariant} ensures 
that $\txid' \in \Tx[\mkvs, \vi] \cup \txidset_{\mathsf{rd}}$. 
Suppose $\txid'' \toEDGE{\SO_{\aexec'}} \txid$, then $\txid'' = \txid_{\cl}^{n}$ for some $n \in \Nat$.
Because $\txid'' \neq \txid$ and $\txidset_{\aexec'} \setminus \txidset_{\aexec} = \Set{ \txid }$, we also 
have that $\txid'' \in \aexec$. By the invariant of $I(\aexec, \cl)$, 
we have that $\VIS^{-1}_{\aexec}(\cl) \subseteq I(\aexec, \cl)$:
because $\txid' \toEDGE{\VIS_{\aexec'}} \txid''$ and $\txid'' \neq \txid$ we have 
that $\txid' \toEDGE{\VIS_{\aexec}} \txid''$ and therefore $\txid' \in I(\aexec, \cl)$. 

\item $I(\aexec', \cl) \subseteq \Tx[\aexec', \vi'] = \Tx[\mkvs', \vi']$. 
    In this case, because $\ET_{\MR} \vdash (\mkvs, \vi) \csat \fp: (\mkvs',\vi')$, 
then it must be the case that $\vi \viewleq \vi'$. 
A trivial consequence of this fact is that $\Tx[\mkvs, \vi] \subseteq \Tx[\mkvs, \vi']$.
Also, because $\aexec' = \extend[\aexec, \txid, \Tx[\mkvs, \vi] \cup \txidset_{\rd}]$, 
we have that $\Tx[\mkvs_{\aexec}, \vi] = \Tx[\mkvs_{\aexec'}, \vi]$. 
\ac{to infer this there should be a Lemma that states that if $\vi \in \Views(\mkvs)$, 
then $\Tx[\updateKV[\mkvs, \vi', \fp, \txid], \vi] = \Tx[\mkvs, \vi]$.}
Finally, note that $\Set{\txid_{\cl}^{n} \in \aexec' }[ n \in \Nat] = 
\Set{ \txid_{\cl}^{n} \in \txidset_{\aexec} }[ n \in \Nat] \cup \txid$, that for any 
$\txid_{\cl}^{n} \in \txidset_{\aexec}$ we have that $\VIS^{-1}_{\aexec'}(\txid_{\cl}^{n}) = 
\VIS^{-1}_{\aexec}(\txid_{\cl}^{n})$, and that 
$\VIS_{\aexec'}^{-1}(\txid) = \Tx[\mkvs, \vi] \cup \txidset_{\mathsf{rd}}$. 
Using all these facts, we obtain 
\begin{align*}
    I(\aexec', \cl) 
    &= \left( \bigcup_{\Set{\txid_{\cl}^{n} \in \aexec' }[ n \in \Nat]} \VIS_{\aexec'}^{-1}(\txid_{\cl}^{n}) \right) \setminus \txidset_\rd \\
    &= \left( \left( \bigcup_{\Set{\txid_{\cl}^{n} \in \aexec }[ n \in \Nat]} \VIS_{\aexec}^{-1}(\txid_{\cl}^{n}) \right) \setminus \txidset_\rd  \right) \cup \left( \VIS^{-1}_{\aexec'}(\txid) \setminus \txidset_\rd  \right) \\
    &= I(\aexec, \cl) \cup \Tx[\mkvs, \vi] \\
    &\stackrel{\eqref{eq:mr_invariant}}{\subseteq} \Tx[\mkvs, \vi] \\
    &= \Tx[\mkvs_\aexec, \vi] \\
    &= \Tx[\mkvs_{\aexec'}, \vi] \\
    &\subseteq \Tx[\mkvs_{\aexec'}, \vi']
\end{align*}
\end{itemize}

We show that the execution test $\ET_{\MR}$ is complete 
with respect to the axiomatic definition 
\[(\RP_{\LWW}, \Set{\lambda \aexec.(\VIS_{\aexec};\SO_{\aexec})})\]
Let $\aexec$ be an abstract execution that satisfies the definition
$\CMa(\RP_{\LWW}, \Set{\lambda \aexec.(\VIS_{\aexec};\SO_{\aexec})})$, 
and consider a transaction $\txid \in \txidset_{\aexec}$. 
Assume i-\emph{th} transaction \( \txid_i \) in the arbitrary order,
and let $\vi_i = \getView[\aexec, \VIS^{-1}_{\aexec}(\txid_{i})]$.
We have two possible cases: 
\begin{itemize}
    \item the transaction $\txid'_{i} = \min_{\SO_{\aexec}}\Set{\txid' }[ \txid_{i} \toEDGE{\SO_{\aexec}} \txid']$ is 
defined. In this case let 
\[\vi'_{i} =\getView[\aexec, (\AR^{-1}_{\aexec})\rflx(\txid_{i}) \cap \VIS^{-1}_{\aexec}(\txid'_{i})]\]
Note that $\txid_{i} \toEDGE{\SO_{\aexec}} \txid'_{i}$, and because $\aexec \models \VIS_{\aexec} ; \SO_{\aexec}$, 
it follows that $\VIS^{-1}_{\aexec}(\txid_{i}) \subseteq \VIS^{-1}_{\aexec}(\txid'_{i})$. 
We also have that $\VIS^{-1}_{\aexec}(\txid_{i}) \subseteq (\AR^{-1}_{\aexec})\rflx(\txid_{i})$ because of 
the definition of abstract execution. It follows that 
\[
\VIS^{-1}_{\aexec}(\txid_{i}) \subseteq (\AR^{-1}_{\aexec})\rflx(\txid_{i}) \cap \VIS^{-1}_{\aexec}(\txid'_{i}),
\]
Recall that  $\vi_i = \getView[\aexec, \VIS^{-1}_{\aexec}(\txid_{i})]$,
and $\vi'_{i} =\getView[\aexec, (\AR^{-1}_{\aexec})\rflx(\txid_{i}) \cap \VIS^{-1}_{\aexec}(\txid'_{i})]$.
Thus we have that $\vi_i \viewleq \vi'_{i}$, and therefore $\ET_{\MR} \vdash (\mkvs_{\cut[\aexec, i]}, \vi_i) 
\csat \TtoOp{T}_{\aexec}(\txid_{i}) : (\mkvs_{\cut[\aexec, i+1]},\vi'_{i})$. 
\item the transaction $\txid'_{i} = \min_{\SO_{\aexec}}\Set{\txid' }[ \txid_{i} \toEDGE{\SO_{\aexec}} \txid_{i}]$ 
is not defined. In this case, let 
\[\vi'_{i} = \getView[\aexec, (\AR^{-1}_{\aexec})\rflx(\txid_{i})]\]
As for the case above, we have that $\vi_i \viewleq \vi'_{i}$, and therefore 
$\ET_{\MR} \vdash (\mkvs_{\cut[\aexec, i]}, \vi_i) \csat \TtoOp{T}_{\aexec}(\txid_{i}) : (\mkvs_{\cut[\aexec, i+1]},\vi'_{i},\vi'_{i})$. 
\end{itemize}

\subsection{Monotonic Write \( \MW \)}
\label{sec:sound-complete-mw}

The execution test $\ET_\MW$ is sound with respect to the axiomatic definition \cite{surech-session-guarantee}
\[(\RP_{\LWW}, \Set{\lambda \aexec. ( \SO_{\aexec} \cap \WW_\aexec ) ; \VIS_{\aexec} })\]
We pick the invariant as empty set given the fact of no constraint on the view after update:
\[ 
    I( \aexec, \cl ) = \emptyset 
\]
Assume a kv-store $\mkvs$, an initial and a final view $\vi, \vi'$  a fingerprint $\fp$ 
such that $\ET_{\MW} \vdash (\mkvs, \vi) \csat \fp: (\mkvs',\vi')$. 
Also choose an arbitrary $\cl$, a transaction identifier $\txid \in \nextTxid(\mkvs, \cl)$, 
and an abstract execution $\aexec$ such that $\mkvs_{\aexec} = \mkvs$ and 
\( I(\aexec, \cl) =  \emptyset \subseteq \Tx[\mkvs, \vi] \).
Note that since the invariant  is empty set, it remains to prove that there exists a set of read-only transactions \( \txidset_\rd \) such that
\( \aexec' = \extend(\aexec, \txid, \Tx[\mkvs, \vi] \cup \txidset_\rd, \fp ) \) and:
\[
    \begin{array}{@{}l@{}}
        \fora{ \txid' }  (\txid' ,\txid)  \in ( \SO_{\aexec'} \cap \WW_{\aexec'} ) ; \VIS_{\aexec'}
        \implies \txid' \in \Tx[\mkvs, \vi] \cup \txidset_\rd
    \end{array}
\]
which can be derived from \cref{thm:view-vis-relation}.

The execution test $\ET_{\MW}$ is complete with respect to 
the axiomatic definition
\[(\RP_{\LWW}, \Set{\lambda \aexec.(( \SO_{\aexec} \cap \WW_{\aexec} ) ; \VIS_{\aexec})})\]
Let $\aexec$ be an abstract execution that satisfies the definition
$\CMa(\RP_{\LWW}, \Set{\lambda \aexec.(( \SO_{\aexec} \cap \WW_{\aexec} ) ; \VIS_{\aexec})})$, 
and consider a transaction $\txid \in \txidset_{\aexec}$. 
Let \( \mkvs = \mkvs_\aexec \).
Assume i-\emph{th} transaction \( \txid_i \) in the arbitrary order,
and let view \( \vi_{i} = \getView[\mkvs, \VIS^{-1}_{\aexec}(\txid_{i})] \).
We also pick any final view such that \( \vi'_{i} \subseteq \getView[\mkvs, (\AR^{-1}_{\aexec})\rflx(\txid_{i})] \).
It suffices to prove \( \ET_\MW \vdash (\mkvs_{\cut[\aexec, i-1]}, \vi_i ) \csat  \TtoOp{T}_{\aexec}(\txid_{i}) : (\mkvs_{\cut[\aexec, i-1]}, \vi'_{i}) \).
It means to prove the following:
\[
    \vi_i = \getView[\mkvs_{\cut[\aexec, i-1]}, \lfpTx[\mkvs_{\cut[\aexec, i-1]},\vi,\SO \cap \WW_{\mkvs_{\cut[\aexec, i-1]}}]]
\]
which can be derived from \cref{thm:view-vis-relation}.

\subsection{Read Your Write \( \RYW \) }

\label{sec:sound-complete-ryw}

The execution test $\ET_\RYW$ is sound with respect to the axiomatic definition \cite{surech-session-guarantee}
$(\RP_{\LWW}, \Set{\lambda \aexec \ldotp \SO_{\aexec} })$.
We pick an invariant for the \( \ET_\RYW \) as the following:
\[
    I(\aexec, \cl) = \left( \bigcup_{\Set{\txid_{\cl}^{n} \in \txidset_{\aexec} }[ n \in \Nat ]} (\SO_{\aexec}^{-1})\rflx(\txid^n_\cl) \right) \setminus \txidset_\rd
\]
where \( \txidset_\rd \) is all the read-only transactions in \( \bigcup_{\Set{\txid_{\cl}^{n} \in \txidset_{\aexec} }[ n \in \Nat ]} (\SO_{\aexec}^{-1})\rflx(\txid^n_\cl) \).
Assume a kv-store $\mkvs$, an initial and a final view $\vi, \vi'$  a fingerprint $\fp$ 
such that $\ET_{\RYW} \vdash (\mkvs, \vi) \csat \fp: (\mkvs',\vi')$. 
Also choose an arbitrary $\cl$, a transaction identifier $\txid_\cl^n \in \nextTxid(\mkvs, \cl)$, 
and an abstract execution $\aexec$ such that $\mkvs_{\aexec} = \mkvs$ and 
\( I(\aexec, \cl) \subseteq \Tx[\mkvs, \vi] \).
Let a new abstract execution \( \aexec' = \extend[\aexec, \txid_\cl^n, \fp, \Tx[\mkvs, \vi] \cup \txidset_\rd] \).
We need to prove that \( \aexec' \) satisfies the constraint and the invariant is preserved:
\begin{itemize}
    \item \( \txid \in \Tx[\mkvs, \vi] \cup \txidset_\rd  \) for all \( \txid \) such that \( \txid \toEDGE{\SO_{\aexec'}} \txid_\cl^n  \). 
    Assume a transaction \( \txid \) such that \( \txid \toEDGE{\SO_{\aexec'}} \txid_\cl^n \).
It immediately implies that \( \txid = \txid_\cl^m\) where \( m < n \) and \( \txid_\cl^m \in \aexec \).
Thus we prove that 
\[ 
    \txid \in \left( \bigcup_{\Set{\txid_{\cl}^{n} \in \txidset_{\aexec} }[ n \in \Nat]} (\SO_{\aexec}^{-1})\rflx(\txid^n_\cl) \right) \subseteq \Tx[\mkvs,\vi] \cup \txidset_\rd
\]
\item \(I(\aexec',\cl) \subseteq \Tx[\mkvs_{\aexec'}, \vi'] \).
Let \( \txidset'_\rd = \txidset_\rd \) if the new transaction \( \txid_\cl^n\) has writes, otherwise \( \txidset'_\rd = \txidset_\rd \cup \Set{\txid_\cl^n}\).
First we have
\[ I(\aexec', \cl) = \left(\bigcup_{\Set{\txid_{\cl}^{m} \in \txidset_{\aexec'} }[ m \in \Nat ]} (\SO_{\aexec'}^{-1})\rflx(\txid^m_\cl) \right) \setminus \txidset'_{\rd} = \left( (\SO_{\aexec'}^{-1})\rflx(\txid^n_\cl) \right) \setminus \txidset'_\rd 
\]
Note that \( \txid^n_\cl \) is the latest transaction committed by the client \( \cl \).
For any transaction \( \txid \in (\SO_{\aexec'}^{-1})\rflx(\txid^n_\cl) \setminus \txidset'_\rd \) that has write,
because execution test requires \( z \in \vi'(\key) \) for any key \( \key \) and index \( z \) such that \( \wtOf( \mkvs_{\aexec'}(\key, z) ) \toEDGE{\SO_\aexec} \txid \),
then \( \txid \in \Tx[\mkvs_{\aexec'}, \vi'] \) as what we wanted.
\end{itemize}

The execution test $\ET_{\RYW}$ is complete with respect to 
the axiomatic definition $(\RP_{\LWW}, \Set{\lambda \aexec.\SO_{\aexec} })$. 
Let $\aexec$ be an abstract execution that satisfies the definition
$\CMa(\RP_{\LWW}, \Set{\lambda \aexec.\SO_{\aexec} })$.
Assume i-\emph{th} transaction \( \txid_i \) in the arbitrary order,
and let view \( \vi_{i} = \getView[\aexec, \VIS^{-1}_{\aexec}(\txid_{i})] \).
We construct the final view \( \vi'_i\) depending on whether \( \txid_i \) is the last transaction from the client.
\begin{itemize}
\item If the transaction \( \txid'_i = \min_{\SO_\aexec}\left(\Set{\txid'}[\txid_i \toEDGE{\SO_\aexec} \txid'] \right) \)  is defined,
then \( \vi'_i = \getView[\aexec, \txidset_i] \) where \( \txidset_i \subseteq (\AR_{\aexec}^{-1})\rflx(\txid_i) \cap \VIS_\aexec^{-1}(\txid'_i) \) for some \( \txidset_i \).
Given the definition \( \lambda \aexec.\SO_{\aexec} \), 
we know \( \SO_\aexec^{-1}(\txid'_i) \subseteq \VIS_\aexec^{-1}(\txid'_i) \),
so \(  (\AR_\aexec^{-1})\rflx(\txid_i) \cap \SO^{-1}(\txid'_i) = (\SO^{-1})\rflx(\txid_i) \subseteq \txidset_i \).
Take \( j,\key \) such that \( \wtOf[\mkvs_{\cut[\aexec, i]}(\key,j)] \toEDGE{\SO\rflx} \txid'_i\).
By the constraint of \( \aexec \), that is \( \SO_{\aexec} \subseteq \VIS_\aexec \), it follows \( \wtOf[\mkvs_{\cut[\aexec, i]}(\key,j)] \in \txidset_i \).
Recall \( \vi'_i = \getView[\mkvs_{\cut[\aexec, i]}, \txidset_i] \).
By the definition of \( \getView \), it follows \( i \in \vi'_i(\key) \).
Therefore \( \ET_\RYW \vdash (\mkvs_{\cut[\aexec, i-1]}, \vi_i) \csat \TtoOp{T}_{\aexec}(\txid_{i}) : (\mkvs_{\cut[\aexec, i]}, \vi'_{i}) \).

\item If there is no other transaction after \( \txid_i \) from the same client,
we pick \( \vi'_i = \getView[\aexec, \txidset_i] \) where \( \txidset_i = (\SO_\aexec^{-1})\rflx(\txid_i) \),
so \( \ET_\RYW \vdash (\mkvs_{\cut[\aexec, i-1]}, \vi_i) \csat \TtoOp{T}_{\aexec}(\txid_{i}) : (\mkvs_{\cut[\aexec, i]}, \vi'_{i}) \).
\end{itemize}

\subsection{Write Following Read \( \WFR \) }
\label{sec:sound-complete-wfr}

The write-read relation  on \( \aexec \) is defined as the following:
\[
    \WR(\aexec, \key) \defeq \Set{ (\txid, \txid') }[ \exsts{\val} (\otW, \key, \val) \in_\aexec \txid \land (\otR, \key, \val) \in_\aexec \txid' \land \txid = \max_\AR(\VIS^{-1}(\txid')) ]
\]
The notation \( \WR_\aexec \) is defined as \( \WR_\aexec \defeq \bigcup_{\key \in \Keys} \WR(\aexec, \key) \).
Note that for a kv-store \( \mkvs \) such that \( \mkvs = \mkvs_\aexec \),
by the definition of  \(  \mkvs = \mkvs_\aexec \), 
the following holds:
\[
    \WR_\aexec = \Set{(\txid, \txid')}[\exsts{\key, i } \mkvs(\key, i) = (\stub, \txid, \txid'\cup \stub)]
\]
Note that such \( \WR_\aexec \) coincides with \( \WR_\Gr \) and \( \WR_\mkvs \).

The execution test $\ET_\WFR$ is sound with respect to the axiomatic definition \cite{surech-session-guarantee}
\[ (\RP_{\LWW}, \Set{\lambda \aexec. \WR_\aexec ; (\SO \cap \RW_{\aexec} )\rflx ; \VIS_{\aexec} })\]
We pick the invariant as \( I( \aexec, \cl ) = \emptyset \), given the fact of no constraint on the view after update.
Assume a kv-store $\mkvs$, an initial and a final view $\vi, \vi'$  a fingerprint $\fp$ 
such that $\ET_{\WFR} \vdash (\mkvs, \vi) \csat \fp: (\mkvs', \vi')$. 
Also choose an arbitrary $\cl$, a transaction identifier $\txid \in \nextTxid(\mkvs, \cl)$, 
and an abstract execution $\aexec$ such that $\mkvs_{\aexec} = \mkvs$ and 
\( I(\aexec, \cl) =  \emptyset \subseteq \Tx[\mkvs, \vi] \).
Note that since the invariant is empty set, it remains to prove there is a set of read-only transactions \( \txidset_\rd \) such that
Let \( \aexec' = \extend[\aexec, \txid, \Tx[\mkvs, \vi] \cup \txidset_\rd, \fp] \) and
\[
    \begin{array}{@{}l@{}}
        \fora{ \txid' } 
        (\txid' ,\txid)  \in \WR_{\aexec'} ; \SO_{\aexec'}\rflx ; \VIS_{\aexec'} 
        \implies \txid' \in \Tx[\mkvs, \vi]
    \end{array}
\]
which can be derived from \cref{thm:view-vis-relation}.

The execution test $\ET_\WFR$ is complete with respect to the axiomatic definition 
\[ (\RP_{\LWW}, \Set{\lambda \aexec. \WR ; \WR_{\mkvs} ; (\SO \cap \RW_{\aexec'} )\rflx ; \VIS_{\aexec'} })\]
Assume i-\emph{th} transaction \( \txid_i \) in the arbitrary order,
and let view \( \vi_{i} = \getView[\mkvs_{\cut[\aexec, i-1]}, \VIS^{-1}_{\aexec}(\txid_{i})] \).
We also pick any final view such that \( \vi'_{i} \subseteq \getView[\mkvs_{\cut[\aexec, i]}, (\AR^{-1}_{\aexec})\rflx(\txid_{i})] \).
Note that there is nothing to prove for \( \vi'_i \),
so it is sufficient to prove the following:
\[
    \vi_i = \getView[\mkvs_{\cut[\aexec, i-1]}, \lfpTx[\mkvs_{\cut[\aexec, i-1]},\vi,\WR_{\mkvs_{\cut[\aexec, i-1]}}; \SO]]
\]
which can be derived from \cref{thm:view-vis-relation}.

\subsection{Causal Consistency \( \CC \)}
\label{sec:sound-complete-cc}

The wildly used definition on abstract executions for causal consistency is that \( \VIS \) is transitive.
Yet it is for the sack of elegant definition, while there is a minimum visibility relation given by \( (\WR_\aexec \cup \SO_\aexec) ; \VIS_\aexec \subseteq \VIS_\aexec \) (\cref{lem:aexec-spec-cc}).

\begin{lemma}
    \label{lem:aexec-spec-cc}
    For any abstract execution \( \aexec \) under last-write-win, if it satisfies the following:
    \[
        (\WR_\aexec \cup \SO_\aexec) ; \VIS_\aexec \subseteq \VIS_\aexec \quad \SO_\aexec \subseteq \VIS_\aexec
    \]
    There exists a new abstract execution \( \aexec' \) where \( \txidset_\aexec = \txidset_{\aexec'} \), \( \AR_\aexec = \AR_{\aexec'} \),
    \( \VIS_{\aexec'} ; \VIS_{\aexec'} \subseteq \VIS_{\aexec'} \), and
    under last-write-win \( \TtoOp{T}_{\aexec}(\txid) = \TtoOp{T}_{\aexec'}(\txid) \) for all transactions \( \txid \).
\end{lemma}
\begin{proof}
    To recall, the write-read relation under a key \( \WR(\aexec, \key) \) is defined as 
    \( \WR(\aexec, \key) \defeq \Set{ (\txid, \txid') }[ \exsts{\val} (\otW, \key, \val) \in_\aexec \txid \land (\otR, \key, \val) \in_\aexec \txid' \land \txid = \max_\AR(\VIS^{-1}(\txid')) ]\).
    Given an \( \aexec \) that satisfies the following
    \[
        (\WR_\aexec \cup \SO_\aexec ) ; \VIS_\aexec \subseteq \VIS_\aexec \quad \SO_\aexec \subseteq \VIS_\aexec
    \]
    we erase some visibility relation for each transaction following the order of arbitration \( \AR \) until the visibility is transitive.
    Assume the i-\emph{th} transaction \( \txid_i \)  with respect to the arbitration order.
    Let \( R_i \) denote a new visibility for transaction \( \txid_i \) such that
    \( R_i\projection{2} = \Set{\txid_i}\)
    and the visibility relation before (including) \( \txid_i \) is transitive.
    Let \( \aexec_i = \mkvs_{\cut[\aexec, i]} \) and \( \VIS_i = \bigcup_{0 \leq k \leq i} R_i \).
    For each step, says i-\emph{th} step, we  preserve the following:
    \begin{gather}
        \VIS_i ; \VIS_i \subseteq \VIS_i \label{equ:vis-i-transitive} \\
        \fora{\txid} (\txid,\txid_i) \in R_i \implies (\txid, \txid_i) \in (\WR_i \cup \SO_i)
        \label{equ:last-read-correct}
    \end{gather}
    
    \begin{itemize}
    \item \caseB{\( i = 1 \) and \( R_1 = \emptyset \)}
    Assume it is from client \( \cl \).
    There is no transaction committed before, so \( \VIS_1 = \emptyset \) and \( \VIS_1 ; \VIS_1 \subseteq \VIS_1 \) as \cref{equ:vis-i-transitive}.

    \item \caseI{i-\emph{th} step}
    Suppose the (i-1)-\emph{th} step satisfies \cref{equ:vis-i-transitive} and \cref{equ:last-read-correct}.
    Let consider i-\emph{th} step and the transaction \( \txid_i \).
    Initially we take \( R_i \) as empty set.
    We first extend \( R_i \) by closing with respect to \( \WR_i \)
    and prove that it does not affect any read from the transaction \( \txid_i \).
    Then we will do the same for \( \SO_i \).
    \begin{itemize}
        \item \( \WR_i\). For any read \( (\otR, \key, \val ) \in \txid_i \),
        there must be a transaction \( \txid_j \) that \( \txid_j \toEDGE{\WR(\aexec_i,\key), \AR} \txid_i \) and \( j < i \).
        We include \( (\txid_j, \txid_i) \in R_i \).
        Let consider all the visible transactions of \( \txid_j \).
        Assume a transaction \( \txid' \in \VIS_{i-1}^{-1}(\txid_j) \), 
        thus \( \txid' \in \VIS_{j}^{-1}(\txid_j) = R_j^{-1}(\txid_j) \).
        It is safe to include \( (\txid', \txid_i) \in R_i \) without affecting the read result,
        because those transaction \( \txid' \) is already visible for \( \txid_i \) in the abstract execution \( \aexec \):
        by \cref{equ:last-read-correct} we know \( R_j \subseteq (\WR_j \cup \SO_j)^{+} \subseteq (\WR_\aexec \cup \SO_\aexec)^{+}\),
        and by the definition of \( \WR(\aexec_i,\key) \) we know \( \WR(\aexec_i,\key) \subseteq \VIS_\aexec\).

        \item Given \( \SO_\aexec \subseteq \VIS_\aexec \), we include \( (\txid_j,\txid_i) \) for some \( \txid_j \)
        such that \( \txid_j \toEDGE{\SO_\aexec} \txid_i\).
        For the similar reason as \( \WR \),
        it is safe to includes all the visible transactions \( \txid' \) for \( \txid_j \), \ie \( \txid' \in R_j^{-1}\).
        \end{itemize}
        
    By the construction, both \cref{equ:vis-i-transitive} and \cref{equ:last-read-correct} are preserved. 
    Thus we have the proof.
    \end{itemize}
\end{proof}
\begin{proposition}
    \label{prop:cc-vis}
    For any abstract execution \( \aexec \) under last-write-win, if it satisfies the following:
    \[
        (\WR_\aexec \cup \SO_\aexec)^{+} ; \VIS_\aexec \subseteq \VIS_\aexec \quad \SO_\aexec \subseteq \VIS_\aexec
    \]
    then
    \[
        \exsts{R \subseteq \AR_\aexec} \VIS = (\WR_\aexec \cup \SO_\aexec \cup R)^{+}
    \]
\end{proposition}

By \cref{lem:aexec-spec-cc}, the execution test $\ET_\CC$ is sound with respect to the axiomatic definition 
\[ (\RP_{\LWW}, \Set{\lambda \aexec. ( \SO_{\aexec} \cup \WR_{\aexec} )^{+} ; \VIS_{\aexec}, \lambda \aexec \ldotp \SO_\aexec })\]
We pick an invariant for the \( \ET_\CC \) as the union of those for \( \MR\) and \( \RYW \) shown in the following:
\begin{align*}
    I_1(\aexec, \cl) & =  \left( \bigcup_{\Set{\txid_{\cl}^{n} \in \txidset_{\aexec} }[ n \in \Nat ]} \VIS_{\aexec}^{-1}(\txid^n_\cl) \right) \setminus \txidset_\rd \\
    I_2(\aexec, \cl) & =  \left( \bigcup_{\Set{\txid_{\cl}^{n} \in \txidset_{\aexec} }[ n \in \Nat ]} (\SO_{\aexec}^{-1})\rflx(\txid^n_\cl) \right) \setminus \txidset_\rd
\end{align*}
where \( \txidset_\rd \) is all the read-only transactions included in both:
\[ \txidset_\rd \in \left( \bigcup_{\Set{\txid_{\cl}^{n} \in \txidset_{\aexec} }[ n \in \Nat ]} \VIS_{\aexec}^{-1}(\txid^n_\cl) \right)\]
and \[ \txidset_\rd \in  \left( \bigcup_{\Set{\txid_{\cl}^{n} \in \txidset_{\aexec} }[ n \in \Nat ]} (\SO_{\aexec}^{-1})\rflx(\txid^n_\cl) \right) \]
Assume a kv-store $\mkvs$, an initial and a final view $\vi, \vi'$  a fingerprint $\fp$ 
such that $\ET_{\CC} \vdash (\mkvs, \vi) \csat \fp: (\mkvs',\vi')$. 
Also choose an arbitrary $\cl$, a transaction identifier $\txid_\cl^n \in \nextTxid(\mkvs, \cl)$, 
and an abstract execution $\aexec$ such that $\mkvs_{\aexec} = \mkvs$ and 
\( I_1(\aexec, \cl) \cup I_2(\aexec, \cl) \subseteq \Tx[\mkvs, \vi] \).
We are about to prove there exists an extra set of read-only transactions \( \txidset'_\rd \) such that
the new abstract execution \( \aexec' = \extend[\aexec, \txid_\cl^n, \fp, \Tx[\mkvs, \vi] \cup \txidset_\rd \cup \txidset'_\rd] \) and:
\begin{gather}
    \fora{\txid} (\txid, \txid_\cl^n) \in \SO_{\aexec'} \implies \txid \in \Tx[\mkvs, \vi] \cup \txidset_\rd \cup \txidset'_\rd \label{equ:cc-sound-update-so}\\
    \fora{\txid} (\txid, \txid_\cl^n) \in ( \SO_{\aexec'} \cup \WR_{\aexec'} ) ; \VIS_{\aexec'} \implies \txid \in \Tx[\mkvs, \vi] \cup \txidset_\rd \cup \txidset'_\rd \label{equ:cc-sound-update-visvis}\\
    I_1(\aexec',\cl) \cup I_2(\aexec',\cl) \subseteq \Tx[\mkvs_{\aexec'}, \vi'] \label{equ:cc-sound-inv} 
\end{gather}

\begin{itemize}
\item The invariant \( I_2 \) implies \cref{equ:cc-sound-update-so} as the same as \( \RYW \) in \cref{sec:sound-complete-ryw}.
\item \cref{equ:cc-sound-update-visvis}.
    Note that \( (\txid, \txid_\cl^n) \in ( \SO_{\aexec'} \cup \WR_{\aexec'} ); \VIS_{\aexec'} \implies (\txid, \txid_\cl^n) \in ( \SO_{\aexec} \cup \WR_{\aexec} ) ; \VIS_{\aexec'}\).
    Also, recall that \( \SO_\aexec = \SO_\mkvs \) and \( \WR_\aexec = \WR_\mkvs \).
    Let \( \txidset'_\rd = \lfpTx[\mkvs,\vi,\SO_{\mkvs} \cup \WR_{\mkvs}] \). 
    This means that \( \aexec' = \extend[\aexec, \txid_\cl^n, \fp, \lfpTx[\mkvs, \vi, \SO_{\mkvs} \cup \WR_{\mkvs}] \cup \txidset_\rd ] \).
    Let assume \( \txid \toEDGE{\SO_{\mkvs} \cup \WR_{\mkvs}} \txid' \) and \( \txid' \in \lfpTx[\mkvs, \vi, \SO_{\mkvs} \cup \WR_{\mkvs}] \cup \txidset_\rd \).
    We have two possible cases:
    \begin{itemize}
        \item If \( \txid' \in \lfpTx[\mkvs, \vi, \SO_{\mkvs} \cup \WR_{\mkvs}] \), by  \cref{thm:view-vis-relation}, we know \( \txid \in \lfpTx[\mkvs, \vi, \SO_{\mkvs} \cup \WR_{\mkvs}] \).
        \item If \( \txid' \in \txidset_\rd \), there are two cases:
        \begin{itemize}
            \item \( \txid' \in  \left( \bigcup_{\Set{\txid_{\cl}^{n} \in \txidset_{\aexec} }[ n \in \Nat ]} \VIS_{\aexec}^{-1}(\txid^n_\cl) \right) \).
                By the property of \( \aexec \) (before update) that \( ( \SO \cup \WR_\aexec ) ; \VIS_\aexec \in \VIS_\aexec \), it is known that \( \txid \in \left( \bigcup_{\Set{\txid_{\cl}^{n} \in \txidset_{\aexec} }[ n \in \Nat ]} \VIS_{\aexec}^{-1}(\txid^n_\cl) \right) \), that is, \( \txid \in \Tx[\mkvs,\vi] \cup \txidset_\rd\).

            \item \( \txid' \in  \left( \bigcup_{\Set{\txid_{\cl}^{n} \in \txidset_{\aexec} }[ n \in \Nat ]} \SO_{\aexec}^{-1}(\txid^n_\cl) \right) \).
                Then we know \( \txid \in (\SO \cup \WR_\aexec)^{-1} \left( \bigcup_{\Set{\txid_{\cl}^{n} \in \txidset_{\aexec} }[ n \in \Nat ]} \SO_{\aexec}^{-1}(\txid^n_\cl) \right) \).
                By the property of \( \aexec \) (before update) that \( \SO \cup \WR_\aexec \in \VIS_\aexec \),
                it follows:
                \begin{align*}
                    \txid & \in VIS_\aexec^{-1} \left( \bigcup_{\Set{\txid_{\cl}^{n} \in \txidset_{\aexec} }[ n \in \Nat ]} \SO_{\aexec}^{-1}(\txid^n_\cl) \right) \\
                          & = \left( \bigcup_{\Set{\txid_{\cl}^{n} \in \txidset_{\aexec} }[ n \in \Nat ]} \VIS_{\aexec}^{-1}(\txid^n_\cl) \right)  \\
                          & = \Tx[\mkvs,\vi] \cup \txidset_\rd
                \end{align*}
                
        \end{itemize}
    \end{itemize}
\item Finally the new abstract execution preserves the invariant \( I_1 \) and \( I_2 \) 
because  \( \CC \) satisfies \( \MW \) and \( \RYW \).
The proofs are the same as those in \cref{sec:sound-complete-mr} and \cref{sec:sound-complete-ryw}.

\end{itemize}

The execution test $\ET_\CC$ is complete with respect to the axiomatic definition 
\[ (\RP_{\LWW}, \Set{\lambda \aexec.  \VIS_\aexec ; \VIS_{\aexec}, \lambda \aexec \ldotp \SO_\aexec })\]
Assume i-\emph{th} transaction \( \txid_i \) in the arbitrary order,
and let view \( \vi_{i} = \getView[\aexec, \VIS^{-1}_{\aexec}(\txid_{i})] \).
We pick final view as \( \vi'_{i} = \getView[\aexec, (\AR^{-1}_{\aexec})\rflx(\txid_{i}) \cap \VIS_{\aexec}^{-1}(\txid'_i)] \),
if \( \txid'_i = \min_{\SO}\Set{\txid'}[\txid_i \toEDGE{\SO} \txid' ]\) is defined,
otherwise  \( \vi'_{i} = \getView[\aexec, (\AR^{-1}_{\aexec})\rflx(\txid_{i})]\).
Let the \( \mkvs = \mkvs_{\cut[\aexec, i-1]} \).
Now we prove the three parts separately.
\begin{itemize}
    \item \( \MR \).  By \cref{prop:cc-vis} since 
    \( \VIS_\aexec ; \SO_\aexec \subseteq \VIS_\aexec ; \VIS_\aexec \subseteq \VIS_\aexec \)
    so it follows as in \cref{sec:sound-complete-mr}.
    \item \( \RYW \). For \( \RYW \), since \( \WR_\aexec ; \SO_\aexec ; \VIS_\aexec \subseteq \VIS_\aexec ; \VIS_\aexec ; \VIS_\aexec \subseteq \VIS_\aexec\), 
    the proof is as the same proof as in \cref{sec:sound-complete-ryw}.
    \item \( \allowed[\WR_\mkvs \cup \SO]\). It is derived from \cref{thm:view-vis-relation} and 
        \( (\WR_\aexec \cup \SO) ; \VIS_\aexec \subseteq \VIS_\aexec ; \VIS_\aexec \subseteq \VIS_\aexec\).
\end{itemize}

\subsection{Update Atomic \( \UA \)}
\label{sec:sound-complete-ua}

Given abstract execution \( \aexec \), we define write-write relation for a key \( \key \) as the following~\cite{framework-concur}:
\[ 
    \WW(\aexec,\key) \defeq \Set{(\txid, \txid')}[\txid \toEDGE{\AR_\aexec} \txid' \land (\otW,\key, \stub ) \in \txid \land (\otW,\key, \stub ) \in \txid' ]
\]
Then, the notation \( \WW_\aexec \defeq \bigcup_{\key \in \Keys} \WW(\aexec, \key) \).
Note that for a kv-store \( \mkvs \) such that \( \mkvs = \mkvs_\aexec \),
by the definition of  \(  \mkvs = \mkvs_\aexec \), 
the following holds:
\[
    \WW_\aexec = \Set{(\txid, \txid')}[\exsts{\key, i,j } \txid = \wtOf(\mkvs(\key, i)) \land \txid' = \wtOf(\mkvs(\key, j)) \land i < j]
\]
Also the \( \WW_\aexec \) coincides with \( \WW_\Gr \) and \( \WW_\mkvs \).

The execution test $\ET_\UA$ is sound with respect to the axiomatic definition \( (\RP_{\LWW}, \Set{\lambda \aexec. \WW_\aexec }) \).
We pick the invariant as \( I( \aexec, \cl ) = \emptyset \), given the fact of no constraint on the final view.
Assume a kv-store $\mkvs$, an initial and a final view $\vi, \vi'$  a fingerprint $\fp$ 
such that $\ET_{\UA} \vdash (\mkvs, \vi) \csat \fp: (\mkvs', \vi')$. 
Also choose an arbitrary $\cl$, a transaction identifier $\txid \in \nextTxid(\mkvs, \cl)$, 
and an abstract execution $\aexec$ such that $\mkvs_{\aexec} = \mkvs$ and 
\( I(\aexec, \cl) =  \emptyset \subseteq \Tx[\mkvs, \vi] \).
Let \( \aexec' = \extend[\aexec, \txid, \Tx[\mkvs, \vi], \fp] \).
Note that since the invariant is empty set, it remains to prove the following:
\[
    \begin{array}{@{}l@{}}
        \fora{ \txid' } \txid' \toEDGE{\WW_{\aexec'}} \txid \implies \txid' \in \Tx[\mkvs, \vi]
    \end{array}
\]
Assume a transaction \( \txid' \) that writes to a key \( \key \) as \( \txid \), \ie \( \txid' \toEDGE{\WW_{\aexec'}} \txid \).
Since that \( \txid' \) is a transaction already existing in \( \mkvs\),
we have \( \wtOf[\mkvs(\key, i)] = \txid' \) for some index \( i \) and key \( \key \).
It means \( (\otW, \key, \valueOf[\mkvs(\key, i)]) \in \fp \).
By the execution test of \( \UA \), we know \( i \in \vi(\key) \) therefore \( \txid' \in \Tx[\mkvs, \vi] \).

The execution test $\ET_\UA$ is complete with respect to 
the axiomatic definition \( (\RP_{\LWW}, \Set{\lambda \aexec. \WW_\aexec }) \).
Assume i-\emph{th} transaction \( \txid_i \) in the arbitrary order,
and let view \( \vi_{i} = \getView[\aexec, \VIS^{-1}_{\aexec}(\txid_{i})] \).
We also pick any final view such that \( \vi'_{i} \subseteq \getView[\aexec, (\AR^{-1}_{\aexec})\rflx(\txid_{i})] \).
Note that there is nothing to prove for \( \vi'_i \),
so it is sufficient to prove the following:
\[
    \fora{\key} (\otW, \key, \stub) \in \TtoOp{T}_{\aexec}(\txid_{i}) 
    \implies 
    \fora{j : 0 \leq j < \abs{\mkvs_{\cut[\aexec, i-1]}(\key)}} j \in \vi_i(\key)
\]
Let consider a key \( \key \) that have been overwritten by the transaction \( \txid_i \).
By the constraint of \( \aexec \) that \( \WW_\aexec \subseteq \VIS_\aexec \),
for any transaction \( \txid \) that writes to the same key \( \key \) and committed before \( \txid_i \), 
they are included in the visible set \(\txid \in \VIS^{-1}_{\aexec}(\txid_{i}) \).
Note that \( \txid \toEDGE{\WW_\aexec} \txid_i \implies \txid \toEDGE{\AR_\aexec} \txid_i \implies \txid \in \mkvs_{\cut(\aexec,i-1)}\).
Since that the transaction \( \txid \) write to the key \( \key \),
it means \( \wtOf(\mkvs_{\cut(\aexec, i-1)}(\key,j)) = \txid \) for some index \( j \).
Then by the definition of \( \getView \), we have \( j \in \vi_i(\key)\).

\subsection{Consistency Prefix \( \CP \) }
\label{sec:sound-complete-cp}

Given abstract execution \( \aexec \), we define read-write read-write relation:
\[
    \RW(\aexec,\key) \defeq \Set{(\txid, \txid')}[\txid \toEDGE{\AR_\aexec} \txid' \land (\otR,\key, \stub ) \in \txid \land (\otW,\key, \stub ) \in \txid' ] 
\]
It is easy to see \( \RW(\aexec,\key) \)  can be derived from \( \WW(\aexec,\key) \) and \( \WR(\aexec, \key ) \) as the following:
\[
    \RW(\aexec,\key) = \Set{(\txid, \txid')}[ \exsts{\txid'' } (\txid'', \txid) \in \WR(\aexec, \key) \land (\txid'', \txid') \in \WW(\aexec, \key) ]
\]
Then, the notation \( \RW_\aexec \defeq \bigcup_{\key \in \Keys} \RW(\aexec, \key) \).
Note that for a kv-store \( \mkvs \) such that \( \mkvs = \mkvs_\aexec \),
by the definition of  \(  \mkvs = \mkvs_\aexec \), 
the following holds:
\[
    \RW_\aexec = \Set{(\txid, \txid')}[\exsts{\key, i,j } \txid \in \rsOf(\mkvs(\key, i)) \land \txid' = \wtOf(\mkvs(\key, j)) \land i < j]
\]
The \( \RW_\aexec \) also coincides with \( \RW_\Gr \) and \( \RW_\mkvs \).

An abstract execution \( \aexec \) satisfies consistency prefix (\(\CP\)), 
if it satisfies \( \AR_\aexec ; \VIS_\aexec \subseteq \VIS_\aexec \) and \( \SO_\aexec \subseteq \VIS_\aexec \).
Given the definition, there is a corresponding definition on dependency graph by solve the following inequalities:
\[
    \begin{array}{@{}l@{}}
        \WR \subseteq \VIS \\
        \WW \subseteq \AR \\
        \VIS \subseteq \AR \\
        \VIS ; \RW \subseteq \AR \\
        \AR ; \AR \subseteq \AR  \\
        \SO \subseteq \VIS \\
        \AR ; \VIS \subseteq \VIS
    \end{array}
\]
By solving the inequalities the visibility and arbitration relations are:
\begin{align*}
    \AR  & \defeq \left( (\SO \cup \WR ) ; \RW\rflx \cup \WW \cup R \right)^+ \\
    \VIS & \defeq \left( (\SO \cup \WR ) ; \RW\rflx \cup \WW \cup R \right)^* ; (\SO \cup \WR )
\end{align*}
for some relation \( R \subseteq \AR \).
When \( R = \emptyset \), it is the smallest solution therefore the minimum visibility required.

\sx{A bit verbal}
\begin{lemma}
    \label{lem:cp-eauiv-spec}
    For any abstract execution \( \aexec \),
    if it satisfies 
    \[
        \left( (\SO \cup \WR ) ; \RW\rflx \cup \WW \right) ; \VIS_\aexec \subseteq \VIS_\aexec 
        \qquad \SO_\aexec \subseteq \VIS_\aexec
    \]
    then there exists a new \( \aexec' \) such that \( \txidset_\aexec = \txidset_{\aexec'} \), 
    under last-write-win \( \TtoOp{T}_{\aexec}(\txid) = \TtoOp{T}_{\aexec'}(\txid) \) for all transactions \( \txid \),
    and the relations satisfy the following:
    \[ 
        \AR_{\aexec'} ; \VIS_{\aexec'} \subseteq \VIS_{\aexec'}  \qquad \SO_{\aexec'} \subseteq \VIS_{\aexec'}
    \]
    and vice versa.
\end{lemma}
\begin{proof}
Assume abstract execution \( \aexec' \) that satisfies \( \AR_{\aexec'} ; \VIS_{\aexec'} \subseteq \VIS_{\aexec'} \)
and  \( \SO_{\aexec'} \subseteq \VIS_{\aexec'} \).
We already show that:
\begin{align*}
    \AR_{\aexec'} & = \left( (\SO_\aexec \cup \WR_\aexec ) ; \RW_\aexec\rflx \cup \WW_\aexec \cup R \right)^+ \\
    \VIS_{\aexec'} & = \left( (\SO_\aexec \cup \WR_\aexec ) ; \RW_\aexec\rflx \cup \WW_\aexec \cup R \right)^* ; (\SO_\aexec \cup \WR_\aexec )
\end{align*}
for some relation \( R \subseteq \AR_{\aexec'} \).
If we take \( R  = \emptyset \), we have the proof for:
\[
        \SO \subseteq \VIS_\aexec \qquad 
        \left( (\SO_\aexec \cup \WR_\aexec ) ; \RW_\aexec\rflx \cup \WW_\aexec \right) ; \VIS_\aexec \subseteq \VIS_\aexec
\]
For another way, we pick the \( R \) that extends
\( \left( (\SO_\aexec \cup \WR_\aexec ) ; \RW_\aexec\rflx \cup \WW_\aexec \cup R \right)^+ \) 
to a total order.
\end{proof}

By \cref{lem:cp-eauiv-spec} to prove soundness and completeness of \( \ET_\CP \), it is sufficient to use the definition:
\[
    (\RP_{\LWW}, \Set{\lambda \aexec. \left( (\SO \cup \WR ) ; \RW\rflx \cup \WW \right) ; \VIS_\aexec, \lambda \aexec \ldotp \SO_\aexec }) 
\]

For the soundness, we pick the invariant as the following:
\begin{align*}
    I_1(\aexec, \cl) & = \left( \bigcup_{\Set{\txid_{\cl}^{i} \in \txidset_{\aexec} }[ i \in \Nat ]} \VIS_{\aexec}^{-1}(\txid^i_\cl) \right) \setminus \txidset_\rd \\
    I_2(\aexec, \cl) & = \left( \bigcup_{\Set{\txid_{\cl}^{i} \in \txidset_{\aexec} }[ i \in \Nat ]} (\SO_{\aexec}^{-1})\rflx(\txid^i_\cl) \right) \setminus \txidset_\rd
\end{align*}
where \( \txidset_\rd \) is all the read-only transactions included in both 
\[ \txidset_\rd \in \left( \bigcup_{\Set{\txid_{\cl}^{i} \in \txidset_{\aexec} }[ i \in \Nat ]} \VIS_{\aexec}^{-1}(\txid^i_\cl) \right)\] 
and \[ \txidset_\rd \in \left( \bigcup_{\Set{\txid_{\cl}^{i} \in \txidset_{\aexec} }[ i \in \Nat ]} (\SO_{\aexec}^{-1})\rflx(\txid^i_\cl) \right) \]
Assume a key-value store $\mkvs$, an initial and a final view $\vi, \vi'$  a fingerprint $\fp$ 
such that $\ET_{\CP} \vdash (\mkvs, \vi) \csat \fp: ( \mkvs',\vi')$. 
Also choose an arbitrary $\cl$, a transaction identifier $\txid_\cl^n \in \nextTxid(\mkvs, \cl)$, 
and an abstract execution $\aexec$ such that $\mkvs_{\aexec} = \mkvs$ and 
\( I_1(\aexec, \cl) \cup I_2(\aexec, \cl) \subseteq \Tx[\mkvs, \vi] \).
We are about to prove that there exists an extra set of read-only transaction \( \txidset'_\rd \) such that
the new abstract execution \( \aexec' = \extend[\aexec, \txid_\cl^n, \fp, \Tx[\mkvs, \vi] \cup \txidset_\rd \cup \txidset'_\rd] \) and
\begin{gather}
    \fora{\txid} (\txid, \txid_\cl^n) \in \SO_{\aexec'} \implies \txid \in \Tx[\mkvs, \vi] \cup \txidset_\rd \cup \txidset'_\rd \label{equ:cp-sound-update-so}\\
    \begin{array}{l}
        \fora{\txid} (\txid, \txid_\cl^n) \in \left( (\SO_{\aexec'} \cup \WR_{\aexec'} ) ; \RW_{\aexec'}\rflx \cup \WW_{\aexec'} \right) ; \VIS_{\aexec'} 
    \implies \txid \in \Tx[\mkvs, \vi] \cup \txidset_\rd \cup \txidset'_\rd 
    \end{array}
    \label{equ:cp-sound-update-arvis}\\
    I_1(\aexec',\cl) \cup I_2(\aexec',\cl) \subseteq \Tx[\mkvs_{\aexec'}, \vi'] \label{equ:cp-sound-inv} 
\end{gather}
\begin{itemize}
\item the invariant \( I_2 \) implies the \cref{equ:cp-sound-update-so} where the proof is the same as \( \RYW \) in \cref{sec:sound-complete-ryw}.

\item \cref{equ:cp-sound-update-arvis}.
    Note that 
    \begin{centermultline} 
        (\txid, \txid_\cl^n) \in \left( (\SO_{\aexec'} \cup \WR_{\aexec'} ) ; \RW_{\aexec'}\rflx \cup \WW_{\aexec'} \right) ; \VIS_{\aexec'} \\
        {} \implies (\txid, \txid_\cl^n) \in \left( (\SO_{\aexec} \cup \WR_{\aexec} ) ; \RW_{\aexec}\rflx \cup \WW_{\aexec} \right) ; \VIS_{\aexec'}
    \end{centermultline}
    Also, recall that \( \rel_\aexec = \rel_\mkvs \) for \( \rel \in \Set{\SO, \WR, \WW, \RW} \).
    Let \[ \txidset'_\rd = \lfpTx[\mkvs,\vi,(\SO_{\mkvs} \cup \WR_{\mkvs} ) ; \RW_{\mkvs}\rflx \cup \WW_{\mkvs} ] \]
    Let assume \( \txid \toEDGE{(\SO_{\mkvs} \cup \WR_{\mkvs} ) ; \RW_{\mkvs}\rflx \cup \WW_{\mkvs}} \txid' \) and \( \txid' \in \lfpTx[\mkvs, \vi, (\SO_{\mkvs} \cup \WR_{\mkvs} ) ; \RW_{\mkvs}\rflx \cup \WW_{\mkvs}] \cup \txidset_\rd \).
    We have two possible cases:
    \begin{itemize}
        \item If \( \txid' \in \lfpTx[\mkvs, \vi, (\SO_{\mkvs} \cup \WR_{\mkvs} ) ; \RW_{\mkvs}\rflx \cup \WW_{\mkvs}] \), by \cref{thm:view-vis-relation}, we know \[ \txid \in \lfpTx[\mkvs, \vi, (\SO_{\mkvs} \cup \WR_{\mkvs} ) ; \RW_{\mkvs}\rflx \cup \WW_{\mkvs}] \]
        \item If \( \txid' \in \txidset_\rd \), there are two cases:
        \begin{itemize}
            \item \( \txid' \in  \left( \bigcup_{\Set{\txid_{\cl}^{n} \in \txidset_{\aexec} }[ n \in \Nat ]} \VIS_{\aexec}^{-1}(\txid^n_\cl) \right) \).
                Note that \( \txid' \) is a read-only transaction,
                which means \( \txid \toEDGE{\SO_{\mkvs} \cup \WR_{\mkvs} } \txid' \).
                By the property of \( \aexec \) (before update) that \( ( \SO \cup \WR_\aexec ) ; \VIS_\aexec  \in \VIS_\aexec \), it is known that \( \txid \in \left( \bigcup_{\Set{\txid_{\cl}^{n} \in \txidset_{\aexec} }[ n \in \Nat ]} \VIS_{\aexec}^{-1}(\txid^n_\cl) \right) \), that is, \( \txid \in \Tx[\mkvs,\vi] \cup \txidset_\rd\).

            \item \( \txid' \in  \left( \bigcup_{\Set{\txid_{\cl}^{n} \in \txidset_{\aexec} }[ n \in \Nat ]} \SO_{\aexec}^{-1}(\txid^n_\cl) \right) \) and it is a read only transaction.
                Then we know \( \txid \in (\SO \cup \WR_\aexec)^{-1} \left( \bigcup_{\Set{\txid_{\cl}^{n} \in \txidset_{\aexec} }[ n \in \Nat ]} \SO_{\aexec}^{-1}(\txid^n_\cl) \right) \).
                By the property of \( \aexec \) (before update) that \( \SO \cup \WR_\aexec \in \VIS_\aexec \),
                it follows:
                \begin{align*}
                    \txid & \in VIS_\aexec^{-1} \left( \bigcup_{\Set{\txid_{\cl}^{n} \in \txidset_{\aexec} }[ n \in \Nat ]} \SO_{\aexec}^{-1}(\txid^n_\cl) \right) \\
                          & = \left( \bigcup_{\Set{\txid_{\cl}^{n} \in \txidset_{\aexec} }[ n \in \Nat ]} \VIS_{\aexec}^{-1}(\txid^n_\cl) \right)  \\
                          & = \Tx[\mkvs,\vi] \cup \txidset_\rd
                \end{align*}
                
        \end{itemize}
    \end{itemize}

\item Since \( \CP \) satisfies \( \RYW \) and \( \MR \), thus invariants \( I_1 \) and  \( I_2 \) are preserved after update.

\end{itemize}

The execution test $\ET_\CP$ is complete with respect to the axiomatic definition 
\[ (\RP_{\LWW}, \Set{\lambda \aexec.  \AR_\aexec ; \VIS_{\aexec}, \lambda \aexec \ldotp \SO_\aexec })\]
Assume i-\emph{th} transaction \( \txid_i \) in the arbitrary order,
and let view \( \vi_{i} = \getView[\aexec, \VIS^{-1}_{\aexec}(\txid_{i})] \).
We pick final view as \( \vi'_{i} = \getView[\aexec, (\AR^{-1}_{\aexec})\rflx(\txid_{i}) \cap \VIS_{\aexec}^{-1}(\txid'_i)] \),
if \( \txid'_i = \min_{\SO}\Set{\txid'}[\txid_i \toEDGE{\SO} \txid' ]\) is defined,
otherwise  \( \vi'_{i} = \getView[\aexec, (\AR^{-1}_{\aexec})\rflx(\txid_{i})]\).
Let the \( \mkvs = \mkvs_{\cut[\aexec, i-1]} \).
Now we prove the three parts separately.

\begin{itemize}
    \item \( \MR \).  By \cref{prop:cc-vis} since 
    \( \VIS_\aexec ; \SO_\aexec \subseteq \AR_\aexec ; \VIS_\aexec \subseteq \VIS_\aexec \)
    so it follows as in \cref{sec:sound-complete-mr}.
    \item \( \RYW \). For \( \RYW \), since \( \WR_\aexec ; \SO_\aexec ; \VIS_\aexec \subseteq \AR_\aexec ; \AR_\aexec ; \VIS_\aexec \subseteq \VIS_\aexec\), 
    the proof is as the same proof as in \cref{sec:sound-complete-ryw}.

    \item \( \allowed[(\SO ; \RW_{\mkvs}\rflx) \cup (\WR_{\mkvs} ; \RW_{\mkvs}\rflx) \cup \WW_\mkvs] \) can be derived 
    from \cref{thm:view-vis-relation} and \[ (\SO ; \RW_{\aexec}\rflx) \cup (\WR_{\aexec} ; \RW_{\aexec}\rflx) \cup \WW_\aexec ; \VIS_\aexec \subseteq \AR_\aexec ; \VIS_\aexec \subseteq \VIS_\aexec \]
\end{itemize}

\subsection{Parallel Snapshot Isolation \(\PSI\)}
\label{sec:sound-complete-psi}

The axiomatic definition for \( \PSI \) is 
\[ 
    (\RP_{\LWW}, \Set{\lambda \aexec. \VIS_{\aexec} ; \VIS_{\aexec}, \lambda \aexec \ldotp \SO_\aexec, \lambda \aexec. \WW_\aexec })
\]
There exist a minimum visibility such that 
\[ 
    (\RP_{\LWW}, \Set{\lambda \aexec. (\WR_{\aexec} \cup \WW_{\aexec} \cup \SO ) ; \VIS_{\aexec}, \lambda \aexec \ldotp \SO_\aexec, \lambda \aexec. \WW_\aexec })
\]
by solve the following inequalities:
\[
    \begin{array}{@{}l@{}}
        \WR \subseteq \VIS \\
        \WW \subseteq \VIS \\
        \SO \subseteq \VIS \\
        \VIS ; \VIS \subseteq \VIS 
    \end{array}
\]
It is easy to see the former implies to later.
For another way round, \cref{lem:aexec-spec-psi}.

\begin{lemma}
    \label{lem:aexec-spec-psi}
    For any abstract execution \( \aexec \) under last-write-win, if it satisfies the following:
    \[
        (\WR_\aexec \cup \WW_\aexec \cup \SO_\aexec ) ; \VIS_\aexec \subseteq \VIS_\aexec \quad \SO_\aexec \subseteq \VIS_\aexec
    \]
    There exists a new abstract execution \( \aexec' \) where \( \txidset_\aexec = \txidset_{\aexec'} \), \( \AR_\aexec = \AR_{\aexec'} \),
    \( \VIS_{\aexec'} ; \VIS_{\aexec'} \subseteq \VIS_{\aexec'} \), and
    under last-write-win \( \TtoOp{T}_{\aexec}(\txid) = \TtoOp{T}_{\aexec'}(\txid) \) for all transactions \( \txid \).
\end{lemma}
\begin{proof}
    we erase some visibility relation for each transaction following the order of arbitration \( \AR \) until the visibility is transitive.
    Assume the i-\emph{th} transaction \( \txid_i \)  with respect to the arbitration order.
    Let \( R_i \) denote a new visibility for transaction \( \txid_i \) such that
    \( R_i\projection{2} = \Set{\txid_i}\)
    and the visibility relation before (including) \( \txid_i \) is transitive.
    Let \( \aexec_i = \mkvs_{\cut[\aexec, i]} \) and \( \VIS_i = \bigcup_{0 \leq k \leq i} R_i \).
    For each step, says i-\emph{th} step, we  preserve the following:
    \begin{gather}
        \VIS_i ; \VIS_i \subseteq \VIS_i \label{equ:vis-i-transitive-psi} \\
        \fora{\txid} (\txid,\txid_i) \in R_i \implies (\txid, \txid_i) \in (\WR_i \cup \WW_i \cup \SO_i)
        \label{equ:last-read-correct-psi}
    \end{gather}
    \begin{itemize}
    \item \caseB{\( i = 1 \) and \( R_1 = \emptyset \)}
    Assume it is from client \( \cl \).
    There is no transaction committed before, so \( \VIS_1 = \emptyset \) and \( \VIS_1 ; \VIS_1 \subseteq \VIS_1 \) as \cref{equ:vis-i-transitive-psi}.

    \item \caseI{i-\emph{th} step}
    Suppose the (i-1)-\emph{th} step satisfies \cref{equ:vis-i-transitive-psi} and \cref{equ:last-read-correct-psi}.
    Let consider i-\emph{th} step and the transaction \( \txid_i \).
    Initially we take \( R_i \) as empty set.
    We first extend \( R_i \) by closing with respect to \( \WR_i \)
    and prove that it does not affect any read from the transaction \( \txid_i \).
    Then we will do the same for \( \SO_i \) and \( \WW_i \).
    \begin{itemize}
        \item \( \WR_i\). For any read \( (\otR, \key, \val ) \in \txid_i \),
        there must be a transaction \( \txid_j \) that \( \txid_j \toEDGE{\WR(\aexec_i,\key), \AR} \txid_i \) and \( j < i \).
        We include \( (\txid_j, \txid_i) \in R_i \).
        Let consider all the visible transactions of \( \txid_j \).
        Assume a transaction \( \txid' \in \VIS_{i-1}^{-1}(\txid_j) \), 
        thus \( \txid' \in \VIS_{j}^{-1}(\txid_j) = R_j^{-1}(\txid_j) \).
        It is safe to include \( (\txid', \txid_i) \in R_i \) without affecting the read result,
        because those transaction \( \txid' \) is already visible for \( \txid_i \) in the abstract execution \( \aexec \):
        by \cref{equ:last-read-correct-psi} we know \( R_j \subseteq (\WR_j \cup \SO_j \cup \WW_j)^{+} \subseteq (\WR_\aexec \cup \SO_\aexec \cup \WW_\aexec)^{+}\),
        and by the definition of \( \WR(\aexec_i,\key) \) we know \( \WR(\aexec_i,\key) \subseteq \VIS_\aexec\).

    \item Given \( \SO_\aexec \subseteq \VIS_\aexec \) (and \( \WW_\aexec \subseteq \VIS_\aexec \) respectively) we include \( (\txid_j,\txid_i) \) for some \( \txid_j \)
        such that \( \txid_j \toEDGE{\SO_\aexec} \txid_i\) (and \( \txid_j \toEDGE{\WW_\aexec} \txid_i \) respectively).
        For the similar reason as \( \WR \),
        it is safe to includes all the visible transactions \( \txid' \) for \( \txid_j \), \ie \( \txid' \in R_j^{-1}\).
        \end{itemize}
        
    By the construction, both \cref{equ:vis-i-transitive} and \cref{equ:last-read-correct} are preserved. 
    Thus we have the proof.
    \end{itemize}
\end{proof}

To prove soundness, we pick an invariant for the \( \ET_\PSI \) as the union of those for \( \MR\) and \( \RYW \) shown in the following:
\begin{align*}
    I_1(\aexec, \cl) & =  \left( \bigcup_{\Set{\txid_{\cl}^{n} \in \txidset_{\aexec} }[ n \in \Nat ]} \VIS_{\aexec}^{-1}(\txid^n_\cl) \right) \setminus \txidset_\rd \\
    I_2(\aexec, \cl) & =  \left( \bigcup_{\Set{\txid_{\cl}^{n} \in \txidset_{\aexec} }[ n \in \Nat ]} (\SO_{\aexec}^{-1})\rflx(\txid^n_\cl) \right) \setminus \txidset_\rd
\end{align*}
where \( \txidset_\rd \) is all the read-only transactions included in both 
\[ \txidset_\rd \in \left( \bigcup_{\Set{\txid_{\cl}^{n} \in \txidset_{\aexec} }[ n \in \Nat ]} \VIS_{\aexec}^{-1}(\txid^n_\cl) \right)\]
and \[ \txidset_\rd  \in \left( \bigcup_{\Set{\txid_{\cl}^{n} \in \txidset_{\aexec} }[ n \in \Nat ]} (\SO_{\aexec}^{-1})\rflx(\txid^n_\cl) \right) \]
Assume a kv-store $\mkvs$, an initial and a final view $\vi, \vi'$  a fingerprint $\fp$ 
such that $\ET_{\PSI} \vdash (\mkvs, \vi) \csat \fp: (\mkvs',\vi')$. 
Also choose an arbitrary $\cl$, a transaction identifier $\txid_\cl^n \in \nextTxid(\mkvs, \cl)$, 
and an abstract execution $\aexec$ such that $\mkvs_{\aexec} = \mkvs$ and 
\( I_1(\aexec, \cl) \cup I_2(\aexec, \cl) \subseteq \Tx[\mkvs, \vi] \).
We are about to prove there exists an extra set of read-only transactions \( \txidset'_\rd \) such that
the new abstract execution \( \aexec' = \extend[\aexec, \txid_\cl^n, \fp, \Tx[\mkvs, \vi] \cup \txidset_\rd \cup \txidset'_\rd] \) and:
\begin{gather}
    \fora{\txid} (\txid, \txid_\cl^n) \in \SO_{\aexec'} \implies \txid \in \Tx[\mkvs, \vi] \cup \txidset_\rd \cup \txidset'_\rd \label{equ:psi-sound-update-so}\\
    \fora{\txid} (\txid, \txid_\cl^n) \in \WW_{\aexec'} \implies \txid \in \Tx[\mkvs, \vi] \cup \txidset_\rd \cup \txidset'_\rd \label{equ:psi-sound-update-ua}\\
    \fora{\txid} (\txid, \txid_\cl^n) \in ( \SO_{\aexec'} \cup \WR_{\aexec'} \cup \WW_{\aexec'} ) ; \VIS_{\aexec'} \implies \txid \in \Tx[\mkvs, \vi] \cup \txidset_\rd \cup \txidset'_\rd \label{equ:psi-sound-update-closure}\\
    I_1(\aexec',\cl) \cup I_2(\aexec',\cl) \subseteq \Tx[\mkvs_{\aexec'}, \vi'] \label{equ:psi-sound-inv} 
\end{gather}
\begin{itemize}
\item The invariant \( I_2 \) implies \cref{equ:psi-sound-update-so} as the same as \( \RYW \) in \cref{sec:sound-complete-ryw}.
\item Since \( \PSI \) also satisfies \( \UA \), the \cref{equ:si-sound-update-ww} can be proven as the same as \( \UA \) in \cref{sec:sound-complete-ua}.
\item \cref{equ:psi-sound-update-closure}.
    Note that \[ (\txid, \txid_\cl^n) \in ( \SO_{\aexec'} \cup \WR_{\aexec'} \cup \WW_{\aexec'}); \VIS_{\aexec'} \implies (\txid, \txid_\cl^n) \in ( \SO_{\aexec} \cup \WR_{\aexec}  \cup \WW_{\aexec} ) ; \VIS_{\aexec'} \]
    Also, recall that \( \SO_\aexec = \SO_\mkvs \), \( \WR_\aexec = \WR_\mkvs \) and  \( \WW_\aexec = \WW_\mkvs \).
    Let \[ \txidset'_\rd = \lfpTx[\mkvs,\vi,\SO_{\mkvs} \cup \WR_{\mkvs} \cup \WW_{\mkvs}] \]
    This means \[ \aexec' = \extend[\aexec, \txid_\cl^n, \fp, \lfpTx[\mkvs, \vi, \SO_{\mkvs} \cup \WR_{\mkvs} \cup \WW_\mkvs] \cup \txidset_\rd ] \]
    Let assume \( \txid \toEDGE{\SO_{\mkvs} \cup \WR_{\mkvs} \cup \WW_{\mkvs}} \txid' \) and \( \txid' \in \lfpTx[\mkvs, \vi, \SO_{\mkvs} \cup \WR_{\mkvs}\cup \WW_\mkvs ] \cup \txidset_\rd \).
    We have two possible cases:
    \begin{itemize}
        \item If \( \txid' \in \lfpTx[\mkvs, \vi, \SO_{\mkvs} \cup \WR_{\mkvs} \cup \WW_{\mkvs}] \), by  \cref{thm:view-vis-relation}, we know \[ \txid \in \lfpTx[\mkvs, \vi, \SO_{\mkvs} \cup \WR_{\mkvs} \cup \WW_{\mkvs}] \]
        \item If \( \txid' \in \txidset_\rd \), there are two cases:
        \begin{itemize}
            \item \( \txid' \in  \left( \bigcup_{\Set{\txid_{\cl}^{n} \in \txidset_{\aexec} }[ n \in \Nat ]} \VIS_{\aexec}^{-1}(\txid^n_\cl) \right) \).
                Since \( \txid' \) is a read-only transaction, it means \( \txid \toEDGE{\SO_{\mkvs} \cup \WR_{\mkvs} } \txid' \).
                By the property of \( \aexec \) (before update) that \( ( \SO \cup \WR_\aexec  ) ; \VIS_\aexec \in \VIS_\aexec \), it is known that \( \txid \in \left( \bigcup_{\Set{\txid_{\cl}^{n} \in \txidset_{\aexec} }[ n \in \Nat ]} \VIS_{\aexec}^{-1}(\txid^n_\cl) \right) \), that is, \( \txid \in \Tx[\mkvs,\vi] \cup \txidset_\rd\).

            \item \( \txid' \in  \left( \bigcup_{\Set{\txid_{\cl}^{n} \in \txidset_{\aexec} }[ n \in \Nat ]} \SO_{\aexec}^{-1}(\txid^n_\cl) \right) \).
                Given that \( \txid' \) is a read only transaction, we know \( \txid \in (\SO \cup \WR_\aexec)^{-1} \left( \bigcup_{\Set{\txid_{\cl}^{n} \in \txidset_{\aexec} }[ n \in \Nat ]} \SO_{\aexec}^{-1}(\txid^n_\cl) \right) \).
                By the property of \( \aexec \) (before update) that \( \SO \cup \WR_\aexec \in \VIS_\aexec \),
                it follows:
                \begin{align*}
                    \txid & \in VIS_\aexec^{-1} \left( \bigcup_{\Set{\txid_{\cl}^{n} \in \txidset_{\aexec} }[ n \in \Nat ]} \SO_{\aexec}^{-1}(\txid^n_\cl) \right) \\
                          & = \left( \bigcup_{\Set{\txid_{\cl}^{n} \in \txidset_{\aexec} }[ n \in \Nat ]} \VIS_{\aexec}^{-1}(\txid^n_\cl) \right)  \\
                          & = \Tx[\mkvs,\vi] \cup \txidset_\rd
                \end{align*}
                
        \end{itemize}
    \end{itemize}
\item Finally the new abstract execution preserves the invariant \( I_1 \) and \( I_2 \) 
because  \( \CC \) satisfies \( \MW \) and \( \RYW \).
\end{itemize}

The execution test $\ET_\PSI$ is complete with respect to the axiomatic definition 
\[ (\RP_{\LWW}, \Set{\lambda \aexec.  \VIS_\aexec ; \VIS_{\aexec}, \lambda \aexec \ldotp \SO_\aexec , \lambda \aexec \ldotp \WW_\aexec })\]
Assume i-\emph{th} transaction \( \txid_i \) in the arbitrary order,
and let view \( \vi_{i} = \getView[\aexec, \VIS^{-1}_{\aexec}(\txid_{i})] \).
We pick final view as \( \vi'_{i} = \getView[\aexec, (\AR^{-1}_{\aexec})\rflx(\txid_{i}) \cap \VIS_{\aexec}^{-1}(\txid'_i)] \),
if \( \txid'_i = \min_{\SO}\Set{\txid'}[\txid_i \toEDGE{\SO} \txid' ]\) is defined,
otherwise  \( \vi'_{i} = \getView[\aexec, (\AR^{-1}_{\aexec})\rflx(\txid_{i})]\).
Let the \( \mkvs = \mkvs_{\cut[\aexec, i-1]} \).
Now we prove the three parts separately.
\begin{itemize}
    \item \( \MR \).  By \cref{prop:cc-vis} since 
    \( \VIS_\aexec ; \SO_\aexec \subseteq \VIS_\aexec ; \VIS_\aexec \subseteq \VIS_\aexec \)
    so it follows as in \cref{sec:sound-complete-mr}.
    \item \( \RYW \). For \( \RYW \), since \( \WR_\aexec ; \SO_\aexec ; \VIS_\aexec \subseteq \VIS_\aexec ; \VIS_\aexec ; \VIS_\aexec \subseteq \VIS_\aexec\), 
    the proof is as the same proof as in \cref{sec:sound-complete-ryw}.
    \item \( \UA \). Since \( \WW_\aexec \subseteq \VIS_\aexec\), 
    the proof is as the same proof as in \cref{sec:sound-complete-ua}.
    \item \( \allowed[\WR_\mkvs \WW_\mkvs \cup \cup \SO]\). It is derived from \cref{thm:view-vis-relation} and 
        \[ (\WR_\aexec \cup \WW_\aexec \cup \SO_\aexec) ; \VIS_\aexec \subseteq \VIS_\aexec ; \VIS_\aexec \subseteq \VIS_\aexec\]
\end{itemize}

\subsection{Snapshot Isolation \( \SI \)}
\label{sec:sound-complete-si}

The axiomatic definition for \( \SI \) is 
\[ 
(\RP_{\LWW}, \Set{\lambda \aexec. \AR_\aexec ; \VIS_\aexec, \lambda \aexec \ldotp \SO_\aexec, \lambda \aexec \ldotp \WW_\aexec }) 
\]
By a lemma proven in \cite{SIanalysis}, for any \( \aexec \) satisfies the \( \SI \)
there exists an equivalent \( \aexec' \) with minimum visibility \( \VIS_{\aexec'} \subseteq \VIS_\aexec \) satisfying 
\[ 
    \left( \RP_{\LWW}, \Set{\lambda \aexec. \left( (\SO_{\aexec'} \cup \WW_{\aexec'} \cup \WR_{\aexec'} ) ; \RW_{\aexec'}\rflx \right) ; \VIS'_{\aexec'}, 
    \lambda \aexec \ldotp (\WW_{\aexec'} \cup \SO_{\aexec'}) } \right) 
\]
Under the minimum visibility \( \VIS \) all the transactions still have the same behaviour as before,
meaning they do not violate last-write-win.

To prove the soundness, we pick the invariant as the following:
\begin{align*}
    I_1(\aexec, \cl) & = \left( \bigcup_{\Set{\txid_{\cl}^{i} \in \txidset_{\aexec} }[ i \in \Nat ]} \VIS_{\aexec}^{-1}(\txid^i_\cl) \right) \setminus \txidset_\rd \\
    I_2(\aexec, \cl) & = \left( \bigcup_{\Set{\txid_{\cl}^{i} \in \txidset_{\aexec} }[ i \in \Nat ]} (\SO_{\aexec}^{-1})\rflx(\txid^i_\cl) \right) \setminus \txidset_\rd
\end{align*}
where \( \txidset_\rd \) is all the read-only transactions included in both 
\[ \txidset_\rd \in \left( \bigcup_{\Set{\txid_{\cl}^{i} \in \txidset_{\aexec} }[ i \in \Nat ]} \VIS_{\aexec}^{-1}(\txid^i_\cl) \right)\]
and \[ \txidset_\rd \in \left( \bigcup_{\Set{\txid_{\cl}^{i} \in \txidset_{\aexec} }[ i \in \Nat ]} (\SO_{\aexec}^{-1})\rflx(\txid^i_\cl) \right) \]
Assume a kv-store $\mkvs$, an initial and a final view $\vi, \vi'$  a fingerprint $\fp$ 
such that $\ET_{\SI} \vdash (\mkvs, \vi) \csat \fp: (\mkvs',\vi')$. 
Also choose an arbitrary $\cl$, a transaction identifier $\txid_\cl^n \in \nextTxid(\mkvs, \cl)$, 
and an abstract execution $\aexec$ such that $\mkvs_{\aexec} = \mkvs$ and 
\( I_1(\aexec, \cl) \cup I_2(\aexec, \cl) \subseteq \Tx[\mkvs, \vi] \).
We are about to prove there exists an extra set of read-only transaction \( \txidset'_\rd \) such that
the new abstract execution \( \aexec' = \extend[\aexec, \txid_\cl^n, \fp, \Tx[\mkvs, \vi] \cup \txidset_\rd \cup \txidset'_\rd]\) and 
\begin{gather}
    \fora{\txid} (\txid, \txid_\cl^n) \in \SO_{\aexec'} \implies \txid \in \Tx[\mkvs, \vi] \cup \txidset_\rd \cup \txidset'_\rd \label{equ:si-sound-update-so}\\
    \fora{\txid} (\txid, \txid_\cl^n) \in \WW_{\aexec'} \implies \txid \in \Tx[\mkvs, \vi] \cup \txidset_\rd \cup \txidset'_\rd \label{equ:si-sound-update-ww}\\
    \begin{array}{l}
    \fora{\txid} (\txid, \txid_\cl^n) \in \left( (\SO_{\aexec'} \cup \WW_{\aexec'} \cup \WR_{\aexec'} ) ; \RW_{\aexec'}\rflx \right) ; \VIS_{\aexec'} 
    \implies \txid \in \Tx[\mkvs, \vi] \cup \txidset_\rd \cup \txidset'_\rd 
    \end{array}
    \label{equ:si-sound-update-arvis}\\
    I_1(\aexec',\cl) \cup I_2(\aexec',\cl) \subseteq \Tx[\mkvs_{\aexec'}, \vi'] \label{equ:si-sound-inv} 
\end{gather}
\begin{itemize}
\item The invariant \( I_2 \) implies \cref{equ:si-sound-update-so} as the same as \( \RYW \) in \cref{sec:sound-complete-ryw}.
\item Since \( \SI \) also satisfies \( \UA \), the \cref{equ:si-sound-update-ww} can be proven as the same as \( \UA \) in \cref{sec:sound-complete-ua}.
\item \cref{equ:si-sound-update-arvis}. 
    Note that 
    \begin{centermultline}
        (\txid, \txid_\cl^n) \in \left( (\SO_{\aexec'} \cup \WW_{\aexec'} \cup \WR_{\aexec'} ) ; \RW_{\aexec'}\rflx \right); \VIS_{\aexec'} \\
        {} \implies (\txid, \txid_\cl^n) \in \left( (\SO_{\aexec} \cup \WW_{\aexec} \cup \WR_{\aexec} ) ; \RW_{\aexec}\rflx \right) ; \VIS_{\aexec'}
    \end{centermultline}
    Also, recall that \( \rel_\aexec = \rel_\mkvs \) for \( \rel \in \Set{\SO, \WR, \WW, \RW} \).
    Let \[ \txidset'_\rd = \lfpTx[\mkvs,\vi, \left( (\SO_{\aexec'} \cup \WW_{\aexec'} \cup \WR_{\aexec'} ) ; \RW_{\aexec'}\rflx \right)] \]
    This means that \[ \aexec' = \extend[\aexec, \txid_\cl^n, \fp, \lfpTx[\mkvs, \vi, \left( (\SO_{\aexec'} \cup \WW_{\aexec'} \cup \WR_{\aexec'} ) ; \RW_{\aexec'}\rflx \right)] \cup \txidset_\rd ] \]
    Let assume \( \txid \toEDGE{\left( (\SO_{\aexec'} \cup \WW_{\aexec'} \cup \WR_{\aexec'} ) ; \RW_{\aexec'}\rflx \right)} \txid' \) and \[ \txid' \in \lfpTx[\mkvs, \vi, \left( (\SO_{\aexec'} \cup \WW_{\aexec'} \cup \WR_{\aexec'} ) ; \RW_{\aexec'}\rflx \right)] \cup \txidset_\rd \]
    We have two possible cases:
    \begin{itemize}
        \item If \( \txid' \in \lfpTx[\mkvs, \vi, \left( (\SO_{\aexec'} \cup \WW_{\aexec'} \cup \WR_{\aexec'} ) ; \RW_{\aexec'}\rflx \right)] \), by  \cref{thm:view-vis-relation}, we know \[ \txid \in \lfpTx[\mkvs, \vi, \left( (\SO_{\aexec'} \cup \WW_{\aexec'} \cup \WR_{\aexec'} ) ; \RW_{\aexec'}\rflx \right)] \]
        \item If \( \txid' \in \txidset_\rd \), there are two cases:
        \begin{itemize}
            \item \( \txid' \in  \left( \bigcup_{\Set{\txid_{\cl}^{n} \in \txidset_{\aexec} }[ n \in \Nat ]} \VIS_{\aexec}^{-1}(\txid^n_\cl) \right) \).
                Since \( \txid' \) is a read-only transaction, it means \( \txid \toEDGE{\SO_{\mkvs} \cup \WR_{\mkvs} } \txid' \).
                By the property of \( \aexec \) (before update) that \( ( \SO \cup \WR_\aexec  ) ; \VIS_\aexec \in \VIS_\aexec \), it is known that \( \txid \in \left( \bigcup_{\Set{\txid_{\cl}^{n} \in \txidset_{\aexec} }[ n \in \Nat ]} \VIS_{\aexec}^{-1}(\txid^n_\cl) \right) \), that is, \( \txid \in \Tx[\mkvs,\vi] \cup \txidset_\rd\).

            \item \( \txid' \in  \left( \bigcup_{\Set{\txid_{\cl}^{n} \in \txidset_{\aexec} }[ n \in \Nat ]} \SO_{\aexec}^{-1}(\txid^n_\cl) \right) \).
                Given that \( \txid' \) is a read only transaction, we know \( \txid \in (\SO \cup \WR_\aexec)^{-1} \left( \bigcup_{\Set{\txid_{\cl}^{n} \in \txidset_{\aexec} }[ n \in \Nat ]} \SO_{\aexec}^{-1}(\txid^n_\cl) \right) \).
                By the property of \( \aexec \) (before update) that \( \SO \cup \WR_\aexec \in \VIS_\aexec \),
                it follows:
                \begin{align*}
                    \txid & \in VIS_\aexec^{-1} \left( \bigcup_{\Set{\txid_{\cl}^{n} \in \txidset_{\aexec} }[ n \in \Nat ]} \SO_{\aexec}^{-1}(\txid^n_\cl) \right) \\
                          & = \left( \bigcup_{\Set{\txid_{\cl}^{n} \in \txidset_{\aexec} }[ n \in \Nat ]} \VIS_{\aexec}^{-1}(\txid^n_\cl) \right)  \\
                          & = \Tx[\mkvs,\vi] \cup \txidset_\rd
                \end{align*}
        \end{itemize}
    \end{itemize}
\item Since \( \SI \) satisfies \( \RYW \) and \( \MR \), thus invariants \( I_1 \) and  \( I_2 \) are preserved, that is, \cref{equ:si-sound-inv}.
\end{itemize}

The execution test $\ET_\SI$ is complete with respect to the axiomatic definition 
\[ (\RP_{\LWW}, \Set{\lambda \aexec.  \AR_\aexec ; \VIS_{\aexec}, \lambda \aexec \ldotp \SO_\aexec , \lambda \aexec \ldotp \WW_\aexec })\]
Assume i-\emph{th} transaction \( \txid_i \) in the arbitrary order,
and let view \( \vi_{i} = \getView[\aexec, \VIS^{-1}_{\aexec}(\txid_{i})] \).
We pick final view as \( \vi'_{i} = \getView[\aexec, (\AR^{-1}_{\aexec})\rflx(\txid_{i}) \cap \VIS_{\aexec}^{-1}(\txid'_i)] \),
if \( \txid'_i = \min_{\SO}\Set{\txid'}[\txid_i \toEDGE{\SO} \txid' ]\) is defined,
otherwise  \( \vi'_{i} = \getView[\aexec, (\AR^{-1}_{\aexec})\rflx(\txid_{i})]\).
Let the \( \mkvs = \mkvs_{\cut[\aexec, i-1]} \).
Now we prove the three parts separately.
\begin{itemize}
    \item \( \MR \).  By \cref{prop:cc-vis} since 
    \( \VIS_\aexec ; \SO_\aexec \subseteq \VIS_\aexec ; \VIS_\aexec \subseteq \VIS_\aexec \)
    so it follows as in \cref{sec:sound-complete-mr}.
    \item \( \RYW \). For \( \RYW \), since \( \WR_\aexec ; \SO_\aexec ; \VIS_\aexec \subseteq \VIS_\aexec ; \VIS_\aexec ; \VIS_\aexec \subseteq \VIS_\aexec\), 
    the proof is as the same proof as in \cref{sec:sound-complete-ryw}.
    \item \( \UA \). Since \( \WW_\aexec \subseteq \VIS_\aexec\), 
    the proof is as the same proof as in \cref{sec:sound-complete-ua}.
    \item \( \allowed[\left( (\SO_{\mkvs} \cup \WW_{\mkvs} \cup \WR_{\mkvs} ) ; \RW_{\mkvs}\rflx \right)]\). It is derived from \cref{thm:view-vis-relation} and 
        \[ \left( (\SO_{\aexec} \cup \WW_{\aexec} \cup \WR_{\aexec} ) ; \RW_{\aexec}\rflx \right) ; \VIS_\aexec \subseteq \AR_\aexec ; \VIS_\aexec \subseteq \VIS_\aexec\]
\end{itemize}

\subsection{Serialisability \( \SER \)}
\label{sec:sound-complete-ser}

The execution test $\ET_\SER$ is sound with respect to the axiomatic definition 
\[ 
    (\RP_{\LWW}, \Set{\lambda \aexec. \AR })
\]
We pick the invariant as \( I( \aexec, \cl ) = \emptyset \), given the fact of no constraint on the view after update.
Assume a kv-store $\mkvs$, an initial and a final view $\vi, \vi'$  a fingerprint $\fp$ 
such that $\ET_{\SER} \vdash (\mkvs, \vi) \csat \fp: (\mkvs',\vi')$. 
Also choose an arbitrary $\cl$, a transaction identifier $\txid \in \nextTxid(\mkvs, \cl)$, 
and an abstract execution $\aexec$ such that $\mkvs_{\aexec} = \mkvs$ and 
\( I(\aexec, \cl) =  \emptyset \subseteq \Tx[\mkvs, \vi] \).
Let \( \aexec' = \extend[\aexec, \txid, \Tx[\mkvs, \vi], \fp] \).
Note that since the invariant is empty set, it remains to prove there exists a set of read-only transactions \( \txidset_\rd \) such that:
\[
    \begin{array}{@{}l@{}}
        \fora{ \txid' } 
        \txid' \toEDGE{\AR_{\aexec'}} \txid \implies \txid' \in \Tx[\mkvs, \vi] \cup \txidset_\rd
    \end{array}
\]
Since the abstract execution satisfies the constraint for \( \SER \), \ie \( \AR \subseteq \VIS \), we know \( \AR = \VIS \).
Since \( \Tx[\mkvs, \vi]  \) contains all transactions that write at least a key, 
we can pick a \( \txidset_\rd \) such that \( \Tx[\mkvs, \vi] \cup \txidset_\rd = \txidset_\aexec\),
which gives us the proof.

The execution test $\ET_\UA$ is complete with respect to the axiomatic definition \( (\RP_{\LWW}, \Set{\lambda \aexec. \AR_\aexec }) \).
Assume i-\emph{th} transaction \( \txid_i \) in the arbitrary order,
and let view \( \vi_{i} = \getView[\aexec, \VIS^{-1}_{\aexec}(\txid_{i})] \).
We also pick any final view such that \( \vi'_{i} \subseteq \getView[\aexec, (\AR^{-1}_{\aexec})\rflx(\txid_{i})] \).
Note that there is nothing to prove for \( \vi'_i \),
Now we need to prove the following:
\[
    \fora{\key, j}  0 \leq j < \abs{\mkvs_{\cut[\aexec, i-1]}(\key)} \implies j \in \vi_i(\key)
\]
Because \( \VIS^{-1}(\txid_i) = \AR^{-1}(\txid_i) = \Set{\txid }[\txid \in \mkvs_{\cut[\aexec, i-1]} ]\),
so for any key \( \key \) and index \( j \) such that \( 0 \leq j < \abs{\mkvs_{\cut[\aexec, i-1]}(\key)} \),
the j-\emph{th} version of the key contains in the view, \ie \( j \in \vi(\key)\).

\section{Program Analysis}
\label{app:robustness}
%
We give applications of our theory aimed at showing the 
robustness of a transactional library against a given consistency 
model. 
The first application considers a single counter library, 
and proves that it is robust against Parallel Snapshot Isolation. 
We present a general robustness conditions for \( \WSI \) and 
then show  multiple-counter example and a banking example \cite{bank-example-wsi} are robust against \( \WSI \).

\subsection{Single counter}
We start by reviewing the transactional code for the 
increment and read operations provided by a counter 
object over a key $\key$, denoted as  $\ctrinc(\key)$ and 
$\ctrread(\key)$, respectively.

\begin{align*}
\ctrinc(\key) & =
\begin{session}
\begin{transaction}
\plookup{\pv{a}}{\key};\\
\pmutate{\key}{\pv{a}+1};
\end{transaction}
\end{session}
&
\ctrread(\key) & =
\begin{session}
\begin{transaction}
\plookup{\pv{a}}{\key};
\end{transaction}
\end{session}
\end{align*}
Clients can interact with the key-value store only by invoking the $\ctrinc(\key)$ and 
$\ctrread(\key)$ operations. A transactional library is a set of transactional operations. 
For a single counter over key $\key$, we define the transactional library $\Counter(\key) = \Set{ \ctrinc(\key), \ctrread(\key)}$,
while for multiple counters over a set of keys $\mathsf{K} = \Set{\key_i}_{i \in I}$, respectively, we define $\Counter(\mathsf{K}) = 
\bigcup_{i \in I} \Counter(\key_i)$.

\mypar{KV-store semantics of a transactional library}
Given the transactional code 
$\ptrans{\trans}$, we define $\fp(\mkvs, \vi, \ptrans{\trans})$ 
to be the fingerprint that would be produced by a client that has view $\vi$ 
over the kv-store $\mkvs$, upon executing $\ptrans{\trans}$.
For the $\ctrinc(\key)$ and $\ctrread(\key)$ operations discussed above, 
we have that 
\[\fp(\mkvs, \vi, \ctrinc(\key)) = \Set{(\otR, \key, n), (\otW, \key, n+1) }[n = \snapshot[\mkvs, \vi](\key) ]\]
and \[\fp(\mkvs, \vi ,\codeFont{read}(\key)) = \Set{(\otR, \key, n) }[ n = \snapshot[\mkvs, \vi](\key) ]\]
Given an execution test $\ET$, and a transactional library $L = \Set{\ptrans{\trans_i}}_{i \in I}$, 
we define the set of valid $\ET$-traces for $L$ as the set 
$\codeFont{Traces}(\ET, \Set{\ptrans{\trans_i}}_{i \in I})$ 
of $\ET$-traces in which only $\ET$-reductions of the form 
\[
(\mkvs_{0}, \vienv_{0}) \toET{(\cl_0, \lambda_0)} (\mkvs_{1}, \vienv_{1}) \toET{(\cl_1, \lambda_1)} \cdots 
\toET{(\cl_{n-1}, \lambda_{n-1})} (\mkvs_{n}, \vienv_{n}),
\]
where for any $j=0,\cdots,n-1$, either $\lambda_{j} = \varepsilon$ or $\lambda_{j} = \fp(\mkvs_{j}, \vienv_{j}(\cl_{j}), \ptrans{\trans_{i}})$ 
for some $i \in I$. Henceforth we commit an abuse of notation and write $(\mkvs, \vienv) \toET{(\cl, \ptrans{\trans})} (\mkvs', \vienv')$ 
in lieu of $(\mkvs, \vienv) \toET{(\cl, \fp(\mkvs,\vienv(\cl), \ptrans{\trans})} (\mkvs', \vienv')$.
We also let $\mathsf{KVStores}(\ET, \Set{\ptrans{\trans_i}}_{i \in I})$ be the set of kv-stores 
that can be obtained when clients can only perform operations from $\Set{\ptrans{\trans_i}}_{i \in I}$ 
under the execution test $\ET$. Specifically, 
\[
    \mathsf{KVStores}(\ET, \Set{\ptrans{\trans_i}}_{i \in I}) \defeq
    \Set{ \mkvs }[%
        \left( (\mkvs_0, \vienv_{0}) \toET{\cdot} \cdots \toET{\cdot} (\mkvs, \stub) \right)
        \in \mathsf{Traces}(\ET, \Set{\ptrans{\trans_{i}}}_{i \in I}) ]
\]

\subsubsection{Anomaly of a single counter under Causal Consistency}
It is well known that the transactional library consisting of a single counter over a single 
key, $\Counter(\key)$, implemented on top of a kv-store guaranteeing Causal Consistency, 
leads to executions over the kv-store that cannot be simulated by the same transactional 
library implemented on top of a serialisable kv-store. 
For simplicity, let us assume that $\Keys = \Set{\key}$.
Let $\mkvs_{0} = [\key \mapsto (0, \txid_{0}, \emptyset)]$,  
$\mkvs_1 = [\key \mapsto (0, \txid_{0}, \Set{\txid_{\cl_1}^{1}} \lcat (0, \txid_{\cl_1}^{1}, \emptyset)$, 
$\mkvs_2 = [\key \mapsto (0, \txid_{0}, \Set{\txid_{\cl_1}^{1}, \txid_{\cl_2}^{1}}) \lcat (0, \txid_{\cl_1}^{1}, \emptyset) 
\lcat (0, \txid_{\cl_2}^{1}, \emptyset)$. Let also
$\vi_{0} = [\key \mapsto {0}]$. Then we have that 
\[
    (\mkvs_{0}, [\cl_1 \mapsto \vi_0, \cl_2 \mapsto \vi_0]) \toET{(\cl_1, \mathsf{inc(\key)})}[\ET_{\CC}]
    (\mkvs_1, [\cl_1 \mapsto \_, \cl_2 \mapsto \vi_0]) \toET{(\cl_1, \mathsf{inc(\key)})}[\ET_{\CC}]
(\mkvs_2, \_).
\]
By looking at the kv-store $\mkvs_{2}$, we immediately find a cycle in the graph induced by 
the relations $\SO_{\mkvs_{2}}, \WR_{\mkvs_{2}}, \WW_{\mkvs_{2}}, \RW_{\mkvs_{2}}$: 
$\txid_{\cl_1}^{1} \toEDGE{\RW} \txid_{\cl_2}^{1} \toEDGE{\RW} \txid_{\cl_1}^{1}$. 
Following from \cref{thm:serialisable_nocycle}, then 
which proves that $\mkvs_2$ is not included in $\CMs(\ET_{\SER})$, i.e. it is 
not serialisable.

\subsubsection{Robustness of a Single counter under Parallel Snapshot Isolation}
Here we show that the  single counter library $\Counter(\key)$ is robust under any consistency model 
that guarantees both write conflict detection (formalised by the execution test 
$\ET_{\UA}$), monotonic reads (formalised by the execution test $\ET_{\MR}$) 
and read your writes (formalised by the execution test $\ET_{\RYW}$). 
Because $\ET_{\PSI}$ guarantees all such consistency guarantees, i.e. 
$\CMs(\ET_{\PSI}) \subseteq \CMs(\ET_{\MR} \cap \ET_{\RYW} \cap \ET_{\UA})$, 
then it also follows that a single counter is robust under Parallel Snapshot Isolation.
\begin{proposition}
\label{prop:counter_hhshape}
Let $\mkvs \in \mathsf{KVStores}(\ET_{\UA} \cap \ET_{\MR} \cap \ET_{\RYW}, \Counter(\key))$. 
Then there exist $\Set{\txid_i}_{i = 1}^{n}$ and $\Set{\txidset_{i}}_{i = 0}^{n}$ such that 
\begin{align}
\mkvs(\key) = \left( (0, \txid_{0}, \txidset_{0} \uplus \Set{\txid_1}) \lcat \cdots \lcat (n-1, \txid_{n-1}, \txidset_{n-1} \uplus \Set{\txid_{n}}) \right) 
\lcat (n, \txid_{n}, \txidset_{n}) \label{eq:psi_counter_shape}\\
\fora{ i : 0 \leq i \leq n } \txidset_{i} \cap \Set{\txid_{i}}_{i=0}^{n} = \emptyset \label{eq:psi_counter_rwtxs}\\
\fora{ \txid, \txid', i,j: 0 \leq i,j \leq n } \txid \toEDGE{\SO} \txid' 
\land \txid \in \Set{\txid_{i}} \cup \txidset_{i} \implies 
\left(\begin{array}{l}
(\txid' = \txid_{j} \implies i < j) \land {} \\
(\txid' \in \txidset_{j} \implies i \leq j) \\
\end{array}\right) \label{eq:psi_counter_so}
\end{align}
%
%
%
\end{proposition}

\begin{proof}
It suffices to prove that the properties \eqref{eq:psi_counter_shape},\eqref{eq:psi_counter_rwtxs}, 
\eqref{eq:psi_counter_so} given in \cref{prop:counter_hhshape}, are invariant under 
$(\ET_{\MR} \cap \ET_{\RYW} \cap \ET_{\UA})$-reductions of the form 
\begin{align}
(\mkvs, \vienv) \toET{(\cl, \func{inc}[\key])}[\ET_{\UA} \cap \ET_{\MR} \cap \ET_{\RYW}] (\mkvs', \vienv') \label{eq:psi_counter_inc}\\
(\mkvs, \vienv) \toET{(\cl, \func{read}[\key])}[\ET_{\UA} \cap \ET_{\MR} \cap \ET_{\RYW}] (\mkvs', \vienv). \label{eq:psi_counter_read}
\end{align}
To this end, we will need the following auxiliary result which holds for any configuration $(\mkvs, \vienv)$ 
that can be obtained under the execution test $\ET_{\RYW} \cap \ET_{\MR}$:


\begin{centermultline}[eq:psi_counter_view]
    \fora{ i, j, n,m, \cl, \key } \txid_{\cl}^{n} \in \Set{\wtOf(\mkvs(\key, i))} \cup \rsOf(\mkvs(\key, i))  \\
    {} \land \txid_{\cl}^{m} \in \Set{\wtOf(\mkvs(\key, j))} \cup \rsOf(\mkvs(\key, j)) \land m < n 
    \land i \in \vienv(\cl)(\key) 
    \implies j \in \vienv(\cl)(\key) 
\end{centermultline} 

Suppose that there exist two sets $\Set{\txid_{i}}_{i=1}^{n}$ and 
$\Set{\txidset_{i}}_{i=0}^{n}$ such that $(\mkvs, \Set{\txid_{i}}_{i=1}^{n}, \Set{\txidset_{i}}_{i=0}^{n})$ 
satisfies the properties \eqref{eq:psi_counter_shape}-\eqref{eq:psi_counter_so}. 
We prove that, for transitions of the form \eqref{eq:psi_counter_inc}-\eqref{eq:psi_counter_read}, 
there exist an index $m$ and two collections $\Set{\txid_{i}}_{i=1}^{m}$, $\Set{\txidset'_{i}}_{i=0}^{m}$ 
such that $(\mkvs', \Set{\txid_{i}}_{i=1}^{m}, \Set{\txidset'_{i}}_{i=0}^{m})$ satisfies the properties 
\eqref{eq:psi_counter_shape}-\eqref{eq:psi_counter_so}. We consider the two transitions separately.

\begin{itemize}
\item 
Assume that
\[
(\mkvs, \vienv) \toET{(\cl, \func{inc}[\key])}[\ET_{\UA} \cap \ET_{\MR} \cap  \ET_{\RYW}] (\mkvs', \vienv')
\]
for some $\cl, \mkvs', \vienv'$. Let $n+1 = \lvert \mkvs(\key) \rvert$. Because of the definition of 
$\ET_{\UA}$, we must have that $\vienv(\cl) = [\key \mapsto \Set{0, \cdots, n}]$. Also, 
because $\mkvs$ satisfies \eqref{eq:psi_counter_shape}, we have that $\snapshot[\mkvs, \vienv(\cl)](\key) = n$. 
In particular, $\fp(\key, \vienv(\cl), \func{inc}[\key]) = \Set{(\otR, \key, n), (\otW, \key, n+1)}$. 
Thus we have that 
\[\mkvs' \in \updateKV[\mkvs, \vienv(\cl), \cl, \Set{(\otR, \key, n), (\otW, \key, n+1)}] \]
Let $\txid_{n+1}$ be the transaction identifier 
chosen to update $\mkvs$, i.e. \[\mkvs' = \updateKV[\mkvs, \vienv(\cl), \txid_{n+1}, \Set{(\otR, \key, n), (\otW, \key, n+1)}]\]
where $\txid_{n+1} \in \nextTxid(\mkvs, \cl)$; 
let also $\txidset_{n+1} = \emptyset$. Then we have the following: 
\begin{itemize}
\item  $(\mkvs', \Set{\txid_{i}}_{i=1}^{n+1}, \Set{\txidset_{i}}_{i=0}^{n+1})$ satisfies Property \eqref{eq:psi_counter_shape}. 
Recall that $(\mkvs, \Set{\txid_{i}}_{i=1}^{n}, \Set{\txidset_{i}}_{i=0}^{n})$ satisfies \eqref{eq:psi_counter_shape}, 
i.e.
\[\mkvs(\key) = \left( (0, \txid_{0}, \txidset_{0} \uplus \Set{\txid_1}) \lcat \cdots \lcat (n-1, \txid_{n-1}, \txidset_{n-1} \uplus \Set{\txid_{n}}) \right) 
\lcat (n, \txid_{n}, \txidset_{n}).
\]
It follows that $\mkvs'(\key) = \left( (0, \txid_{0}, \txidset_{0} \uplus \Set{\txid_1}) \lcat \cdots \lcat (n-1, \txid_{n-1}, \txidset_{n} \uplus \Set{\txid_{n+1}}) \right) 
\lcat (n+1, \txid_{n+1}, \txidset_{n+1})$, 
where we recall that $\txidset_{n+1} = \emptyset$.

\item $(\mkvs', \Set{\txid_{i}}_{i=1}^{n+1}, \Set{\txidset_{i}}_{i=0}^{n+1})$ 
satisfies Property \eqref{eq:psi_counter_rwtxs}. Let $i =0, \cdots, n+1$. If $i = n+1$, then 
$\txidset_{i} = \emptyset$, from which $\txidset_{i} \cap \Set{\txid_{j}}_{j=0}^{n+1} = \emptyset$ follows. If $i < n+1$, then 
because $(\mkvs, \Set{\txid_{i}}_{i=1}^{n}, \Set{\txidset_{i}}_{i=0}^{n})$ 
satisfies Property \eqref{eq:psi_counter_rwtxs}, then $\txidset_{i} \cap \Set{\txid_{j}}_{j=0}^{n} = \emptyset$. 
Finally, because $\txid_{n+1}$ was chosen to be fresh with respect to the transaction identifiers appearing in 
$\mkvs$, and $\txidset_{i} \subseteq \rsOf(\mkvs(\key, i))$, then  we also have that $\txidset_{i} \cap \Set{\txid_{n+1}} = \emptyset$. 
\item $(\mkvs', \Set{\txid_{i}}_{i=1}^{n+1}, \Set{\txidset_{i}}_{i=0}^{n+1})$ satisfies Property \eqref{eq:psi_counter_so}. Let 
$\txid, \txid'$ be such that $\txid \toEDGE{\SO} \txid'$. Choose two arbitrary indexes $i,j=0,\cdots, n+1$, 
and assume that $\txid \in \Set{\txid_{i}} \cup \txidset_{i}$. Note that if $i \leq n$, $j \leq n$, then 
because $(\mkvs, \Set{\txid_{i}}_{i=1}^{n}, \Set{\txidset_{i}}_{i=0}^{n})$ satisfies Property $\eqref{eq:psi_counter_so}$, then 
if $\txid' = \txid_{j}$ it follows that $i < j$, and if $\txid' \in \txidset_{j}$ it follows that $i \leq j$, as 
we wanted to prove. 
If $\txid \in \Set{\txid_{n+1}} \cup \txidset_{n+1}$, then it must be $\txid = \txid_{n+1}$ because 
$\txidset_{n+1} = \emptyset$. Recall that $\txid_{n+1}$ is the transaction identifier that was used 
to update $\mkvs$ to $\mkvs'$, i.e. $\mkvs' = \updateKV[\mkvs, \vienv(\cl), \txid_{n+1}, \stub]$. By 
definition of $\updateKV$, it follows that $\txid_{n+1} \in \nextTxid(\mkvs, \cl)$, 
and because $\txid_{n+1} \toEDGE{\SO} \txid'$, then $\txid'$ cannot appear in $\mkvs'$. 
In particular, 
$\txid' \notin \Set{\txid_{j}}_{j=0}^{n+1} \cup \bigcup \Set{\txidset_{j}}_{j=0}^{n+1}$, hence in this case there is nothing to prove. 
Finally, if $\txid' \in \Set{\txid_{n+1}} \cup \txidset_{n+1}$, then 
it must be the case that $\txid' = \txid_{n+1}$. If $\txid = \txid_{j}$, because 
$\txid \toEDGE{\SO} \txid'$ and $\txid' = \txid_{n+1}$, it cannot be $\txid = \txid_{n+1}$, 
hence it must be $i \leq n < n+1$. 
\end{itemize}

\item Suppose that 
\[
(\mkvs, \vienv) \toET{(\cl, \func{read}[\key])}[\ET_{\UA} \cap \ET_{\MR} \cap \ET_{\RYW}] (\mkvs', \vienv').
\]
As in the previous case, we have that $\mkvs' = \updateKV[\mkvs, \vienv(\cl), \txid, \Set{(\otR, \key, i)}]$, where 
$m = \snapshot[\mkvs, \vienv(\cl)](\key)$  - 
in particular, because $(\mkvs, \Set{\txid_{i}}_{i=1}^{n}, \Set{\txidset_{i}}_{i=0}^{n}$ satisfies 
Property \eqref{eq:psi_counter_shape}, then it must be the case that $m = \max_{<}(\vienv(\cl)(\key))$ - 
and $\txid \in \nextTxid(\mkvs, \cl)$. 
For $i=0,\cdots, n$, let $\txidset'_{i} := \txidset_{i}$ if $i \neq m$, $\txidset'_{i} = \txidset_{i} \cup \Set{\txid}$ if 
$i = m$. Then we have that $(\mkvs', \Set{\txid_{i}}_{i=0}^{n}, \Set{\txidset'_{i}}_{i=0}^{n})$ satisfies 
properties \eqref{eq:psi_counter_shape}-\eqref{eq:psi_counter_so}.
Putting all these facts together, we obtain the following: 
\begin{itemize}
\item $(\mkvs', \Set{\txid_{i}}_{i=0}^{n}, \Set{\txidset'_{i}}_{i=0}^{n})$ satisfies Property \eqref{eq:psi_counter_shape}. 
WIthout loss of generality, suppose that $m < n$. 
Because $(\mkvs,  \Set{\txid_{i}}_{i=0}^{n}, \Set{\txidset_{i}}_{i=0}^{n})$ satisfies Property \eqref{eq:psi_counter_shape}, 
we have that 
\[
\mkvs(\key) = \left( (0, \txid_{0}, \txidset_{0} \uplus \Set{\txid_1}) \lcat \cdots \lcat (m, \txid_{m}, \txidset_{m} \uplus \Set{\txid_{m+1}}) 
\lcat \cdots \lcat (n-1, \txid_{n-1}, \txidset_{n-1} \uplus \Set{\txid_{n}}) \right) \lcat (n, \txid_{n}, \txidset_{n}),
\] 
and from the definition of $\updateKV$ it follows that 
\[
\begin{array}{ll}
\mkvs(\key) &= \left( (0, \txid_{0}, \txidset_{0} \uplus \Set{\txid_1}) \lcat \cdots \lcat (m, \txid_{m}, \txidset_{m} \cup \Set{\txid} \uplus \Set{\txid_{m+1}}) 
\lcat \cdots \lcat (n-1, \txid_{n-1}, \txidset_{n-1} \uplus \Set{\txid_{n}}) \right) \lcat (n, \txid_{n}, \txidset_{n}) \\
& = \left( (0, \txid_{0}, \txidset'_{0} \uplus \Set{\txid_1}) \lcat \cdots \lcat (m, \txid_{m}, \txidset'_{m} \uplus \Set{\txid_{m+1}}) 
\lcat \cdots \lcat (n-1, \txid_{n-1}, \txidset_{n-1} \uplus \Set{\txid_{n}}) \right) \lcat (n, \txid_{n}, \txidset_{n})
\end{array}
\]
%

\item $(\mkvs',  \Set{\txid_{i}}_{i=0}^{n}, \Set{\txidset'_{i}}_{i=0}^{n})$ satisfies Property \eqref{eq:psi_counter_rwtxs}. 
Recall that $m = \max_{<}(\vienv(\cl)(\key))$; let $i=0,\cdots,n$.

Let again $i = \max_{<}\vienv(\cl)(\key)$ . 
If $i \neq m$, then $\txidset'_{i} = \txidset_{i}$, and because $(\mkvs,  \Set{\txid_{i}}_{i=0}^{n}, \Set{\txidset_{i}}_{i=0}^{n})$ 
satisfies Property \eqref{eq:psi_counter_rwtxs} 
we have that $\txidset'_{i} \cap \Set{\txid_{i}}_{i=0}^{n} = \emptyset$. If $i = m$, then 
we have that $\txidset'_{i} = \txidset'_{m} = \txidset_{m} \cup \Set{\txid}$, where we recall that $\txid \in \nextTxid(\mkvs, \cl)$. 
Because $(\mkvs,  \Set{\txid_{i}}_{i=0}^{n}, \Set{\txidset_{i}}_{i=0}^{n})$ 
satisfies Property \eqref{eq:psi_counter_rwtxs}, we have that $\txidset_{m} \cap \Set{\txid_{i}}_{i=0}^{n} 
= \emptyset$. Finally, because $\txid \in \nextTxid(\mkvs,\cl)$, then it must be the case that 
for any $i = 0,\cdots, n$, $\txid \notin \Set{\wtOf(\mkvs'(\key,i))}_{i=0}^{m} = \Set{\txid_{i}}_{i=0}^{m}$,  
where the last equality follows because we have already proved that $(\mkvs',  \Set{\txid_{i}}_{i=0}^{n}, \Set{\txidset'_{i}}_{i=0}^{n})$ 
satisfies Property \eqref{eq:psi_counter_shape}.

\item $(\mkvs',  \Set{\txid_{i}}_{i=0}^{n}, \Set{\txidset'_{i}}_{i=0}^{n})$ satisfies Property \eqref{eq:psi_counter_so}. 
Let $\txid', \txid''$ be such that $\txid' \toEDGE{\SO} \txid''$. 
Suppose also that $\txid' \in \Set{\txid_{i}} \cup \txidset'_{i}$ for some $i = 0,\cdots, n$. We consider two different cases:
\begin{itemize}
\item $\txid' = \txid_{i}$. Suppose then that $\txid'' = \txid_{j}$ for some $j = 0, \cdots, n$. Because 
$(\mkvs,  \Set{\txid_{i}}_{i=0}^{n}, \Set{\txidset_{i}}_{i=0}^{n})$ satisfies Property 
\eqref{eq:psi_counter_so}, then it must be the case that $i < j$. Otherwise, 
suppose that $\txid'' \in \txidset'_{j}$ for some $j=0,\cdots,n$. If $j \neq m$, then $\txidset'_{j} = \txidset_{j}$, 
and because $(\mkvs,  \Set{\txid_{i}}_{i=0}^{n}, \Set{\txidset_{i}}_{i=0}^{n})$ satisfies Property \eqref{eq:psi_counter_so}, we have that $i \leq j$. 
Otherwise, $\txidset'_{j} = \txidset'_{m} =  \txidset_{m} \cup \Set{\txid}$. Without loss of generality, in this case 
we can assume that $\txid'' = \txid$ (we have already shown that if $\txid'' \in \txidset_{j}$, then 
it must be $i \leq j$. Recall that $j = m = \max(\vienv(\cl)(\key))$, by the Definition of 
$\ET_{\UA}$ it must be the case that $\vienv(\cl) = [\key \mapsto \Set{0,\cdots, j}]$. 
It also follows that $\txid = \txid_{\cl}^{p}$ for some $p \geq 0$, and because $\txid' \toEDGE{\SO} \txid'' = \txid$, 
then $\txid' = \txid_{\cl}^{q}$ for some $q < p$. Because of Property 
\eqref{eq:psi_counter_view}, and because $\txid' = \txid_{i} = \wtOf(\mkvs(\key, i))$, then it must be the case the 
case that $i \in \vienv(\cl)(\key)$, hence  $i \leq m = j$.

\item $\txid' \in \txidset'_{i}$. We need to distinguish the cases $i \neq m$, leading to $\txidset'_{i} = \txidset_{i}$, 
or $i = m$, in which case $\txidset'_{i} = \txidset'_{m} = \txidset_{m} \cup \Set{\txid}$. If either $i \neq m$, or $i = m$ and $\txid \in 
\txidset_{m}$, then we can proceed as in the case $\txid' = \txid_{i}$. Otherwise, suppose that $i = m$ and 
$\txid' = \txid$. Then, because $\txid' \toEDGE{\SO} \txid''$, and $\txid \in \nextTxid(\mkvs,\cl)$, 
it must be the case that $\txid = \txid_{\cl}^{p}$ for some $p \geq 0$, and whenever 
$\txid_{\cl}^{\cdot} \in \key$, then $\txid_{\cl}^{\cdot} \toEDGE{\SO} \txid$. In particular 
we cannot have that $\txid'' \in \key$, because $\txid \toEDGE{\SO} \txid''$, which 
concludes the proof.
\end{itemize}

\item $(\mkvs', \vienv')$ satisfies Property \eqref{eq:psi_counter_view}.

\end{itemize}

\end{itemize}

\end{proof}

\begin{corollary}
\label{cor:psi_counter_acyclic}
Given $\mkvs \in \mathsf{KVStores}(\ET_{\UA} \cap \ET_{\MR} \cap \ET_{\RYW}, \mathsf{Counter})$, 
then $\graphOf[\mkvs]$ is acyclic.
\end{corollary}

\begin{proof}
Let $\Set{\txid_{i}}_{i=1}^{n}$, $\Set{\txidset_{i}}_{i=0}^{n}$ 
be such that $(\Set{\txid_{i}}_{i=1}^{n}, \Set{\txidset_{i}}_{i=0}^{n})$ 
satisfies properties \eqref{eq:psi_counter_shape}-\eqref{eq:psi_counter_so}. 
First, we define a partial order between transactions appearing in $\mkvs$ 
as the smallest relation $\dashrightarrow$ such that for any $\txid, \txid', \txid''$ and 
$i,j = 0,\cdots, n$
\[
\begin{array}{ll}
\txid \in \txidset_{i} &\implies \txid_{i} \dashrightarrow \txid,\\
i < j &\implies \txid_{i} \dashrightarrow \txid_{j},\\
\txid \in \txidset_{i} \land i < j & \implies \txid \dashrightarrow \txid_{j}\\
\txid, \txid' \in \txidset_{i} \land \txid \toEDGE{\SO} \txid' &\implies \txid \dashrightarrow \txid'\\
\txid \dashrightarrow \txid' \rightarrow \txid'' &\implies \txid \dashrightarrow \txid''
\end{array}
\]
It is immediate that if $\txid \dashrightarrow \txid'$ then either $\txid \in \Set{\txid_{i}} \cup \txidset_i$, 
$\txid' = \Set{\txid_{j}} \cup \txidset_{j}$ for some $i,j$ such that $i < j$, or $\txid = \txid_{i}$, $\txid' \in \txidset_i$, 
or $\txid, \txid' \in \txidset_{i}$ and $\txid \toEDGE{\SO_{\mkvs}} \txid'$. A consequence of this fact, 
is that $\dashrightarrow$ is irreflexive.

Next, observe that we have the following: 
\begin{itemize}
\item whenever $\txid \toEDGE{\WR_{\mkvs}} \txid'$, then 
there exists an index $i = 0,\cdots, n$ such that $\txid = \txid_{i}$, 
and either $i < n$ and $\txid' \in \txidset_{i} \cup \Set{ \txid_{i+1} }$, 
or $i = n$ and $\txid' \in \txidset_{i}$: by definition, we have that $\txid \dashrightarrow \txid'$;
\item whenever $\txid, \toEDGE{\WW_{\mkvs}} \txid'$, 
then there exist two indexes $i, j: 0 \leq i < j \leq n$ such that 
$\txid = \txid_{i}, \txid' = \txid_{j}$; again, we have that $\txid \dashrightarrow \txid'$, 
\item whenever $\txid \toEDGE{\RW_{\mkvs}} \txid'$, then 
there exist two indexes $i, j: 0 \leq i < j \leq n$ such that either 
$\txid \in \txidset_{i}$ and $\txid' = \txid_{j}$, or $\txid = \txid_{i+1}$, 
$i+1 < j$ and $\txid' = \txid_{j}$; in both cases, we obtain that $\txid \dashrightarrow \txid'$,
\item whenever $\txid \toEDGE{\SO_{\mkvs}} \txid'$, then 
$\txid \in \Set{\txid_{i}} \cup \txidset_{i}$ for some $i=0,\cdots,n$, 
and either $\txid' = \txid_{j}$ for some $i < j$,  or $\txid' \in \txidset_{j}$ for 
some $i \leq j$; it follows that $\txid \dashrightarrow \txid'$.
\end{itemize}

We have proved that $\dashrightarrow$ is an irreflexive relation, and it contains $(\SO_{\mkvs} \cup \WR_{\mkvs} \cup \WW_{\mkvs} \cup \RW_{\mkvs})^+$; 
because any subset of an irreflexive relation is itself irreflexive, we obtain that $\graphOf[\mkvs]$ is acyclic.
\end{proof}

\begin{corollary}
$\mathsf{KVStores}(\ET_{\PSI}, \mathsf{Counter}(\key)) \subseteq \mathsf{KVStores}(\ET_{\SER}, \mathsf{Counter}(\key))$. 
\end{corollary}

\begin{proof}
Let $\mkvs \in \mathsf{KVStores}(\ET_{\PSI}, \mathsf{Counter}(\key))$. Because $\ET_{\PSI} \supseteq \ET_{\MR} \cap \ET_{\RYW} \cap \ET_{\UA}$, 
we have that $\mkvs \in \mathsf{KVStores}(\ET_{\MR} \cap \ET_{\RYW} \cap \ET_{\UA}, \mathsf{Counter}(\key))$. 
By \cref{cor:psi_counter_acyclic} we have that $\graphOf[\mkvs]$ is acyclic. We can now employ the construction 
outlined in \cite{laws} to recover an abstract execution $\aexec = (\txidset_{\mkvs}, \VIS, \AR)$ such that $\SO \subseteq \VIS$ and $\AR \subseteq \VIS$, 
and $\graphOf[\aexec] = \graphOf[\mkvs]$.
Finally, the results from \cref{sec:sound-complete-ser} establish that, from $\aexec$ we can recover a $\ET_{\SER}$-trace in 
$\mathsf{Traces}(\ET_{\SER}, \mathsf{Counter}(\key))$ 
whose last configuration is $(\mkvs', \_)$, and 
$\graphOf[\mkvs'] = \graphOf[\aexec] = \graphOf[\mkvs]$, leading to $\mkvs' = \mkvs$. It follows that $\mkvs \in 
\mathsf{KVStores}(\ET_{\SER}, \mathsf{Counter}(\key))$.
\end{proof}

\subsection{Robust against \WSI}
\begin{definition}
    \label{def:wsi-safe}
    A key-value store \( \mkvs \) is \(\WSI\) safe if \( \mkvs \) is 
    reachable from executing an program \( \prog \) from an initial configuration \( \conf_0 \),
    \ie \( \ET_\WSI \vdash \conf_0, \prog \toPROG{} (\mkvs, \vienv), \prog' \),
    and \( \mkvs \) satisfies the following:
    \begin{align}
         & \fora{\txid,\key, \key',i,j} ( \txid \in \rsOf[\mkvs(\key,i)] \implies \txid \neq \wtOf[\mkvs(\key,i)] ) \implies \txid \neq \wtOf[\mkvs(\key',j)] \label{equ:wsi-safe-read-only} \\
         & \fora{\txid,\key,i} \txid \neq \txid_0 \exsts{j} \txid = \wtOf[\mkvs(\key,i)] \implies \txid \in \rsOf[\mkvs(\key,j)] \label{equ:wsi-safe-write-must-read} \\
         & \fora{\txid,\key,\key',i, j} \txid \neq \txid_0  \exsts{k} \txid = \wtOf[\mkvs(\key,i)] \land \txid \in \rsOf[\mkvs(\key',j)] \implies \txid = \wtOf[\mkvs(\key',j)] \label{equ:wsi-safe-all-write}
    \end{align}
\end{definition}

\begin{theorem}
    If a key-value store \( \mkvs \) is \(\WSI\) safe, it is robust against \(\WSI\).
\end{theorem}
\begin{proof}
    Assume a kv-store \( \mkvs \) that is \( \WSI\) safe.
    Given \cref{def:wsi-safe} that \( \mkvs \) is reachable under \(\WSI\) 
    therefore \( \CC \) and \( \UA \) since \( \CC \cup \UA \subseteq \WSI \), 
    it is easy to derive the following properties:
    \begin{align}
        & \fora{\txid, \txid'} \txid \toEDGE{\WR \cup \SO \cup \WW}  \txid' \implies \txid \neq \txid' \label{equ:acyclic-wsi}\\
        & \fora{\key,i,j} \txid = \wtOf[\mkvs(\key,i)] \land \txid \in \rsOf[\mkvs(\key,j)] \implies i = j + 1 \label{equ:wsi-sat-ua} 
    \end{align}
    \Cref{equ:acyclic-wsi} 
    To prove the robustness, it is sufficient to prove that 
    the relation \( (\WW \cup \WR \cup \RW \cup \SO)^+ \) is irreflexive,
    that is, for any transactions \( \txid \) and \( \txid' \):
    \[
        \txid \toEDGE{(\WW \cup \WR \cup \RW \cup \SO)^+} \txid' \implies \txid \neq \txid'
    \]
    We prove that by contradiction.
    Let assume \( \txid = \txid' \).
    By \cref{equ:acyclic-wsi}, it must be the case that the cycle contains \( \RW \), 
    which means there exists \( \txid_1 \) to \( \txid_n \)  such that
    \[
        \txid = \txid_1 \toEDGE{\rel^*} \txid_2 \toEDGE{\RW} \txid_3 \toEDGE{\rel^*} \cdots \toEDGE{\rel^*} \txid_{n-2} \toEDGE{\RW} \txid_{n-1} \toEDGE{\rel^*}  \txid_n = \txid' 
    \]
    where \( \rel  = \WR \cup \SO \cup \WW \).
    We replace some edges from the cycle.
    \begin{itemize}
        \item First, let consider transactions \( \txid_i \) such that \( \txid_i \toEDGE{\RW} \txid_{i+1}\).
    This means \( \txid_i \in \rsOf[ \mkvs(\key, x ) ]\) 
    and \( \txid_{i+1} = \wtOf[ \mkvs(\key, y ) ] \) for some key \( \key \) and two indexes \( x,y \) such that \( x < y \).
    There are two possible cases depending on if \( \txid_i\) wrote the key \( \key \).
        \begin{itemize}
            \item if \( \txid_i \) also wrote any key \( \key' \), 
                by \cref{lem:wsi-rw-to-ww}, it also wrote the key \( \key \) 
                and we can replace the edge with a \( \WW \) edge, that is \( \txid_i \toEDGE{\WW} \txid_{i+1}\).
            \item if \( \txid_i \) did not wrote any key, we leave the edge the same as before.
        \end{itemize}
        After the first step, any \( \RW \) edge in the cycle must start from a read only transaction.
        \begin{centermultline}[equ:wsi-rw-start-read-only]
            \txid_i \toEDGE{\RW} \txid_{i+1} \implies \fora{\key,i} \txid \neq \wtOf[\mkvs(\key,i)]
        \end{centermultline}
    \item Second, let now consider transactions \( \txid_i \) such that \( \cdots \toEDGE{\RW} \txid_i \toEDGE{\rel^*} \txid_{i+1} \toEDGE{\RW} \cdots \).  
        Transaction \( \txid_i \) at least wrote a key but \( \txid_{i+1}\) is a read-only transaction,
        thus \( \txid_i \neq \txid_{i+1}\).
        This means \( \cdots \toEDGE{\RW} \txid_i \toEDGE{\rel^+} \txid_{i+1} \toEDGE{\RW} \cdots \).
    \item Last, by \cref{lem:wsi-ww-to-wr} we replace all the \( \WW \) with \( \WR^* \).
    \end{itemize}
    Let \( \rel' = \WR \cup \SO\).
    Now we have cycle in the following form:
    \[
        \txid = \txid'_1 \toEDGE{{\rel'}^*} \txid'_2 \toEDGE{\RW} \txid'_3 \toEDGE{{\rel'}^+} \cdot \toEDGE{{\rel'}^+} \txid'_{m-2} \toEDGE{\RW} \txid_{m-1} \toEDGE{{\rel'}^*}  \txid'_m = \txid' 
    \]
    for some transactions \( \txid'_1 \) to \( \txid'_m \) and \( m \leq n \).
    This means \( \txid \toEDGE{((\WR \cup \SO ); \RW^?)^*} \txid' \).
    Because \( \mkvs \) is reachable under \( \WSI \) and so \( \CP \), it must the case that \( \txid \neq \txid' \),
    which contradicts with the assumption.
    Therefore, the relation \( (\WW \cup \WR \cup \RW \cup \SO)^+ \) is irreflexive.
\end{proof}

\begin{lemma}
    \label{lem:wsi-rw-to-ww}
    If a key-value store \( \mkvs \) is \WSI safe, then for any transactions \( \txid, \txid' \)
    \[
        \txid \toEDGE{\RW} \txid' \land \exsts{\key,i} \txid = \wtOf[\mkvs(\key,i)] \implies \txid \toEDGE{\WW} \txid' 
    \]
\end{lemma}
\begin{proof}
    Assume \( \txid \toEDGE{\RW} \txid' \), which means \( \txid  \in \rsOf[\mkvs(\key,i)]\) and \( \txid' = \wtOf[\mkvs(\key,j)]\)
    for a key \( \key \) and two indexes \(i,j \) such that \( i < j \).
    Assume the transaction \( \txid \) also wrote some key \( \key' \).
    Since that \( \mkvs \) is \( \WSI \) safe, \( \txid \) must write key \( \key \) too,
    \ie \( \txid  = \wtOf[\mkvs(\key,z)] \) for some index \(z \).
    Because the \( \mkvs \) is reachable under \( \WSI \) and therefore \( \UA \), this means \( z = i + 1\).
    Since that each version can only have one writer,
    we have \( i < z = i + 1 < j\), therefore \( \txid \toEDGE{\WW} \txid' \).
\end{proof}

\begin{lemma}
    \label{lem:wsi-ww-to-wr}
    If a key-value store \( \mkvs \) is \(\WSI\) safe, then for any transactions \( \txid, \txid' \)
    \[
        \txid \toEDGE{\WW} \txid' \implies \txid \toEDGE{(\WR)^*} \txid' 
    \]
\end{lemma}
\begin{proof}
    Assume a kv-store \( \mkvs \).
    Assume a key \( \key \) and two versions of it, \( i \) and \( j \) respectively with \( i< j\).
    Assume \(\txid = \wtOf[\mkvs(\key, i)] \) and \( \txid' = \wtOf[\mkvs(\key, j)] \).
    We prove \( \txid \toEDGE{(\WR)^*} \txid' \) by induction on the distance of the two versions.
    \begin{itemize}
    \item \caseB{ \(j - i = 1\) }
    By \( \WSI \) safe (\cref{def:wsi-safe}), \( \txid' \) must also read the key \( \key \),
    that is, \( \txid' \in \rsOf[\mkvs(\key,z)]\) for some \( z \).
    Because the \( \mkvs \) is reachable under \( \WSI \) and therefore \( \UA \),
    this means that if \( \txid' \) read and writes key \( \key \), 
    it must read the immediate predecessor.
    This means \( z = i\) and then \( \txid \toEDGE{\WR} \txid' \).
    \item \caseI{\(j - i > 1\)}
    By the \cref{def:wsi-safe}, \( \txid' \) must also read the key \( \key \),
    that is, \( \txid' \in \rsOf[\mkvs(\key,z)]\) for some \( z \).
    Because the \( \mkvs \) is reachable under \( \WSI \) and therefore \( \UA \),
    this means if \( \txid' \) read and writes key \( \key \), 
    it must read the immediate previous version with respect to the version it wrote.
    This means \( z = j - 1\).
    Assume the writer of the \(z\)-th version is \( \txid'' = \wtOf[\mkvs(\key, j-1)]\).
    We have \( \txid'' \toEDGE{\WR} \txid' \).
    Applying \ih,  we get \( \txid \toEDGE{(\WR)^*} \txid'' \).
    Thus we have \( \txid \toEDGE{(\WR)^*} \txid' \).
    \end{itemize}
\end{proof}

\subsection{Multiple counters}
\label{sec:app-multi-counter-robust}
We define a multi-counter library on a set of keys \( \codeFont{\keyset} \) as the following:
\[
    \mathsf{Counters(\keyset)} \defeq \bigcup_{\codeFont{\key} \in \codeFont{\keyset} } \Counter(\key)
\]
\subsubsection{Anomaly of multiple counters under Parallel Snapshot Isolation}
Suppose that the kv-store contains only two keys $\key, \key'$, each of which 
can be accessed and modified by clients using the code of transactional libraries 
$\mathsf{Counter}(\key), \mathsf{Counter}(\key'), \cdots$. We show that in this
 case it is possible to have the interactions of two client with the kv-store result 
 in  a non-serialisable final configuration. 
 
 More formally, suppose that $\Keys = \Set{\key_1, \key_2}$, and let $\mathsf{Counter} = \bigcup_{\key \in \Keys} \mathsf{Counter}(\key)$. 
 Let also 
 \[
 \begin{array}{lcl}
 \mkvs_{0} &=& [\key_1 \mapsto (0, \txid_{0}, \emptyset), \key_2 \mapsto (0, \txid_0, \emptyset)]\\
 \mkvs_{1} &=& [\key_1 \mapsto (0, \txid_{0}, \Set{\txid_{\cl_1}^{1}}) \lcat (1, \txid_{\cl_1}^{1}, \emptyset), \key_2 \mapsto (0, \txid_0, \emptyset)]\\
 \mkvs_{2} &=& [\key_1 \mapsto (0, \txid_{0}, \Set{\txid_{\cl_1}^{1}}) \lcat (1, \txid_{\cl_1}^{1}, \emptyset), \key_2 \mapsto (0, \txid_0, \Set{\txid_{\cl_2}}) \lcat(1, \txid_{\cl_2}^{1}, \emptyset)\\
 \mkvs_{3} &=& [\key_1 \mapsto (0, \txid_{0}, \Set{\txid_{\cl_1}^{1}}) \lcat (1, \txid_{\cl_1}^{1}, \emptyset), \key_2 \mapsto (0, \txid_0, \Set{\txid_{\cl_2}}) \lcat(1, \txid_{\cl_2}^{1}, \Set{\txid_{\cl_1}^{2}})\\
  \mkvs_{4} &=& [\key_1 \mapsto (0, \txid_{0}, \Set{\txid_{\cl_1}^{1}}) \lcat (1, \txid_{\cl_1}^{1}, \Set{\txid_{\cl_2}^{2}}), \key_2 \mapsto (0, \txid_0, \Set{\txid_{\cl_2}}) \lcat(1, \txid_{\cl_2}^{1}, \Set{\txid_{\cl_1}^{2}})\\
 && \\
 \vienv_0 &=& [\cl_1 \mapsto [\key_1 \mapsto \Set{0},  \key_2 \mapsto  \Set{0}], \cl_2 \mapsto [\key_1 \mapsto \Set{0}, \key_2 \mapsto \Set{0}]]\\
 \vienv_1 &=& [\cl_1 \mapsto [\key_1 \mapsto \Set{0,1}, \key_2 \mapsto \Set{0}], \cl_2 \mapsto [\key_1 \mapsto \Set{0}, \key_2 \mapsto \Set{0}]]\\
 \vienv_2 &=& [\cl_1 \mapsto [\key_1 \mapsto \Set{0,1}, \key_2 \mapsto \Set{0}], \cl_2 \mapsto [\key_1 \mapsto \Set{0}, \key_2 \mapsto \Set{0,1}]]\\
 \vienv_3 &=& \vienv_2\\
 \vienv_4 &=& \vienv_2\\
\end{array}
\]
Observe that we have the sequence of $\ET_{\PSI}$-reductions 
 \begin{centermultline}
     (\mkvs_0, \vienv_0) \toET{\cl_1, \mathsf{inc}(\key_1)}[\ET_{\PSI}] (\mkvs_1, \vienv_1) \toET{(\cl_2, \mathsf{inc}(\key_2)}[\ET_{\PSI}] {} \\
 (\mkvs_2, \vienv_2) \toET{(\cl_1, \mathsf{read}(\key_2)}[\ET_{\PSI}] (\mkvs_3, \vienv_3) \toET{(\cl_2, \mathsf{read}(\key_1)}[\ET_{\PSI}] 
 (\mkvs_4, \vienv_4)
 \end{centermultline}
and therefore $\mkvs_4 \in \mathsf{KVStores}(\ET_{\PSI}, \mathsf{Counter})$. 
On the other hand, for $\graphOf[\mkvs_4]$ we have the following cycle, which proves that 
$\mkvs_4 \notin \mathsf{KVStrores}(\ET_{\SER}, \mathsf{Counter})$: 
\[
\txid_{\cl_1}^{1} \toEDGE{\SO_{\mkvs_4}} \txid_{\cl_1}^{2} \toEDGE{\RW_{\mkvs_4}} \txid_{\cl_2}^{1} \toEDGE{\SO_{\mkvs_4}} 
\txid_{\cl_2}^{2} \toEDGE{\RW_{\mkvs_4}} \txid_{\cl_1}^{1}.
\]

\subsubsection{Robustness under Weak Snapshot Isolation}
It is easy to see a multi-counter libraries is \( \WSI \) safe, therefore robust under \( \WSI \).
\begin{theorem}
    Mulit-counter libraries \( \mathsf{Counters(\keyset)}  \) are \( \WSI \) safe.
\end{theorem}
\begin{proof}
    Assume an initial configuration \( \conf_0 = (\mkvs_0, \vienv_0 \) 
    and some \( \prog_0 \) where \( \dom[prog] \subseteq \dom[\vienv_0] \).
    Under \( \WSI \), we prove any reachable kv-store \( \mkvs_i \) satisfies \cref{equ:wsi-safe-read-only,equ:wsi-safe-write-must-read,equ:wsi-safe-all-write} by induction on the length of trace.
    \begin{itemize}
        \item \caseB{\( i = 0\) } 
            The formulae \cref{equ:wsi-safe-read-only,equ:wsi-safe-write-must-read,equ:wsi-safe-all-write} trivially hold given \( \mkvs_0 \) contains only the initial transaction \( \txid_0 \).
        \item \caseI{\( i > 0 \)}
            Let \( \conf_i = (\mkvs_i, \vienv_i)\) be the result of running \( \prog_0 \) for \( i \) steps.
            We perform case analysis for the next transaction \( \txid_{i+1} \).
            \begin{itemize}
                \item If \( \txid_{i+1}\) reads a key \( \pv{\key} \), \ie \ctrread(\pv{\key}),
                    it must start from a view that is closed to the relation \( (  (\WW \cup \WR \cup \SO) \cup \WR;\RW \cup \SO;\RW )^* \).
                    Let \( \mkvs_i(\pv{\key},j) = (\val, \txid', \txidset') \) be the latest version included in the view.
                    Thus the new kv-store \( \mkvs_{i+1} = \mkvs_{i}\rmto{\pv{\key}}{\mkvs_i(\pv{\key})\rmto{j}{(\val, \txid', \txidset' \uplus \Set{\txid_{i+1}})}} \).
                    Given \( \txid_{i+1} \) only read the key \( \pv{\key} \) without writing, \cref{equ:wsi-safe-read-only,equ:wsi-safe-all-write,equ:wsi-safe-write-must-read} trivially holds.
                    For other transactions \( \txid \) that are different from \( \txid_{i+1} \), they must exist in \( \mkvs_i \).
                    By \ih, then we prove that \( mkvs_{i+1} \) is \( \WSI \) safe.
                \item If \( \txid_{i+1}\) increments a key \( \pv{\key} \), \ie \ctrinc(\pv{\key}),
                    it means that all versions of \( \key \) must be included in the view.
                    Let \( \mkvs_i(\pv{\key},j) = (\val, \txid', \txidset') \) be the latest version of key \( \pv{\key} \).
                    Thus the new kv-store \( \mkvs_{i+1} = \mkvs_{i}\rmto{\pv{\key}}{( \mkvs_i(\pv{\key})\rmto{j}{(\val, \txid', \txidset' \uplus \Set{\txid_{i+1}})} ) \lcat (\val+1, \txid_{i+1}, \emptyset) } \).
                    Given \( \txid_{i+1} \) only read and then rewrites the key \( \pv{\key} \), \cref{equ:wsi-safe-read-only,equ:wsi-safe-all-write,equ:wsi-safe-write-must-read} trivially holds.
                    For other transactions \( \txid \) that are different from \( \txid_{i+1} \), they must exist in \( \mkvs_i \).
                    By \ih, then we prove that \( mkvs_{i+1} \) is \( \WSI \) safe.
            \end{itemize}
    \end{itemize}
\end{proof}

\subsection{Bank Example}
\citet{bank-example-wsi} presented a bank example
and claimed that this example is robust against  \( \SI \).
We find out that the bank example is also robust against \( \WSI \).
The example bases on relational database with three tables, account, saving and checking.
The account table maps customer names to customer IDs (\( \codeFont{Account(\underline{Name}, CustomerID )} \))
and saving and checking map customer IDs to saving balances (\( \codeFont{Saving(\underline{CustomerID}, Balance )} \)) 
and checking balances (\( \codeFont{Checking(\underline{CustomerID}, Balance )} \)) respectively.
We ignore the account table since it is an immutable lookup table.
We encode the saving and checking tables together as a kv-store.
Each customer is represent as an integer \( n \), that is,
\( (\stub, n) \in \codeFont{Account(\underline{Name}, CustomerID )} \),
its checking balance is associated with 
key \( n_s = 2 \times n \) and saving with \( n_c = 2 \times n + 1 \).
\begin{align*}
    n_c &\defeq 2 \times n &
    n_s &\defeq 2 \times n + 1 &
    \Keys &\defeq \bigcup_{n \in \Nat} \Set{n_c, n_s}
\end{align*}
If \( n \) is a customer, then 
\[ (n, \valueOf[\mkvs(n_s, \abs{\mkvs(n_s)})]) \in \codeFont{Saving(\underline{CustomerID}, Balance )} \]
\noindent and 
\[ (n, \valueOf[\mkvs(n_s, \abs{\mkvs(n_c)})]) \in \codeFont{Checking(\underline{CustomerID}, Balance )} \]
To interact with tables, there are five types of transactions.
For brevity we assume balances are integers.
\begin{align*}
    \codeFont{balance(n)} & \defeq
    \begin{transaction}
    \plookup{\pv{x}}{\pv{n}_c}; \ 
    \plookup{\pv{y}}{\pv{n}_s}; \ 
    \passign{\ret}{\pv{x}+\pv{y}}
    \end{transaction} \\
    \codeFont{depositChecking(n,v)} & \defeq
    \begin{transaction}
    \pifs{\pv{v} \geq 0} \ 
    \plookup{\pv{x}}{\pv{n}_c}; \ 
    \pmutate{\pv{n}_c}{\pv{x} + \pv{v}}; \ 
    \pife
    \end{transaction}  \\
    \codeFont{transactSaving(n,v)} & \defeq
    \begin{transaction}
    \plookup{\pv{x}}{\pv{n}_s}; \ 
    \pifs{\pv{v} + \pv{x} \geq 0} \ 
    \pmutate{\pv{n}_s}{\pv{x} + \pv{v}}; \ 
    \pife
    \end{transaction}
\end{align*}
\( \codeFont{balance(n)} \) returns customer \( n \) total balance.
\( \codeFont{depositChecking(n,v)} \) deposits \( v \) to the checking account of customer \( n \),
if \( v  \) is non-negative, otherwise the transaction does nothing.
While \( \codeFont{transactSaving(n,v)} \) allows a consumer \( n \) to deposit or withdraw money
from the saving account as long as the saving account afterwards is non-negative.
\begin{align*}
    \codeFont{amalgamate(n,n')} & \defeq
    \begin{transaction}
    \plookup{\pv{x}}{\pv{n}_s}; \ 
    \plookup{\pv{y}}{\pv{n}_c}; \ 
    \plookup{\pv{z}}{\pv{n'}_c}; \\
    \pmutate{\pv{n}_s}{0}; \ 
    \pmutate{\pv{n}_c}{0}; \ 
    \pmutate{\pv{n'}_c}{\pv{x} + \pv{y} + \pv{z}}; 
    \end{transaction} \\
    \codeFont{writeCheck(n,v)} & \defeq
    \begin{transaction}
    \plookup{\pv{x}}{\pv{n}_s}; \ 
    \plookup{\pv{y}}{\pv{n}_c}; \\
    \pifs{\pv{x} + \pv{y} < \pv{v} } \
        \pmutate{\pv{n}_c}{\pv{y} - \pv{v} - 1 }; \
    \pifm \
        \pmutate{\pv{n}_c}{\pv{y} - \pv{v} }; \ 
    \pife \\
    \pmutate{\pv{n}_s}{\pv{x}}; 
    \end{transaction} 
\end{align*}
\( \codeFont{amalgamate(n,n')} \) represents moving all funds from consumer \( n \) to
the checking account of customer \( n'\).
Last, \( \codeFont{writeCheck(n,v)} \) updates the checking account of \( n \).
If funds, both saving and checking, from \( n \) is greater than the \( v \),
the transaction deduct \( v \) from the checking account of \( n \).
If funds are not enough, the transaction further deducts one pounds as penalty.
\citet{bank-example-wsi} argued that, to make this example robust against \( \SI \),
\( \codeFont{writeCheck(n,v)} \) must be strengthened by writing back the balance to the saving account 
(the last line, \(\pmutate{\pv{n}_s}{\pv{x}} \)),
even thought the saving balance is unchanged.
The bank \( \codeFont{bank} \) libraries are defined by
\[ 
    \codeFont{Bank} \defeq \Set{\codeFont{balance(n)}, \codeFont{depositChecking(n,v)}, 
    \codeFont{amalgamate(n,n')}, \\ \codeFont{writeCheck(n,v)}, \codeFont{writeCheck(n,v)} }
    [ n,n' \in \Nat \land v \in \mathbb{Z} ] 
\]
\begin{theorem}
    The bank libraries \codeFont{Bank} are \( \WSI \) safe.
\end{theorem}
\begin{proof}
    Assume an initial configuration \( \conf_0 = (\mkvs_0, \vienv_0 ) \) 
    and some \( \prog_0 \) where \( \dom(\prog) \subseteq \dom(\vienv_0) \).
    Under \( \WSI \), we prove any reachable kv-store \( \mkvs_i \) satisfies \cref{equ:wsi-safe-read-only,equ:wsi-safe-write-must-read,equ:wsi-safe-all-write} by induction on the length of trace.
    \begin{itemize}
        \item \caseB{\( i = 0\) } 
            The formulae \cref{equ:wsi-safe-read-only,equ:wsi-safe-write-must-read,equ:wsi-safe-all-write} trivially hold given \( \mkvs_0 \) contains only the initial transaction \( \txid_0 \).
        \item \caseI{\( i > 0 \)}
            Let \( \conf_i = (\mkvs_i, \vienv_i)\) be the result of running \( \prog_0 \) for \( i \) steps.
            We perform case analysis for the next transaction \( \txid_{i+1} \).
            \begin{itemize}
                \item If \( \txid_{i+1} \) is \( \codeFont{balance(n)} \), 
                the only possible fingerprint is \( \Set{(\otR, n_c, \val_c), (\otR, n_s, \val_s)} \) 
                for some values \( \val_c \) and \( \val_s \).
                Since it is a read-only transaction, 
                \cref{equ:wsi-safe-read-only,equ:wsi-safe-write-must-read,equ:wsi-safe-all-write} trivially hold.
                \item If  \( \txid_{i+1} \) is \( \codeFont{depositChecking(n,v)} \), 
                in the cases of \( v < 0 \), the fingerprint is empty and 
                \cref{equ:wsi-safe-read-only,equ:wsi-safe-write-must-read,equ:wsi-safe-all-write} trivially hold.
                However, in the case of \( v \geq 0 \), the fingerprint is \( \Set{(\otR, n_c, \val_c), (\otW, n_c, \val_c + v)} \).
                Because it read and wrote only one key, \( n_c \),
                \cref{equ:wsi-safe-read-only,equ:wsi-safe-write-must-read,equ:wsi-safe-all-write} hold.
                \item If \( \txid_{i+1} \) is \( \codeFont{transactSaving(n,v)} \), 
                    there are two cases:
                    either a read-only fingerprint \( \Set{(\otR, n_s, \val_s)} \) 
                    when saving account has insufficient funds, or a read and write on key \( n_s \), 
                    that is \( \Set{(\otR, n_s, \val_s), (\otW, n_s, \val_s + v) }\).
                    For both cases it is easy to see 
                    \cref{equ:wsi-safe-read-only,equ:wsi-safe-write-must-read,equ:wsi-safe-all-write} hold.
                \item If  \( \txid_{i+1} \) is \( \codeFont{amalgamate(n,n')} \),
                    the fingerprint is
                    \[
                        \Set{(\otR, n_s, \val_s), (\otW, n_s, 0), (\otR, n_c, \val_c), (\otW, n_c, 0), (\otR, n'_c, \val'_c), (\otW, n'_c, \val'_c + \val_s + \val_c)} 
                    \]
                    Because the transaction always read and then wrote keys it touched, namely \( n_s, n_c \) and \( n'_c \), 
                    \cref{equ:wsi-safe-read-only,equ:wsi-safe-write-must-read,equ:wsi-safe-all-write} hold.
                \item Last, if \( \txid_{i+1} \) is \( \codeFont{writeCheck(n,v)} \),
                the fingerprint is
                \[
                    \Set{(\otR, n_s, \val_s), (\otW, n_s, \val_s), (\otR, n_c, \val_c), (\otW, n_c, \val'_c)} 
                \]
                where \( \val'_c \) can be either \( \val_c - v\) or \( \val_c - v - 1 \).
                Similar to \(  \codeFont{amalgamate(n,n')} \),
                the transaction always read and then wrote keys it touched,
                so \cref{equ:wsi-safe-read-only,equ:wsi-safe-write-must-read,equ:wsi-safe-all-write} hold.
            \end{itemize}
    \end{itemize}
\end{proof}

\section{Verification of implementations}
\label{sec:implementation-verification}
\label{app:implementation-verification}
We verify two protocols, COPS and Clock-SI, that the former is a full replicated implementation for causal consistency and the latter is a shard implementation for snapshot isolation.
\subsection{COPS}
\label{sec:cops}
\subsubsection{Code}
\paragraph{\bf Structure}
COPS is a fully replicated database protocol for causal consistency.
There are two versions, 
that the simplified version only supports either a single read or a single write per transaction, 
and the full version supports either multiple reads or a single write pe transactions.
Here we verify the full version.

The overall database is modelled as a key-value store.
Each key, instead of a single value, is associated with a list of versions,
consisting of value, version number \verb|VersionNo| and dependencies \verb|Dep|.
COPS relies on the version number to resolve conflict,
that is, the write with greater version number wins.
Version number composite by time (higher bits) and replica identifier (lower bits).
Since the replica identifiers are full ordered, therefore version numbers are full ordered.
The dependencies is a set of versions (pairs of keys and versions numbers)
that the current version depends on.

\begin{lstlisting}[caption={COPS Structure},label={lst:cops-structure}]
VersionNo :: LocalTime + ID
Dep :: Set(Key,VersionNo)
KV :: Key -> List(Val,VersionNo,Dep)
\end{lstlisting}

Under the hood, there are many replicas,
where each replica has a unique identifier
and contains the full key-value store yet might be out of data.
Replica also tracks its local time.

\begin{lstlisting}[caption={COPS Replicas},label={lst:cops-replica}]
Replicas :: ID -> (KV,LocalTime)
\end{lstlisting}

To track session order, Client interact with a replica via certain API.
To call those API,
client has the responsibility to provide context \verb|Ctx| 
which contains versions that has been read from or written to the replica from the same client.

\begin{lstlisting}[caption={Client context},label={lst:cops-client-ctx}]
Ctx :: {
    readers : List(Key,VersionNo,Dep)
    writers : List(Key,VersionNo,Dep)
}
\end{lstlisting}

\paragraph{\bf Write}
The client call \verb|put| to commit a new write to a key \verb|k| with value \verb|v| with the context \verb|ctx|.
It extracts the dependencies from the context, by unioning all the versions inside the context,
then calls the replica API \verb|ver = put_after(k,v,deps,dep_to_nearest(deps),dep)|,
which require to provide the dependencies so that the replica can check whether they are exists.
Note \verb|dep_to_nearest(deps)| is for performance by providing the latest versions.
The replica returns a version number allocated for the new version for key \verb|k|,
and then the version is added into the context.

\begin{lstlisting}[caption={Client API for write},label={lst:cops-client-api-write}]
Ctx put(k,v,ctx) {
    // add up all the read and write dependency
    deps = ctx_to_dep(ctx);
    // call replica API
    ver = put_after(k,v,deps,dep_to_nearest(deps));
    // update context
    ctx.writers += (k,ver,deps);
    return ctx;
}

Dep ctx_to_dep(ctx) { 
    return { (k,ver) | (k,ver,_) (*$\in$*) ctx.readers (*$\lor$*) (k,ver,_) (*$\in$*) ctx.writers } 
}

Dep dep_to_nearest(deps) {
    return { (k,ver) | (*$\forall$*)k',ver',deps'. (k',ver',deps') (*$\in$*) deps (*$\implies$*) (k,ver) (*$\notin$*) deps' };
}
\end{lstlisting}

The \verb|put_after| waits until all the versions contained in \verb|nearest| exists,
consequently all the versions contained in \verb|deps| exists.
The replica increments the local time and insert the new version with version numbers \verb|time ++ id| (local time concatenating replica identifier) to the key \verb|k|.
At this point, replica returns the new version number to client and 
later on it will broad case to other replica\footnote{It uses message queue for broad-casting}.

\begin{lstlisting}[caption={Replica API for write},label={lst:cops-replica-api-write}]
VersionNo put_after(k,v,deps,nearest,vers){
    for (k,ver) in nearest { wait until (_,ver,_) (*$\in$*) kv(k); }

    time = inc(local_time);

    // appending local kv with a new version.
    list_isnert(kv[k], (v, (time ++ id), deps) );

    asyn_brordcase(k, v, (time ++ id), deps);
    return (local_time + id);
}
\end{lstlisting}

When a replica receives a update message, it checks the existence of versions included in the dependencies and then adds the new version to the replica.
Last, the replica updates the local time if the new version's time is greater than the local time.

\begin{lstlisting}[caption={Receive update message},label={lst:cops-replica-receive-msg}]
on_receive(k,v,ver,deps) {
    for (k',ver') in deps { wait until (_,ver',_) (*$\in$*) kv(k'); }

    list_isnert(kv[k],(v,ver,deps));
    (remote_local_time + id) = ver;
    local_time = max(remote_local_time, local_time);
}
\end{lstlisting}

\paragraph{\bf Read}
To read multiple keys \verb|ks| in a transaction, client calls the \verb|get_trans(ks,ctx)|.
Note that between two reads for different keys, 
the replica might be interleave and schedule other transactions.
The challenge here is to ensure all the values are consistent, \ie
overall they satisfy the causal consistency.
That is, if the transaction fetches a version \( \ver \) for key \( \key \),
and this version \( \ver \) depends on anther version \( \ver' \) for another key \( \key' \), 
then the transaction should at least fetches \( \ver' \) for the key \( \key' \),
or any later version  for the key \( \key' \).

The algorithm, in the first phase, optimistically reads the current latest version for each key from the replica via the replica API \verb|rst[k]=get_by_version(k,LATEST)|.
In the second phase, it computes the maximum version \verb|ccv[k]| from any dependencies \verb|rst[k].deps| read in the first phase.
Such \verb|ccv[k]| is the minimum version that should be fetched.
Therefore at the end, it only needs to re-fetched the specifically version \verb|ccv[k]|,
if the version fetched in the first phase is older than \verb|ccv[k]|.

\begin{lstlisting}[caption={Reads},label={lst:cops-client-read}]
List(Val) get_trans(ks,ctx) {
    // only guarantee to read up-to-date value 
    // the moment reading the individual key
    for k in ks { rst[k] = get_by_version(k,LATEST); }

    for k in ks {
        ccv[k] = max (ccv[k],rst[k].ver);
        for dep in rst[k].deps
            if ( dep.key (*$\in$*) ks ) ccv[k] = max (ccv[dep.key],dep.vers);
    }

    for k in ks 
        if ( ccv[k] > rst[k].vers ) rst[k] = get_by_version(k,ccv[k]);

    // update the ctx
    for (k,ver,deps) in rst { ctx.readers += (k,ver,deps); }

    return to_vals(rst);
}                                   
\end{lstlisting}

The client API \verb|get_by_version(k,ver)| returns the version \verb|ver| for key \verb|k|.

\begin{lstlisting}[caption={Replica API for read},label={lst:cops-replica-read}]
(Val,Version,Dep) get_by_version(k,ver) {
    if (ver  = LATEST){ ver = max(kv[k].vers); }

    wait until (_,ver,_) (*$\in$*) kv(k);
    
    let (val,ver,deps) from kv[k];
    return (val,ver,deps);
}
\end{lstlisting}

\subsubsection{Verification}
\paragraph{\bf Semantics the code}
Let \( \repl \in \Repls \) denotes the set of totally ordered replicates.
Each replicate can have multiple clients, and 
each clients can commit a sequence of either read-only transitions or single-write transactions.
To model these, we annotate the transaction identifier with replicate \( \repl \), client \( \cl \), 
local time of the replicate \( n \) and read-only transactions counter \( n' \), \ie \( \txidCOPS{\repl}{\cl}{n}{n'} \).
Note that the \( (n, \repl, n') \) can be treated as a single number that \( n \) are the higher bits, 
\( \repl \) the middle bits and \( n' \) the lower bits.
For a new single-write transaction, it is allocated with a transaction identifier with larger local time,
and for a read-only transactions, it is allocated with a transaction identifier with larger read-only counter.
There is a total order among transitions from the same replica and from the same client.

To model the dependencies of each version,
We extend version from \cref{def:his_heap} 
with the set of all versions it dependencies on, \( \dep \in \pset{\Keys \times \TxID} \).
The function \( \depOf{\ver} \) denotes the dependencies set of the version.
We use \( \mkvsCOPS \) for key-value store whose versions contain the dependencies.

We use view to model the client context,
that is, a version is included in a context if and only if such version is in the view of the client.
We also use view to model a replica state,
that is, if a replica contains a version if and only if such version in the view of the replica.
For readability, we annotate view with either a replica, \( \viREPL \), or a client, \( \viCL \).
The view environment is extended with replicas and their views, \( \vienvCOPS : (\Repls \times \Clients ) \parfinfun \Views \).
We give the following semantics to capture the behaviours of the code.

\paragraph{\bf Write}
For purpose of verification, we eliminate code for performance, and put the client API and replica API in the same function (\cref{lst:simplified-put}).
\begin{lstlisting}[caption={put},label={lst:simplified-put}]
// mixing the client API and system API
put(repl,k,v,ctx) {

    // Dependency for previous reads and writes
    deps = ctx_to_dep(ctx);(*\label{line:put-ctx-to-deps}*)

    // increase local time and appending local kv with a new version.
    inc(repl.local_time);(*\label{line:put-inc-local}*) 
    list_isnert(repl.kv[k],(v, (local_time + id), deps));(*\label{line:put-update-kv}*)

    // update dependency for writes
    ctx.writers += (k,(local_time + id),deps);(*\label{line:put-update-ctx}*)

    // broad case
    asyn_brordcase(k, v, (time ++ id), deps);
    return (local_time + id);
}
\end{lstlisting}

The following is the rule corresponds to \cref{lst:simplified-put}:

\begin{mathpar}
    \inferrule[Put]{%
        ( \stk, \ss, \emptyset ), \pmutate{\key}{\vx} \toTRANS
        ( \stk', \stub, \Set{(\otW, \key, \val )} ), \pskip
        \\\\
        \dep = \Set{(\key', \txid)}[\exsts{i} i \in \viCL(\key') \land \txid = \wtOf(\mkvsCOPS(\key', i)) \lor (\key', \txid) \in \depOf[\mkvsCOPS(\key', i)] ] \texttt{ ---> \cref{lst:simplified-put}, \cref{line:put-ctx-to-deps}} 
        \\\\
        \txid = \min\Set{%
        \txidCOPS{\repl}{\cl}{n'}{0}
        }[%
            \fora{\key', i \in \viREPL(\key'), n} 
            \txidCOPS{\stub}{\stub}{n}{\stub} = \wtOf(\mkvsCOPS(\key',i))
            \implies n' > n 
        ] \texttt{ ---> \cref{lst:simplified-put}, \cref{line:put-inc-local}}
        \\\\
        \mkvsCOPS' = \mkvsCOPS\rmto{\key}{\mkvsCOPS(\key) \lcat (\key, \txid, \emptyset, \dep)} \texttt{ ---> \cref{lst:simplified-put}, \cref{line:put-update-kv}}
        \\\\
        \viREPL' = \viREPL\rmto{\key}{\viREPL(\key) \uplus \Set{\abs{\mkvsCOPS'(\key)} - 1}} \texttt{ ---> \cref{lst:simplified-put}, \cref{line:put-update-kv}}
        \\\\
        \viCL' = \viCL\rmto{\key}{\viREPL(\key) \uplus \Set{\abs{\mkvsCOPS'(\key)} - 1}} \texttt{ ---> \cref{lst:simplified-put}, \cref{line:put-update-ctx}}
    }{%
    \repl, \cl \vdash 
    \mkvsCOPS, \viREPL, \viCL, \stk, \ptrans{\pmutate{\key}{\vx};} \toCMD{\viCL, \Set{(\otW, \key, \val )} }
    \mkvsCOPS', \viREPL', \viCL', \stk', \pskip
    }
\end{mathpar}

The first premiss is to execute the transaction locally (\cref{fig:semantics-trans}).
Since there is only a write, the snapshot \( \ss \) can be any snapshot.
The second line computes the dependency set for the new write operation,
by collecting all the writers of versions included in the view \( \viCL \).
The third line simulates the increment of local time.
Even thought we do not directly track the local time of a replica, 
yet the local time can compute as the maximum time contained in the replica's view \( \viREPL \).
The forth and fifth simulates the updates of the replica's key-value store,
and the last premiss simulates the update of client context.

\paragraph{\bf Read}
The following is a simplified algorithm by directly taking a list of versions 
\verb|ccv| satisfies causal consistency constraint,
\ie the second phase of \cref{lst:cops-client-read},
and then read the versions indicated by \verb|ccv|.

\begin{lstlisting}[caption={get\_trans},label={lst:get-trans}]
List(Val) get_trans(ks,ctx) {
    // only guarantee to read up-to-date value 
    // the moment reading the individual key
    for k in ks { rst[k] = get_by_version(k,LATEST); } (*\label{line:get-trans-pick-ccv-1}*)

    for k in ks { 
        ccv[k] = max (ccv[k],rst[k].ver); (*\label{line:get-trans-ccv-1}*)
        for dep in rst[k].deps (*\label{line:get-trans-ccv-2}*)
            if ( dep.key (*$\in$*) ks ) ccv[k] = max (ccv[dep.key],dep.vers); (*\label{line:get-trans-ccv-3}*) 
    } (*\label{line:get-trans-ccv-4}*)
    for k in ks   (*\label{line:get-trans-ccv-5}*)
        if ( ccv[k] > rst[k].vers ) rst[k] = get_by_version(k,ccv[k]); (*\label{line:get-trans-ccv-6}*)

    // update the ctx
    for (k,ver,deps) in rst { ctx.readers += (k,ver,deps); } (*\label{line:get-trans-update-ctx-1}*)

    return to_vals(rst); (*\label{line:get-trans-read-1}*)
}                                   
\end{lstlisting}

The following is the rule for read-only transaction:

\begin{mathpar}
    \inferrule[GetTrans]{%
        \trans =  \plookup{\vx_1}{\key_1}; \dots; \plookup{\vx_j}{\key_j};
        \\
        \vi_0 = \viCL
        \\\\
        {\begin{array}{@{}l@{}}
            \mathsf{for} \ i \ \mathsf{in} \Set{1, \dots, j} \\
            \quad m_i \in \viREPL(\key_i) \\
            \quad \vi'_i = \vi_{i-1}\rmto{\key_i}{\vi_{i-1}(\key_i) \cup \Set{m_i}} 
            \texttt{ ---> \cref{lst:get-trans}, \cref{line:get-trans-pick-ccv-1}} \\
            \quad \vi_i = \lambda \key \ldotp \vi'_i(\key) \cup \Set{x}[(\key,\wtOf[\mkvsCOPS(\key,x)]) \in \depOf[\mkvsCOPS(\key_i,m_i)]]  \\
            \qquad \texttt{ ---> \cref{lst:get-trans}, \cref{line:get-trans-ccv-1,line:get-trans-ccv-2,line:get-trans-ccv-3,line:get-trans-ccv-4,line:get-trans-ccv-5,line:get-trans-ccv-6,line:get-trans-update-ctx-1}} \\
        \end{array}}
        \\\\
        \\\\
        \viCL' = \vi_j
        \\
        ( \stk, \func{getMax}[\mkvsCOPS, \viCL'], \emptyset ), \trans \toTRANS
        ( \stk', \stub, \fp ), \pskip \texttt{ ---> \cref{lst:get-trans}, \cref{line:get-trans-read-1}}
        \\\\
        \txidCOPS{\repl}{\cl}{n'}{n} = \max\Set{\txidCOPS{\repl}{\cl}{z'}{z}}[\txidCOPS{\repl}{\cl}{z'}{z} \in \mkvsCOPS ]
        \\
        \mkvsCOPS' = \updateKV[\mkvsCOPS, \viCL', \fp, \txidCOPS{\repl}{\cl}{n'}{n+1}] 
    }{%
        \repl, \cl \vdash 
        \mkvsCOPS, \viREPL, \viCL, \stk, \ptrans{\plookup{\vx_1}{\key_1}; \dots; \plookup{\vx_j}{\key_j}; } \toCMD{\viCL', \fp}
        \mkvsCOPS', \viREPL, \viCL', \stk', \pskip
    }
    \and
    \inferrule[ClientCommit]{%
        \repl, \cl \vdash 
        \mkvsCOPS, \vienvCOPS(\repl), \vienvCOPS(\cl), \stk, \prog(\cl) \toCMD{\viCL'', \fp}
        \mkvsCOPS', \viREPL', \viCL', \stk', \cmd'
    }{%
        \mkvsCOPS, \vienvCOPS, \thdenv, \prog \toPROG{\viCL'', \fp}
        \mkvsCOPS', \vienvCOPS\rmto{\repl}{\viREPL'}\rmto{\cl}{\viCL'}, \thdenv\rmto{\cl}{\stk'}, \prog\rmto{\cl}{\cmd'}
    }
\end{mathpar}

The for-loop in the premiss picks a version for key \( \key_i \) from \( \viREPL \) and adds to client view, which corresponds to the initial \verb|ccv|\((\key_i)\).
Then it add all the new dependencies \( \depOf[\mkvsCOPS(\key_i,m_i)] \) to the view that yields \( \vi_i \).
The view \( \vi_i \) corresponds to \verb|ccv|\((\key_i)\) after the update against the dependencies.
Given the view \( \viCL \), the client fetches the version with the maximum writer it can observed for each key, which is computed by \( \func{getMax} \) function. 
It is different from \( \snapshot \) because \( \snapshot \) fetches the latest version with respect to the position in the list, but later one we will prove \( \func{getMax} \) and \( \snapshot \) are equivalent.
\begin{align*}
    \func{getMax}[\mkvsCOPS, \viCL] & \defeq 
    \lambda \key \ldotp \left( \max\Set{(\val, \txid, \txidset, \dep)}[\exsts{i} (\val, \txid, \txidset, \dep) = \mkvsCOPS(\key, i) \land i \in \viCL(\key)] \right)\projection{1} \\
    (\val, \txidCOPS{\repl}{\cl}{n}{n'}, \txidset, \dep) & > (\val', \txidCOPS{\repl'}{\cl'}{n''}{n'''}, \txidset', \dep') \defiff (n, \repl, n') > (n'', \repl, n''') 
\end{align*}
The rest part are trivial be picking a new transaction identifier with larger read counter and committing to key-value store.

Last the rule for receiving a update.
A replica updates its local state only if all the dependencies has been receive as shown in \cref{lst:cops-replica-receive-msg}.
\begin{mathpar}
    \inferrule[sync]{%
        \viREPL = \vienvCOPS(\repl)\rmto{\key}{\vienvCOPS(\repl)(\key) \uplus i} 
        \\\\
        {\left(\begin{array}{@{}l@{}}
        \fora{\key', m, \ver} 
        \ver = \mkvsCOPS(\key, i) \land {}
        (\key', \wtOf(\mkvsCOPS(\key', m))) \in \ver\projection{4}
        \implies m \in \viREPL'(\key') 
        \end{array}\right)} 
    }{%
        \mkvsCOPS, \vienvCOPS, \thdenv, \prog \toPROG{}
        \mkvsCOPS, \vienvCOPS\rmto{\repl}{\viREPL}, \thdenv, \prog
    }
\end{mathpar}
The premiss says, the first line, the replica \( \repl \) receive a new version \( i \) for key \( \key \),
and, the second line, only if all the dependencies of the new update already in the replica's view.

\paragraph{\bf COPS key-value store}
To verify the algorithm, we prove that for any COPS trace produced by the algorithm,
there exists a corresponding causal consistency trace.
First we prove that the key-value stores from COPS trace satisfy \cref{def:mkvs}.

\begin{theorem}[Well-formed COPS key-value]
    \label{thm:cops-key-value-well-form}
    Let ignore the dependencies of versions from \( \mkvsCOPS \).
    Given the initial key-value store \( \mkvsCOPS_0 \), initial views \( \vienvCOPS_0 \) and some programs \( \prog_0 \), for any \( \mkvsCOPS_i \) and \( \vienvCOPS_i \)  such that: 
    \[
        \mkvsCOPS_0, \vienvCOPS_0, \thdenv_0, \prog_0 {\toPROG{}}^* \mkvsCOPS_i, \vienvCOPS_i, \thdenv_i, \prog_i
    \]
    The key-value store \( \mkvsCOPS_i \) satisfies \cref{def:mkvs} and any replica or client view \( \vi \) from \( \vienvCOPS_i \) is a valid view of the key-value store, \ie \( \vi \in \Views(\mkvsCOPS_i) \).
\end{theorem}
\begin{proof}
    We need to prove the  \( \mkvsCOPS_i \) satisfies the well-formed conditions,
    and any view \( \vi_i \Views(\mkvsCOPS_i) \).
    We prove it by introduction on the length \( i \).
    \begin{itemize}
    \item \caseB{\(i = 0\)}
        It holds trivially since each key only has the initial version \( (\val_0,\txid_0,\emptyset, \emptyset) \).
        Since there is only the initial version for each key, it is easy to see that any view \( \vi_0 \) satisfying the well-formed conditions in \cref{def:views}.
    \item \caseI{\(i+1\)}
        We perform case analysis on the possible \((i+1)\)-\emph{th} step:
        \begin{itemize}
            \item \rl{Put}
                Assume the client \( \cl \) of a replica \( \repl \) commits a single-write transaction \( \txid \) that installs a new version for key \( \key \).
                By the premiss of \rl{Put}, the new transaction identifier \( \txid = \txidCOPS{\repl}{\cl}{n'}{0} \) where for some \( n' \) that is greater than any \( n \) from any writers \( \txidCOPS{\stub}{\stub}{n}{\stub} \) that are observable by the replica \( \repl \).
                Since the new transaction \( \txid = \txidCOPS{\repl}{\cl}{n'}{0} \) is a single-write transaction which is always installed at the end of the list associated to \( \key \), it is sufficient to prove the following:
                \begin{gather}
                    \fora{j} 0 \leq j < \abs{ \mkvsCOPS_i(\key) } \implies \wtOf(\mkvsCOPS_{i}(\key, j)) \neq \txid \label{equ:write-trans-unique} \\
                    \fora{j, n} \txidCOPS{\repl}{\cl}{n}{\stub} \in \Set{\wtOf(\mkvsCOPS_{i}(\key,j))} \cup \rsOf(\mkvsCOPS_{i}(\key, j)) \implies n < n' \label{equ:replica-time-monotonic-inc}
                \end{gather}
                \cref{lem:mono-local-time} implies \cref{equ:write-trans-unique} and \cref{equ:replica-time-monotonic-inc}.
                Thus the new key-value store \( \mkvsCOPS_{i+1} \) satisfies the well-formed conditions.
                Now let consider the views, especially the views of the replica \( \viREPL' \) and the client \( \viCL' \).
                Because views different replicas or clients remain unchanged, by \ih they satisfy \( \vi' \in \Views(\mkvsCOPS_{i+1}) \).
                The new view for replica \( \viREPL' = \viREPL\rmto{\key}{\abs{\mkvsCOPS_{i+1}(\key)} - 1} \)
                where \( \viREPL \) is the replica's view before updating and the writer of the last version of \( \key \) is \( \txid \).
                Because \( \txid \) is a single-write transaction, so the new view \( \viREPL' \) still satisfies the atomic read.
                For similar reason, the new view for client \( \viCL' \) till satisfies atomic read.
                Therefore we have \( \viREPL', \viCL' \in \Views(\mkvsCOPS_{i+1}) \).

            \item \rl{GetTrans}
                Assume the client \( \cl \) of a replica \( \repl \) commits a read-only transaction \( \txid \).
                Since it is a read-only transaction, it suffice to prove the following:
                \begin{gather}
                    \fora{\key,j} 0 \leq j < \abs{ \mkvsCOPS_i(\key) } \implies \txid \notin \rsOf(\mkvsCOPS_{i}(\key, j)) \label{equ:read-trans-unique} 
                \end{gather}
                \cref{lem:mono-local-time} implies \cref{equ:read-trans-unique}.
                Thus the new key-value store \( \mkvsCOPS_{i+1} \) satisfies the well-formed conditions.
                Now let consider the views.
                Since only the client view has changed, it is sufficient to  prove that \( \viCL' \in \Views(\mkvsCOPS_{i+1}) \), 
                where \( \viCL' \) is the new client view.
                By \cref{lem:write-unique}, the new view \( \viCL' \) satisfies the atomic read constraint.
                Therefore \( \viCL' \in \Views(\mkvsCOPS_{i+1})  \).
        \end{itemize}
    \end{itemize}
\end{proof}

\begin{lemma}[Monotonic local time]
    \label{lem:mono-local-time}
    Given a reduction step such that: 
    \[
        \mkvsCOPS, \viREPL, \viCL, \stk, \cmd {\toCMD{}}^* \mkvsCOPS', \viREPL', \viCL', \stk', \cmd'
    \]
    let \( \txidCOPS{\repl}{\cl}{n}{n'} \) be the new transaction, \ie \( \txidCOPS{\repl}{\cl}{n}{n'} \in \mkvsCOPS' \land  \txidCOPS{\repl}{\cl}{n}{n'} \notin \mkvsCOPS \).
    It implies the new transaction is greater than any transaction committed by the same client view \( \viCL \), \ie
    \[ 
        \txidCOPS{\repl}{\cl}{n''}{n'''} \in \mkvsCOPS \implies (n,\repl,n') > (n'', \repl, n''')
    \]
\end{lemma}
\begin{proof}
    We perform case analysis.
    \begin{itemize}
        \item \rl{Put}
            By the premiss of the rule, 
            the new transaction is in the form \( \txidCOPS{\repl}{\cl}{n}{0} \),
            and for any existing transaction \( \txidCOPS{\repl}{\cl}{n'}{\stub} \),
            \[
                \txidCOPS{\repl}{\cl}{n'}{\stub} \in \mkvsCOPS \implies n > n'
            \]
        \item \rl{GetTrans}
            Let \( \txidCOPS{\repl}{\cl}{n}{n'} \) be the new transaction.
            By the premiss we have that for any existing transaction \( \txidCOPS{\repl}{\cl}{n''}{n'''} \), 
            \[
                \txidCOPS{\repl}{\cl}{n''}{n'''} \in \mkvsCOPS \implies n = n'' \land n' > n'''
            \]
    \end{itemize}
\end{proof}

\begin{lemma}[Unique writer]
    \label{lem:write-unique}
    Each version has a unique writer.
\end{lemma}
\begin{proof}
    Because a transaction can write to at most one key, and 
    a client at least observes its own writes (the premiss of \rl{Put})
    we have the proof.
\end{proof}

\mypar{COPS normal trace}
We define COPS semi-normal trace then normal trace.
Later, we prove for any COPS trace, there exists an equivalent normal trace.

The dependency of a transaction \( \depOf[\mkvsCOPS, \txid] \) is defined as:
\begin{itemize}
    \item if \( \txid \) is a single-write transaction:
    \begin{align*}
        \depOf[\mkvsCOPS, \txid] & \defeq 
        \begin{multlined}
            \dep  \ \text{where} \ 
            \exsts{\key,j} \lastConf(\tr)\projection{1}(\key,j) = (\stub, \txid, \stub, \dep)
        \end{multlined}
    \end{align*}
    \item if \( \txid \) is a read-only transaction:
    \begin{align*}
        \depOf[\mkvsCOPS, \txid] & \defeq 
        \begin{multlined}
            \bigcup_{\Set{\ver}[\exsts{\key,j} \ver = \mkvsCOPS(\key,j) \land \txid \in \rsOf(\ver)]} \depOf[\ver] \\
        \end{multlined}
    \end{align*}
\end{itemize}
Given a trace \( \tr \), 
the function \( \func{maxVersion}[\tr, \txid] \) returns the version that the transaction \( \txid \) depends on
and is the last one that appears in the trace:
\begin{align*}
    \func{maxVersion}[\tr, \txid] & \defeq \txid_i
    \quad \begin{multlined}[t]
        \text{the \( \txid_i \) is the last one appears in \( \tr \) such that} \\
        (\key',\txid_i) \in \depOf[\lastConf(\tr)\projection{1}, \txid] \land \txid_i \notin \mkvsCOPS_i \land  \txid_i \in \mkvsCOPS_{i+1}
    \end{multlined}
\end{align*}
Give two COPS's traces \( \tr \) and \( \tr' \), \( \mkvsCOPS \) being the final state of \( \tr \) and \( \mkvsCOPS' \) for \(\tr'\),
if the two traces are equivalent, if and only if,
\[ 
    \begin{array}{@{}l@{}}
    \fora{\txid,\fp,\dep} \txid \in \mkvsCOPS
    \land \fp = \func{RWset}[\mkvsCOPS,\txid] \land \dep =  \depOf[\mkvsCOPS,\txid] \\
    \quad {} \iff
    \txid \in \mkvsCOPS'
    \land \fp = \func{RWset}[\mkvsCOPS',\txid] \land \dep =  \depOf[\mkvsCOPS',\txid] \\
    \end{array}
\]
where \( \func{RWset}[\mkvsCOPS,\txid] \) is the read-write set of \( \txid \).
Note that \( \func{RWset} \) is well-defined by \cref{thm:cops-key-value-well-form}.

A COPS's semi-normal trace is a trace \( \tr \) if it is in the form that
read-only transactions \( \txid_\rd \) directly follows its \( \func{maxVersion}[\tr, \txid_\rd] \) or another read-only transaction \( \txid_{\rd}' \) such that \( \func{maxVersion}[\tr, \txid_\rd]  =  \func{maxVersion}[\tr, \txid_{\rd}']\).

\begin{corollary}
    \label{lem:cops-semi-normal-trace}
    \label{cor:cops-semi-normal-trace}
    For any COPS's trace \( \tr \), there exists a equivalent semi-normal trace \( \tr' \) 
    such that \( \lastConf(\tr) = \lastConf(\tr') \).
\end{corollary}
\begin{proof}
    It is easy to prove by induction on the numbers of read-only transactions that are not in the wanted position.
    Let take the first one for those read-only transactions \( \txid_\rd \) who does not follows its \( \func{maxVersion}[\tr, \txid_i] \).
    It is safe to move the reduction step to the right in the trace,
    until it directly follows its \( \func{maxVersion}[\tr, \txid_\rd] \) or another read-only transaction \( \txid_{\rd}' \) such that \( \func{maxVersion}[\tr, \txid_\rd] = \func{maxVersion}[\tr, \txid_{\rd}']\). 
    Assume it is \(i\)-\emph{th} step, and assume  \( (i-1)\)-\emph{th} is not  \( \func{maxVersion}[\tr, \txid_\rd] \) nor another read-only transaction \( \txid_{\rd}' \) such that \( \func{maxVersion}[\tr, \txid_\rd] = \func{maxVersion}[\tr, \txid_{\rd}']\).
    It means the \( (i-1)\)-\emph{th} step must be a write that \( \txid_\rd \) does not depend on.
    Because if \( (i-1)\)-\emph{th} step is a read only transaction \( \txid''_\rd \) 
    it must be the case 
    \( \func{maxVersion}[\tr, \txid_\rd] = \func{maxVersion}[\tr, \txid_{\rd}'']\), 
    otherwise \( \txid_\rd \) is not the first read-only transaction that is not in the wanted place.
\end{proof}

A COPS's normal trace is a trace \( \tr \) if it is semi-normal trace and the single-write transactions step appears in the trace in the order of the writers of those transactions.

\begin{theorem}[normal trace]
    \label{thm:cops-normal-trace}
    \label{lem:cops-normal-trace}
    \label{cor:cops-normal-trace}
    For any COPS's trace \( \tr \), there exists an equivalent normal trace \( \tr' \).
\end{theorem}
\begin{proof}
    By \cref{cor:cops-semi-normal-trace}, let consider a semi-normal trace \( \tr \).
    It is easy to prove by induction on the single-write transactions that are out of order.
    Let take the first single-write transaction \( \txid = \txidCOPS{\repl}{\cl}{n}{n'} \) that is out of order.
    Suppose it is the i-\emph{step}.
    Assume \( \txid \) write to key \( \key \) with value \( \val \).
    Assume the preview write-only transaction \( \txid'  = \txidCOPS{\repl'}{\cl'}{n''}{n'''} \) and it follows \( (n,\repl,n') < (n'', \repl', n''') \).
    The intuition here is to swap the reduction steps of these two transactions and those read-only transactions following them.
    We assume \( \txid \)  and \( \txid' \) write to the same key, otherwise it is safe to swap.
    For any read-only transactions \( \txid_{\rd} \in \txidset_{\rd} \) following \( \txid \), we knows \( \txid' \notin \depOf[\mkvsCOPS_i, \txid_{\rd}] \), otherwise it violate \( \func{maxVersion}[\tr, \txid_\rd] = \txid \).
    Thus it is safe to swap the reduction steps and alter views that includes either or both versions.
    Note that if a view \( \vi \) includes both \( \txid \)  and \( \txid' \), and if \( \txid' \) also write to the same key \( \key \) with value \( \val' \), the \( \vi \) will fetch the value from \( \txid' \):
    \( \func{getMax}[\mkvsCOPS_i,\vi](\key) = \val' \)
\end{proof}

\begin{corollary}
    \label{cor:get-max-to-get-snap}
    For the \rl{Put} and \rl{GetTrans}, it is safe to replace the function \( \func{getMax} \) with the function \( \snapshot \).
\end{corollary}
\begin{proof}
    By \cref{thm:cops-normal-trace}, each trace \( \tr \) has an equivalent normal trace \( \tr' \).
    Assume i-\emph{th} step in the trace \( \tr' \).
    Assume the key-value store before is \( \mkvsCOPS_i \).
    For any versions \( m \) and \( j \) from the same key in the \( \tr' \):
    \[
        0 < m < j < \abs{\mkvsCOPS_i(\key)} \implies \wtOf(\mkvsCOPS_i(\key,m)) <  \wtOf(\mkvsCOPS_i(\key,j))
    \]
    thus it is safe to use \( \snapshot \).
\end{proof}

\paragraph{\bf COPS normal trace to \( \ET_\CC \) trace}
Given \cref{thm:cops-key-value-well-form,thm:cops-normal-trace,cor:get-max-to-get-snap}, for any COPS's trace \( \tr \), 
there exists a trace \( \tr' \) that is a normal trace and satisfies \( \ET_\top \).
Then by the following theorem (\cref{thm:cops-cc}), we proof the COPS trace satisfies \( \ET_\CC \).

\begin{theorem}[COPS satisfying causal consistency]
    \label{thm:cops-cc}
    Given a trace starting from the initial key-value store \( \mkvsCOPS_0 \), initial views \( \vienvCOPS_0 \) and some programs \( \prog_0 \), for any \( \mkvsCOPS_i \) and \( \vienvCOPS_i \)  such that: 
    \[
        \mkvsCOPS_i, \vienvCOPS_i, \thdenv_i, \prog_i \toPROG{\viCL,\fp} \mkvsCOPS_{i+1}, \vienvCOPS_{i+1}, \thdenv_{i+1}, \prog_{i+1} 
    \]
    then the i-\emph{th} step satisfies \( \ET_\CC \), \ie
    \[
        \ET_\CC \vdash ( \mkvsCOPS_i, \viCL ) \csat \fp : (\mkvsCOPS_{i+1},\vienvCOPS_{i+1}(\cl) )
    \]
\end{theorem}
\begin{proof}
    We prove this by induction on the length of trace \( i \).
    We introduce two invariants \( I_1, I_2 \).
    The \( I_1 \) states that for any view \( \vi \), if the view \( \vi \) includes a version, 
    it also includes all the version it depends on, that is,
    \[
        \begin{array}{@{}l@{}}
        \fora{\cl, \vi,\key,\key',j,m,\dep} \\
        \quad \vi = \func{co-dom}[\vienvCOPS_i] \land j \in \viCL(\key') 
        \land (\stub,\stub,\stub,\dep) = \mkvsCOPS_i(\key,j)
        \land (\key',\wtOf(\mkvsCOPS_i(\key',m))) \in \dep \\
        \qquad \implies m \in \vi(\key')
        \end{array}
    \]
    The \( I_2 \) states that for any transaction \( \txid \), 
    \( \depOf[\mkvsCOPS_i,\txid] \) includes transactions that \( \txid \) depends on
    with respect to \( (\SO \cup \WR) \).
    \[
        \begin{array}{@{}l@{}}
        \fora{\txid,\key,j} 
        \txid = \wtOf[\mkvsCOPS_i(\key,j)]
        \implies 
        ((\SO \cup \WR)^{-1})^*(\txid) \subseteq \depOf[\mkvsCOPS_i(\key,j)]
        \end{array}
    \]
    if the view \( \vi \) includes a version, 
    Since \( \ET_\CC = \ET_\MR \cap \ET_\RYW \cap \allowed[\SO \cup \WR] \), we prove the three constraints separately.
    In each case we need to consider \rl{Put}, \rl{GetTrans} and \rl{Sync}
    Let \( \txid \) be the new transaction committed to replica \( \repl \) by the client \( \cl \).
    Let \( \viCL \) and \( \viCL' \) be the client views before and after respectively,
    and \( \viREPL \) and \( \viREPL' \) be the replica views before and after respectively.
    \begin{itemize}
        \item \( \ET_\MR \). 
            For \( \rl{Put} \), it is easy to see \( \viCL \viewleq \viCL' \) and \( \viREPL \viewleq \viREPL' \).
            For \( \rl{GetTrans} \), it is known that \( \viREPL = \viREPL' \). 
            In the premiss of \( \rl{GetTrans} \),
            it fetches a version, \( m_i \)-\emph{th} version, from the replica for each key \( \key_i \) and adds all dependencies, which are associated with version, in the client view.
            This means \( \viCL \viewleq \viCL' \).

        \item \( \ET_\RYW \). 
            By \( \ET_\MR \), it suffices to only consider 
            if the newly committed transaction is included in the view.
            We only need to consider \rl{Put}.
            The new views \( \viCL' = \viCL\rmto{\key}{j} \) and \( \viREPL' = \viREPL\rmto{\key}{j} \)
            where the new version \( \mkvsCOPS_{i+1}(\key,j)\) is written by the client \( \cl \).
            
        \item \( \allowed[\SO \cup \WR] \).
            Given the invariants \( I_1 \) and \( I_2 \),
            let consider \rl{Put}, \rl{GetTrans} and \rl{Sync}.
            \begin{itemize}
                \item \rl{Put}.
                By invariants \( I_1 \) and \( I_2 \),
                we know \( \viCL = \getView[\mkvsCOPS_i,((\SO \cup \WR)^{-1})^{*}(\Tx[\mkvsCOPS_i,\viCL])]\).
                Now we need to re-establish the \( I_1 \) for the new views \( \viCL' \) and \( \vienvCOPS' \) and \( I_2 \) for the new transaction \( \txid \).
                \begin{itemize}
                    \item Since it is a single-write transaction,
                    assume it appends \( j\)-\emph{th} version to key \( \key \).
                    This means the new client views \( \viCL' = \viCL\rmto{\key}{ \viCL(\key) \cup \Set{j} } \),
                    Let \( \dep \) be the dependency for the new version, \ie \( \dep = \depOf(\mkvsCOPS_{i+1}(\key,j)) \).
                    By the premiss, the \( \dep \) includes all versions included in \( \viCL \).
                    This means for any key \( \key' \) and index \( m \):
                    \[
                        (\key',\wtOf[\mkvsCOPS_i(\key',m)]) \in \dep \implies m \in \viCL'(\key')
                    \]
                    Note that \( \viCL' \viewleq \viREPL' \),
                    therefore the invariant \( I_1 \) holds.
                    \item 
                    It is enough to only consider the newly committed transaction \( \txid \).
                    Since it is a write transaction,
                    let assume any transaction \( \txid' \) such that \( \txid' \toEDGE{\SO} \txid \).
                    By \( \MR \) and \( \RYW \), transaction \( \txid' \) is in the view \( \viCL \),
                    that is, for some \(\key',m \):
                    \[
                        \txid' \in \Set{\wtOf[\mkvsCOPS_i(\key',m)]} \cup \rsOf[\mkvsCOPS_i(\key',m)] \implies m \in \viCL(\key')
                    \]
                    Given the definition of dependencies the new version written by \( \txid \):
                    \[ 
                        \dep = \Set{(\key'', \txid'')}[
                            \exsts{z} z \in \viCL(\key'') \land \txid'' = \wtOf(\mkvsCOPS_i(\key'', z)) \lor (\key'', \txid'') \in \depOf[\mkvsCOPS_i(\key'', z)] 
                        ]
                    \] 
                    it follows \( (\key',m) \in \dep \).
                    By \ih of invariant \( I_2 \), we know that  \( \dep \) is close with respect to \( \SO \cup \WR \).
                    Thus we know for the new transaction \( ((\SO \cup \WR)^{-1})^*(\txid) \subseteq \depOf[\mkvsCOPS_i,\txid] \),
                    that is, we re-establish invariant for \( i+1 \).
                \end{itemize}
                \item 
                For \rl{GetTrans}, the client \( \cl \) commits a read-only transaction.
                By the premiss of the rule, the client will pick a new view \( \viCL' \)
                such that \( \vi_0 = \viCL \viewleq \viCL' = \vi_j \) and:
                \[
                    {\begin{array}{@{}l@{}}
                        \mathsf{for} \ z \ \mathsf{in} \Set{1, \dots, j} \\
                        \quad m_z \in \viREPL(\key_z) \\
                        \quad \vi'_z = \vi_{z-1}\rmto{\key_z}{\vi_{z-1}(\key_z) \cup \Set{m_z}}  \\
                        \quad \vi_z = \lambda \key \ldotp \vi'_z(\key) \cup \Set{x}[(\key,\wtOf[\mkvsCOPS(\key,x)]) \in \depOf[\mkvsCOPS(\key_z,m_z)]]  \\
                    \end{array}}
                \]
                We prove each iteration preserves the invariant \( I_1 \).
                Suppose the invariants holds after \((z-1)\)-\emph{th}, let consider \(z\)-\emph{th}.
                By construction, \( \vi'_z = \vi_{z-1}\rmto{\key_z}{\vi_{z-1}(\key_z) \cup \Set{m_z}} \),
                the \( m_z \) version of some key \( \key_z \) is included in \( \vi'_z \).
                Then all the versions \( \mkvsCOPS_i(\key_z, m_z) \) is is included in \( \vi_z \) as 
                \[
                     \vi_z = \lambda \key \ldotp \vi'_z(\key) \cup \Set{x}[(\key,\wtOf[\mkvsCOPS(\key,x)]) \in \depOf[\mkvsCOPS(\key_z,m_z)]] 
                \]
                This implies \( I_1 \) for \( \vi_z \).
                For \( I_2 \), we apply \ih of invariant \( I_2 \),
                since the new transaction is a read-only transaction.
            \item For \rl{Sync}, the premiss of the rule implies that \( I_1 \) holds under the new replica view \( \viREPL' \).
            For \( I_2 \), we apply \ih of invariant \( I_2 \).
            \end{itemize}
    \end{itemize}
\end{proof}

\subsection{Clock-SI}
\label{sec:clock-si}
\subsubsection{Code}

\paragraph{\bf Structure}
Clock-SI is a partitioned distributed NoSQL database, which means 
each server, also called shard, contains part of keys and does not overlap with any other servers.
Clock-SI implements snapshot isolation.
To achieve that, each shard tracks the physical time.
Note that times between shards do not match,
but there is a upper bound of the difference.
\begin{lstlisting}[caption={Shard},label={lst:clock-si-shard}]
Shard :: ID -> ( clockTime )
\end{lstlisting}
A key maintains a list of values and their versions.
A version is the time when such value is committed.
\begin{lstlisting}[caption={Key-value store},label={lst:clock-si-key-value-store}]
VersionNo :: Time
KV :: Keys -> List( Val, VersionNo )
(each key is asscoaited with a shard)
\end{lstlisting}
The idea behind Clock-SI is that
a client starts a transaction in a shard, 
and the shard is responsible for fetching value from other shards
if keys are not stored in the local shard.
During execution, a transaction tracks the write set.
\begin{lstlisting}[caption={Write set},label={lst:clock-si-write-set}]
WS :: Key -> Val
\end{lstlisting}
At the end, the transaction commits all the update in the write set,
and the local shard acts as coordinator to update keys either locally or remotely.
To commit a transaction, Clock-SI use two-phase commits protocol.
A transaction has four states:
\begin{itemize}
    \item \verb|active|, the transaction is still running;
    \item \verb|prepared|, shards receive the update requests from the coordinator;
    \item \verb|committing|, shards receive the update confirmations from the coordinator;
    \item \verb|committed|, the transaction commits successfully.
\end{itemize}
To implement SI, a transaction also tracks its snapshot time 
so it knows which version should be fetched.
Also a transaction tracks the prepared and committing times,
which are used to postpone other transactions' reads 
if those transactions' snapshots time are greater.
\begin{lstlisting}[caption={Transaction runtime},label={lst:clock-si-state}]
State :: { active, prepared, committing, committed }
Trans :: ( state, snapTime, prepareTime, commitTime, ws )
\end{lstlisting}

\paragraph{\bf Start Transaction}
Clock-SI proposes two versions, with or without session order.
Here we verify the one with session order.
To start a transaction, the client contacts a shard 
and provides the previous committing time.
The shard will return a snapshot time, 
which is greater than the committing time provided, for the new transaction.
Note that client might connects to a different shard from last time,
which means that 
the shard might have to wait until the shard local time is greater than the committing time.
\begin{lstlisting}[caption={Transaction runtime},label={lst:clock-si-trans-runtime}]
startTransaction( Trans t, Time ts )
    wait until ts < getClockTime();
    t.snapshotTime = getClockTime();
    t.state = active;
\end{lstlisting}
From this point, such transaction will always interact with the shard and
the shard will act as coordinator if necessary.

\paragraph{\bf Read}
A transaction \verb|t| might read within the transaction if the key has been updated by the same transaction before,
that is, read from the write set \verb|ws|.
A transaction \verb|t| might read from the original shard if the key store in the shard,
but it has to wait until any other transactions \verb|t'| commit successfully
who are supposed to commit before the current transaction's snapshot time,
\ie \verb|t'| are in \verb|prepared| or \verb|committing| stage 
and the corresponding time is less the \verb|t| snapshot time.
\begin{lstlisting}[caption={Read from original shard},label={lst:clock-si-read-original}]
Read( Trans t, key k )
    if ( k in t.ws ) return ws(k);
    if ( k is updated by t' and t'.state = committing 
                and t.snapshotTime > t'.committingTime)
        wait until t'.state == committed;
    if ( k is updated by t' 
                    and t'.state = prepared and t.snapshotTime > t'.preparedTime 
                    and t.snapshotTime > t'.committingTime )
        wait until t'.state == committed;
    return K(k,i), where i is the latest version before t.snapshotTime;
\end{lstlisting}
If the key is not stored in the original shard, 
the original shard sends a read request to the shard containing the key.
Because of time difference, the remote shard's time might before the snapshot time of the transaction.
In this case, the shard wait until the time catches up.
\begin{lstlisting}[caption={Read from original shard},label={lst:clock-si-read-remote}]
On read k request from a remote transaction t
    wait until t.snapshotTime < getClockTime() 
    return read(t,k);
\end{lstlisting}

\paragraph{\bf Commit Write Set}
If all the keys in the write set are hosted in the original shard that the transaction first connected,
the write set only needs to commit local.
\begin{lstlisting}[caption={Local Commit},label={lst:clock-si-local-commit}]
localCommit( Trans t )
    if noConcurrentWrite(t) {
        t.state = committing;
        t.commitTime = getClockTime();
        log t.commitTime;
        log t.ws;
        t.state = committed;
    }
\end{lstlisting}
To commit local, it first checks, by \verb|noConcurrentWrite(t)|,
if there is any transaction \verb|t'| 
that writes to the same key as the transaction new transaction \verb|t|,
and the transaction \verb|t'| commit after the snapshot of \verb|t|.
Since writing database needs time,
it sets the transaction state to \verb|committing| and
log the \verb|commitTime|, before the updating really happens.
During \verb|committing| state, other transactions will be pending, 
if they want to read the keys being updated.
Last, the state of transaction is set to \verb|committed|.

To commit to several shards,
Clock-SI uses two-phase protocol.
\begin{lstlisting}[caption={Distributed Commit},label={lst:clock-si-distributed-commit}]
distributedCommit( Trans t )
    for p in t.updatedPartitions { send ``prepare t'' to p; }
    wait receiving ``t prepared'' from all participants, store into prep;
    t.state = committing;
    t.commitTime = max(prep);
    log t.commitTime;
    t.state = committed;
    for p in t.updatedPartitions { send ``commit t'' to p; }

On receiving ``prepare t''
    if noConcurrentWrite(t) {
        log t.ws and t.coordinator ID
        t.state = prepared;
        t.prepareTime = getClockTime();
        send ``t prepared'' to t.coordinator
    }

On receiving ``commit t''
    log t.commitTime
    t.state = committed
\end{lstlisting}
The original shard, who acts as the coordinator,
sends \verb|''prepare t''| to shards that will be updated.
Any shard receiving \verb|''prepare t''| checks, similarly, 
if there is any transaction write to the same key committing after the snapshot time.
If the check passes, 
the shard logs the write set and the coordinator shard ID,
set the state to \verb|prepared|,
and sends the local time to the coordinator.
Once the coordinator receives all the \verb|prepared| messages,
it starts the second phase by setting the state to \verb|committing|.
Then the coordinator picks the largest time from 
all the \verb|prepared| messages as the commit time for the new transaction.
Since the write set has been logged in the first phase, 
so here it can immediately set the state to be \verb|committed|.
Last, the coordinator needs to send \verb|commit t| to other shards 
so they will log the commit time and set the state to \verb|committed|.
Note that participants have different view for the new transaction from the coordinator,
but it guarantees eventually they all updated to \verb|committed| with the same commit time.

\subsubsection{Verification}

\paragraph{\bf Structure}
We model the database use key-value store from \cref{def:mkvs},
yet here it is necessary to satisfy the well-formed conditions.
Transaction identifier \( \txid_\cl^n \) are labelled with the committing time \( n \).
Sometime we also write \( \txid_\cl^\ct \) or omit the client label, \ie \( \txid^n\) and \( \txid^\ct \).

Database is partitioned into several \emph{shards}.
A shard \(\sd \in \Shards \)  contains some keys which are disjointed from keys in other shards.
The \( \func{shardOf}[\key] \) denotes the shard where the key \( \key \) locates.

Shards and clients are associated with clock times, \(  \ct \in \ClockTimes \defeq \Nat \), which represent the current times of shards and clients.
We use notation \( \cts \in (\Shards \cup \Clients ) \parfinfun \ClockTimes\).

We will use notation \( \ptrans{\trans} \) to denote the static code of a transaction,
and \( \runtrans{\trans}{\ct}{\fp} \) for the runtime of a transaction,
where \( \ct \)  is the snapshot time and \( \fp \) is the read-write set.
Note that in the model, we only distinguishes \verb|active| and \verb|committed| state,
since the \verb|prepared| and \verb|committing| are only for better performance.

\paragraph{\bf Start Transaction}
To start a transaction, it picks a random shard \( \sd \) as the coordinator,
reads the local time \( \cts(\sd) \) as the snapshot,
and sets the initial read-write set to be an empty set.
Also the client time is updated to the snapshot time.
For technical reason,
we also update the shard time to avoid time collision to other transaction about to commit.
Note that in real life, all the operations running in a shard take many time cycles,
so it is impossible to have time collision.
\[
    \inferrule[StartTrans]{ 
        \ct < \cts(\sd) \and 
        \cts' = \cts\rmto{\sd}{\cts(\sd) + 1} \quad \texttt{--->  simulate time elapses}
        }{%
            \cl \vdash \mkvs, \ct, \cts, \stk, \ptrans{\trans} \toCMD{\cts(\sd), \ct, \emptyset, \perp}
            \mkvs, \cts(\sd), \cts', \stk, \runtrans{\trans}{\cts(\sd)}{\emptyset}
        }
\]

\paragraph{\bf Read}
The clock-SI protocol includes some codes related to performance which does not affect the correctness.
Clock-SI distinguishes a local read/commit and a remote read/commit,
yet it is sufficient to assume all the read and commit are ``remote'',
while the local read and commit can be treated as self communication.
Similarly we assume a transaction always commits to several shards.
\begin{lstlisting}[caption={simplified read},label={lst:simplified-read}]
On receive ``read(t,k)'' {
    if ( k in t.ws ) return ws(k); (*\label{line:read-from-local}*)

    asssert( t.snapshotTime < getClockTime() ) (*\label{line:snapshot-time-grt-than-shard}*)
    for t' that writes to k:
        if(t.snapshotTime > t'.preparedTime 
                    || t.snapshotTime > t'.committingTime) 
            asssert( t.state == committed )

    return K(k,i), where i is the latest version before t.snapshotTime; (*\label{line:read-i-from-shard}*)
}
\end{lstlisting}

If the key exists in the write set \( \fp \),
the transaction read from the write set immediately.
\[
    \inferrule[ReadTrans]{ 
            \key = \evalE{\expr} \\
            (\otW, \key, \val ) \in \fp \quad \texttt{---> \cref{lst:simplified-read}, \cref{line:read-from-local}}\\
        }{%
        \cl \vdash \mkvs, \ct, \cts, \stk, \runtrans{\plookup{\vx}{\expr}}{\ct}{\fp} \toCMD{\cl,\ct,\fp \addO (\otR, \key, \val),\perp}
            \mkvs, \ct, \cts, \stk\rmto{\vx}{\val}, \runtrans{\pskip}{\ct}{\fp \addO (\otR, \key, \val)}
        }
\]

Otherwise, the transaction needs to fetch the value from the shard.
The first premiss says the transaction must wait until the shard local time \( \cts(\func{shardOf}[\key]) \) is greater than the snapshot time \( \ct \).
If so, by the second line, the transaction fetches the latest version for key \( \key \) before the snapshot time \( \ct \).
\[
    \inferrule[ReadRemote]{ 
            \key = \evalE{\expr} \\
            (\otW, \key, \stub ) \notin \fp \\
            \ct < \cts(\func{shardOf}[\key]) \quad \texttt{---> \cref{lst:simplified-read}, \cref{line:snapshot-time-grt-than-shard}} \\\\
            n = \max\Set{n'}[\exsts{j} \txid^{n'} = \wtOf(\mkvs(\key, j)) \land n' < \ct] \quad \texttt{---> \cref{lst:simplified-read}, \cref{line:read-i-from-shard}} \\\\ 
            \mkvs(\key, i) = (\val, \txid^n, \stub) 
        }{%
        \cl \vdash \mkvs, \ct, \cts, \stk, \runtrans{\plookup{\vx}{\expr}}{\ct}{\fp} \toCMD{\cl,\ct,\fp \addO (\otR, \key, \val), \perp}
            \mkvs, \ct, \cts', \stk\rmto{\vx}{\val}, \runtrans{\pskip}{\ct}{\fp \addO (\otR, \key, \val)}
        }
\]

\paragraph{\bf Write}
Write will not go to the shard until committing time.
Before it only log it in the write set.
\[
    \inferrule[Write]{ 
            \key = \evalE{\expr_1} \\
            \val = \evalE{\expr_2} \\
        }{%
            \cl \vdash \mkvs, \ct, \cts, \stk, \runtrans{\pmutate{\expr_1}{\expr_2}}{\ct}{\fp} \toCMD{\cl,\ct,\fp \addO (\otW, \key, \val), \perp}
            \mkvs, \ct, \cts, \stk, \runtrans{\pskip}{\ct}{\fp \addO (\otW, \key, \val)}
        }
\]

\paragraph{\bf Commit}
We also assume transaction always commit to several shards and the local commit is treated as self-communication.

\begin{lstlisting}[caption={simplified commit},label={lst:simplified-commit}]
commit( Trans t )
    for p in t.updatedPartitions
        send ``prepare t'' to p;
    wait receiving ``t prepared'' from all participants, store into prep; (*\label{line:pick-time-1}*)
    t.state = committing; (*\label{line:pick-time-2}*)
    t.commitTime = max(prep); (*\label{line:pick-time-3}*)
    log t.commitTime;
    t.state = committed;
    for p in t.updatedPartitions
        send ``commit t'' to p;

On receiving ``prepare t''
    if noConcurrentWrite(t) (*\label{line:check-concur-write}*)
        log t.ws to t.coordinator ID
        t.state = prepared;
        t.prepareTime = getClockTime();
        send ``t prepared'' to t.coordinator

On receiving ``commit t''
    log t.commitTime (*\label{line:clock-si-commit-1}*)
    t.state = committed  (*\label{line:clock-si-commit-2}*)
\end{lstlisting}

Note that Clock-SI uses two phase commit:
the coordinator (the shard that the client directly connects to) distinguishes ``committing'' state  and ``committed'' state, where in between the coordinator pick the committing time and log the write set,
and the participants distinguishes ``prepared'' state and ``committed'' state.
Such operations are for possible network partition or single shard errors, and allowed a more fine-grain implementations which do not affect the correctness,
therefore it suffices to assume they are one atomic step.

\[
    \inferrule[Commit]{ 
        \fora{\key,i} (\otW, \key, \stub) \in \fp \land \wtOf(\mkvs(\key,i)) < \ct \quad \texttt{---> \cref{lst:simplified-commit}, \cref{line:check-concur-write}} \\\\  
        n = \max\left( \Set{\ct'}[\exsts{\key} (\stub, \key, \stub) \in \fp \land \ct' = \cts(\func{shardOf}[\key])] \cup \Set{\ct} \right) \quad \texttt{---> \cref{lst:simplified-commit}, \cref{line:pick-time-1,line:pick-time-2,line:pick-time-3}} \\\\
        \mkvs' =  \func{commitKV}[\mkvs,\ct,\txid_\cl^n,\fp] \quad \texttt{---> \cref{lst:simplified-commit}, \cref{line:clock-si-commit-1,line:clock-si-commit-2}} \\\\
        \fora{\sd}
        {\left( \begin{array}{@{}l@{}}
            \sd \in \Set{\func{shardOf}[\key]}[(\stub, \key, \stub) \in \fp] \\
            \quad \implies \cts'(\sd) = \cts(\sd) + 1 
        \end{array} \right)} \lor (\cts'(\sd) = \cts(\sd)) \quad \texttt{--->  simulate time elapses}
        }{%
            \sd, \cl \vdash \mkvs, \ct, \cts, \stk, \runtrans{\pskip}{\ct}{\fp} \toCMD{\cl,\ct,\fp,n}
            \mkvs', n+1, \cts', \stk, \pskip
        }
\]

To commit the new transaction, it needs to check, by the first premiss,
there is no other transactions writing to the same keys after the snapshot time.
If it passes, by the second line it picks the maximum time \( n \) among all participants
as the commit time.
The new key-value store \( \mkvs' =  \func{commitKV}[\mkvs,\ct,\txid_\cl^n,\fp] \),
where 
\begin{align*}
    \func{commitKV}[\mkvs,\ct,\txid,\fp \uplus \Set{(\otR,\key,\val)}] & \defeq 
    \begin{multlined}[t]
    \texttt{let} \ n = \max\Set{n'}[\exsts{j} \txid^{n'} = \wtOf(\mkvs(\key, j)) \land n' < \ct] \\
    \texttt{and} \ \mkvs(\key,i) = (\val, \txid^n, \txidset) \\
    \texttt{and} \ \mkvs' = \func{commitKV}[\mkvs,\ct,\txid,\fp] \\
    \texttt{in} \ \mkvs'\rmto{\key}{\mkvs'(\key)\rmto{i}{(\val, \txid^n, \txidset \cup \Set{\txid})}}
    \end{multlined} \\
    \func{commitKV}[\mkvs,\ct,\txid,\fp \uplus \Set{(\otW,\key,\val)}] & \defeq 
    \begin{multlined}[t]
    \texttt{let} \ \mkvs' = \func{commitKV}[\mkvs,\ct,\txid,\fp] \\
    \texttt{in} \ \mkvs'\rmto{\key}{\mkvs'(\key) \lcat (\val, \txid, \emptyset)}
    \end{multlined} 
\end{align*}
Note that \( \func{commitKV} \) is similar to \( \updateKV \) by appending the new version to the end of a list.
The \( \func{commitKV} \) also updates versions read by the new transaction 
using the snapshot time of the transaction.
Last, like \rl{StartTrans} 
we update the client time after the commit time, \ie \( n + 1 \) 
and simulate time elapses for all shards updated.

\paragraph{\bf Time Tick}
For technical reasoning, we have non-deterministic time elapses.
\[
    \inferrule[TimeTick]{ }{%
        \mkvs, \cts, \cts', \thdenv, \prog \toPROG{\sd,\cts(\sd) + 1}
        \mkvs, \cts, \cts'\rmto{\sd}{\cts(\sd) + 1}, \thdenv, \prog
    }
\]

\[
    \inferrule[ClientStep]{ 
            \cl \vdash 
            \mkvs, \cts(\cl), \cts', \thdenv(\cl), \prog(\cl) \toCMD{\cl,\ct',\fp,\ct''}
            \mkvs', \ct, \cts'', \stk, \cmd
        }{%
            \mkvs, \cts, \cts', \thdenv, \prog \toPROG{\cl,\ct',\fp,\ct''}
            \mkvs, \cts\rmto{\cl}{\ct}, \cts'', \thdenv\rmto{\cl}{\stk}, \prog\rmto{\cl}{\cmd}
        }
\]

\paragraph{\bf Verification}
Clock-SI allows interleaving, 
yet for any clock-si trace \( \tr \) there exists a equivalent trace \( \tr' \)
where transactions do not interleave with others (\cref{thm:clock-si-normal-trace}).
Furthermore, in such trace \( \tr' \), transactions are reduced in their commit order.
\begin{theorem}[Normal clock-SI trace]
\label{thm:clock-si-normal-trace}
A clock-SI trace \( \tr \) is a clock-SI normal trace if it satisfies the following:
there is no interleaving of a transaction,
\begin{centermultline}[equ:clock-si-no-interleaving]
        \fora{\cl,\ct, \mkvs_i, \cts_i, \cts'_i, \thdenv_i, \prog_i} \\
        \tr = \cdots \toPROG{\cl,\ct,\stub, \perp} \mkvs_i, \cts_i, \cts'_i, \thdenv_i, \prog_i \toPROG{\stub} \cdots \\
        {} \implies \exsts{\ct', \mkvs_j, \cts_j, \cts'_j, \thdenv_j, \prog_j}  \\
        \tr = \cdots \toPROG{\cl,\ct,\stub, \perp} \mkvs_i, \cts_i, \cts'_i, \thdenv_i, \prog_i \toPROG{\cl,\ct,\stub, \perp} \stub \toPROG{\cl,\ct,\stub, \perp} 
        \cdots \toPROG{\cl,\ct,\stub,\ct'} \mkvs_j, \cts_j, \cts'_j, \thdenv_j, \prog_j
\end{centermultline}
and transactions in the trace appear in the committing order, 
\begin{equation}
    \label{equ:clock-si-commit-order}
    \begin{array}{@{}l@{}}
        \fora{\cl_i,\cl_j,\ct_i, \ct_j, \ct'_i, \cl'_j, \fp_i, \fp_j \mkvs_i, \mkvs_j, \cts_i, \cts_j, \cts'_i, \cts'_j, \thdenv_i, \thdenv_j,  \prog_i, \prog_j} \\
        \tr = \cdots \toPROG{\cl_i,\ct_i,\fp_i,\ct'_i} \mkvs_i, \cts_i, \cts'_i, \thdenv_i, \prog_i \toPROG{\stub} 
        \cdots \toPROG{\cl_j,\ct_j,\fp_j,\ct'_j} \mkvs_j, \cts_j, \cts'_j, \thdenv_j, \prog_j 
        \implies \ct'_i < \ct'_j
    \end{array}
\end{equation}
For any clock-SI trace \( \tr \), there exists an equivalent normal trace \( \tr' \) which has the same final configuration as \( \tr \).
\end{theorem}
\begin{proof}
    Given a trace \( \tr \), we first construct a trace \( \tr' \)  that satisfies \cref{equ:clock-si-commit-order}, by swapping steps.
    Let take the first two transactions \( \txid_{\cl_i}^n \) and \( \txid_{\cl_j}^m \) that are out of order, \ie \( n > m \) and 
    \[
    \begin{array}{@{}l@{}}
        \tr = \cdots \toPROG{\cl_i,\ct_i,\fp_i,n} \mkvs_i, \cts_i, \cts'_i, \thdenv_i, \prog_i \toPROG{\stub}  \cdots \toPROG{\cl_j,\ct_j,\fp_j,m} \mkvs_j, \cts_j, \cts'_j, \thdenv_j, \prog_j \\
    \end{array}
    \]
    By \cref{lem:mono-client-clock-time}, the two clients are different \( \cl_i \neq \cl_j \) and thus two steps are unique in the trace.
    We will construct a trace that \( \txid_{\cl_i}^n \) commits after \( \txid_{\cl_j}^m  \).
    \begin{itemize}
    \item First, it is important to prove that \( \txid_{\cl_j}^m \) does not read any version written by \( \txid_{\cl_i}^n\).
    By \cref{lem:commit-after-snapshot-time}, the snapshot time \( \ct_j \) of \( \txid_{\cl_j}^m \) is less than the commit time, 
    \ie \( \ct_j < m \), therefore \( \ct_j < n \).                                                                                  
    By the \rl{read} rule, \( \ct_j < n \) implies the transaction \( \txid_{\cl_j}^m \) never read any version written by \( \txid_{\cl_i}^n \).

    \item Let consider any possible time tick for those shard \( \sd \) that has been updated by \( \txid_{\cl_j}^m \),
    that is, \( \sd = \func{shardOf}[\key]\) for some key \( \key \)  that \( (\otW, \key, \stub) \in \fp_i \) and
    \begin{equation}
    \label{equ:time-tick-move-afterwards}
    \begin{array}{@{}l@{}}
        \tr = \cdots \toPROG{\cl_i,\ct_i,\fp_i,n} \mkvs_i, \cts_i, \cts'_i, \thdenv_i, \prog_i \toPROG{\stub} 
        \cdots \toPROG{\sd,\ct} \stub \toPROG{\stub} \cdots \toPROG{\cl_j,\ct_j,\fp_j,m} \mkvs_j, \cts_j, \cts'_j, \thdenv_j, \prog_j \\
    \end{array}
    \end{equation}
    Since \( \ct_j < m < n < \ct \), therefore such time tick will not affect the transaction \( \txid_{\cl_j}^m \),
    which means it is safe to move the time tick step after the \( \txid_{\cl_j}^m \).
    \end{itemize}
    Now we can move the commit of \( \txid_{\cl_i}^n \) and time tick steps similar to \cref{equ:time-tick-move-afterwards} after the commit of \( \txid_{\cl_j}^m \),
    \[
    \begin{array}{@{}l@{}}
        \tr' = \cdots \toPROG{\cl_j,\ct_j,\fp_j,m} \mkvs_j, \cts_j, \cts'_j, \thdenv_j, \prog_j \toPROG{\cl_i,\ct_i,\fp_i,n} \mkvs_i, \cts_i, \cts'_i, \thdenv_i, \prog_i \toPROG{\sd,\ct} \cdots \\
    \end{array}
    \]
    We continually swap the out of order transaction until the newly constructed trace \( \tr' \)  satisfying \cref{equ:clock-si-commit-order}.

    Now let consider \cref{equ:clock-si-no-interleaving}.
    Let take the first transaction \( \txid \) whose read has been interleaved by other transaction or a time tick.
    \begin{itemize}
        \item If it is a step that the transaction \( \txid \) read from local state,
        \[
        \begin{array}{@{}l@{}}
            \tr = \cdots \toPROG{\cl,\ct,\fp \addO (\otR, \key, \val ),\perp} \stub  \toPROG{\alpha} \cdots  \toPROG{\cl,\ct,\fp'',n} \cdots \\
        \end{array}
        \]
        then by \rl{ReadTrans} we know \( \fp \addO (\otR, \key, \val ) = \fp\), and it is safe to swap the two steps as the following
        \[
        \begin{array}{@{}l@{}}
            \tr' = \cdots \toPROG{\alpha} \stub \toPROG{\cl,\ct,\fp \addO (\otR, \key, \val ),\perp} \cdots \toPROG{\cl,\ct,\fp'',n} \cdots \\
        \end{array}
        \]
        \item If it is a step that the transaction \( \txid \) read from remote, 
            the step might be interleaved by a step from other transaction or time tick step.
        \begin{itemize}                                                                       
            \item if it is interleaved by the commit of other transaction \( \txid' = \txid_{\cl'}^m \), that is
        \[
        \begin{array}{@{}l@{}}
            \tr = \cdots \toPROG{\cl,\ct,\fp,\perp} \mkvs, \cts, \cts', \thdenv, \prog  \toPROG{\cl',\ct',\fp',m} \cdots \toPROG{\cl,\ct,\fp'',n} \cdots \\
        \end{array}
        \]
        where \( \cl' \neq \cl \).
        \begin{itemize}
            \item if the transaction \( \txid' \) does not write to any key \( \key \) that is read by \( \txid \),
                \[
                    \fora{\key} (\otR, \key, \stub) \in \fp \implies (\otW, \key, \stub) \notin \fp'
                \]
            In this case, it is safe to swap the two steps
            \[
            \begin{array}{@{}l@{}}
                \tr' = \cdots \toPROG{\cl',\ct',\fp',m} \stub \toPROG{\cl,\ct,\fp,\perp} \cdots \toPROG{\cl,\ct,\fp'',n} \cdots \\
            \end{array}
            \]
            \item if the transaction \( \txid' \) write to a key \( \key \) that is read by \( \txid \),
                \[
                    (\otR, \key, \stub) \in \fp \land (\otW, \key, \stub) \in \fp'
                \]
                Let \( \sd = \func{shardOf}[\key] \).
                By the \rl{ReadRemote}, we know the current clock time for the shard \( \sd \) is greater than \( \ct \) which is the snapshot time of \( \txid \), 
                that is, \( \cts'(\sd) > \ct \).
                Then by \rl{commit}, the commit time of \( \txid' \) is picked as the maximum of the shards it touched, 
                \ie \( m \geq \cts'(\sd) \).
                Now by the \rl{ReadRemote} and \( m \geq \ct \), it is safe to swap the two steps since the new version of \( \key \) does not affect the \( \txid \).
        \end{itemize}
        \item if it is interleaved by the read of other transaction \( \txid' \), that is
        \[
        \begin{array}{@{}l@{}}
            \tr = \cdots \toPROG{\cl,\ct,\fp,\perp} \mkvs, \cts, \cts', \thdenv, \prog  \toPROG{\cl',\ct',\fp',\perp} \cdots \toPROG{\cl,\ct,\fp'',n} \cdots \\
        \end{array}
        \]
        Because reads have no side effect to any shard by \rl{readRemote},
        it is safe to swap the two steps
        \[
        \begin{array}{@{}l@{}}
            \tr' = \cdots \toPROG{\cl',\ct',\fp', \perp} \stub \toPROG{\cl,\ct,\fp,\perp} \cdots \toPROG{\cl,\ct,\fp'',n} \cdots \\
        \end{array}
        \]
        \item if it is interleaved by a time tick step,
        \[
        \begin{array}{@{}l@{}}
            \tr = \cdots \toPROG{\cl,\ct,\fp \addO (\otR, \key, \val),\perp} \mkvs, \cts, \cts', \thdenv, \prog  \toPROG{\sd, \ct'} \cdots \toPROG{\cl,\ct,\fp'',n} \cdots \\
        \end{array}
        \]
        \begin{itemize}
            \item if the transaction \( \txid \) does not read from the shard \( \sd \), it means for any key \( \key \),
                \[
                    \func{shardOf}[\key] \neq \sd
                \]
            In this case, it is safe to swap the two steps
            \[
            \begin{array}{@{}l@{}}
            \tr' = \cdots \toPROG{\sd, \ct'} \stub \toPROG{\cl,\ct,\fp \addO (\otR, \key, \val),\perp} \cdots \toPROG{\cl,\ct,\fp'',n} \cdots \\
            \end{array}
            \]
            \item if the transaction \( \txid \) read from the shard \( \sd \), it means that there exists a key \( \key \)
            \[
                \func{shardOf}[\key] = \sd
            \]
            By the \rl{ReadRemote}, we know the current clock time for the shard \( \sd \) is greater than the snapshot time of \( \txid \), 
            that is, \( \cts'(\sd) > \ct \).
            Then by \rl{TimeTick},  we have \( \ct' > \cts'(\sd) \).
            Now by the \rl{ReadRemote} and \( \ct' > \ct \), it is safe to swap the two steps.
        \end{itemize}
    \end{itemize}
    \end{itemize}
\end{proof}

\begin{lemma}[Monotonic client clock time]
    \label{lem:mono-client-clock-time}
    The clock time associated with a client monotonically increases,
    That is, given a step
    \[
        \mkvs, \cts, \cts', \thdenv, \prog \toPROG{\stub} \mkvs', \cts'', \cts''', \thdenv', \prog'
    \]
    then for any clients \( \cl \),
    \[
        \cts(\cl) \leq \cts(\cl')
    \]
\end{lemma}
\begin{proof}
    It suffices to only check the \rl{ClientStep} rule which is the only rule updates the client clock time,
    especially, it is enough to check the client \( \cl \) that who starts or commits a new transaction.
    \begin{itemize}
        \item \rl{Commit}.
            Let \( \ct \) be the clock time before committing, \( \ct = \cts(\cl)\).
            By the premiss of the rule, the new client time \( n + 1 \) satisfies that, 
            \[
                n = \max\left( \Set{\ct'}[\exsts{\key} (\stub, \key, \stub) \in \fp \land \ct' = \cts(\func{shardOf}[\key])] \cup \Set{\ct} \right)
            \]
            It means \( \ct < (n + 1)\).
        \item \rl{StartTrans}.
            Let \( \ct \) be the clock time before taking snapshot, \( \ct = \cts(\cl)\).
            By the premiss of the rule the new  client time \( \cts'(\sd) \) for a shard \( \sd \), 
            such that \( \ct < \cts'(\sd)  \).
    \end{itemize}
\end{proof}

\begin{lemma}[No side effect local operation]
    \label{lem:no-side-effect-local-operation}
    Any transactional operation has no side effect to the shard and key-value store,
    \[
        \mkvs, \cts, \cts', \thdenv, \prog \toPROG{\cl, \ct, \fp, \perp} \mkvs', \cts'', \cts''', \thdenv', \prog' \implies \mkvs = \mkvs'
    \]
\end{lemma}
\begin{proof}
    It is easy to see that 
    \rl{StartTrans}, \rl{ReadTrans}, \rl{ReadRemote} and \rl{Write} do not change the state of key-value store.
\end{proof}

Clock-SI also has a notion view which corresponds the snapshot time.
The following definition \( \func{viewOf}[\mkvs,\ct] \) extracts the view from snapshot time.
\begin{definition}
    \label{clock-si-view}
    \label{def:clock-si-view}
    Given a normal clock-SI trace \( \tr \) and a transaction \( \txid_\cl \), such that
    \[
        \tr = \cdots \toPROG{\cl, \ct, \emptyset, \perp} \cdots \toPROG{\cl, \ct, \fp, \perp} \mkvs, \cts, \cts', \thdenv, \prog  \toPROG{\cl, \ct, \fp, \ct'} \cdots
    \] 
    the initial view of the transaction is defined as the following:
    \begin{align*}
        \func{viewOf}[\mkvs,\ct] & \defeq \lambda \key \ldotp \Set{i}[\exsts{\txid^{n} } \wtOf(\mkvs(\key,i)) = \txid^{n} \land n < \ct]
    \end{align*}
\end{definition}

Given the view \( \func{viewOf}[\mkvs,\ct] \) for each transaction, 
we first prove that clock-si produces a well-formed key-value store (\cref{def:mkvs}).
\begin{lemma}
    \label{lem:well-formed-clock-si-view}
    Given any key-value store \( \mkvs \) and snapshot time \( \ct \) from a clock-SI trace \( \tr \),
    \[
        \tr = \cdots \toPROG{\stub} \mkvs, \cts, \cts', \thdenv, \prog \to{\cl, \ct, \fp, \ct'} \cdots
    \]
    \func{viewOf}[\mkvs, \ct] and \func{viewOf}[\mkvs, \ct'] (\cref{clock-si-view}) produce well-formed views.
\end{lemma}
\begin{proof}                     
    It suffices to prove that \cref{eq:view.atomic} in \cref{def:view}.
    Assume a key-value store \( \mkvs \) and a snapshot time \( \ct \).
    Suppose a version \( i \) in the view \( i \in \func{viewOf}[\mkvs,\ct](\key)\) for some key \( \key \).
    By \cref{def:clock-si-view}, the version is committed before the snapshot time,
    \ie \( \txid^n = \wtOf(\mkvs(\key, i)) \land n < \ct\).
    Assume another version \( \txid^n = \wtOf(\mkvs(\key', j)) \) for some key \( \key' \) and index \( j \).
    By \cref{def:clock-si-view} we have \( j \in \func{viewOf}[\mkvs,\ct](\key') \).
    Similarly \( \func{viewOf}[\mkvs, \ct'] \) is a well-formed view.
\end{proof}

Second, given the view \( \func{viewOf}[\mkvs,\ct] \) for each transaction, 
both \( \func{commitKV} \) and \( \updateKV \) produce the same result.
\begin{lemma}
    Given a normal clock-SI trace \( \tr \) and a transaction \( \txid_\cl \), such that
    \[
        \tr = \cdots \toPROG{\cl, \ct, \emptyset, \perp} \cdots \toPROG{\cl, \ct, \fp, \perp} \mkvs, \cts, \cts', \thdenv, \prog  \toPROG{\cl, \ct, \fp, \ct'} \cdots
    \] 
    the following holds:
    \[
        \func{commitKV}[\mkvs,\ct,\txid_\cl^{\ct'},\fp] = \updateKV[\mkvs, \func{viewOf}[\mkvs, \ct],\fp, \txid_\cl^{\ct'}] 
    \]
\end{lemma}
\begin{proof}
    We prove by induction on \( \fp \).
    \begin{itemize}
        \item \caseB{\( \fp = \emptyset \)}
            It is easy to see that 
            \[ \func{commitKV}[\mkvs,\ct,\txid_\cl^{\ct'},\emptyset] = \mkvs =  \updateKV{\mkvs, \func{viewOf}[\mkvs, \ct],\fp, \txid_\cl^{\ct'}} \]
        \item \caseI{\( \fp \uplus (\otW, \key, \val) \)}
            Because in both functions, the new version is installed at the tail of the list associated with \( \key \),
            \[
                \begin{array}{@{}l@{}}
                \func{commitKV}[\mkvs,\ct,\txid_\cl^{\ct'},\fp \uplus (\otW, \key, \val)]  \\
                \quad \begin{array}[t]{@{}c l@{}}
                = &
                \func{commitKV}[\mkvs,\ct,\txid_\cl^{\ct'},\fp]\rmto{\key}{\mkvs(\key) \lcat (\val, \txid, \emptyset)} \\
                = & 
                \updateKV[\mkvs,\func{viewOf}[\mkvs,\ct],\fp,\txid_\cl^{\ct'}]\rmto{\key}{\mkvs(\key) \lcat (\val, \txid, \emptyset)} \\
                = & 
                \updateKV[\mkvs,\func{viewOf}[\mkvs,\ct],\fp \uplus (\otW, \key, \val),\txid_\cl^{\ct'}]
                \end{array}
                \end{array}
            \]
        \item \caseI{\( \fp \uplus (\otR, \key, \val) \)}
            Let \( \mkvs(\key,i) \) be the version being read.
            That is, the writer \( \txid^n = \wtOf(\mkvs(\key,i)) \)
            is the latest transaction written to the key \( \key \) before the snapshot time \( \ct \),
            \[
                n = \max\Set{n'}[%
                    \exsts{j} 
                    \txid^{n'} = \wtOf(\mkvs(\key, j)) 
                    \land n' < \ct%
                ] 
            \]
            Let the new version \( \ver = \left( \valueOf(\mkvs(\key,i)), \wtOf(\mkvs(\key,i)), \rsOf(\mkvs(\key,i)) \uplus \Set{\txid_\cl^{\ct'}} \right) \).
            By \cref{lem:well-formed-clock-si-view}, it follows \( i \in \func{viewOf}[\mkvs,\ct](\key) \), then by \cref{lem:clock-si-mono-writer}, the version is the latest one \( i = \max(\func{viewOf}[\mkvs,\ct](\key)) \).
            Therefore we have,
            \[
                \begin{array}{@{}l@{}}
                \func{commitKV}[\mkvs,\ct,\txid_\cl^{\ct'},\fp \uplus (\otR, \key, \val)]  \\
                \quad \begin{array}[t]{@{}c l@{}}
                = &
                \func{commitKV}[\mkvs,\ct,\txid_\cl^{\ct'},\fp]\rmto{\key}{\mkvs(\key)\rmto{i}{\ver}} \\
                = & 
                \updateKV[\mkvs,\func{viewOf}[\mkvs,\ct],\fp,\txid_\cl^{\ct'}]\rmto{\key}{\mkvs(\key)\rmto{i}{\ver}} \\
                = & 
                \updateKV[\mkvs,\func{viewOf}[\mkvs,\ct],\fp \uplus (\otR, \key, \val),\txid_\cl^{\ct'}]
                \end{array}
                \end{array}
            \]
    \end{itemize}
\end{proof}

\begin{lemma}[Strictly monotonic writers]
    \label{lem:clock-si-mono-writer}
    Each version for a key has a writer with strictly greater clock time than any versions before:
    \[
        \begin{array}{@{}l@{}}
            \fora{\mkvs,\key,i,j,\txid^n,\txid^m} 
            \wtOf(\mkvs(\key,i)) = \txid^n 
            \land \wtOf(\mkvs(\key,j)) = \txid^m 
            \land i < j
            \implies 
            n < m
        \end{array}
    \]
\end{lemma}

By \cref{thm:clock-si-normal-trace}, it is sufficient to only consider normal clock-SI trace.
Since transactions do not interleave in a normal clock-SI trace,
all transactional execution can be replaced by \cref{fig:semantics-trans}.
\begin{theorem}[Simulation]
    \label{thm:clock-si-transaction-to-atomic}
    Given a clock-SI normal trace \( \tr \), a transaction \( \txid_\cl^n \) from the trace,
    and the following transactional internal steps
    \[
        \mkvs_0, \ct_0, \cts_0, \stk_0, \ptrans{\trans} \toCMD{\cl,\ct,\emptyset,\perp} \cdots  \toPROG{\cl,\ct,\fp,n} \mkvs_i, \ct_i, \cts_i, \stk_i, \ptrans{\pskip}
    \]
    for some \( i \), there exists a trace
    \[
        (\stk_0, \snapshot[\mkvs_0,\func{viewOf}[\mkvs_0,\ct]], \emptyset), \trans \toTRANS^*
        (\stk_i, \sn_i, \fp_i), \pskip
    \]
    that produces the same final fingerprint in the end.
\end{theorem}
\begin{proof}
    Given the internal steps of a transaction
    \[
        \mkvs_0, \ct_0, \cts_0, \stk_0, \ptrans{\trans_0}_\ct^{\fp_0} \toCMD{\stub} 
        \cdots  \toPROG{\stub} \mkvs_i, \ct_i, \cts_i, \stk_i, \ptrans{\trans_i}_\ct^{\fp_i}
    \]
    We construct the following trace,
    \[
        (\stk_0, \snapshot[\mkvs_0,\func{viewOf}[\mkvs_0,\ct]], \fp_0), \trans_0 \toTRANS^*
        (\stk_i, \sn_i, \fp_i), \trans_i
    \]
    Let consider how many transactional internal steps.
    \begin{itemize}
        \item \caseB{i = 0}
        In this case, 
        \[
            \mkvs_0, \ct_0, \cts_0, \stk_0, \ptrans{\trans}_\ct^{\fp_0}
        \]
        It is easy to construct the following
        \[
            (\stk_0, \snapshot[\mkvs,\func{viewOf}[\mkvs,\ct]], \fp_0), \trans_0 
        \]
        \item \caseI{i + 1}
        Suppose  a trace with \(i\) steps,
        \[
            \mkvs_0, \ct_0, \cts_0, \stk_0, \ptrans{\trans_0}_\ct^{\fp_0} \toCMD{\stub} 
            \cdots  \toPROG{\stub} \mkvs_i, \ct_i, \cts_i, \stk_i, \ptrans{\trans_i}_\ct^{\fp_i}
        \]
        and a trace
        \[
            (\stk_0, \snapshot[\mkvs_0,\func{viewOf}[\mkvs_0,\ct]], \fp_0), \trans_0 \toTRANS^*
            (\stk_i, \sn_i, \fp_i), \trans_i
        \]
        Now let consider the next step.
        \begin{itemize}
            \item \rl{ReadTrans}.
                In this case
                \[
                    \mkvs_i, \ct_i, \cts_i, \stk_i, \ptrans{\trans_i}_\ct^{\fp_i}
                    \toPROG{\stub}
                    \mkvs_{i+1}, \ct_{i+1}, \cts_{i+1}, \stk_{i+1}, \ptrans{\trans_{i+1}}_\ct^{\fp_{i+1}}
                \]
                such that
                \[
                    \fp_{i+1} = \fp_i \addO (\otR, \key, \val)  = \fp_i
                    \land (\otW,\key, \val) \in \fp_i
                \]
                for some key \( \key \) and value \( \val \), and
                \[
                    \trans_{i} \equiv \plookup{\vx}{\expr} ; \trans 
                    \land \evalE[\stk_i]{\expr} = \key 
                    \land \stk_{i+1} = \stk_i\rmto{\pv}{\val}
                    \land  \trans_{i+1} \equiv \pskip ; \trans
                \]
                for some variable \( \vx \), expression \( \expr \) and continuation \( \trans \).
                Since \( ( \otW, \key ,\val ) \in \fp_i  \), it means \( \ss_i(\key) = \val \) for the local snapshot.
                By the \rl{TPrimitive}, we have 
                \[
                    (\stk_i, \sn_i, \fp_i), \trans_i \toTRANS \plookup{\vx}{\expr} ; \trans 
                    \toTRANS (\stk_{i+1}, \sn_i, \fp_{i+1} ), \trans_{i+1}
                \]
            \item \rl{ReadRemote}.
                In this case
                \[
                    \mkvs_i, \ct_i, \cts_i, \stk_i, \ptrans{\trans_i}_\ct^{\fp_i}
                    \toPROG{\stub}
                    \mkvs_{i+1}, \ct_{i+1}, \cts_{i+1}, \stk_{i+1}, \ptrans{\trans_{i+1}}_\ct^{\fp_{i+1}}
                \]
                such that
                \[
                    \fp_{i+1} = \fp_i \addO (\otR, \key, \val) = \fp_i \uplus \Set{(\otR, \key, \val)}
                    \land \fora{\val} (\otW,\key, \val') \notin \fp_i
                \]
                for some key \( \key \) and value \( \val \), and
                \[
                    \trans_{i} \equiv \plookup{\vx}{\expr} ; \trans 
                    \land \evalE[\stk_i]{\expr} = \key 
                    \land \stk_{i+1} = \stk_i\rmto{\pv}{\val}
                    \land  \trans_{i+1} \equiv \pskip ; \trans
                \]
                for some variable \( \vx \), expression \( \expr \) and continuation \( \trans \).
                By the premiss of the \rl{ReadRemote}, the value read is from the last version before the snapshot time:
                \[
                    n = \max\Set{n'}[%
                        \exsts{j} \txid^{n'} = \wtOf(\mkvs(\key, j)) \land n' < \ct%
                    ]
                    \land \valueOf(\mkvs_i(\key,n) ) = \val
                \]                 
                By the definition of \( \vi_0 = \func{viewOf}[\mkvs_0,\ct] \) and \( \snapshot[\mkvs_0,\vi_0] \) and the fact that there is no write to the key \( \key \),
                it follows \( \sn_i(\key) = \val \).
                Thus, by the \rl{TPrimitive}, we have 
                \[
                    (\stk_i, \sn_i, \fp_i), \trans_i \toTRANS \plookup{\vx}{\expr} ; \trans 
                    \toTRANS (\stk_{i+1}, \sn_i, \fp_{i+1} ), \trans_{i+1}
                \]
            \item \rl{Write}.
                In this case
                \[
                    \mkvs_i, \ct_i, \cts_i, \stk_i, \ptrans{\trans_i}_\ct^{\fp_i}
                    \toPROG{\stub}
                    \mkvs_{i+1}, \ct_{i+1}, \cts_{i+1}, \stk_{i+1}, \ptrans{\trans_{i+1}}_\ct^{\fp_{i+1}}
                \]
                such that
                \[
                    \fp_{i+1} = \fp_i \addO (\otW, \key, \val) = \fp_i \setminus \Set{(\otW, \key, \val')}[\val' \in \Val] \uplus \Set{(\otR, \key, \val)}
                \]
                for some key \( \key \) and value \( \val \), and
                \[
                    \trans_{i} \equiv \pmutate{\expr_1}{\expr_2} ; \trans 
                    \land \evalE[\stk_i]{\expr_1} = \key 
                    \land \evalE[\stk_i]{\expr_2} = \val 
                    \land  \trans_{i+1} \equiv \pskip ; \trans
                \]
                for some expressions \( \expr_1 \) and \( \expr_2 \), and continuation \( \trans \).
                By the \rl{TPrimitive}, it is easy to see: 
                \[
                    (\stk_i, \sn_i, \fp_i), \trans_i \toTRANS \pmutate{\expr_1}{\expr_2} ; \trans 
                    \toTRANS (\stk_{i+1}, \sn_i, \fp_{i+1} ), \trans_{i+1}
                \]
        \end{itemize}
    \end{itemize}
\end{proof}

By \cref{def:clock-si-view,lem:well-formed-clock-si-view,thm:clock-si-transaction-to-atomic},
we know for each clock-SI trace, there exists a trace that satisfies \( \ET_\perp \).
Last, we prove such trace also satisfies \( \ET_\SI \).
\begin{theorem}[Clock-SI satisfying SI]
    For any normal trace clock-SI trace \( \tr \), and transaction \( \txid_\cl^{n} \) such that
    \[
        \tr = \cdots \toPROG{\cl, \ct, \fp, \perp} \mkvs, \cts, \cts', \thdenv, \prog  \toPROG{\cl, \ct, \fp, n} \mkvs', \cts'', \cts''', \thdenv', \prog' \toPROG{\stub} \cdots
    \]
    the transaction satisfies \( \ET_\SI \), \ie \( \ET_\SI \vdash (\mkvs, \func{viewOf}[\mkvs,\ct]) \csat \fp : \func{viewOf}[\mkvs,\cts''(\cl)] \)
\end{theorem}
\begin{proof}
    Recall \( \ET_\SI  = \Set{(\mkvs, \vi, \fp, \vi')}[\dagger] \cap \ET_\MR \cap \ET_\RYW  \cap \ET_\UA \)
    Note that final view of the client, \( \cts''(\cl) = n + 1 \).
    We prove the four parts separately.
    \begin{itemize}
        \item \( \Set{(\mkvs, \func{viewOf}[\mkvs,\ct], \fp, \func{viewOf}[\mkvs,\cts''(\cl)])}[\dagger] \).
            Assume a version \( i \in \func{viewOf}[\mkvs,\ct](\key) \) for some key \( \key \).
            Suppose a version \( \mkvs(\key',j)\) such that 
            \[ 
                \wtOf(\mkvs(\key',j)) \toEDGE{((\SO \cup \WR_{\mkvs} \cup \WW_{\mkvs}) ; \RW_{\mkvs}\rflx)^+} \wtOf(\mkvs(\key,i))
            \]
            Let \( \txid^n = \wtOf(\mkvs(\key',j)) \) and \( \txid^m = \wtOf(\mkvs(\key,i)) \).
            By \cref{lem:clock-si-rw,lem:clock-si-wr-ww-so}, we know \( n < m \) then \( j  \in \func{viewOf}[\mkvs,\ct](\key')\).
        \item \( \ET_\MR \).
            By \rl{Commit}, we know \( \ct \leq n < \cts''(\cl) \) then \( \func{viewOf}[\mkvs,\ct] \viewleq \func{viewOf}[\mkvs,\cts''(\cl)] \).
        \item \( \ET_\MW \).
            By \rl{Commit}, for any write \( (\otW, \key, \val) \in \fp \), there is a new version written by the client \( \cl \) in the \( \mkvs'  \),
            \[
                \wtOf(\mkvs'(\key,\abs{\mkvs'(\key)} - 1)) = \txid_\cl^{n}
            \]
            Since \( n < \cts''(\cl)\), it follows \( \abs{\mkvs'(\key)} - 1 \in \func{viewOf}[\mkvs,\cts''(\cl)](\key) \).
        \item \( \ET_\UA \).
            By the premiss of \rl{Commit}, for any write \( (\otW, \key, \val) \in \fp \), any existed versions of the key \( \key \)
            must be installed by some transactions before the snapshot time of \( \ct \),
            \[
                \fora{\key,i} (\otW, \key, \stub) \in \fp \land \wtOf(\mkvs(\key,i)) < \ct 
            \]
            It implies that 
            \[ 
                \fora{i} i \in \dom(\mkvs(\key)) \implies i \in \func{viewOf}[\mkvs,\ct](\key) 
            \]
    \end{itemize}
\end{proof}

\begin{lemma}[\( \RW_\mkvs \)]
    \label{lem:clock-si-rw}
    Given a normal clock-SI trace \( \tr \), and two transactions \( \txid_\cl^n \) and \( \txid_{\cl'}^m \) from the trace
    \[
        \begin{array}{@{}l@{}}
            \tr = \cdots \toPROG{\stub} \mkvs,\cts,\cts',\thdenv, \prog \toPROG{\cl,\ct,\fp,n} \cdots \land 
            \tr = \cdot \toPROG{\stub} \mkvs',\cts'',\cts''',\thdenv', \prog' \toPROG{\cl',\ct',\fp',m} \cdots
        \end{array}
    \]
    Suppose the final state of the trace \( \tr \) is \( \mkvs'' \).
    , if \( \txid_\cl^n \toEDGE{\RW_{\mkvs''}\rflx} \txid_{\cl'}^m \) then the snapshot time of \( \txid_{\cl}^n \) took snapshot before the commit time of \( \txid_{\cl'}^m \), \ie \( \ct \leq m \).
\end{lemma}
\begin{proof}
    By definition of \( \txid_\cl^n \toEDGE{\RW_{\mkvs''}\rflx} \txid_{\cl'}^m \),
    it follows that
    \[
        \txid_\cl^n \in \rsOf(\mkvs''(\key,i)) 
        \land \txid_{\cl'}^m = \wtOf(\mkvs''(\key,j)) 
        \land i < j
    \]                              
    for some key \( \key \) and indexes \( i,j \).
    There are two cases depending on the commit order.
    \begin{itemize}
        \item If \( \txid_\cl^n \) commits after \( \txid_{\cl'}^m  \),
            we have,
            \[
                \begin{array}{@{}l@{}}
                \tr = \cdot \toPROG{\stub} \mkvs',\cts'',\cts''',\thdenv', \prog' \toPROG{\cl',\ct',\fp',m} \cdots
                \toPROG{\stub} \mkvs,\cts,\cts',\thdenv, \prog \toPROG{\cl,\ct,\fp,n} \cdots \\
                \end{array}
            \]
            We prove by contradiction.
            Assume \( \ct > m \).
            Since it is a normal trace \( \tr \) (\cref{thm:clock-si-normal-trace}), 
            it follows \( n > m \).
            Note that both transactions access the key \( \key \), and then by \cref{lem:mono-shard-clock-time,lem:commit-after-snapshot-time}, we have \( n > \ct > m  \).
            Given \( \ct > m \), by \rl{ReadRemote} the transaction \( \txid_\cl^n \) should at least read the version written by \( \txid_{\cl'}^m \) for the key \( \key \).
            That is,
            \[
                \txid_\cl^n \in \rsOf(\mkvs''(\key,i)) 
                \land \txid_{\cl'}^m = \wtOf(\mkvs''(\key,j)) 
                \land i > j
            \]
            which contradict  \( \txid_\cl^n \toEDGE{\RW_{\mkvs''}\rflx} \txid_{\cl'}^m  \).
        \item If \( \txid_\cl^n \) commits before \( \txid_{\cl'}^m  \),
            \[
                \begin{array}{@{}l@{}}
                    \tr = \cdots \toPROG{\stub} \mkvs,\cts,\cts',\thdenv, \prog \toPROG{\cl,\ct,\fp,n} \cdots
                    \toPROG{\stub} \mkvs',\cts'',\cts''',\thdenv', \prog' \toPROG{\cl',\ct',\fp',m} \cdots
                \end{array}
            \]
            It is trivial that \( \ct \leq m \) by \cref{lem:mono-shard-clock-time,lem:commit-after-snapshot-time}.
    \end{itemize}
\end{proof}

\begin{lemma}[\( \WR_\mkvs \), \( \WW_\mkvs  \) and \( \SO_\mkvs \)]
    \label{lem:clock-si-wr-ww-so}
    Given a normal clock-SI trace \( \tr \), and two transactions \( \txid_\cl^n \) and \( \txid_{\cl'}^m \) from the trace
    \[
        \begin{array}{@{}l@{}}
            \tr = \cdots \toPROG{\stub} \mkvs,\cts,\cts',\thdenv, \prog \toPROG{\cl,\ct,\fp,n} \cdots \land 
            \tr = \cdots \toPROG{\stub} \mkvs',\cts'',\cts''',\thdenv', \prog' \toPROG{\cl',\ct',\fp',m} \cdots
        \end{array}
    \]
    Suppose the final state of the trace \( \tr \) is \( \mkvs'' \).
    , if \( \txid_\cl^n \toEDGE{\WR_{\mkvs''}\rflx} \txid_{\cl'}^m \) then the transaction \( \txid_{\cl}^n \) commit before the commit time of \( \txid_{\cl'}^m \), \ie \( n < m \).
    Similarly, \( n <  m \) for the relations \( \WW_\mkvs \) and \( \SO_\mkvs \).
\end{lemma}
\begin{proof}
    \begin{itemize}
        \item \( \WR_{\mkvs''} \).
            Since  \( \txid_\cl^n \toEDGE{\WR_{\mkvs''}\rflx} \txid_{\cl'}^m \),
            it is only possible that the later commits after the former,
            \[
                \begin{array}{@{}l@{}}
                    \tr = \cdots \toPROG{\stub} \mkvs,\cts,\cts',\thdenv, \prog \toPROG{\cl,\ct,\fp,n} \cdots \land
                    \toPROG{\stub} \mkvs',\cts'',\cts''',\thdenv', \prog' \toPROG{\cl',\ct',\fp',m} \cdots
                \end{array}
            \]
            By \cref{lem:clock-si-reader-greater-than-writer},we know \( n < m \).
        \item \( \WW_{\mkvs''} \).
            By the definition of  \( \WW_{\mkvs''} \) and \cref{lem:clock-si-mono-writer},we know \( n < m \).
        \item \( \SO_{\mkvs''} \).
            By the definition of  \( \SO_{\mkvs''} \) and \cref{lem:mono-client-clock-time},we know \( n < m \).
    \end{itemize}
\end{proof}

\begin{lemma}[Reader greater than writer]
    \label{lem:clock-si-reader-greater-than-writer}
    Assume a trace \( \tr \) and two transactions \( \txid_{\cl}^n\) and \( \txid_{\cl'}^m \),
        \[
            \begin{array}{@{}l@{}}
                \tr = \cdots \toPROG{\stub} \mkvs,\cts,\cts',\thdenv, \prog \toPROG{\cl,\ct,\fp,n} \cdots \land
                \toPROG{\stub} \mkvs',\cts'',\cts''',\thdenv', \prog' \toPROG{\cl',\ct',\fp',m} \cdots
            \end{array}
        \]
    Assume the final state of key-value store of the trace is \( \mkvs'' \).
    If  \( \txid_{\cl'}^m \) reads a version written \ by \( \txid_{\cl}^n\)
    \[
        \begin{array}{@{}l@{}}
            \wtOf(\mkvs''(\key,i)) = \txid^n 
            \land \txid^m \in \rsOf(\mkvs''(\key,j))
        \end{array}
    \]
    Then, the snapshot times of readers of a version is greater then the commit time of the writer \( n < \ct' \)
\end{lemma}
\begin{proof}
    Trivially, \( \wtOf(\mkvs'(\key,i)) = \txid^n \).
    By the \rl{ReadRemote}, it follows \[ n  = \max\Set{n'}[\exsts{j} \txid^{n'} = \wtOf(\mkvs(\key, j)) \land n' < \ct'] \] which implies \( n < \ct' \).
\end{proof}

\begin{lemma}[Commit time after snapshot time]
    \label{lem:commit-after-snapshot-time}
    The commit time of a transaction is after the snapshot time.
    Suppose the following step,
    \[
        \mkvs,\cts,\cts',\thdenv, \prog \toPROG{\cl,\ct,\fp,n} \mkvs',\cts'',\cts''',\thdenv', \prog'
    \]
    then \( \ct < n \).
\end{lemma}
\begin{proof}
    It is easy to see by \rl{ClientStep} and then \rl{Commit} that 
    \[
        n > n - 1 \max\left( \Set{\ct'}[\exsts{\key} (\stub, \key, \stub) \in \fp \land \ct' = \cts'(\func{shardOf}[\key])] \cup \Set{\ct} \right)
    \]
    so \( \ct < n \).
\end{proof}

\begin{lemma}[Monotonic shard clock time]
    \label{lem:mono-shard-clock-time}
    The clock time associated with a shard monotonically increases,
    Suppose the following step,
    \[
        \mkvs,\cts,\cts',\thdenv, \prog \toPROG{\cl,\ct,\fp,n} \mkvs',\cts'',\cts''',\thdenv', \prog'
    \]
    then 
    \[
        \fora{\sd \in \dom(\cts')} \cts'(\sd) \leq \cts'''(\sd)
    \]
\end{lemma}
\begin{proof}
    We perform case analysis on rules.
    \begin{itemize}
        \item \rl{TimeTick}
            By the rule there is one shard \( \sd' \) ticks time \( \cts'''(\sd') = \cts'(\sd) + 1 > \cts'(\sd) \).
        \item \rl{ClientStep}.
            There are further five cases, yet only \rl{StartTrans} and \rl{Commit} change the shard's clock times.
            \begin{itemize}
                \item \rl{StartTrans}
                    By the rule a new transaction starts in a shard \( \sd' \) and triggering the shard \( \sd' \) ticks time \( \cts'''(\sd') = \cts'(\sd) + 1 > \cts'(\sd) \).
                \item \rl{Commit}
                    By the rule the transaction commits their fingerprint \( \fp \) to those shards \( \sd' \) it read or write, and triggering the shard \( \sd' \) ticks time \( \cts'''(\sd') = \cts'(\sd) + 1 > \cts'(\sd) \).
            \end{itemize}
    \end{itemize}
\end{proof}

\end{appendices}
\end{document}